# A compilation of redshifts for Compact Groups of Galaxies

G.C. Baiesi Pillastrini[1]
*Sezione Spettroscopia Astronomica - U.A.I.*
email: gcbp@it.packardbell.org

[1] permanent address: via Garzoni, 2/2 40138 Bologna Italy

Abstract

We present a compilation of redshifts for 4,911 Compact Groups of galaxies (CG) referenced to the heliocentric reference frame, where 694 of them are available from the literature as of 1992. An unpublished redshift list for 4,217 CGs identified in the SDSS data release 4 by Deng et al. (2006) has been added too. We present an additional list of 157 CGs with published redshifts for one or more member galaxies and 13 “apparent” CGs with discrepant memberships. The electronic version of the compilation is provided by email request.

Keywords: catalog: redshift: galaxies: clusters

## 1. Introduction

The purpose of this paper is to make available to the Astronomical community a compilation of published redshifts for Compact Groups of Galaxies (CGs). The many new measurements now available in literature allow us to collect sparse published lists and provide a substantially complete compilation of redshifts for these peculiar objects. An unpublished list of 4,217 redshifts for CGs has been included too. This very important list has been obtained from the SDSS data release 4 by Deng et al.(2006) and provided by Deng (2006). The compilation begins with the first list of redhifts measured by Hickson et al. (1992) for a complete sample of 100 CGs. Note that also some older references have been added for completeness.

## 2. The list of redshifts

We list 5,081 heliocentric redshifts for CGs. For 157 of them we list redshifts obtained for one or more member galaxies of the group. For the remaining 13 objects with discordant memberships, redshifts for individual galaxies are given.
The list contains the following information:
Col.(1) Compact Group name (as given by the Authors).
Col.(2) Right Ascension J2000 in decimal numerical format. The assigned celestial position for each CG of ref.(27) has been calculated by averaging positions of member galaxies.
Col.(3) Declination J2000 as above (blank rows mean repeated coordinates).
Col.(4) Number of galaxies in the group (when specified by the Authors).
Col.(5) Heliocentric redshift. For 17 CGs with redshifts available for 4 or more member galaxies, a mean redshift $\langle z \rangle$ has been computed [a note “mean value” appears in Col.(10)]. The following procedure has been applied: if $\sigma \geq 0.003$, a discordant $z$ is removed when $\langle z \rangle + 3\sigma \leq z \leq \langle z \rangle - 3\sigma$. If the computed new $\langle z \rangle$ has $\sigma \leq 0.003$, then it is accepted and the number of redshifts used in the calculation is reported in Col.(4). Note that redshifts available from ref.(2) were published corrected for the CMB dipole moment. In the present form, they have been recomputed with respect to the heliocentric reference frame.
Col.(6) Redshifts for individual member galaxies.
Col.(7) ID member galaxy code given by the Author(s).
Col.(8) Other name/identification of the CG.
Col.(9) Reference code of the published redshift.
Col.(10) Additional notes.
References are given at the bottom of the list.

| (1) | (2) | (3) | (4) | (5) | (6) | (7) | (8) | (9) | (10) |
|---|---|---|---|---|---|---|---|---|---|
| | 0,1730 | -10,4853 | 4 | 0.077 | | | | 27 | |
| HCG 099 | 0,1833 | 28,3889 | 4 | 0,029 | | | U12896 | 1 | |
| | | | 3 | 0,029 | | | UZCCG001 | 5 | |
| | | | 3 | 0,029 | | | RSCG 89 | 9 | |
| HCG 100 | 0,3375 | 13,1325 | 4 | 0,018 | | | N7803 | 1 | |
| LCCG 76 | 0,5750 | -44,0019 | 7 | 0,040 | | | | 2 | |
| | 0,9216 | 0,4894 | 5 | 0.1 | | | | 27 | |
| | 1,1743 | 15,8519 | 4 | 0.037 | | | | 27 | |
| | 1,2083 | 0,9991 | 5 | 0.1 | | | | 27 | |
| | 1,2105 | -10,4694 | 5 | 0.077 | | | | 27 | |
| UZCCG002 | 1,3250 | 5,1739 | 3 | 0,018 | | | | 5 | |
| | 1,3607 | 16,0145 | 5 | 0.116 | | | | 27 | |
| | 1,5340 | 0,5542 | 5 | 0.062 | | | | 27 | |
| | 1,6325 | -9,8679 | 4 | 0.084 | | | | 27 | |
| | 1,6924 | -9,7422 | 6 | 0.112 | | | | 27 | |
| | 1,8237 | -0,7980 | 7 | 0.074 | | | | 27 | |
| | 1,8261 | 15,8495 | 4 | 0.125 | | | | 27 | |
| PCG000843-004404 | 2,1833 | 0,7347 | 4 | | 0.1540 | C | | 7 | group member/coord.from Ref.(6) |
| | 2,2926 | -0,1857 | 5 | 0.076 | | | | 27 | |
| | 2,3014 | -0,3023 | 5 | 0.075 | | | | 27 | |
| | 2,3302 | 0,3321 | 5 | 0.063 | | | | 27 | |
| | 2,3725 | -10,1173 | 4 | 0.107 | | | | 27 | |
| | 2,5194 | -0,8343 | 7 | 0.039 | | | | 27 | |
| | 2,7718 | -10,0343 | 6 | 0.084 | | | | 27 | |
| | 3,1625 | -0,0013 | 4 | 0.04 | | | | 27 | |
| | 3,1822 | 0,0669 | 4 | 0.039 | | | | 27 | |
| | 3,2228 | 1,1492 | 4 | 0.039 | | | | 27 | |
| UZCCG003 | 3,2750 | 30,9528 | 3 | 0,016 | | | | 5 | |
| UZCCG004 | 3,5417 | 48,2028 | 4 | 0,017 | | | | 5 | |
| | 3,6119 | -0,0903 | 4 | 0.064 | | | | 27 | |
| | 3,6471 | 0,8431 | 4 | 0.064 | | | | 27 | |
| | 3,6675 | -10,1601 | 8 | 0.084 | | | | 27 | |
| | 3,6824 | -0,0489 | 6 | 0.065 | | | | 27 | |
| SCG 20 | 3,7333 | -24,0903 | | 0,026 | | | | 12 | |
| | 3,8955 | 14,5983 | 4 | 0.082 | | | | 27 | |
| | 3,8987 | 14,6578 | 4 | 0.083 | | | | 27 | |
| | 3,9854 | 14,7565 | 5 | 0.086 | | | | 27 | |
| | 4,0730 | -0,3209 | 5 | 0.06 | | | | 27 | |
| | 4,1680 | -0,4569 | 6 | 0.065 | | | | 27 | |
| | 4,2058 | -0,3823 | 5 | 0.039 | | | | 27 | |
| | 4,2264 | -0,4824 | 4 | 0.066 | | | | 27 | |

| | | | | | | |
|---|---|---|---|---|---|---|
| | 4,3130 | -10,2692 | 7 | 0.085 | | 27 |
| | 4,4839 | -0,7461 | 4 | 0.064 | | 27 |
| | 4,5129 | -9,9481 | 4 | 0.1 | | 27 |
| | 4,5587 | -10,4850 | 4 | 0.076 | | 27 |
| RSCG 01 | 4,6125 | 30,0536 | 5 | 0,023 | | 9 |
| | 5,2658 | -0,9713 | 4 | 0.107 | | 27 |
| UZCCG005 | 5,3750 | 22,4736 | 7 | 0,020 | | 5 |
| | 5,4130 | -0,2391 | 4 | 0.063 | | 27 |
| | 5,4180 | -1,0551 | 4 | 0.085 | | 27 |
| | 5,5524 | -9,0337 | 4 | 0.108 | | 27 |
| | 5,5681 | -0,9372 | 5 | 0.058 | | 27 |
| | 5,5786 | -0,8246 | 4 | 0.059 | | 27 |
| | 5,7244 | 14,2346 | 4 | 0.114 | | 27 |
| | 5,8744 | -0,7617 | 7 | 0.064 | | 27 |
| | 5,8842 | -0,9752 | 5 | 0.063 | | 27 |
| | 5,9458 | -0,8875 | 6 | 0.065 | | 27 |
| LCCG 01 | 6,1292 | -45,2941 | 3 | 0,067 | | 2 |
| | 6,3952 | -9,5268 | 4 | 0.054 | | 27 |
| HCG 001 | 6,5000 | 25,7181 | 4 | 0,034 | U248 | 1 |
| UZCCG006 | 6,8625 | -1,7033 | 3 | 0,013 | | 5 |
| UZCCG007 | 7,1792 | 2,7283 | 5 | 0,015 | | 5 |
| | 7,2195 | -9,5881 | 4 | 0.074 | | 27 |
| | 7,2625 | -0,1366 | 5 | 0.058 | | 27 |
| | 7,2831 | -0,8050 | 5 | 0.107 | | 27 |
| | 7,3541 | -0,2432 | 5 | 0.061 | | 27 |
| | 7,3820 | -0,2752 | 7 | 0.059 | | 27 |
| HCG 002 | 7,8750 | 8,4314 | 4 | 0,014 | U312 | 1 |
| | | | 3 | 0,014 | UZCCG008 | 5 |
| SCG 15 | 8,1708 | -25,6130 | | 0,010 | | 12 |
| SCG 51 | 8,4000 | -27,7833 | | 0,005 | | 12 |
| RSCG 02 | 8,4708 | 7,2506 | 3 | 0,018 | | 9 |
| HCG 003 | 8,5458 | -7,5930 | 4 | 0,026 | | 1 |
| HCG 004 | 8,5667 | -21,4466 | 5 | 0,028 | | 1 |
| | | | 5 | 0,027 | HCG 004 | 12 |
| | 8,9953 | -10,2286 | 4 | 0.021 | | 27 |
| | 9,1015 | -9,0860 | 4 | 0.055 | | 27 |
| SCG 07 | 9,4292 | -33,6900 | | 0,030 | | 12 |
| | 9,4510 | 15,5393 | 5 | 0.08 | | 27 |
| | 9,6883 | 0,0901 | 5 | 0.071 | | 27 |
| HCG 005 | 9,7250 | 7,0636 | 4 | 0,041 | N190 | 1 |
| HCG 006 | 9,7917 | -8,3953 | 4 | 0,038 | | 1 |
| RSCG 03 | 9,8167 | 0,8883 | 3 | 0,014 | HCG7 a | 9 |
| HCG 007 | 9,8500 | 0,8781 | 4 | 0,014 | N192 | 1 |
| | | | 3 | 0,014 | UZCCG009 | 5 |
| | | | 4 | 0,015 | HCG 007 | 12 |
| | 10,4111 | -9,1714 | 4 | 0.06 | | 27 |

| | | | | | | | | |
|---|---|---|---|---|---|---|---|---|
| | 10,4569 | -10,7884 | 4 | 0.053 | | | 27 | |
| | 10,4590 | -9,0452 | 5 | 0.053 | | | 27 | |
| | 10,5106 | -9,8677 | 8 | 0.055 | | | 27 | |
| UZCCG010 | 10,5250 | 29,6186 | 3 | 0,017 | | | 5 | |
| | 10,5589 | -9,5320 | 5 | 0.053 | | | 27 | |
| | 10,5659 | -9,1607 | 6 | 0.05 | | | 27 | |
| | 10,6443 | -9,7388 | 5 | 0.051 | | | 27 | |
| RSCG 04 | 10,7042 | -23,5530 | 3 | 0,022 | | | 9 | |
| | 10,7653 | -11,0300 | 4 | 0.036 | | | 27 | |
| | 10,7933 | -9,7051 | 6 | 0.05 | | | 27 | |
| | 10,7945 | 15,2550 | 9 | 0.078 | | | 27 | |
| UZCCG011 | 10,8083 | 23,4961 | 3 | 0,024 | | | 5 | |
| | 11,0006 | -9,0690 | 4 | 0.055 | | | 27 | |
| | 11,3721 | 15,2705 | 4 | 0.056 | | | 27 | |
| | 11,5699 | -9,5935 | 4 | 0.02 | | | 27 | |
| | 11,6517 | 0,0760 | 6 | 0.113 | | | 27 | |
| | 11,6960 | -0,9482 | 4 | 0.12 | | | 27 | |
| | 11,8394 | -0,8542 | 4 | 0.119 | | | 27 | |
| SCG 28 | 11,8833 | -21,4878 | | 0,022 | | | 12 | |
| | 11,9442 | 15,5372 | 4 | 0.079 | | | 27 | |
| LCCG 02 | 12,0458 | -42,0128 | 4 | 0,079 | | | 2 | |
| HCG 008 | 12,4042 | 23,5808 | 4 | 0,055 | | | 1 | |
| UZCCG012 | 12,5167 | -1,8775 | 3 | 0,013 | | | 5 | |
| | 12,7035 | 15,9379 | 5 | 0.077 | | | 27 | |
| UZCCG013 | 12,9083 | 29,6436 | 3 | 0,017 | | | 5 | |
| | 13,1759 | -9,8927 | 4 | 0.079 | | | 27 | |
| | 13,2560 | -0,4142 | 4 | 0.116 | | | 27 | |
| | 13,5613 | -10,6061 | 4 | 0.12 | | | 27 | |
| HCG009 | 13,5750 | -23,5511 | 4 | | 0.0672 | A | 1 | not CG / discordant redshifts |
| | | | | | 0.0314 | B | 1 | |
| | | | | | 0.0344 | C | 1 | |
| | | | | | 0.0591 | D | 1 | |
| | 13,9425 | -10,9434 | 5 | 0.056 | | | 27 | |
| | 13,9785 | -10,3397 | 5 | 0.057 | | | 27 | |
| | 14,0461 | -9,6969 | 4 | 0.051 | | | 27 | |
| | 14,0488 | -10,0848 | 4 | 0.058 | | | 27 | |
| | 14,0708 | -1,0237 | 5 | 0.044 | | | 27 | |
| | 14,1258 | 13,8716 | 5 | 0.082 | | | 27 | |
| | 14,1270 | 15,5452 | 5 | 0.076 | | | 27 | |
| | 14,1409 | -0,9864 | 8 | 0.048 | | | 27 | |
| | 14,1687 | 0,0966 | 4 | 0.071 | | | 27 | |
| | 14,2077 | 0,4541 | 5 | 0.067 | | | 27 | |
| | 14,2245 | -0,7570 | 9 | 0.042 | | | 27 | |
| | 14,2322 | 15,2224 | 5 | 0.077 | | | 27 | |
| | 14,3069 | -9,2641 | 4 | 0.05 | | | 27 | |
| | 14,3455 | -1,0338 | 6 | 0.046 | | | 27 | |
| UZCCG014 | 14,4542 | 30,3528 | 3 | 0,017 | | | 5 | |
| | 14,4834 | 14,6272 | 4 | 0.078 | | | 27 | |
| SHK 031 | 14,5583 | 13,9031 | 7 | 0,187 | | | 20 | |
| PCG005817+135444 | 14,5583 | 13,9031 | | | 0.1840 | B | 7 | group member |
| | 15,2455 | 0,5371 | 4 | 0.11 | | | 27 | |
| | 15,4050 | -0,2292 | 4 | 0.112 | | | 27 | |
| | 15,5337 | 0,6447 | 4 | 0.042 | | | 27 | |
| | 15,9591 | -0,3930 | 5 | 0.05 | | | 27 | |
| | 16,0005 | -0,8544 | 5 | 0.065 | | | 27 | |
| | 16,2319 | 14,8711 | 6 | 0.04 | | | 27 | |
| | 16,3237 | -0,9296 | 9 | 0.067 | | | 27 | |
| | 16,5707 | 14,4924 | 5 | 0.041 | | | 27 | |
| LCCG 03 | 16,5958 | -44,3194 | 3 | 0,139 | | | 2 | |
| RSCG 05 | 16,8500 | 32,3567 | 5 | 0,017 | | | 9 | |
| | 16,9662 | -0,7490 | 6 | 0.046 | | | 27 | |
| | 17,0363 | -9,5590 | 4 | 0.075 | | | 27 | |
| UZCCG015 | 17,0542 | 33,2589 | 4 | 0,015 | | | 5 | |
| | 17,1725 | 14,2663 | 8 | 0.06 | | | 27 | |
| | 17,2523 | 0,1876 | 4 | 0.051 | | | 27 | |
| | 17,3836 | 0,2392 | 6 | 0.045 | | | 27 | |
| | 17,4795 | 14,1551 | 6 | 0.062 | | | 27 | |
| | 17,4888 | 13,9211 | 4 | 0.057 | | | 27 | |
| | 17,5418 | 14,1193 | 4 | 0.054 | | | 27 | |
| | 17,6159 | 13,9911 | 4 | 0.056 | | | 27 | |
| SHK 038 | 17,7292 | 8,3156 | 6 | 0,088 | | | 20 | |
| UZCCG016 | 17,7417 | 33,1303 | 3 | 0,017 | | | 5 | |
| | 17,7707 | 14,8742 | 4 | 0.06 | | | 27 | |
| | 18,0326 | 15,7730 | 4 | 0.061 | | | 27 | |
| | 18,1161 | 15,5506 | 7 | 0.039 | | | 27 | |
| UZCCG017 | 18,1625 | 0,3178 | 3 | 0,018 | | | 5 | |
| | 18,2056 | -9,8326 | 4 | 0.124 | | | 27 | |
| | 18,2247 | 15,4694 | 9 | 0.044 | | | 27 | |
| | 18,2535 | 15,2623 | 4 | 0.045 | | | 27 | |
| | 18,2720 | 15,4967 | 9 | 0.043 | | | 27 | |
| | 18,2887 | 14,3903 | 4 | 0.056 | | | 27 | |
| | 18,5162 | 15,7443 | 6 | 0.038 | | | 27 | |
| | 18,5520 | -0,9306 | 4 | 0.034 | | | 27 | |
| PCG011428-005123 | 18,6167 | 0,8567 | 4 | | 0.0810 | A | 7 | group member/coord.from Ref.(6) |
| | 18,6776 | 13,5717 | 8 | 0.059 | | | 27 | |
| | 18,7716 | 0,3650 | 7 | 0.048 | | | 27 | |
| | 18,7783 | -10,8760 | 4 | 0.12 | | | 27 | |
| | 18,7840 | 15,3812 | 6 | 0.068 | | | 27 | |
| | 18,8140 | 14,3275 | 4 | 0.068 | | | 27 | |
| UZCCG018 | 18,9083 | 5,2933 | 3 | 0,017 | | | 5 | |
| | 18,9752 | 13,5237 | 4 | 0.06 | | | 27 | |
| SCG 43 | 18,9875 | -29,7666 | | 0,032 | | | 12 | |

| | | | | | | | | | |
|---|---|---|---|---|---|---|---|---|---|
| UZCCG019 | 18,9958 | 33,0725 | 3 | 0,017 | | | | 5 | |
| RSCG 06 | 19,0458 | 46,7386 | 3 | 0,017 | | | | 9 | |
| | | | 3 | 0,018 | | | UZCCG020 | 5 | |
| | 19,2467 | 14,4406 | 4 | 0.06 | | | | 27 | |
| | 19,2875 | -9,0721 | 5 | 0.065 | | | | 27 | |
| | 19,3073 | -0,4744 | 5 | 0.092 | | | | 27 | |
| | 19,3788 | 0,5972 | 4 | 0.059 | | | | 27 | |
| | 19,6167 | -0,3405 | 4 | 0.047 | | | | 27 | |
| | 19,6684 | -1,0823 | 4 | 0.044 | | | | 27 | |
| | 19,6813 | -0,9230 | 6 | 0.045 | | | | 27 | |
| | 19,7897 | -0,9744 | 5 | 0.046 | | | | 27 | |
| | 19,7931 | 14,9414 | 4 | 0.125 | | | | 27 | |
| | 19,8453 | 1,0265 | 4 | 0.089 | | | | 27 | |
| | 19,8995 | -8,4865 | 4 | 0.075 | | | | 27 | |
| | 19,9723 | -0,3485 | 4 | 0.054 | | | | 27 | |
| | 19,9757 | 14,8513 | 5 | 0.053 | | | | 27 | |
| | 19,9775 | -10,6191 | 4 | 0.101 | | | | 27 | |
| | 20,0663 | -0,1112 | 4 | 0.077 | | | | 27 | |
| UZCCG021 | 20,2208 | 1,3739 | 3 | 0,018 | | | | 5 | |
| UZCCG022 | 20,2917 | 40,4686 | 3 | 0,020 | | | | 5 | |
| | 20,3258 | 0,0608 | 5 | 0.078 | | | | 27 | |
| | 20,4290 | -0,8275 | 4 | 0.054 | | | | 27 | |
| | 20,4539 | 14,2047 | 7 | 0.053 | | | | 27 | |
| | 20,5367 | -10,1083 | 5 | 0.105 | | | | 27 | |
| | 20,6208 | -9,6967 | 4 | 0.042 | | | | 27 | |
| PCG012244+134854 | 20,6875 | 13,8153 | 4 | | 0.0450 | A | | 7 | group member/coord.from Ref.(6) |
| | 20,7055 | -10,4190 | 4 | 0.105 | | | | 27 | |
| RSCG 07 | 20,7875 | 33,4531 | 3 | 0,015 | | | | 9 | |
| | 20,8188 | 0,2844 | 4 | 0.094 | | | | 27 | |
| RSCG 08 | 20,8750 | 33,2714 | 5 | 0,017 | | | | 9 | |
| LCCG 04 | 20,8958 | -39,2369 | 3 | 0,122 | | | | 2 | |
| | 20,9302 | -9,8948 | 4 | 0.103 | | | | 27 | |
| | 20,9723 | -9,5515 | 4 | 0.044 | | | | 27 | |
| | 20,9985 | 15,6058 | 5 | 0.133 | | | | 27 | |
| UZCCG023 | 21,0375 | 33,8578 | 3 | 0,018 | | | | 5 | |
| SHK 040 | 21,2542 | 8,4767 | 60 | 0,048 | | | | 11 | mean value/coord.from Ref.(10) |
| RSCG 09 | 21,3167 | 14,8683 | 3 | 0,021 | | | | 9 | |
| | | | 3 | 0,022 | | | UZCCG024 | 5 | |
| RSCG 10 | 21,3583 | -1,5183 | 3 | 0,017 | | | | 9 | |
| RSCG 11 | 21,4750 | -1,3266 | 5 | 0,018 | | | | 9 | |
| HCG 010 | 21,5292 | 34,6908 | 4 | 0,016 | | | N536 | 1 | |
| | | | 4 | 0,016 | | | UZCCG025 | 5 | |
| | | | 3 | 0,016 | | | RSCG 12 | 9 | |
| HCG011 | 21,6417 | -23,2311 | 4 | | 0.0184 | A | | 1 | not CG / discordant redshifts |
| | | | | | 0.0443 | B | | 1 | |
| | | | | | 0.0430 | C | | 1 | |

| | | | | | | | | |
|---|---|---|---|---|---|---|---|---|
| | | | | | 0.0323 | D | 1 | |
| | 21,8643 | -9,8402 | 8 | 0.078 | | | 27 | |
| HCG 012 | 21,8917 | -4,6705 | 5 | 0,049 | | | 1 | |
| PCG012735+145913 | 21,8958 | 14,9872 | 4 | | 0.1100 | A | 7 | group member/coord.from Ref.(6) |
| | 22,8970 | 0,6292 | 5 | 0.078 | | | 27 | |
| HCG 013 | 23,0917 | -7,8811 | 5 | 0,041 | | | 1 | |
| UZCCG026 | 23,4792 | -1,0661 | 3 | 0,016 | | | 5 | |
| | 23,4847 | -1,1147 | 4 | 0.081 | | | 27 | |
| | 23,5248 | -9,8503 | 4 | 0.11 | | | 27 | |
| | 23,7063 | -0,6472 | 8 | 0.08 | | | 27 | |
| | 23,7592 | -0,6363 | 4 | 0.083 | | | 27 | |
| | 23,8887 | 15,2911 | 4 | 0.071 | | | 27 | |
| | 23,9153 | 13,1711 | 5 | 0.121 | | | 27 | |
| | 23,9448 | -9,9819 | 4 | 0.114 | | | 27 | |
| | 24,2033 | -10,5590 | 4 | 0.099 | | | 27 | |
| | 24,3497 | -0,4085 | 4 | 0.055 | | | 27 | |
| | 24,4368 | -9,2537 | 9 | 0.039 | | | 27 | |
| | 24,4456 | 0,9063 | 6 | 0.043 | | | 27 | |
| SHK 043 | 24,6083 | 8,5206 | 8 | 0,124 | | | 20 | |
| | 24,7728 | -0,0749 | 4 | 0.057 | | | 27 | |
| | 24,8331 | -8,9895 | 5 | 0.051 | | | 27 | |
| | 25,0493 | 14,8653 | 5 | 0.028 | | | 27 | |
| | 25,2893 | -0,2361 | 4 | 0.056 | | | 27 | |
| LCCG 05 | 25,3042 | -44,5983 | 3 | 0,095 | | | 2 | |
| | 25,5941 | -10,2088 | 6 | 0.112 | | | 27 | |
| | 25,6982 | -8,8839 | 4 | 0.037 | | | 27 | |
| RSCG 13 | 25,7875 | -34,2105 | 3 | 0,013 | | | 9 | |
| UZCCG027 | 25,8250 | 4,2339 | 3 | 0,018 | | | 5 | |
| | 26,2224 | 0,2621 | 4 | 0.06 | | | 27 | |
| | 26,2415 | -8,3688 | 6 | 0.052 | | | 27 | |
| UZCCG028 | 26,3083 | 10,4828 | 4 | 0,018 | | | 5 | |
| | 26,3730 | -10,1496 | 5 | 0.052 | | | 27 | |
| | 26,4189 | 0,0424 | 6 | 0.081 | | | 27 | |
| | 26,5323 | -9,8994 | 7 | 0.053 | | | 27 | |
| | 26,5598 | -9,9557 | 4 | 0.051 | | | 27 | |
| | 26,6029 | -0,5334 | 4 | 0.083 | | | 27 | |
| | 26,6305 | -9,6444 | 4 | 0.052 | | | 27 | |
| | 26,7149 | -0,0710 | 4 | 0.085 | | | 27 | |
| | 26,7199 | -0,4963 | 5 | 0.056 | | | 27 | |
| | 26,8022 | -9,4929 | 4 | 0.051 | | | 27 | |
| | 27,0751 | 14,1418 | 9 | 0.07 | | | 27 | |
| | 27,1099 | 14,4877 | 4 | 0.071 | | | 27 | |
| UZCCG029 | 27,2125 | 10,5367 | 3 | 0,018 | | | 5 | |
| | 27,3136 | 13,9927 | 5 | 0.069 | | | 27 | |
| | 27,3173 | 13,8967 | 7 | 0.071 | | | 27 | |
| | 27,3508 | -10,0013 | 6 | 0.051 | | | 27 | |

| | | | | | | | | | |
|---|---|---|---|---|---|---|---|---|---|
| RSCG 14 | 27,4250 | 22,0058 | 3 | 0,010 | | | | 9 | |
| | 27,4962 | -0,0750 | 5 | 0.087 | | | | 27 | |
| | 27,5356 | 13,7967 | 5 | 0.044 | | | | 27 | |
| | 28,1615 | 1,0265 | 8 | 0.059 | | | | 27 | |
| | 28,1616 | 1,0757 | 5 | 0.06 | | | | 27 | |
| RSCG 15 | 28,1708 | 36,1461 | 4 | 0,016 | | | | 9 | |
| PCG015254-001033 | 28,2250 | 0,1761 | 4 | | 0.0810 | C | | 7 | group member/coord.from Ref.(6) |
| | 28,2286 | -0,1961 | 5 | 0.081 | | | | 27 | |
| | 28,3067 | -1,0670 | 4 | 0.06 | | | | 27 | |
| | 28,6028 | 14,6513 | 5 | 0.085 | | | | 27 | |
| | 28,6524 | -9,7941 | 5 | 0.053 | | | | 27 | |
| RSCG 16 | 28,7542 | 36,9033 | 3 | 0,018 | | | | 9 | |
| | 28,9768 | -8,1253 | 4 | 0.051 | | | | 27 | |
| | 28,9969 | 14,8267 | 4 | 0.044 | | | | 27 | |
| | 29,0581 | 0,9940 | 4 | 0.077 | | | | 27 | |
| | 29,0807 | 1,0161 | 4 | 0.078 | | | | 27 | |
| RSCG 17 | 29,0917 | 5,6436 | 3 | 0,019 | | | | 9 | |
| | 29,0954 | 1,0221 | 4 | 0.079 | | | | 27 | |
| | 29,0987 | -9,5338 | 7 | 0.053 | | | | 27 | |
| | 29,1135 | 1,0208 | 8 | 0.08 | | | | 27 | |
| | 29,1600 | 13,2268 | 5 | 0.045 | | | | 27 | |
| RSCG 18 | 29,2000 | 33,0319 | 3 | 0,015 | | | | 9 | |
| | 29,3083 | 13,0856 | 5 | 0.033 | | | | 27 | |
| | 29,3234 | -9,3535 | 4 | 0.083 | | | | 27 | |
| | 29,4473 | -8,9145 | 4 | 0.082 | | | | 27 | |
| | 29,4486 | -9,0305 | 4 | 0.081 | | | | 27 | |
| | 29,4985 | -9,5115 | 5 | 0.079 | | | | 27 | |
| | 29,7211 | 13,0861 | 6 | 0.044 | | | | 27 | |
| HCG 014 | 29,9500 | -7,0286 | 4 | 0,018 | | | | 1 | |
| UZCCG030 | 30,0792 | 24,4356 | 3 | 0,016 | | | | 5 | |
| | 30,0898 | -8,3021 | 4 | 0.079 | | | | 27 | |
| | 30,1032 | -0,5944 | 4 | 0.043 | | | | 27 | |
| | 30,1594 | 13,0349 | 5 | 0.045 | | | | 27 | |
| LCCG 06 | 30,6042 | -44,7816 | 3 | 0,063 | | | | 2 | |
| | 30,6116 | 14,2177 | 4 | 0.044 | | | | 27 | |
| | 30,6343 | -1,0029 | 5 | 0.044 | | | | 27 | |
| | 30,6820 | -1,0821 | 4 | 0.041 | | | | 27 | |
| | 30,8496 | 0,1872 | 4 | 0.076 | | | | 27 | |
| UZCCG031 | 31,1458 | 28,8183 | 3 | 0,016 | | | | 5 | |
| | 31,2937 | 0,1786 | 4 | 0.077 | | | | 27 | |
| | 31,3955 | -0,6696 | 6 | 0.043 | | | | 27 | |
| UZCCG032 | 31,5667 | 29,9172 | 3 | 0,016 | | | | 5 | |
| | 31,6581 | 0,7055 | 6 | 0.041 | | | | 27 | |
| HCG 015 | 31,9125 | 2,1383 | 6 | 0,023 | | | U1624 | 1 | |
| | | | | 0,023 | | | UZCCG033 | 5 | |
| | 31,9308 | 13,6314 | 4 | 0.047 | | | | 27 | |
| | 31,9430 | -0,1488 | 4 | 0.075 | | | | 27 | |
| | 32,2202 | 12,4554 | 5 | 0.061 | | | | 27 | |
| RSCG 19 | 32,3792 | -10,1586 | 4 | 0,013 | | | HCG 16 | 9 | |
| HCG 016 | 32,3875 | -10,1630 | 4 | 0,013 | | | ARP318 | 1 | |
| | | | | 0,013 | | | HCG 016 | 12 | |
| | | | 4 | 0,013 | | | RSCG 19 | 9 | |
| PCG021053-010929 | 32,7250 | -1,1583 | 4 | | 0.1730 | B | | 7 | group member/coord.from Ref.(6) |
| | 32,8898 | 13,0045 | 7 | 0.061 | | | | 27 | |
| | 33,0144 | 14,1420 | 4 | 0.026 | | | | 27 | |
| | 33,1011 | -9,2567 | 6 | 0.136 | | | | 27 | |
| HCG 017 | 33,5250 | 13,3133 | 5 | 0,060 | | | | 1 | |
| | 33,6441 | 13,3174 | 4 | 0.147 | | | | 27 | |
| | 34,6740 | -8,2828 | 4 | 0.11 | | | | 27 | |
| | 36,1543 | -9,1156 | 4 | 0.052 | | | | 27 | |
| | 36,3441 | -8,6237 | 5 | 0.055 | | | | 27 | |
| | 36,4099 | -8,0711 | 5 | 0.054 | | | | 27 | |
| | 36,4136 | -8,6965 | 4 | 0.111 | | | | 27 | |
| | 36,9808 | -1,0961 | 4 | 0.069 | | | | 27 | |
| | 37,0168 | 1,0654 | 4 | 0.129 | | | | 27 | |
| | 37,0300 | -1,1055 | 5 | 0.068 | | | | 27 | |
| | 37,0373 | -0,2371 | 4 | 0.059 | | | | 27 | |
| UZCCG034 | 37,1792 | -1,1905 | 4 | 0,005 | | | | 5 | |
| LCCG 07 | 37,2333 | -44,4561 | 3 | 0,098 | | | | 2 | |
| | 37,2827 | -8,5547 | 4 | 0.119 | | | | 27 | |
| UZCCG035 | 37,3917 | 31,5222 | 3 | 0,017 | | | | 5 | |
| | 38,0368 | 1,0646 | 9 | 0.022 | | | | 27 | |
| | 38,0448 | 0,8789 | 4 | 0.023 | | | | 27 | |
| | 38,0875 | 0,8252 | 4 | 0.021 | | | | 27 | |
| | 38,3498 | 0,0446 | 6 | 0.054 | | | | 27 | |
| | 38,5650 | -1,0668 | 4 | 0.049 | | | | 27 | |
| UZCCG036 | 38,6125 | 32,9156 | 3 | 0,015 | | | | 5 | |
| | 38,7167 | -8,7395 | 4 | 0.043 | | | | 27 | |
| | 39,3702 | -0,6149 | 4 | 0.037 | | | | 27 | |
| | 39,6564 | 0,4077 | 6 | 0.047 | | | | 27 | |
| HCG 018 | 39,7792 | 18,3831 | 3 | | 0.0334 | A | | 1 | not CG / discordant redshifts |
| | | | | | 0.0138 | C | | 1 | |
| | | | | | 0.0136 | D | | 1 | |
| | 39,8937 | -7,3137 | 4 | 0.024 | | | | 27 | |
| | 40,1951 | -8,4903 | 5 | 0.038 | | | | 27 | |
| HCG 019 | 40,6875 | -12,4119 | 2 | | 0.0143 | A | | 1 | group member |
| | | | | | 0.0140 | B | | 1 | group member |
| UZCCG037 | 40,8833 | 32,4778 | 4 | 0,015 | | | | 5 | |
| | 40,9159 | 1,0840 | 4 | 0.024 | | | | 27 | |
| HCG 020 | 41,0625 | 26,1031 | 6 | 0,048 | | | | 1 | |
| | 41,0745 | -8,1891 | 5 | 0.03 | | | | 27 | |
| HCG 021 | 41,3250 | -17,6194 | 5 | 0,025 | | | N1099 | 1 | |

| | | | | | | |
|---|---|---|---|---|---|---|
| | | | 6 | 0,024 | HCG 021 | 12 |
| HCG 019 | 41,4375 | -12,4116 | | 0,014 | | 12 |
| | 41,6695 | -0,6039 | 4 | 0.042 | | 27 |
| | 41,7054 | -0,4321 | 6 | 0.044 | | 27 |
| | 41,7656 | 0,6520 | 4 | 0.076 | | 27 |
| CG J0247+44.9 | 41,9000 | 44,8608 | 6 | 0,039 | | 14 |
| | 42,6250 | -0,0652 | 4 | 0.029 | | 27 |
| | 42,6872 | -0,9760 | 5 | 0.024 | | 27 |
| LCCG 08 | 43,0375 | -38,9069 | 3 | 0,106 | | 2 |
| | 43,3141 | -0,2815 | 6 | 0.044 | | 27 |
| UZCCG038 | 43,3958 | 12,9650 | 3 | 0,012 | | 5 |
| LCCG 09 | 43,8042 | -39,2408 | 3 | 0,065 | | 2 |
| | 43,8575 | 0,1343 | 4 | 0.041 | | 27 |
| | 45,0621 | -0,4221 | 4 | 0.031 | | 27 |
| | 45,2990 | 1,1429 | 4 | 0.072 | | 27 |
| HCG 022 | 45,8792 | -15,6758 | 5 | 0,009 | N1199 | 1 |
| | | | 5 | 0,009 | HCG 022 | 12 |
| | | | 3 | 0,009 | RSCG 20 | 9 |
| | 45,9939 | -0,9038 | 4 | 0.066 | | 27 |
| | 46,0678 | 1,0189 | 4 | 0.155 | | 27 |
| | 46,3270 | -6,4452 | 5 | 0.087 | | 27 |
| | 46,3530 | -0,2573 | 7 | 0.029 | | 27 |
| | 46,3674 | -0,7222 | 6 | 0.028 | | 27 |
| | 46,4189 | -0,9779 | 4 | 0.11 | | 27 |
| | 46,5573 | -0,4740 | 4 | 0.025 | | 27 |
| HCG 023 | 46,7750 | -9,5855 | 5 | 0,016 | N1214 | 1 |
| | | | 5 | 0,016 | HCG 023 | 12 |
| | 46,8176 | -0,9739 | 4 | 0.084 | | 27 |
| | 46,8653 | -7,5320 | 5 | 0.036 | | 27 |
| | 46,8797 | 0,4421 | 4 | 0.074 | | 27 |
| | 46,9284 | -0,8436 | 7 | 0.038 | | 27 |
| | 47,8511 | -0,2282 | 8 | 0.037 | | 27 |
| | 48,0011 | 0,0006 | 8 | 0.038 | | 27 |
| | 48,6210 | -0,5723 | 4 | 0.117 | | 27 |
| | 49,8419 | -7,9029 | 4 | 0.033 | | 27 |
| RSCG 21 | 49,9042 | 41,5608 | 3 | 0,017 | | 9 |
| HCG 024 | 50,0792 | -10,8647 | 4 | 0,031 | | 1 |
| LCCG 10 | 50,0833 | -42,2633 | 3 | 0,114 | | 2 |
| HCG 025 | 50,1833 | -1,0519 | 7 | 0,021 | | 1 |
| | | | 7 | 0,021 | HCG 025 | 12 |
| HCG 026 | 50,4750 | -13,6458 | 4 | 0,032 | | 1 |
| | | | 4 | 0,032 | HCG 026 | 12 |
| LCCG 11 | 50,5667 | -41,3561 | 4 | 0,061 | | 2 |
| | 50,7239 | -6,6993 | 5 | 0.042 | | 27 |
| | 50,9567 | -0,5179 | 7 | 0.036 | | 27 |
| UZCCG039 | 51,3250 | 40,6589 | 3 | 0,013 | | 5 |
| | 51,5011 | -0,5624 | 6 | 0.038 | | 27 |
| RSCG 22 | 51,6208 | -21,2858 | 3 | 0,005 | | 9 |
| | 52,4925 | -0,8962 | 4 | 0.107 | | 27 |
| | 52,4951 | -6,8603 | 4 | 0.146 | | 27 |
| | 52,5904 | -6,9014 | 4 | 0.145 | | 27 |
| LCCG 12 | 53,0750 | -38,1336 | 3 | 0,059 | | 2 |
| | 53,1981 | -0,1594 | 4 | 0.085 | | 27 |
| | 53,5628 | 1,0919 | 4 | 0.048 | | 27 |
| UZCCG040 | 53,5792 | 39,4986 | 3 | 0,020 | | 5 |
| LCCG 13 | 53,8042 | -38,6711 | 4 | 0,061 | | 2 |
| RSCG 23 | 53,8125 | -32,5525 | 3 | 0,005 | | 9 |
| RSCG 24 | 54,1375 | -35,5253 | 13 | 0,005 | | 9 |
| | 54,2190 | -6,1316 | 4 | 0.114 | | 27 |
| | 54,4428 | 1,0219 | 5 | 0.042 | | 27 |
| RSCG 25 | 55,0167 | -18,6016 | 4 | 0,005 | | 9 |
| | 56,3367 | 0,3062 | 4 | 0.107 | | 27 |
| | 56,5381 | 0,0278 | 4 | 0.038 | | 27 |
| | 56,6174 | -7,1125 | 4 | 0.034 | | 27 |
| | 58,1072 | 0,8140 | 5 | 0.038 | | 27 |
| | 58,2874 | -4,9704 | 5 | 0.113 | | 27 |
| | 59,6658 | -5,3628 | 6 | 0.037 | | 27 |
| LCCG 14 | 59,8792 | -41,3261 | 3 | 0,058 | | 2 |
| RSCG 26 | 63,1042 | -32,9264 | 4 | 0,004 | | 9 |
| RSCG 27 | 63,7958 | -28,4839 | 3 | 0,018 | | 9 |
| LCCG 15 | 64,6708 | -42,1666 | 3 | 0,056 | | 2 |
| LCCG 16 | 64,7000 | -42,2014 | 3 | 0,054 | | 2 |
| HCG 027 | 64,8375 | -11,7097 | 5 | 0,087 | | 1 |
| LCCG 17 | 65,2792 | -45,2522 | 3 | 0,139 | | 2 |
| HCG 028 | 66,8333 | -10,3166 | 4 | 0,038 | | 1 |
| UZCCG041 | 67,6792 | 0,7300 | 3 | 0,012 | | 5 |
| UZCCG042 | 68,6375 | 73,2553 | 3 | 0,014 | | 5 |
| HCG 029 | 68,6917 | -30,5472 | 4 | 0,105 | | 1 |
| HCG 030 | 69,1208 | -2,8325 | 4 | 0,015 | | 1 |
| | 72,0948 | 14,6184 | 5 | 0.091 | | 27 |
| | 72,1507 | 15,9056 | 5 | 0.115 | | 27 |
| HCG 031 | 75,4042 | -4,2566 | 4 | 0,014 | | 1 |
| | | | 4 | 0,013 | HCG 031 | 12 |
| HCG 032 | 75,4292 | -15,4200 | 4 | 0,041 | | 1 |
| HCG 033 | 77,7000 | 18,0347 | 4 | 0,026 | | 1 |
| HCG 034 | 80,4458 | 6,6769 | 4 | 0,031 | | 1 |
| UZCCG043 | 80,4875 | 3,4369 | 3 | 0,014 | | 5 |
| UZCCG044 | 82,9833 | 67,6844 | 3 | 0,017 | | 5 |
| UZCCG045 | 93,6000 | 79,9994 | 3 | 0,015 | | 5 |
| UZCCG046 | 96,2000 | 64,6264 | 3 | 0,014 | | 5 |
| UZCCG047 | 100,2667 | 50,1731 | 3 | 0,019 | | 5 |
| UZCCG048 | 101,7792 | 33,5950 | 3 | 0,017 | | 5 |

| | | | | | |
|---|---|---|---|---|---|
| UZCCG049 | 102,5000 | 25,6706 | 3 | 0,016 | 5 |
| UZCCG050 | 102,7417 | 33,4861 | 5 | 0,017 | 5 |
| UZCCG051 | 103,0125 | 15,2881 | 3 | 0,015 | 5 |
| UZCCG052 | 104,6542 | 45,1075 | 3 | 0,020 | 5 |
| UZCCG053 | 106,4750 | 50,6156 | 3 | 0,020 | 5 |
| UZCCG054 | 107,1375 | 75,3744 | 3 | 0,013 | 5 |
| UZCCG055 | 107,2583 | 46,0983 | 3 | 0,019 | 5 |
| UZCCG056 | 107,9667 | 71,9186 | 3 | 0,010 | 5 |
| UZCCG057 | 108,4917 | 35,2581 | 3 | 0,015 | 5 |
| UZCCG058 | 109,3042 | 34,0136 | 3 | 0,013 | 5 |
| | 110,7068 | 38,0530 | 4 | 0.153 | 27 |
| UZCCG059 | 110,7208 | 58,0842 | 4 | 0,011 | 5 |
| | 111,1334 | 38,8269 | 4 | 0.056 | 27 |
| | 111,9904 | 38,4851 | 6 | 0.089 | 27 |
| | 112,0418 | 39,5915 | 8 | 0.076 | 27 |
| | 112,2685 | 40,1015 | 4 | 0.049 | 27 |
| | 112,9125 | 40,0555 | 4 | 0.12 | 27 |
| | 112,9690 | 40,1928 | 4 | 0.119 | 27 |
| | 112,9990 | 39,1444 | 4 | 0.089 | 27 |
| | 113,0123 | 36,5543 | 6 | 0.085 | 27 |
| | 113,2055 | 39,5157 | 4 | 0.085 | 27 |
| | 113,5797 | 26,9247 | 4 | 0.08 | 27 |
| | 113,6457 | 26,8721 | 5 | 0.079 | 27 |
| | 113,6706 | 26,9569 | 5 | 0.081 | 27 |
| | 113,7834 | 42,4963 | 4 | 0.088 | 27 |
| | 113,8304 | 41,9184 | 5 | 0.087 | 27 |
| | 113,8304 | 42,1903 | 4 | 0.088 | 27 |
| | 113,8730 | 39,7200 | 4 | 0.087 | 27 |
| | 113,9867 | 39,5519 | 4 | 0.082 | 27 |
| | 114,2797 | 38,2429 | 4 | 0.074 | 27 |
| | 114,2990 | 27,2125 | 5 | 0.027 | 27 |
| | 114,4510 | 27,1601 | 4 | 0.092 | 27 |
| | 114,4980 | 23,9224 | 8 | 0.081 | 27 |
| | 114,7027 | 31,2377 | 4 | 0.041 | 27 |
| | 114,7742 | 27,6684 | 4 | 0.081 | 27 |
| | 114,8587 | 41,8266 | 4 | 0.086 | 27 |
| | 115,0967 | 44,9704 | 5 | 0.043 | 27 |
| | 115,2748 | 39,8424 | 5 | 0.072 | 27 |
| | 115,7307 | 37,5253 | 4 | 0.132 | 27 |
| | 116,2305 | 41,6753 | 5 | 0.094 | 27 |
| | 116,3309 | 33,9713 | 6 | 0.063 | 27 |
| | 116,3454 | 34,0568 | 4 | 0.06 | 27 |
| | 116,3971 | 40,2172 | 7 | 0.07 | 27 |
| | 116,4029 | 41,7623 | 8 | 0.029 | 27 |
| | 116,4297 | 38,0047 | 4 | 0.075 | 27 |
| | 116,4552 | 33,1975 | 4 | 0.126 | 27 |
| | 116,4996 | 21,9160 | 4 | 0.045 | 27 |
| | 116,5010 | 30,4316 | 4 | 0.057 | 27 |
| | 116,5065 | 18,2579 | 9 | 0.053 | 27 |
| | 116,5330 | 33,9269 | 8 | 0.062 | 27 |
| UZCCG060 | 116,5500 | 57,0389 | 3 | 0,020 | 5 |
| | 116,6212 | 21,9074 | 4 | 0.046 | 27 |
| | 116,6617 | 39,5688 | 4 | 0.086 | 27 |
| | 116,7540 | 18,4936 | 8 | 0.049 | 27 |
| | 116,7942 | 41,6535 | 5 | 0.028 | 27 |
| UZCCG061 | 116,8167 | 26,9431 | 4 | 0,016 | 5 |
| | 116,8515 | 39,2678 | 5 | 0.085 | 27 |
| | 116,8517 | 32,9741 | 4 | 0.109 | 27 |
| | 116,8544 | 26,6655 | 4 | 0.027 | 27 |
| | 116,8975 | 39,6809 | 4 | 0.097 | 27 |
| | 116,9550 | 18,6749 | 8 | 0.045 | 27 |
| | 116,9950 | 22,8600 | 4 | 0.075 | 27 |
| | 117,0562 | 18,4806 | 4 | 0.042 | 27 |
| | 117,1825 | 19,3773 | 4 | 0.049 | 27 |
| | 117,2377 | 18,7179 | 6 | 0.048 | 27 |
| | 117,2883 | 44,1287 | 5 | 0.032 | 27 |
| | 117,3045 | 38,7119 | 6 | 0.061 | 27 |
| | 117,3145 | 33,0866 | 4 | 0.062 | 27 |
| | 117,3408 | 20,3577 | 4 | 0.05 | 27 |
| | 117,3481 | 30,3706 | 5 | 0.059 | 27 |
| | 117,3992 | 29,3788 | 6 | 0.063 | 27 |
| | 117,4148 | 35,8369 | 4 | 0.135 | 27 |
| | 117,4183 | 31,9325 | 6 | 0.055 | 27 |
| | 117,4401 | 34,8174 | 4 | 0.132 | 27 |
| | 117,4742 | 29,4598 | 8 | 0.064 | 27 |
| | 117,5060 | 30,3222 | 5 | 0.06 | 27 |
| | 117,5094 | 37,3790 | 4 | 0.076 | 27 |
| | 117,5133 | 29,3822 | 5 | 0.062 | 27 |
| | 117,6545 | 19,6453 | 6 | 0.046 | 27 |
| | 117,7003 | 19,4512 | 4 | 0.117 | 27 |
| | 117,7181 | 29,9048 | 4 | 0.027 | 27 |
| | 117,7277 | 33,8018 | 4 | 0.062 | 27 |
| | 117,7621 | 19,9216 | 5 | 0.062 | 27 |
| | 117,7912 | 44,8013 | 5 | 0.049 | 27 |
| | 117,8065 | 43,9843 | 4 | 0.066 | 27 |
| | 117,8437 | 37,6578 | 7 | 0.084 | 27 |
| UZCCG062 | 117,8625 | 74,3592 | 3 | 0,013 | 5 |
| | 117,9002 | 23,2171 | 5 | 0.046 | 27 |
| | 117,9516 | 23,4383 | 4 | 0.044 | 27 |
| | 117,9965 | 42,2003 | 5 | 0.041 | 27 |
| | 118,0739 | 24,3683 | 5 | 0.05 | 27 |
| | 118,1196 | 28,0257 | 4 | 0.068 | 27 |
| | 118,1385 | 26,4526 | 5 | 0.122 | 27 |
| | 118,2104 | 28,8076 | 7 | 0.076 | 27 |
| | 118,2243 | 35,9881 | 4 | 0.112 | 27 |
| | 118,2847 | 23,1356 | 4 | 0.075 | 27 |
| | 118,3026 | 29,5963 | 5 | 0.059 | 27 |
| | 118,3199 | 38,8231 | 5 | 0.041 | 27 |
| | 118,3200 | 29,3334 | 7 | 0.058 | 27 |
| | 118,3783 | 26,4608 | 4 | 0.092 | 27 |
| | 118,4255 | 29,3600 | 4 | 0.057 | 27 |
| | 118,4295 | 19,4149 | 4 | 0.155 | 27 |
| | 118,5108 | 29,3488 | 5 | 0.062 | 27 |
| | 118,5307 | 25,9095 | 4 | 0.061 | 27 |
| | 118,5656 | 30,6880 | 4 | 0.08 | 27 |
| UZCCG063 | 118,7417 | 14,4317 | 3 | 0,029 | 5 |
| | 118,7746 | 22,9309 | 4 | 0.029 | 27 |
| | 118,9295 | 45,2977 | 8 | 0.051 | 27 |
| | 118,9350 | 41,0963 | 6 | 0.075 | 27 |
| | 118,9599 | 45,7372 | 5 | 0.054 | 27 |
| | 119,0386 | 33,7715 | 7 | 0.075 | 27 |
| | 119,0543 | 27,2759 | 6 | 0.065 | 27 |
| | 119,0562 | 33,6536 | 6 | 0.074 | 27 |
| | 119,0832 | 36,9967 | 5 | 0.078 | 27 |
| | 119,1194 | 37,0656 | 4 | 0.077 | 27 |
| | 119,2153 | 39,7240 | 5 | 0.097 | 27 |
| | 119,2757 | 27,3155 | 6 | 0.027 | 27 |
| | 119,3073 | 28,1801 | 5 | 0.069 | 27 |
| | 119,3105 | 39,2390 | 5 | 0.095 | 27 |
| | 119,3481 | 39,3951 | 7 | 0.096 | 27 |
| | 119,4666 | 33,1892 | 5 | 0.07 | 27 |
| | 119,5570 | 37,7714 | 7 | 0.042 | 27 |
| | 119,6130 | 37,7738 | 4 | 0.039 | 27 |
| UZCCG064 | 119,6375 | 59,1122 | 3 | 0,020 | 5 |
| | 119,6420 | 37,7925 | 9 | 0.041 | 27 |
| | 119,6755 | 25,5610 | 5 | 0.042 | 27 |
| | 119,6978 | 27,1778 | 5 | 0.098 | 27 |
| | 119,7375 | 27,1889 | 5 | 0.097 | 27 |
| | 119,7609 | 26,3416 | 4 | 0.06 | 27 |
| | 119,8524 | 46,1657 | 6 | 0.061 | 27 |
| | 119,9103 | 27,1022 | 4 | 0.023 | 27 |
| | 119,9321 | 42,2388 | 4 | 0.044 | 27 |
| | 119,9413 | 32,4469 | 4 | 0.074 | 27 |
| | 119,9866 | 38,2448 | 5 | 0.076 | 27 |
| | 120,0631 | 42,1741 | 5 | 0.032 | 27 |
| | 120,1431 | 46,5073 | 4 | 0.069 | 27 |
| | 120,1491 | 36,1735 | 4 | 0.079 | 27 |
| | 120,1686 | 30,3766 | 5 | 0.046 | 27 |
| | 120,1890 | 27,2172 | 7 | 0.047 | 27 |
| | 120,2438 | 34,6852 | 6 | 0.082 | 27 |
| | 120,2493 | 26,4641 | 4 | 0.042 | 27 |
| | 120,2676 | 39,4372 | 4 | 0.07 | 27 |
| | 120,2877 | 37,1874 | 7 | 0.075 | 27 |
| | 120,3271 | 27,2952 | 4 | 0.048 | 27 |
| | 120,3589 | 26,1684 | 4 | 0.068 | 27 |
| | 120,3869 | 31,2995 | 6 | 0.074 | 27 |
| | 120,4172 | 31,3317 | 4 | 0.075 | 27 |
| | 120,5216 | 31,0524 | 4 | 0.046 | 27 |
| | 120,5386 | 30,4496 | 4 | 0.078 | 27 |
| UZCCG065 | 120,5417 | 9,5383 | 7 | 0,015 | 5 |
| | 120,5956 | 36,9583 | 4 | 0.122 | 27 |
| | 120,5971 | 29,3269 | 4 | 0.13 | 27 |
| | 120,6319 | 46,4415 | 4 | 0.131 | 27 |
| | 120,6604 | 26,6441 | 5 | 0.048 | 27 |
| UZCCG066 | 120,6875 | 61,3828 | 3 | 0,005 | 5 |
| | 120,6892 | 23,9438 | 4 | 0.047 | 27 |
| | 120,7197 | 36,4399 | 4 | 0.073 | 27 |
| | 120,7284 | 36,5076 | 4 | 0.089 | 27 |
| | 120,7594 | 44,5802 | 5 | 0.075 | 27 |
| | 120,8961 | 30,0103 | 5 | 0.04 | 27 |
| | 120,9600 | 36,0663 | 4 | 0.044 | 27 |
| | 121,2044 | 37,4794 | 4 | 0.088 | 27 |
| | 121,2535 | 36,4828 | 7 | 0.089 | 27 |
| | 121,2755 | 33,3297 | 4 | 0.083 | 27 |
| | 121,2933 | 39,7117 | 5 | 0.095 | 27 |
| | 121,3111 | 21,3066 | 4 | 0.124 | 27 |
| | 121,3566 | 22,8969 | 4 | 0.08 | 27 |
| | 121,4715 | 34,7559 | 7 | 0.08 | 27 |
| | 121,4736 | 24,1169 | 4 | 0.059 | 27 |
| | 121,5490 | 29,4745 | 6 | 0.127 | 27 |
| | 121,5984 | 7,2001 | 5 | 0.101 | 27 |
| | 121,7050 | 33,2603 | 6 | 0.084 | 27 |
| | 121,8145 | 31,1150 | 5 | 0.077 | 27 |
| UZCCG067 | 121,8208 | 18,8044 | 3 | 0,016 | 5 |
| | 121,8799 | 37,9223 | 4 | 0.022 | 27 |
| | 122,0011 | 42,7949 | 4 | 0.1 | 27 |
| | 122,0263 | 38,8719 | 7 | 0.042 | 27 |
| | 122,0376 | 39,0744 | 4 | 0.074 | 27 |
| | 122,0531 | 35,2302 | 4 | 0.083 | 27 |
| | 122,1213 | 30,5160 | 4 | 0.051 | 27 |
| | 122,1307 | 38,8627 | 5 | 0.041 | 27 |
| | 122,1380 | 38,8749 | 5 | 0.039 | 27 |
| | 122,1801 | 7,2437 | 5 | 0.099 | 27 |
| | 122,1806 | 5,4446 | 4 | 0.031 | 27 |
| | 122,1860 | 40,1066 | 5 | 0.021 | 27 |
| | 122,2354 | 38,9452 | 4 | 0.04 | 27 |
| | 122,2407 | 34,8909 | 8 | 0.083 | 27 |
| | 122,2997 | 48,2065 | 4 | 0.056 | 27 |
| | 122,3241 | 38,8330 | 4 | 0.032 | 27 |
| | 122,3643 | 42,1906 | 5 | 0.063 | 27 |
| | 122,4296 | 46,6217 | 4 | 0.032 | 27 |
| | 122,4627 | 35,0776 | 6 | 0.082 | 27 |
| | 122,4887 | 24,9511 | 4 | 0.082 | 27 |
| | 122,5740 | 40,3939 | 4 | 0.1 | 27 |
| | 122,5918 | 35,2205 | 5 | 0.087 | 27 |
| | 122,5944 | 5,8915 | 4 | 0.074 | 27 |
| | 122,6230 | 35,0353 | 5 | 0.062 | 27 |
| | 122,6290 | 42,2586 | 6 | 0.062 | 27 |
| | 122,6768 | 35,9725 | 8 | 0.081 | 27 |
| | 122,6973 | 35,9855 | 6 | 0.083 | 27 |
| | 122,7103 | 5,9230 | 4 | 0.072 | 27 |
| | 122,7884 | 36,0176 | 5 | 0.085 | 27 |
| | 122,7978 | 36,0124 | 7 | 0.082 | 27 |
| | 122,8293 | 42,3824 | 6 | 0.064 | 27 |
| | 122,8646 | 42,3593 | 5 | 0.063 | 27 |
| | 123,0473 | 36,9807 | 5 | 0.03 | 27 |
| | 123,1976 | 46,7361 | 5 | 0.054 | 27 |
| | 123,1999 | 39,9108 | 6 | 0.068 | 27 |
| | 123,2917 | 38,3149 | 5 | 0.08 | 27 |
| | 123,3047 | 48,2924 | 8 | 0.055 | 27 |
| | 123,3246 | 39,6611 | 4 | 0.121 | 27 |
| | 123,3612 | 22,9763 | 6 | 0.097 | 27 |
| | 123,3727 | 42,2845 | 5 | 0.065 | 27 |
| | 123,3856 | 3,8410 | 4 | 0.087 | 27 |
| | 123,4418 | 40,9329 | 4 | 0.1 | 27 |
| | 123,5211 | 30,2575 | 4 | 0.075 | 27 |
| | 123,7206 | 45,7857 | 5 | 0.041 | 27 |
| | 123,7265 | 26,2725 | 5 | 0.04 | 27 |
| | 123,7281 | 36,5097 | 5 | 0.082 | 27 |
| | 123,7372 | 6,9799 | 7 | 0.111 | 27 |
| | 123,7869 | 35,7219 | 6 | 0.12 | 27 |
| | 123,8070 | 44,1654 | 4 | 0.183 | 27 |
| UZCCG068 | 123,8083 | 73,4864 | 3 | 0,012 | 5 |
| | 123,8227 | 7,8940 | 4 | 0.112 | 27 |
| | 123,8502 | 4,0999 | 4 | 0.087 | 27 |
| | 123,8949 | 3,7142 | 5 | 0.087 | 27 |
| | 123,9134 | 3,8479 | 5 | 0.088 | 27 |
| | 123,9134 | 25,2581 | 5 | 0.074 | 27 |
| | 124,0067 | 25,1286 | 4 | 0.073 | 27 |
| | 124,0174 | 24,6525 | 4 | 0.083 | 27 |
| | 124,0679 | 34,5316 | 5 | 0.062 | 27 |
| | 124,1242 | 33,3804 | 4 | 0.109 | 27 |
| | 124,2338 | 37,4286 | 6 | 0.081 | 27 |
| | 124,2522 | 3,8914 | 4 | 0.076 | 27 |
| | 124,2618 | 37,3678 | 4 | 0.082 | 27 |
| | 124,3046 | 4,1034 | 5 | 0.075 | 27 |
| | 124,4411 | 51,0440 | 4 | 0.073 | 27 |
| | 124,4453 | 49,8047 | 4 | 0.096 | 27 |
| | 124,4721 | 40,8956 | 6 | 0.066 | 27 |
| | 124,4822 | 33,3623 | 4 | 0.089 | 27 |
| | 124,5116 | 3,5777 | 5 | 0.087 | 27 |
| | 124,5203 | 40,7227 | 5 | 0.061 | 27 |
| | 124,5340 | 40,7448 | 4 | 0.062 | 27 |
| | 124,6146 | 44,4967 | 6 | 0.056 | 27 |
| | 124,6188 | 49,7925 | 4 | 0.095 | 27 |
| | 124,6217 | 37,6156 | 5 | 0.062 | 27 |
| | 124,6291 | 4,2497 | 4 | 0.086 | 27 |
| | 124,6678 | 7,4060 | 5 | 0.076 | 27 |
| UZCCG069 | 124,8833 | 21,9544 | 3 | 0,013 | 5 |
| | 124,9923 | 48,0533 | 4 | 0.072 | 27 |
| | 125,0193 | 50,8458 | 4 | 0.126 | 27 |
| UZCCG070 | 125,0375 | 4,6439 | 3 | 0,014 | 5 |
| | 125,0624 | 36,4170 | 5 | 0.063 | 27 |
| | 125,1302 | 49,1718 | 4 | 0.076 | 27 |
| | 125,1736 | 47,2521 | 4 | 0.055 | 27 |
| | 125,1917 | 7,8625 | 6 | 0.11 | 27 |
| | 125,2105 | 23,9366 | 5 | 0.094 | 27 |
| | 125,3232 | 45,3928 | 4 | 0.054 | 27 |
| | 125,3535 | 25,9820 | 4 | 0.094 | 27 |
| | 125,3929 | 35,8754 | 4 | 0.061 | 27 |
| | 125,4509 | 28,7192 | 4 | 0.077 | 27 |
| | 125,5104 | 47,1974 | 4 | 0.075 | 27 |
| | 125,5224 | 29,8431 | 7 | 0.049 | 27 |
| | 125,5271 | 45,5055 | 4 | 0.048 | 27 |
| | 125,6435 | 50,2573 | 5 | 0.073 | 27 |
| UZCCG071 | 125,6708 | 66,9406 | 3 | 0,014 | 5 |
| | 125,7026 | 37,3820 | 4 | 0.065 | 27 |
| | 125,7133 | 6,8547 | 4 | 0.082 | 27 |
| | 125,7408 | 6,5194 | 4 | 0.086 | 27 |
| UZCCG072 | 125,7500 | 4,3619 | 3 | 0,030 | 5 |
| | 125,7885 | 4,3212 | 4 | 0.028 | 27 |
| | 125,8082 | 3,7407 | 4 | 0.035 | 27 |
| | 125,8753 | 24,9261 | 4 | 0.095 | 27 |
| | 125,9020 | 4,4105 | 5 | 0.032 | 27 |
| | 125,9176 | 4,3359 | 6 | 0.031 | 27 |
| | 125,9378 | 4,3208 | 9 | 0.029 | 27 |

| | | | | | | |
|---|---|---|---|---|---|---|
| | 125,9474 | 7,1191 | 4 | 0.081 | 27 | |
| | 126,0268 | 42,1936 | 4 | 0.058 | 27 | |
| | 126,0804 | 7,3807 | 7 | 0.082 | 27 | |
| | 126,2064 | 4,3794 | 4 | 0.099 | 27 | |
| | 126,2273 | 37,2599 | 7 | 0.083 | 27 | |
| | 126,2539 | 35,6564 | 4 | 0.108 | 27 | |
| | 126,2747 | 28,7133 | 4 | 0.077 | 27 | |
| | 126,2768 | 30,6694 | 4 | 0.09 | 27 | |
| | 126,3024 | 47,6609 | 4 | 0.049 | 27 | |
| | 126,3028 | 47,1824 | 4 | 0.127 | 27 | |
| | 126,3640 | 7,1170 | 4 | 0.111 | 27 | |
| | 126,3758 | 8,5017 | 5 | 0.082 | 27 | |
| | 126,3867 | 47,0934 | 5 | 0.131 | 27 | |
| | 126,4820 | 4,4893 | 5 | 0.102 | 27 | |
| | 126,4847 | 41,0271 | 8 | 0.057 | 27 | |
| | 126,5332 | 36,9611 | 7 | 0.065 | 27 | |
| | 126,5877 | 53,1101 | 4 | 0.064 | 27 | |
| | 126,5945 | 6,7395 | 4 | 0.11 | 27 | |
| UZCCG073 | 126,7500 | 22,8900 | 3 | 0,020 | 5 | |
| | 126,7514 | 30,1444 | 6 | 0.05 | 27 | |
| | 126,7792 | 7,5548 | 4 | 0.031 | 27 | |
| | 127,0125 | 28,2759 | 4 | 0.093 | 27 | |
| SHK 181 | 127,0125 | 28,2672 | 15 | 0,093 | 22 | mean value from NED data |
| | 127,0186 | 28,2621 | 4 | 0.092 | 27 | |
| | 127,0753 | 8,4094 | 4 | 0.066 | 27 | |
| | 127,1120 | 36,8900 | 4 | 0.082 | 27 | |
| | 127,1334 | 26,9761 | 9 | 0.095 | 27 | |
| | 127,1409 | 32,8667 | 7 | 0.092 | 27 | |
| | 127,1556 | 32,8906 | 4 | 0.093 | 27 | |
| | 127,1636 | 47,2697 | 4 | 0.053 | 27 | |
| | 127,1690 | 30,4211 | 4 | 0.053 | 27 | |
| | 127,2061 | 8,3383 | 4 | 0.097 | 27 | |
| | 127,2088 | 30,3828 | 8 | 0.051 | 27 | |
| | 127,2344 | 30,5050 | 6 | 0.046 | 27 | |
| | 127,3047 | 37,2802 | 4 | 0.058 | 27 | |
| | 127,3809 | 41,6886 | 4 | 0.132 | 27 | |
| | 127,4128 | 32,6036 | 4 | 0.094 | 27 | |
| | 127,5051 | 53,3437 | 4 | 0.119 | 27 | |
| | 127,5216 | 36,7245 | 5 | 0.064 | 27 | |
| | 127,5265 | 26,9854 | 5 | 0.094 | 27 | |
| | 127,6001 | 44,3562 | 4 | 0.118 | 27 | |
| | 127,6011 | 48,3966 | 4 | 0.055 | 27 | |
| | 127,7099 | 35,7575 | 4 | 0.054 | 27 | |
| | 127,7433 | 50,6234 | 4 | 0.097 | 27 | |
| | 127,7533 | 37,2483 | 5 | 0.099 | 27 | |
| | 127,7831 | 31,4003 | 4 | 0.048 | 27 | |

| | | | | | |
|---|---|---|---|---|---|
| | 127,8716 | 6,3700 | 5 | 0.111 | 27 |
| | 127,9385 | 36,3833 | 5 | 0.065 | 27 |
| | 128,1663 | 6,6679 | 5 | 0.111 | 27 |
| | 128,1671 | 32,2016 | 6 | 0.085 | 27 |
| | 128,1825 | 41,1652 | 7 | 0.024 | 27 |
| | 128,1953 | 4,3596 | 5 | 0.064 | 27 |
| UZCCG074 | 128,2125 | 52,5269 | 3 | 0,017 | 5 |
| | 128,2181 | 49,7264 | 9 | 0.053 | 27 |
| | 128,2570 | 4,4006 | 4 | 0.065 | 27 |
| | 128,2733 | 51,8910 | 7 | 0.082 | 27 |
| | 128,3680 | 47,0191 | 4 | 0.051 | 27 |
| | 128,4345 | 47,7905 | 4 | 0.133 | 27 |
| | 128,5157 | 51,9257 | 4 | 0.05 | 27 |
| | 128,5158 | 49,7456 | 9 | 0.052 | 27 |
| | 128,5199 | 35,0465 | 6 | 0.053 | 27 |
| | 128,5212 | 46,3718 | 4 | 0.048 | 27 |
| | 128,5401 | 40,7661 | 6 | 0.084 | 27 |
| | 128,7141 | 6,5038 | 4 | 0.03 | 27 |
| | 128,7956 | 38,4622 | 4 | 0.055 | 27 |
| | 128,8540 | 52,4432 | 7 | 0.043 | 27 |
| | 128,9742 | 34,4910 | 4 | 0.054 | 27 |
| | 129,0486 | 39,8815 | 6 | 0.057 | 27 |
| | 129,0628 | 4,3918 | 5 | 0.062 | 27 |
| UZCCG075 | 129,0750 | 1,8025 | 3 | 0,014 | 5 |
| | 129,1086 | 33,4131 | 4 | 0.055 | 27 |
| | 129,1097 | 27,3852 | 4 | 0.094 | 27 |
| | 129,1119 | 54,1361 | 6 | 0.031 | 27 |
| | 129,1172 | 52,4352 | 6 | 0.048 | 27 |
| | 129,1268 | 44,8580 | 8 | 0.038 | 27 |
| | 129,2098 | 28,2397 | 8 | 0.052 | 27 |
| | 129,3865 | 30,7606 | 5 | 0.074 | 27 |
| | 129,3865 | 29,7770 | 4 | 0.051 | 27 |
| | 129,3969 | 44,2158 | 4 | 0.145 | 27 |
| | 129,4113 | 28,2686 | 4 | 0.05 | 27 |
| | 129,4996 | 39,9964 | 8 | 0.025 | 27 |
| | 129,5744 | 27,6083 | 6 | 0.098 | 27 |
| UZCCG076 | 129,5792 | 19,6786 | 3 | 0,015 | 5 |
| | 129,5816 | 52,8125 | 5 | 0.044 | 27 |
| | 129,6070 | 52,0510 | 4 | 0.114 | 27 |
| | 129,6350 | 36,3042 | 4 | 0.055 | 27 |
| | 129,6688 | 4,2814 | 6 | 0.048 | 27 |
| | 129,6772 | 28,7109 | 5 | 0.081 | 27 |
| | 129,6954 | 29,9164 | 6 | 0.05 | 27 |
| | 129,7065 | 28,8336 | 5 | 0.078 | 27 |
| | 129,7759 | 51,1138 | 4 | 0.047 | 27 |
| | 129,8222 | 28,9628 | 5 | 0.082 | 27 |
| | 129,8899 | 28,8424 | 8 | 0.079 | 27 |
| | 129,8981 | 28,7845 | 5 | 0.08 | 27 |
| | 130,1000 | 38,2012 | 5 | 0.041 | 27 |
| | 130,1066 | 29,2387 | 4 | 0.08 | 27 |
| | 130,1141 | 51,9058 | 4 | 0.098 | 27 |
| | 130,1417 | 36,8149 | 7 | 0.055 | 27 |
| | 130,1440 | 28,5206 | 6 | 0.083 | 27 |
| | 130,1652 | 5,5901 | 4 | 0.062 | 27 |
| | 130,1912 | 32,5109 | 8 | 0.068 | 27 |
| | 130,1971 | 31,4275 | 5 | 0.068 | 27 |
| | 130,3081 | 52,9868 | 8 | 0.045 | 27 |
| | 130,4014 | 33,6450 | 6 | 0.109 | 27 |
| | 130,5729 | 51,9609 | 4 | 0.056 | 27 |
| | 130,6348 | 3,9489 | 6 | 0.028 | 27 |
| | 130,6594 | 5,9173 | 4 | 0.073 | 27 |
| | 130,7150 | 36,0156 | 4 | 0.056 | 27 |
| | 130,7887 | 36,1673 | 9 | 0.054 | 27 |
| | 130,8984 | 38,9775 | 5 | 0.121 | 27 |
| | 130,9061 | 32,6508 | 5 | 0.067 | 27 |
| | 131,0470 | 37,0024 | 4 | 0.111 | 27 |
| | 131,0761 | 37,1206 | 4 | 0.112 | 27 |
| | 131,0892 | 37,3575 | 4 | 0.115 | 27 |
| | 131,0991 | 42,9249 | 5 | 0.054 | 27 |
| | 131,1023 | 43,0104 | 5 | 0.055 | 27 |
| | 131,1053 | 53,4373 | 6 | 0.031 | 27 |
| | 131,1334 | 2,1373 | 7 | 0.078 | 27 |
| | 131,2023 | 27,6617 | 5 | 0.084 | 27 |
| UZCCG077 | 131,2167 | 55,0294 | 4 | 0,025 | 5 |
| | 131,2239 | 29,2112 | 5 | 0.073 | 27 |
| | 131,2315 | 51,1649 | 4 | 0.098 | 27 |
| | 131,2449 | 43,5093 | 4 | 0.054 | 27 |
| | 131,3183 | 51,2586 | 5 | 0.08 | 27 |
| | 131,3301 | 29,3219 | 4 | 0.099 | 27 |
| HCG 035 | 131,3333 | 44,5217 | 6 | 0,054 | 1 |
| | 131,3395 | 30,0010 | 4 | 0.065 | 27 |
| | 131,3658 | 27,7338 | 6 | 0.085 | 27 |
| | 131,4113 | 30,2884 | 5 | 0.105 | 27 |
| | 131,4138 | 44,5678 | 5 | 0.054 | 27 |
| | 131,4274 | 29,2969 | 4 | 0.1 | 27 |
| UZCCG078 | 131,4417 | 72,9686 | 4 | 0,012 | 5 |
| | 131,4823 | 1,9216 | 9 | 0.051 | 27 |
| | 131,5176 | 2,4012 | 5 | 0.077 | 27 |
| | 131,5463 | 54,4181 | 5 | 0.031 | 27 |
| | 131,5517 | 4,5183 | 4 | 0.114 | 27 |
| | 131,5740 | 29,6563 | 7 | 0.07 | 27 |
| | 131,5923 | 53,4464 | 4 | 0.113 | 27 |

| | | | | | | |
|---|---|---|---|---|---|---|
| | 131,6025 | 28,1208 | 4 | 0.021 | 27 | |
| | 131,6587 | 29,6802 | 6 | 0.071 | 27 | |
| | 131,6837 | 53,8962 | 5 | 0.046 | 27 | |
| | 131,6853 | 35,5293 | 5 | 0.055 | 27 | |
| | 131,7073 | 30,5558 | 4 | 0.07 | 27 | |
| | 131,7282 | 1,8655 | 5 | 0.05 | 27 | |
| | 131,7348 | 2,1138 | 5 | 0.06 | 27 | |
| | 131,7388 | 53,8241 | 4 | 0.047 | 27 | |
| UZCCG079 | 131,7625 | 13,4464 | 3 | 0,007 | 5 | |
| | 131,7682 | 9,6313 | 4 | 0.063 | 27 | |
| | 131,7749 | 35,4721 | 8 | 0.053 | 27 | |
| | 131,7774 | 53,7462 | 4 | 0.044 | 27 | |
| | 131,8002 | 51,8409 | 5 | 0.118 | 27 | |
| | 131,8010 | 35,5265 | 7 | 0.054 | 27 | |
| | 131,8053 | 52,8454 | 5 | 0.046 | 27 | |
| | 131,8072 | 5,0008 | 4 | 0.097 | 27 | |
| | 131,8118 | 2,2580 | 5 | 0.118 | 27 | |
| | 131,8446 | 33,8725 | 4 | 0.074 | 27 | |
| | 131,8632 | 29,9997 | 4 | 0.104 | 27 | |
| | 131,8653 | 32,0599 | 5 | 0.068 | 27 | |
| SHK 344 | 131,8792 | 3,7153 | 8 | 0,078 | 22 | mean value |
| | 131,8894 | 3,6801 | 4 | 0.078 | 27 | |
| | 131,9335 | 51,0120 | 4 | 0.13 | 27 | |
| | 131,9527 | 36,2725 | 4 | 0.055 | 27 | |
| | 131,9629 | 30,0458 | 4 | 0.106 | 27 | |
| | 131,9640 | 31,9082 | 6 | 0.069 | 27 | |
| | 131,9646 | 34,1618 | 4 | 0.073 | 27 | |
| | 132,0153 | 54,9919 | 4 | 0.026 | 27 | |
| | 132,0184 | 37,7890 | 4 | 0.041 | 27 | |
| | 132,0728 | 32,2501 | 4 | 0.065 | 27 | |
| | 132,0839 | 4,2966 | 4 | 0.059 | 27 | |
| | 132,1313 | 9,0985 | 6 | 0.063 | 27 | |
| | 132,1531 | 3,1723 | 4 | 0.06 | 27 | |
| | 132,2275 | 34,3857 | 5 | 0.059 | 27 | |
| | 132,3050 | 34,8922 | 4 | 0.053 | 27 | |
| | 132,3454 | 29,4568 | 5 | 0.104 | 27 | |
| UZCCG080 | 132,3667 | 70,1908 | 3 | 0,012 | 5 | |
| | 132,4002 | 32,7968 | 4 | 0.068 | 27 | |
| | 132,4345 | 57,1630 | 4 | 0.041 | 27 | |
| | 132,4377 | 46,3231 | 4 | 0.041 | 27 | |
| | 132,4461 | 57,1460 | 4 | 0.042 | 27 | |
| | 132,4654 | 2,7869 | 4 | 0.06 | 27 | |
| | 132,4945 | 51,5921 | 5 | 0.09 | 27 | |
| | 132,5035 | 29,5449 | 5 | 0.105 | 27 | |
| | 132,5864 | 29,5324 | 6 | 0.103 | 27 | |
| | 132,5927 | 32,7556 | 7 | 0.066 | 27 | |

| | | | | | | | | |
|---|---|---|---|---|---|---|---|---|
| | 132,6207 | 29,5292 | 4 | 0.106 | | | 27 | |
| | 132,6532 | 32,8203 | 9 | 0.067 | | | 27 | |
| | 132,6781 | 36,4522 | 4 | 0.125 | | | 27 | |
| | 132,7082 | 55,4074 | 4 | 0.031 | | | 27 | |
| | 132,7401 | 0,3419 | 5 | 0.052 | | | 27 | |
| | 132,7408 | 40,0236 | 4 | 0.041 | | | 27 | |
| | 132,7465 | 30,0105 | 8 | 0.099 | | | 27 | |
| | 132,8074 | 29,3349 | 9 | 0.027 | | | 27 | |
| | 132,8160 | 36,4258 | 4 | 0.025 | | | 27 | |
| | 132,8294 | 9,5739 | 7 | 0.063 | | | 27 | |
| | 132,8546 | 32,4119 | 6 | 0.078 | | | 27 | |
| | 132,8546 | 9,5921 | 6 | 0.064 | | | 27 | |
| | 132,8574 | 55,0745 | 4 | 0.1 | | | 27 | |
| | 132,8777 | 53,2324 | 5 | 0.031 | | | 27 | |
| | 132,9043 | 4,8465 | 5 | 0.077 | | | 27 | |
| | 132,9351 | 29,5099 | 5 | 0.084 | | | 27 | |
| | 132,9466 | 57,1007 | 4 | 0.089 | | | 27 | |
| | 132,9483 | 40,1408 | 5 | 0.024 | | | 27 | |
| | 133,0191 | 33,1398 | 4 | 0.066 | | | 27 | |
| | 133,0381 | 56,0852 | 5 | 0.045 | | | 27 | |
| | 133,0915 | 29,7131 | 4 | 0.104 | | | 27 | |
| | 133,0919 | 32,3993 | 6 | 0.074 | | | 27 | |
| | 133,2067 | 40,8439 | 7 | 0.03 | | | 27 | |
| | 133,2081 | 40,1810 | 5 | 0.088 | | | 27 | |
| | 133,2502 | 34,8726 | 6 | 0.053 | | | 27 | |
| | 133,2970 | 52,1295 | 4 | 0.088 | | | 27 | |
| | 133,3181 | 57,0486 | 4 | 0.026 | | | 27 | |
| | 133,3707 | 34,2021 | 4 | 0.079 | | | 27 | |
| | 133,3849 | 40,7430 | 5 | 0.029 | | | 27 | |
| | 133,3988 | 40,2833 | 8 | 0.086 | | | 27 | |
| | 133,4284 | 29,1409 | 7 | 0.085 | | | 27 | |
| | 133,4498 | 29,0266 | 4 | 0.088 | | | 27 | |
| | 133,4973 | 39,6788 | 7 | 0.057 | | | 27 | |
| | 133,5208 | 55,1315 | 5 | 0.045 | | | 27 | |
| | 133,5391 | 3,5816 | 6 | 0.028 | | | 27 | |
| | 133,5483 | 40,4089 | 6 | 0.088 | | | 27 | |
| | 133,6122 | 35,9292 | 4 | 0.078 | | | 27 | |
| | 133,6473 | 4,4598 | 4 | 0.071 | | | 27 | |
| | 133,6742 | 0,8620 | 4 | 0.107 | | | 27 | |
| | 133,6800 | 41,9352 | 4 | 0.136 | | | 27 | |
| PCG085449+491911 | 133,7042 | 49,3200 | 4 | | 0.1180 | A | 7 | group member/coord.from Ref.(6) |
| | 133,7540 | 36,7764 | 4 | 0.089 | | | 27 | |
| | 133,7633 | 52,6070 | 4 | 0.063 | | | 27 | |
| | 133,7789 | 9,2561 | 4 | 0.068 | | | 27 | |
| | 133,8285 | 32,4858 | 4 | 0.076 | | | 27 | |
| | 133,8331 | 37,0530 | 4 | 0.087 | | | 27 | |
| | 133,8381 | 36,4143 | 8 | 0.088 | | | 27 | |
| | 133,8477 | 34,9662 | 5 | 0.086 | | | 27 | |
| | 133,8485 | 34,8651 | 7 | 0.087 | | | 27 | |
| | 133,8685 | 8,4963 | 4 | 0.063 | | | 27 | |
| | 133,8830 | 38,1371 | 4 | 0.087 | | | 27 | |
| UZCCG081 | 133,8958 | -3,0686 | 7 | 0,006 | | | 5 | |
| | 133,9729 | 34,1297 | 4 | 0.085 | | | 27 | |
| | 134,0274 | 0,7968 | 4 | 0.052 | | | 27 | |
| | 134,0305 | 39,0501 | 7 | 0.095 | | | 27 | |
| | 134,0390 | 48,4378 | 4 | 0.117 | | | 27 | |
| | 134,0661 | 53,3686 | 6 | 0.085 | | | 27 | |
| | 134,0702 | 32,6398 | 5 | 0.066 | | | 27 | |
| | 134,0739 | 53,3448 | 4 | 0.086 | | | 27 | |
| UZCCG082 | 134,1125 | 13,2036 | 3 | 0,014 | | | 5 | |
| | 134,1449 | 32,7146 | 6 | 0.064 | | | 27 | |
| | 134,1455 | 5,8702 | 6 | 0.095 | | | 27 | |
| | 134,1642 | 37,7493 | 4 | 0.091 | | | 27 | |
| | 134,1702 | 5,9391 | 4 | 0.093 | | | 27 | |
| | 134,1916 | 49,9249 | 6 | 0.093 | | | 27 | |
| PCG085650+492633 | 134,2083 | 49,4425 | 4 | | 0.1830 | B | 7 | group member/coord.from Ref.(6) |
| | 134,2447 | 30,6484 | 5 | 0.085 | | | 27 | |
| | 134,2878 | 57,4294 | 5 | 0.04 | | | 27 | |
| | 134,3066 | 37,4493 | 4 | 0.094 | | | 27 | |
| | 134,3124 | 55,6742 | 4 | 0.038 | | | 27 | |
| | 134,3527 | 34,4958 | 4 | 0.052 | | | 27 | |
| | 134,3807 | 39,4571 | 6 | 0.028 | | | 27 | |
| | 134,3928 | 42,1947 | 4 | 0.139 | | | 27 | |
| | 134,4160 | 30,2851 | 6 | 0.085 | | | 27 | |
| | 134,4347 | 32,6263 | 4 | 0.065 | | | 27 | |
| UZCCG083 | 134,4542 | 2,9789 | 3 | 0,013 | | | 5 | |
| PCG085756+383340 | 134,4875 | 38,5611 | 5 | | 0.0930 | C | 7 | group member/coord.from Ref.(6) |
| | 134,5601 | 38,4710 | 6 | 0.093 | | | 27 | |
| | 134,5729 | -0,4713 | 4 | 0.071 | | | 27 | |
| | 134,5778 | 39,3810 | 4 | 0.098 | | | 27 | |
| | 134,5811 | 38,5162 | 8 | 0.094 | | | 27 | |
| | 134,5822 | 48,0975 | 4 | 0.12 | | | 27 | |
| | 134,6077 | 38,5872 | 6 | 0.095 | | | 27 | |
| | 134,6166 | 38,5403 | 8 | 0.092 | | | 27 | |
| | 134,6173 | 1,0575 | 4 | 0.072 | | | 27 | |
| PCG085829+373159 | 134,6250 | 37,5333 | 4 | | 0.0950 | D | 7 | group member/coord.from Ref.(6) |
| PCG085832+382948 | 134,6375 | 38,4967 | 4 | | 0.0900 | A | 7 | group member/coord.from Ref.(6) |
| | 134,6749 | 54,2560 | 4 | 0.085 | | | 27 | |
| | 134,6772 | 38,4405 | 4 | 0.09 | | | 27 | |
| | 134,6918 | 43,4356 | 4 | 0.055 | | | 27 | |
| | 134,6983 | 44,4215 | 4 | 0.088 | | | 27 | |
| | 134,7068 | 54,0119 | 4 | 0.039 | | | 27 | |
| | 134,7083 | 38,9759 | 6 | 0.029 | | | 27 | |
| | 134,7211 | 29,7963 | 4 | 0.049 | | | 27 | |
| | 134,7230 | 49,5978 | 5 | 0.038 | | | 27 | |
| | 134,7711 | 49,7814 | 4 | 0.051 | | | 27 | |
| | 134,7854 | 45,1983 | 7 | 0.086 | | | 27 | |
| | 134,8067 | 37,4549 | 4 | 0.092 | | | 27 | |
| | 134,8109 | 8,0427 | 4 | 0.153 | | | 27 | |
| | 134,8905 | 46,0163 | 5 | 0.089 | | | 27 | |
| | 134,9011 | 39,4400 | 6 | 0.094 | | | 27 | |
| | 134,9029 | 45,1874 | 9 | 0.087 | | | 27 | |
| PCG085943+392602 | 134,9292 | 39,4342 | 4 | | 0.0950 | A | 7 | group member/coord.from Ref.(6) |
| | 134,9896 | 39,3659 | 8 | 0.097 | | | 27 | |
| | 134,9934 | 45,6846 | 6 | 0.052 | | | 27 | |
| | 135,0287 | 39,3357 | 4 | 0.096 | | | 27 | |
| | 135,0760 | 51,4674 | 4 | 0.126 | | | 27 | |
| | 135,0795 | 31,8160 | 4 | 0.065 | | | 27 | |
| | 135,1043 | 51,7384 | 6 | 0.058 | | | 27 | |
| | 135,1334 | 54,9441 | 5 | 0.082 | | | 27 | |
| | 135,1940 | 44,9890 | 4 | 0.037 | | | 27 | |
| | 135,2295 | 58,2325 | 5 | 0.095 | | | 27 | |
| | 135,2457 | 58,2219 | 4 | 0.098 | | | 27 | |
| | 135,3344 | 40,0992 | 4 | 0.096 | | | 27 | |
| | 135,3568 | 2,1401 | 5 | 0.058 | | | 27 | |
| | 135,3573 | 54,9196 | 5 | 0.047 | | | 27 | |
| | 135,3664 | 8,8264 | 5 | 0.084 | | | 27 | |
| | 135,3715 | 9,9019 | 4 | 0.063 | | | 27 | |
| | 135,3980 | 37,3717 | 4 | 0.096 | | | 27 | |
| | 135,4084 | 33,1292 | 9 | 0.064 | | | 27 | |
| | 135,4086 | 2,0068 | 4 | 0.057 | | | 27 | |
| | 135,4820 | 31,0046 | 4 | 0.04 | | | 27 | |
| | 135,4964 | 32,5376 | 7 | 0.054 | | | 27 | |
| | 135,5060 | 38,4073 | 4 | 0.059 | | | 27 | |
| UZCCG084 | 135,5667 | 3,4594 | 3 | 0,027 | | | 5 | |
| | 135,6016 | 52,6065 | 4 | 0.092 | | | 27 | |
| | 135,6277 | 51,9500 | 4 | 0.1 | | | 27 | |
| | 135,6937 | 4,6807 | 4 | 0.058 | | | 27 | |
| | 135,7339 | 1,7554 | 4 | 0.118 | | | 27 | |
| | 135,7686 | 3,3784 | 6 | 0.027 | | | 27 | |
| | 135,8086 | 8,6342 | 4 | 0.084 | | | 27 | |
| UZCCG085 | 135,8500 | 13,6272 | 4 | 0,029 | | | 5 | |
| | 135,8716 | 3,2377 | 4 | 0.099 | | | 27 | |
| | 135,8856 | 40,4257 | 4 | 0.127 | | | 27 | |
| | 135,9169 | 44,6375 | 4 | 0.05 | | | 27 | |
| | 135,9717 | 55,9600 | 6 | 0.082 | | | 27 | |
| | 136,0075 | 36,9762 | 4 | 0.104 | | | 27 | |
| RSCG 28 | 136,0083 | 21,9758 | 3 | 0,010 | | | 9 | |

| | | | | | | | | | |
|---|---|---|---|---|---|---|---|---|---|
| | | | 3 | 0,010 | | | UZCCG086 | 5 | |
| | 136,0442 | 3,9511 | 5 | 0.057 | | | | 27 | |
| | 136,0755 | 55,5246 | 5 | 0.037 | | | | 27 | |
| | 136,0914 | 52,4661 | 4 | 0.083 | | | | 27 | |
| PCG090426+452347 | 136,1125 | 45,3964 | 4 | | 0.1370 | A | | 7 | group member/coord.from Ref.(6) |
| | 136,2583 | 30,6573 | 8 | 0.064 | | | | 27 | |
| UZCCG087 | 136,3000 | 18,3464 | 6 | 0,013 | | | | 5 | |
| | 136,3803 | 9,4626 | 5 | 0.063 | | | | 27 | |
| | 136,3818 | 7,9565 | 5 | 0.063 | | | | 27 | |
| | 136,3832 | 33,0126 | 8 | 0.063 | | | | 27 | |
| PCG090534+470550 | 136,3917 | 47,0972 | 4 | | 0.1740 | C | | 7 | group member/coord.from Ref.(6) |
| | 136,4893 | 50,9783 | 4 | 0.074 | | | | 27 | |
| | 136,5018 | 52,4653 | 5 | 0.058 | | | | 27 | |
| UZCCG088 | 136,5333 | 25,4431 | 3 | 0,009 | | | | 5 | |
| | 136,5617 | 47,8166 | 4 | 0.117 | | | | 27 | |
| | 136,6049 | 41,6648 | 4 | 0.028 | | | | 27 | |
| | 136,6277 | 3,7900 | 5 | 0.057 | | | | 27 | |
| PCG090632+420742 | 136,6375 | 42,1286 | 4 | | 0.1550 | B | | 7 | group member/coord.from Ref.(6) |
| | 136,6469 | 32,6633 | 4 | 0.05 | | | | 27 | |
| | 136,6472 | 57,5874 | 5 | 0.046 | | | | 27 | |
| | 136,6543 | 50,6635 | 5 | 0.037 | | | | 27 | |
| | 136,6884 | 10,3311 | 4 | 0.136 | | | | 27 | |
| | 136,6981 | 49,4584 | 6 | 0.036 | | | | 27 | |
| | 136,7025 | 4,8090 | 6 | 0.125 | | | | 27 | |
| UZCCG089 | 136,7167 | 41,3525 | 3 | 0,026 | | | | 5 | |
| | 136,7676 | 52,4810 | 7 | 0.101 | | | | 27 | |
| | 136,8687 | 4,2248 | 4 | 0.111 | | | | 27 | |
| | 136,8874 | 49,1032 | 4 | 0.082 | | | | 27 | |
| | 136,8927 | 52,0644 | 5 | 0.06 | | | | 27 | |
| | 136,9051 | 37,0729 | 4 | 0.057 | | | | 27 | |
| | 136,9095 | 52,2380 | 4 | 0.057 | | | | 27 | |
| | 136,9498 | 41,3360 | 9 | 0.027 | | | | 27 | |
| | 136,9551 | 1,4918 | 6 | 0.102 | | | | 27 | |
| | 136,9641 | 52,8156 | 4 | 0.099 | | | | 27 | |
| | 136,9928 | 49,7536 | 6 | 0.034 | | | | 27 | |
| | 137,0187 | 37,4471 | 4 | 0.024 | | | | 27 | |
| | 137,0772 | 55,3322 | 6 | 0.046 | | | | 27 | |
| | 137,0901 | 52,3087 | 5 | 0.076 | | | | 27 | |
| | 137,1076 | 59,6696 | 4 | 0.045 | | | | 27 | |
| | 137,1232 | 43,3279 | 4 | 0.042 | | | | 27 | |
| | 137,1363 | 32,1885 | 4 | 0.05 | | | | 27 | |
| | 137,2484 | 47,0977 | 4 | 0.051 | | | | 27 | |
| | 137,2722 | 37,2546 | 4 | 0.03 | | | | 27 | |
| | 137,3061 | 52,6101 | 4 | 0.097 | | | | 27 | |
| HCG 036 | 137,3500 | 15,7956 | 4 | | 0.0127 | A | | 1 | not CG / discordant redshifts |
| | | | | | 0.0211 | B | | 1 | |
| | | | | | 0.0288 | C | | 1 | |
| | | | | | 0.0523 | D | | 1 | |
| | 137,3823 | 35,9310 | 5 | 0.056 | | | | 27 | |
| PCG090938+490321 | 137,4125 | 49,0561 | 4 | | 0.1500 | A | | 7 | group member/coord.from Ref.(6) |
| | 137,4748 | 53,0680 | 4 | 0.1 | | | | 27 | |
| | 137,5084 | 38,7395 | 4 | 0.057 | | | | 27 | |
| RSCG 29 | 137,5250 | 22,8464 | 4 | 0,038 | | | | 9 | |
| | 137,5381 | 48,9810 | 4 | 0.036 | | | | 27 | |
| | 137,5489 | 45,6348 | 5 | 0.027 | | | | 27 | |
| UZCCG090 | 137,6167 | 50,4047 | 3 | 0,017 | | | | 5 | |
| | 137,6236 | 4,4727 | 4 | 0.075 | | | | 27 | |
| | 137,7772 | 38,6140 | 4 | 0.055 | | | | 27 | |
| | 137,7945 | 51,8190 | 4 | 0.098 | | | | 27 | |
| | 137,8154 | 37,1265 | 4 | 0.059 | | | | 27 | |
| | 137,8291 | 7,4499 | 5 | 0.069 | | | | 27 | |
| | 137,8675 | 5,0151 | 4 | 0.125 | | | | 27 | |
| | 137,9892 | 47,6908 | 5 | 0.05 | | | | 27 | |
| | 138,0135 | 56,6488 | 4 | 0.109 | | | | 27 | |
| | 138,0250 | 49,0715 | 4 | 0.081 | | | | 27 | |
| | 138,0698 | 47,7541 | 6 | 0.053 | | | | 27 | |
| | 138,1240 | 0,2705 | 6 | 0.054 | | | | 27 | |
| RSCG 30 | 138,1458 | 34,9850 | 3 | 0,007 | | | | 9 | |
| | | | 4 | 0,007 | | | UZCCG091 | 5 | |
| | 138,1482 | 53,5858 | 7 | 0.101 | | | | 27 | |
| | 138,1609 | 47,7759 | 7 | 0.052 | | | | 27 | |
| | 138,2728 | 4,0851 | 4 | 0.088 | | | | 27 | |
| | 138,3467 | 4,3822 | 4 | 0.07 | | | | 27 | |
| | 138,3524 | 6,4846 | 4 | 0.078 | | | | 27 | |
| | 138,3668 | 7,5516 | 4 | 0.077 | | | | 27 | |
| | 138,3734 | 3,5366 | 4 | 0.1 | | | | 27 | |
| HCG 037 | 138,4000 | 30,0142 | 5 | 0,022 | | | | 1 | |
| | 138,4380 | 2,6987 | 5 | 0.154 | | | | 27 | |
| | 138,4509 | 35,2712 | 4 | 0.061 | | | | 27 | |
| | 138,5020 | 51,5288 | 5 | 0.064 | | | | 27 | |
| | 138,5039 | 51,5665 | 4 | 0.072 | | | | 27 | |
| | 138,5683 | 7,9676 | 5 | 0.031 | | | | 27 | |
| | 138,6988 | 7,1362 | 5 | 0.056 | | | | 27 | |
| | 138,8427 | 34,1240 | 4 | 0.047 | | | | 27 | |
| | 138,9043 | 38,7459 | 4 | 0.069 | | | | 27 | |
| | 138,9322 | 50,0595 | 6 | 0.034 | | | | 27 | |
| | 138,9993 | 39,8988 | 7 | 0.028 | | | | 27 | |
| | 139,0225 | 7,9524 | 4 | 0.112 | | | | 27 | |
| | 139,0232 | 7,3515 | 6 | 0.08 | | | | 27 | |
| | 139,0827 | 41,3206 | 5 | 0.025 | | | | 27 | |
| | 139,2045 | 6,9245 | 6 | 0.081 | | | | 27 | |
| | 139,2082 | 49,7608 | 4 | 0.028 | | | | 27 | |
| UZCCG092 | 139,2625 | 20,1069 | 4 | 0,029 | | | | 5 | |
| | 139,2709 | 50,3576 | 4 | 0.049 | | | | 27 | |
| | 139,3042 | 52,8602 | 4 | 0.039 | | | | 27 | |
| RSCG 31 | 139,3417 | 41,9567 | 3 | 0,006 | | | | 9 | |
| | | | 3 | 0,006 | | | UZCCG093 | 5 | |
| UZCCG094 | 139,3667 | 53,0592 | 3 | 0,008 | | | | 5 | |
| | 139,3811 | 40,3733 | 4 | 0.093 | | | | 27 | |
| | 139,3847 | 6,0561 | 4 | 0.072 | | | | 27 | |
| | 139,4620 | 3,3331 | 4 | 0.057 | | | | 27 | |
| | 139,5220 | 7,3666 | 6 | 0.092 | | | | 27 | |
| | 139,5747 | 33,6512 | 4 | 0.042 | | | | 27 | |
| | 139,7107 | 1,1256 | 7 | 0.087 | | | | 27 | |
| | 139,7168 | 31,8749 | 8 | 0.062 | | | | 27 | |
| | 139,7272 | 1,0986 | 6 | 0.088 | | | | 27 | |
| | 139,7278 | 6,1990 | 5 | 0.072 | | | | 27 | |
| | 139,8654 | 54,8113 | 7 | 0.046 | | | | 27 | |
| RSCG 32 | 139,9375 | 33,7431 | 3 | 0,020 | | | | 9 | |
| | 139,9762 | 36,5994 | 4 | 0.047 | | | | 27 | |
| UZCCG095 | 139,9958 | 0,9714 | 3 | 0,018 | | | | 5 | |
| PCG091959+475900 | 140,0000 | 47,9836 | 4 | | 0.1240 | D | | 7 | group member/coord.from Ref.(6) |
| | 140,0420 | 42,1694 | 4 | 0.041 | | | | 27 | |
| | 140,0423 | 9,2483 | 4 | 0.093 | | | | 27 | |
| | 140,1109 | 40,6599 | 4 | 0.073 | | | | 27 | |
| | 140,1258 | 33,4467 | 4 | 0.025 | | | | 27 | |
| | 140,1510 | 51,2674 | 4 | 0.067 | | | | 27 | |
| | 140,1520 | 54,8182 | 8 | 0.044 | | | | 27 | |
| | 140,2377 | 6,8273 | 5 | 0.078 | | | | 27 | |
| UZCCG096 | 140,3042 | 64,2133 | 4 | 0,005 | | | | 5 | |
| | 140,3167 | 35,2108 | 4 | 0.025 | | | | 27 | |
| | 140,3215 | 6,9698 | 4 | 0.081 | | | | 27 | |
| | 140,3459 | 40,7756 | 9 | 0.074 | | | | 27 | |
| | 140,3865 | 40,8255 | 4 | 0.092 | | | | 27 | |
| | 140,4103 | 51,1018 | 4 | 0.066 | | | | 27 | |
| | 140,4714 | 50,3879 | 4 | 0.085 | | | | 27 | |
| PCG092200+515541 | 140,5000 | 51,9283 | 4 | | 0.2040 | A | | 7 | group member/coord.from Ref.(6) |
| | 140,5318 | 4,1402 | 6 | 0.057 | | | | 27 | |
| | 140,6766 | 50,6381 | 4 | 0.034 | | | | 27 | |
| UZCCG097 | 140,6958 | 40,1714 | 3 | 0,006 | | | | 5 | |
| | 140,7005 | 56,1390 | 6 | 0.049 | | | | 27 | |
| PCG092252+285518 | 140,7208 | 28,9217 | 4 | | 0.0760 | A | | 7 | group member/coord.from Ref.(6) |
| | 140,7853 | 52,9852 | 4 | 0.074 | | | | 27 | |
| | 140,8010 | 57,5776 | 4 | 0.046 | | | | 27 | |
| | 140,9059 | 8,6615 | 4 | 0.127 | | | | 27 | |
| | 141,0210 | 2,4189 | 4 | 0.089 | | | | 27 | |
| | 141,0226 | 54,9016 | 5 | 0.046 | | | | 27 | |
| | 141,0896 | 38,3759 | 4 | 0.027 | | | | 27 | |

| | | | | | | | | |
|---|---|---|---|---|---|---|---|---|
| | 141,1068 | 7,9655 | 4 | 0.105 | | | 27 | |
| | 141,2079 | 3,3678 | 4 | 0.087 | | | 27 | |
| | 141,2105 | 38,2244 | 4 | 0.028 | | | 27 | |
| | 141,2160 | 7,0156 | 7 | 0.08 | | | 27 | |
| | 141,2361 | 3,7215 | 5 | 0.096 | | | 27 | |
| | 141,2377 | 50,1227 | 4 | 0.076 | | | 27 | |
| | 141,3229 | 36,5053 | 4 | 0.112 | | | 27 | |
| UZCCG098 | 141,3708 | 11,4658 | 3 | 0,012 | | | 5 | |
| | 141,4423 | 32,7784 | 4 | 0.052 | | | 27 | |
| | 141,6005 | 52,2183 | 7 | 0.025 | | | 27 | |
| SHK 348 | 141,6583 | 3,4422 | 8 | 0,088 | | | 24 | |
| SHK 348 | 141,6583 | 3,4422 | | 0,090 | | | 11 | mean value |
| | 141,6641 | 3,3423 | 4 | 0.088 | | | 27 | |
| | 141,6763 | 8,7341 | 4 | 0.131 | | | 27 | |
| | 141,7276 | 10,3894 | 5 | 0.079 | | | 27 | |
| | 141,7813 | 1,8557 | 4 | 0.058 | | | 27 | |
| SHK 104 | 141,7958 | 52,9825 | 5 | 0,168 | | | 25 | |
| PCG092726+634929 | 141,8625 | 63,8247 | 4 | | 0.1430 | C | 7 | group member/coord.from Ref.(6) |
| | 141,8663 | 55,1695 | 4 | 0.126 | | | 27 | |
| | 141,8958 | 8,9918 | 4 | 0.076 | | | 27 | |
| | 141,9188 | 2,4139 | 4 | 0.091 | | | 27 | |
| | 141,9376 | 38,9828 | 4 | 0.103 | | | 27 | |
| | 141,9723 | 55,4375 | 4 | 0.046 | | | 27 | |
| | 142,0296 | 7,5348 | 6 | 0.112 | | | 27 | |
| | 142,0454 | 4,1051 | 5 | 0.063 | | | 27 | |
| | 142,0762 | 3,5074 | 4 | 0.088 | | | 27 | |
| | 142,1902 | 0,3732 | 4 | 0.074 | | | 27 | |
| | 142,3524 | 10,7103 | 4 | 0.119 | | | 27 | |
| HCG 039 | 142,3708 | -1,3444 | 4 | 0,070 | | | 1 | |
| | 142,3933 | 55,4259 | 5 | 0.026 | | | 27 | |
| | 142,4907 | 35,0927 | 4 | 0.113 | | | 27 | |
| | 142,6431 | 3,5412 | 4 | 0.101 | | | 27 | |
| | 142,6700 | 7,4612 | 4 | 0.135 | | | 27 | |
| | 142,7755 | 3,8363 | 7 | 0.088 | | | 27 | |
| | 142,7775 | 9,8609 | 6 | 0.076 | | | 27 | |
| | 142,7937 | 39,3549 | 4 | 0.064 | | | 27 | |
| | 142,8247 | 3,8503 | 4 | 0.089 | | | 27 | |
| | 142,8813 | 34,6531 | 4 | 0.113 | | | 27 | |
| | 142,9217 | 0,5395 | 6 | 0.072 | | | 27 | |
| | 142,9357 | 48,1765 | 4 | 0.093 | | | 27 | |
| | 142,9739 | 5,3640 | 4 | 0.076 | | | 27 | |
| UZCCG099 | 142,9833 | 67,6744 | 3 | 0,022 | | | 5 | |
| | 143,0154 | 41,2905 | 4 | 0.028 | | | 27 | |
| | 143,0792 | 9,6262 | 6 | 0.076 | | | 27 | |
| | 143,0894 | 9,1570 | 6 | 0.077 | | | 27 | |
| | 143,0963 | 34,5493 | 7 | 0.028 | | | 27 | |

| | | | | | | | |
|---|---|---|---|---|---|---|---|
| | 143,1088 | 35,4598 | 4 | 0.027 | | 27 | |
| | 143,1378 | 9,5104 | 7 | 0.078 | | 27 | |
| | 143,1392 | 2,5099 | 6 | 0.051 | | 27 | |
| | 143,2685 | 12,0706 | 5 | 0.071 | | 27 | |
| | 143,2743 | 33,8876 | 6 | 0.027 | | 27 | |
| | 143,3481 | 3,8957 | 5 | 0.059 | | 27 | |
| | 143,4146 | 1,6417 | 4 | 0.073 | | 27 | |
| RSCG 33 | 143,4708 | 10,0856 | 3 | 0,011 | | 9 | |
| | | | 4 | 0,010 | UZCCG100 | 5 | |
| | 143,4919 | 61,6859 | 4 | 0.123 | | 27 | |
| | 143,5702 | 60,5993 | 4 | 0.101 | | 27 | |
| | 143,6179 | 9,9470 | 5 | 0.05 | | 27 | |
| | 143,7303 | 54,5310 | 4 | 0.048 | | 27 | |
| | 143,8578 | 3,7300 | 5 | 0.159 | | 27 | |
| | 143,8749 | 10,2886 | 4 | 0.06 | | 27 | |
| | 143,9845 | 34,2978 | 4 | 0.05 | | 27 | |
| | 144,1161 | 53,2336 | 5 | 0.046 | | 27 | |
| | 144,1884 | 10,0107 | 4 | 0.129 | | 27 | |
| | 144,2246 | 54,6176 | 6 | 0.046 | | 27 | |
| | 144,2417 | 10,4290 | 4 | 0.129 | | 27 | |
| | 144,3203 | 41,4939 | 6 | 0.041 | | 27 | |
| | 144,3757 | 4,4411 | 4 | 0.109 | | 27 | |
| | 144,3902 | 10,3374 | 4 | 0.075 | | 27 | |
| | 144,4082 | 48,4907 | 8 | 0.025 | | 27 | |
| UZCCG101 | 144,4208 | 23,1847 | 3 | 0,025 | | 5 | |
| | 144,4867 | 34,9263 | 8 | 0.042 | | 27 | |
| | 144,6690 | 51,9985 | 4 | 0.048 | | 27 | |
| | 144,6701 | 38,9331 | 4 | 0.092 | | 27 | |
| HCG 040 | 144,7292 | -4,8519 | 6 | 0,022 | | 1 | |
| | 144,8152 | 8,9566 | 4 | 0.078 | | 27 | |
| UZCCG102 | 144,8333 | 32,3281 | 3 | 0,022 | | 5 | |
| | 144,9237 | 34,1754 | 4 | 0.081 | | 27 | |
| | 145,0418 | 3,1062 | 4 | 0.094 | | 27 | |
| | 145,0939 | 35,7766 | 4 | 0.02 | | 27 | |
| | 145,0956 | 40,0789 | 4 | 0.041 | | 27 | |
| | 145,2225 | 2,8013 | 7 | 0.046 | | 27 | |
| | 145,2413 | 11,5330 | 7 | 0.022 | | 27 | |
| UZCCG103 | 145,2625 | 21,2467 | 3 | 0,025 | | 5 | |
| SDSS 002 | 145,2708 | 0,4722 | 4 | 0,148 | | 8 | |
| | 145,3484 | 60,6528 | 4 | 0.076 | | 27 | |
| | 145,3516 | 60,7259 | 4 | 0.074 | | 27 | |
| | 145,4073 | 42,5023 | 6 | 0.075 | | 27 | |
| UZCCG104 | 145,4208 | 11,5050 | 3 | 0,019 | | 5 | |
| | 145,4322 | 12,0801 | 4 | 0.023 | | 27 | |
| | 145,4583 | -20,3125 | 5 | 0,715 | | 13 | ass. to MRC B0941-200 quasar |
| | 145,5202 | 61,1772 | 4 | 0.124 | | 27 | |

| | | | | | | | | | |
|---|---|---|---|---|---|---|---|---|---|
| | 145,5232 | 53,2877 | 5 | 0.045 | | | | 27 | |
| | 145,5465 | 5,4107 | 5 | 0.12 | | | | 27 | |
| | 145,6142 | 42,2499 | 4 | 0.072 | | | | 27 | |
| UZCCG105 | 145,6292 | 4,3100 | 3 | 0,029 | | | | 5 | |
| | 145,6621 | 9,7961 | 8 | 0.059 | | | | 27 | |
| | 145,6815 | 55,9936 | 4 | 0.06 | | | | 27 | |
| | 145,7340 | 42,7176 | 4 | 0.074 | | | | 27 | |
| | 145,7650 | 7,6150 | 4 | 0.086 | | | | 27 | |
| | 145,7927 | 39,2892 | 5 | 0.043 | | | | 27 | |
| RSCG 34 | 145,8000 | 31,9131 | 3 | 0,005 | | | | 9 | |
| | | | 3 | 0,005 | | | UZCCG106 | 5 | |
| PCG094316+392308 | 145,8208 | 39,3856 | 5 | | 0.1510 | A | | 7 | group member/coord.from Ref.(6) |
| | 145,8212 | 1,8139 | 6 | 0.1 | | | | 27 | |
| | 145,8445 | 35,3149 | 8 | 0.043 | | | | 27 | |
| | 145,8699 | -0,0804 | 5 | 0.063 | | | | 27 | |
| | 145,8908 | 5,8047 | 4 | 0.091 | | | | 27 | |
| | 145,9098 | 39,3624 | 5 | 0.042 | | | | 27 | |
| | 145,9206 | 41,6960 | 4 | 0.072 | | | | 27 | |
| UZCCG107 | 145,9792 | 42,5264 | 3 | 0,018 | | | | 5 | |
| | 145,9880 | 52,0807 | 4 | 0.046 | | | | 27 | |
| | 146,0847 | 5,3909 | 4 | 0.089 | | | | 27 | |
| | 146,1422 | 36,0633 | 4 | 0.022 | | | | 27 | |
| UZCCG108 | 146,1708 | 0,5239 | 3 | 0,005 | | | | 5 | |
| | 146,2800 | 52,0426 | 4 | 0.047 | | | | 27 | |
| | 146,4485 | 54,4621 | 4 | 0.044 | | | | 27 | |
| | 146,4628 | 42,9366 | 9 | 0.074 | | | | 27 | |
| UZCCG109 | 146,4750 | 9,7019 | 3 | 0,012 | | | | 5 | |
| | 146,4774 | 34,8101 | 7 | 0.041 | | | | 27 | |
| | 146,4777 | 39,2575 | 8 | 0.073 | | | | 27 | |
| | 146,4847 | 60,3785 | 4 | 0.102 | | | | 27 | |
| | 146,4893 | 54,5447 | 5 | 0.049 | | | | 27 | |
| UZCCG110 | 146,5542 | 3,0872 | 3 | 0,020 | | | | 5 | |
| | 146,5764 | -0,8321 | 4 | 0.065 | | | | 27 | |
| | 146,6499 | 34,6369 | 4 | 0.052 | | | | 27 | |
| | 146,6964 | 10,0831 | 5 | 0.056 | | | | 27 | |
| | 146,7003 | 54,4681 | 9 | 0.046 | | | | 27 | |
| | 146,7358 | 43,3945 | 5 | 0.073 | | | | 27 | |
| | 146,7585 | 60,3194 | 4 | 0.094 | | | | 27 | |
| | 146,8182 | 1,7521 | 4 | 0.062 | | | | 27 | |
| | 146,8332 | 6,6735 | 5 | 0.09 | | | | 27 | |
| | 146,8585 | 59,4621 | 4 | 0.075 | | | | 27 | |
| | 146,9334 | 7,5029 | 6 | 0.127 | | | | 27 | |
| | 146,9446 | 42,1446 | 5 | 0.073 | | | | 27 | |
| | 146,9641 | 6,1948 | 4 | 0.09 | | | | 27 | |
| PCG094756+073010 | 146,9875 | 7,5028 | 4 | | 0.1240 | B | | 7 | group member/coord.from Ref.(6) |
| | 147,0186 | 53,8107 | 4 | 0.047 | | | | 27 | |

| | | | | | | | | |
|---|---|---|---|---|---|---|---|---|
| | 147,1407 | 54,3753 | 8 | 0.045 | | | 27 | |
| | 147,1892 | 54,0860 | 7 | 0.048 | | | 27 | |
| | 147,1894 | 37,0962 | 7 | 0.06 | | | 27 | |
| | 147,1940 | 6,1555 | 7 | 0.091 | | | 27 | |
| | 147,2571 | 40,6050 | 5 | 0.082 | | | 27 | |
| PCG094929+481808 | 147,3750 | 48,3025 | 4 | | 0.2030 | A | 7 | group member/coord.from Ref.(6) |
| | 147,4154 | 42,0737 | 4 | 0.072 | | | 27 | |
| UZCCG111 | 147,4500 | 9,0881 | 3 | 0,018 | | | 5 | |
| SDSS 003 | 147,4917 | 0,8825 | 4 | 0,092 | | | 8 | |
| | 147,5489 | 4,2322 | 4 | 0.069 | | | 27 | |
| | 147,5585 | 57,8136 | 4 | 0.083 | | | 27 | |
| | 147,6061 | 38,7596 | 7 | 0.041 | | | 27 | |
| | 147,6665 | -0,4192 | 5 | 0.037 | | | 27 | |
| | 147,6912 | 3,5852 | 4 | 0.089 | | | 27 | |
| | 147,7007 | 2,0301 | 5 | 0.116 | | | 27 | |
| PCG095052+050403 | 147,7167 | 5,0678 | 4 | | 0.1630 | A | 7 | group member/coord.from Ref.(6) |
| | 147,7234 | 51,9014 | 6 | 0.049 | | | 27 | |
| UZCCG112 | 147,7708 | 28,6000 | 3 | 0,005 | | | 5 | |
| | 147,7993 | 4,9008 | 4 | 0.091 | | | 27 | |
| | 147,8161 | 5,4926 | 5 | 0.072 | | | 27 | |
| | 147,8964 | 6,0703 | 6 | 0.092 | | | 27 | |
| | 147,9095 | 35,7368 | 4 | 0.044 | | | 27 | |
| | 147,9673 | 11,9116 | 4 | 0.108 | | | 27 | |
| | 147,9752 | 49,0830 | 4 | 0.051 | | | 27 | |
| | 148,0777 | 52,4138 | 4 | 0.035 | | | 27 | |
| | 148,1409 | 9,7510 | 5 | 0.03 | | | 27 | |
| | 148,2348 | 48,5113 | 5 | 0.115 | | | 27 | |
| | 148,2648 | 8,4609 | 4 | 0.04 | | | 27 | |
| | 148,2819 | 12,3546 | 4 | 0.03 | | | 27 | |
| | 148,3644 | 11,0090 | 4 | 0.053 | | | 27 | |
| | 148,3661 | 0,6491 | 4 | 0.094 | | | 27 | |
| | 148,3722 | 6,3918 | 5 | 0.08 | | | 27 | |
| | 148,4596 | 12,1980 | 7 | 0.128 | | | 27 | |
| PCG095352+571046 | 148,4708 | 57,1797 | 4 | | 0.0810 | B | 7 | group member/coord.from Ref.(6) |
| | 148,4927 | 3,0093 | 4 | 0.071 | | | 27 | |
| | 148,5196 | 10,6465 | 6 | 0.041 | | | 27 | |
| | 148,5504 | 1,9029 | 4 | 0.1 | | | 27 | |
| | 148,5598 | 10,5203 | 6 | 0.04 | | | 27 | |
| | 148,5759 | 57,1706 | 7 | 0.081 | | | 27 | |
| | 148,6353 | 57,3132 | 4 | 0.08 | | | 27 | |
| | 148,6619 | 6,8886 | 4 | 0.074 | | | 27 | |
| | 148,6784 | 48,2451 | 5 | 0.131 | | | 27 | |
| | 148,6972 | 55,1975 | 5 | 0.071 | | | 27 | |
| | 148,7412 | 39,4318 | 4 | 0.048 | | | 27 | |
| | 148,8376 | 37,9867 | 5 | 0.041 | | | 27 | |
| | 148,8469 | 6,8979 | 4 | 0.077 | | | 27 | |
| PCG095527+034508 | 148,8625 | 3,7522 | 4 | | 0.0910 | A | 7 | group member/coord.from Ref.(6) |
| | 148,8714 | -0,0068 | 5 | 0.085 | | | 27 | |
| | 148,9611 | 36,3112 | 6 | 0.052 | | | 27 | |
| | 148,9623 | 0,9661 | 4 | 0.047 | | | 27 | |
| | 148,9722 | 36,1733 | 6 | 0.051 | | | 27 | |
| | 149,0016 | 36,5204 | 8 | 0.05 | | | 27 | |
| SDSS 004 | 149,0375 | 0,3267 | 4 | 0,035 | | | 8 | |
| | 149,0436 | 41,1373 | 4 | 0.069 | | | 27 | |
| | 149,0446 | -0,1921 | 4 | 0.084 | | | 27 | |
| | 149,1310 | 2,1415 | 6 | 0.032 | | | 27 | |
| | 149,1732 | 2,8798 | 4 | 0.092 | | | 27 | |
| | 149,1821 | 36,0603 | 6 | 0.042 | | | 27 | |
| | 149,2029 | -0,5344 | 7 | 0.085 | | | 27 | |
| | 149,2109 | 1,2964 | 4 | 0.064 | | | 27 | |
| | 149,2431 | 12,6271 | 4 | 0.063 | | | 27 | |
| | 149,2506 | -0,4033 | 4 | 0.089 | | | 27 | |
| | 149,2540 | -0,2967 | 5 | 0.087 | | | 27 | |
| SHK 188 | 149,2542 | 26,1736 | 7 | 0,080 | | | 24 | |
| | 149,2645 | 36,8165 | 4 | 0.053 | | | 27 | |
| | 149,2923 | 1,1977 | 5 | 0.062 | | | 27 | |
| | 149,3217 | 44,9557 | 4 | 0.074 | | | 27 | |
| | 149,3293 | 10,1745 | 5 | 0.052 | | | 27 | |
| | 149,3418 | 44,1433 | 4 | 0.074 | | | 27 | |
| | 149,3581 | -0,2823 | 7 | 0.088 | | | 27 | |
| | 149,3784 | 40,9937 | 6 | 0.042 | | | 27 | |
| SDSS 005 | 149,3875 | 0,6678 | 4 | 0,087 | | | 8 | |
| HCG 041 | 149,4167 | 45,2394 | 4 | | 0.0125 | A | 1 | not CG / discordant redshifts |
| | | | | | 0.0242 | B | 1 | |
| | | | | | 0.0324 | C | 1 | |
| | | | | | 0.0148 | D | 1 | |
| | 149,4309 | 37,4171 | 4 | 0.062 | | | 27 | |
| | 149,5182 | 36,7605 | 7 | 0.041 | | | 27 | |
| | 149,5389 | 36,8453 | 4 | 0.051 | | | 27 | |
| | 149,5577 | -0,4099 | 4 | 0.033 | | | 27 | |
| | 149,6754 | 2,3803 | 4 | 0.125 | | | 27 | |
| | 149,7071 | 0,0474 | 4 | 0.073 | | | 27 | |
| | 149,7449 | 6,6750 | 4 | 0.062 | | | 27 | |
| | 149,7909 | 1,0789 | 4 | 0.079 | | | 27 | |
| | 149,7925 | 1,0787 | 6 | 0.082 | | | 27 | |
| | 149,7993 | 2,6817 | 4 | 0.079 | | | 27 | |
| | 149,8118 | 11,6514 | 4 | 0.076 | | | 27 | |
| | 149,8497 | 11,5125 | 5 | 0.075 | | | 27 | |
| | 149,9098 | 53,5973 | 4 | 0.039 | | | 27 | |
| | 149,9111 | 11,5146 | 7 | 0.078 | | | 27 | |
| | 149,9207 | 11,4483 | 4 | 0.08 | | | 27 | |
| | 149,9860 | 11,3737 | 4 | 0.077 | | | 27 | |

| | | | | | | | | | |
|---|---|---|---|---|---|---|---|---|---|
| | 150,0603 | 40,6215 | 6 | 0.068 | | | | 27 | |
| HCG 042 | 150,0917 | -19,6491 | 4 | 0,013 | | | N3091 | 1 | |
| SDSS 006 | 150,1042 | 0,2503 | 5 | 0,220 | | | | 8 | |
| | 150,1778 | 11,8625 | 4 | 0.078 | | | | 27 | |
| | 150,2008 | 4,9291 | 5 | 0.123 | | | | 27 | |
| | 150,2294 | 37,5951 | 4 | 0.051 | | | | 27 | |
| PCG100102-001342 | 150,2625 | 0,2286 | 4 | | 0.0920 | A | | 7 | group member/coord.from Ref.(6) |
| | 150,2634 | 40,5969 | 4 | 0.071 | | | | 27 | |
| | 150,2848 | -0,1016 | 4 | 0.092 | | | | 27 | |
| | 150,3154 | 6,7158 | 5 | 0.094 | | | | 27 | |
| | 150,3160 | 9,5626 | 7 | 0.059 | | | | 27 | |
| SDSS 008 | 150,3625 | 0,3456 | 5 | 0,232 | | | | 8 | |
| | 150,3674 | 0,4128 | 6 | 0.046 | | | | 27 | |
| | 150,3874 | 40,7464 | 4 | 0.09 | | | | 27 | |
| | 150,5022 | 10,8777 | 6 | 0.078 | | | | 27 | |
| | 150,5047 | 9,8788 | 6 | 0.057 | | | | 27 | |
| PCG100237+063626 | 150,6542 | 6,6072 | 4 | | 0.0750 | C | | 7 | group member/coord.from Ref.(6) |
| | 150,7299 | 6,6898 | 4 | 0.09 | | | | 27 | |
| | 150,7397 | 36,9924 | 6 | 0.041 | | | | 27 | |
| | 150,7650 | 54,7724 | 6 | 0.049 | | | | 27 | |
| | 150,7656 | 50,9257 | 5 | 0.05 | | | | 27 | |
| | 150,8299 | 0,6042 | 8 | 0.045 | | | | 27 | |
| | 150,8433 | 55,9679 | 4 | 0.073 | | | | 27 | |
| | 150,8547 | 2,7822 | 4 | 0.103 | | | | 27 | |
| | 150,9092 | 51,6187 | 4 | 0.047 | | | | 27 | |
| | 150,9201 | 1,3260 | 4 | 0.045 | | | | 27 | |
| | 150,9259 | -0,6040 | 4 | 0.095 | | | | 27 | |
| | 150,9605 | 37,9259 | 5 | 0.052 | | | | 27 | |
| | 150,9759 | 0,6078 | 4 | 0.047 | | | | 27 | |
| J1003+1904 | 150,9792 | 19,0817 | 4 | 0,108 | | | | 4 | group coord.from ref.(3) |
| | 150,9960 | 6,6235 | 4 | 0.119 | | | | 27 | |
| SDSS 009 | 151,0292 | 0,3006 | 4 | 0,140 | | | | 8 | |
| | 151,0410 | 0,7066 | 9 | 0.046 | | | | 27 | |
| | 151,0851 | 51,6560 | 4 | 0.049 | | | | 27 | |
| | 151,1621 | 36,8231 | 4 | 0.042 | | | | 27 | |
| | 151,1666 | 2,9865 | 4 | 0.104 | | | | 27 | |
| | 151,2472 | 10,4105 | 4 | 0.056 | | | | 27 | |
| | 151,2727 | 52,8256 | 7 | 0.044 | | | | 27 | |
| | 151,2868 | 44,4732 | 4 | 0.072 | | | | 27 | |
| | 151,2993 | 7,4817 | 5 | 0.062 | | | | 27 | |
| | 151,3500 | 40,7431 | 4 | 0.075 | | | | 27 | |
| | 151,3739 | 53,2286 | 4 | 0.046 | | | | 27 | |
| | 151,3750 | 50,1842 | 6 | 0.05 | | | | 27 | |
| | 151,3885 | 48,7935 | 4 | 0.067 | | | | 27 | |
| | 151,3964 | 48,8184 | 6 | 0.066 | | | | 27 | |
| | 151,5177 | 50,2244 | 7 | 0.051 | | | | 27 | |

| | | | | | | | | |
|---|---|---|---|---|---|---|---|---|
| | 151,5446 | 53,0125 | 5 | 0.043 | | | 27 | |
| | 151,5596 | 11,3296 | 4 | 0.077 | | | 27 | |
| | 151,6163 | 53,8279 | 5 | 0.045 | | | 27 | |
| | 151,6226 | 10,5647 | 4 | 0.043 | | | 27 | |
| | 151,6365 | 1,2121 | 7 | 0.046 | | | 27 | |
| | 151,6693 | 2,9463 | 4 | 0.023 | | | 27 | |
| | 151,6978 | 48,7367 | 5 | 0.065 | | | 27 | |
| | 151,7387 | 39,3602 | 4 | 0.089 | | | 27 | |
| | 151,7409 | 2,6672 | 4 | 0.022 | | | 27 | |
| | 151,7454 | 42,9296 | 4 | 0.072 | | | 27 | |
| LCCG 18 | 151,7542 | -5,6233 | 3 | 0,059 | | | 2 | |
| | 151,7694 | 11,4692 | 4 | 0.078 | | | 27 | |
| | 151,8470 | -0,5786 | 5 | 0.088 | | | 27 | |
| LCCG 19 | 151,8500 | -5,6880 | 3 | 0,084 | | | 2 | |
| | 151,8793 | 37,1996 | 5 | 0.034 | | | 27 | |
| | 151,9004 | 2,2906 | 4 | 0.046 | | | 27 | |
| | 151,9030 | 43,0915 | 5 | 0.073 | | | 27 | |
| | 151,9562 | 9,4452 | 4 | 0.069 | | | 27 | |
| UZCCG113 | 151,9917 | -2,5178 | 3 | 0,021 | | | 5 | |
| | 151,9988 | 0,6777 | 5 | 0.095 | | | 27 | |
| | 152,0207 | 37,8171 | 7 | 0.053 | | | 27 | |
| | 152,0322 | 42,9236 | 7 | 0.074 | | | 27 | |
| | 152,0791 | 55,5556 | 4 | 0.095 | | | 27 | |
| SDSS 011 | 152,0833 | 0,2150 | 4 | 0,094 | | | 8 | |
| | 152,1098 | 39,6392 | 4 | 0.082 | | | 27 | |
| SDSS 013 | 152,1542 | 0,0744 | 4 | 0,184 | | | 8 | |
| PCG100837+171547 | 152,1583 | 17,2633 | 4 | | 0.1210 | B | 7 | group member/coord.from Ref.(6) |
| | 152,2765 | -0,3002 | 5 | 0.07 | | | 27 | |
| | 152,3008 | 62,5002 | 6 | 0.053 | | | 27 | |
| | 152,3889 | -0,3094 | 8 | 0.069 | | | 27 | |
| | 152,4665 | 1,2106 | 4 | 0.045 | | | 27 | |
| | 152,4996 | 11,7536 | 4 | 0.077 | | | 27 | |
| | 152,5342 | 54,4414 | 8 | 0.045 | | | 27 | |
| | 152,5765 | 11,2149 | 4 | 0.078 | | | 27 | |
| | 152,6195 | 11,9518 | 7 | 0.078 | | | 27 | |
| | 152,6716 | 8,6721 | 4 | 0.097 | | | 27 | |
| PCG101053+034611 | 152,7250 | 3,7703 | 4 | | 0.0280 | A | 7 | group member/coord.from Ref.(6) |
| | 152,7368 | 49,9132 | 5 | 0.051 | | | 27 | |
| SDSS 014 | 152,7375 | 0,2531 | 4 | 0,186 | | | 8 | |
| PCG101113+084127 | 152,8042 | 8,6908 | 4 | | 0.0970 | A | 7 | group member/coord.from Ref.(6) |
| HCG 043 | 152,8083 | 0,0317 | 5 | 0,033 | | | 1 | |
| | 152,8229 | 43,2987 | 4 | 0.071 | | | 27 | |
| | 152,8808 | 6,6388 | 4 | 0.104 | | | 27 | |
| | 152,9039 | 38,3041 | 4 | 0.053 | | | 27 | |
| | 152,9241 | 7,9932 | 4 | 0.101 | | | 27 | |
| | 152,9382 | 43,2219 | 4 | 0.072 | | | 27 | |
| | 152,9737 | 41,6505 | 4 | 0.122 | | | 27 | |
| | 153,0170 | 10,4442 | 4 | 0.052 | | | 27 | |
| | 153,0247 | 51,7262 | 6 | 0.042 | | | 27 | |
| | 153,0796 | 8,7295 | 4 | 0.098 | | | 27 | |
| | 153,0811 | 50,0304 | 6 | 0.05 | | | 27 | |
| | 153,1430 | 4,9103 | 6 | 0.032 | | | 27 | |
| | 153,1531 | 59,9991 | 4 | 0.022 | | | 27 | |
| | 153,1703 | 12,4843 | 8 | 0.061 | | | 27 | |
| PCG101241-010609 | 153,1708 | -1,1028 | 4 | | 0.0840 | C | 7 | group member/coord.from Ref.(6) |
| | 153,2090 | 12,1853 | 4 | 0.077 | | | 27 | |
| | 153,2488 | 10,3093 | 8 | 0.07 | | | 27 | |
| | 153,3665 | -0,0205 | 6 | 0.096 | | | 27 | |
| | 153,3960 | 43,4956 | 7 | 0.071 | | | 27 | |
| | 153,4026 | 38,8725 | 6 | 0.022 | | | 27 | |
| | 153,4336 | 38,9253 | 9 | 0.023 | | | 27 | |
| J1013+1945 | 153,4375 | 19,7614 | 4 | 0,112 | | | 4 | group coord.from ref.(3) |
| | 153,4381 | -0,0819 | 5 | 0.092 | | | 27 | |
| | 153,4574 | -0,9119 | 6 | 0.046 | | | 27 | |
| | 153,4636 | 40,6768 | 4 | 0.127 | | | 27 | |
| | 153,4981 | -0,8512 | 6 | 0.042 | | | 27 | |
| | 153,5007 | -0,7729 | 5 | 0.047 | | | 27 | |
| | 153,5400 | -0,8815 | 4 | 0.048 | | | 27 | |
| | 153,5454 | 51,0859 | 4 | 0.068 | | | 27 | |
| | 153,5506 | 45,6260 | 7 | 0.084 | | | 27 | |
| | 153,6269 | -0,8448 | 6 | 0.044 | | | 27 | |
| | 153,6305 | 8,6659 | 6 | 0.069 | | | 27 | |
| | 153,6396 | 4,9180 | 5 | 0.046 | | | 27 | |
| | 153,6540 | 50,8359 | 4 | 0.079 | | | 27 | |
| | 153,7413 | 57,0019 | 4 | 0.124 | | | 27 | |
| | 153,8103 | 51,4973 | 5 | 0.066 | | | 27 | |
| | 153,8587 | 0,5950 | 5 | 0.071 | | | 27 | |
| | 153,8760 | 56,9427 | 7 | 0.046 | | | 27 | |
| SDSS 016 | 153,9167 | 0,8653 | 5 | 0,178 | | | 8 | |
| | 153,9396 | 45,7481 | 4 | 0.068 | | | 27 | |
| | 153,9489 | 1,2934 | 4 | 0.107 | | | 27 | |
| | 153,9547 | 7,0394 | 4 | 0.029 | | | 27 | |
| | 153,9727 | 6,8993 | 8 | 0.028 | | | 27 | |
| | 153,9798 | 46,7667 | 5 | 0.025 | | | 27 | |
| | 154,1146 | 8,9798 | 5 | 0.104 | | | 27 | |
| | 154,1649 | 13,8692 | 6 | 0.083 | | | 27 | |
| | 154,2311 | 60,2194 | 6 | 0.031 | | | 27 | |
| | 154,2574 | 1,0846 | 4 | 0.078 | | | 27 | |
| | 154,2721 | 8,6827 | 5 | 0.104 | | | 27 | |
| | 154,2968 | 6,3462 | 4 | 0.074 | | | 27 | |
| | 154,3116 | 38,1597 | 8 | 0.055 | | | 27 | |
| | 154,3240 | 13,7266 | 4 | 0.081 | | | 27 | |

| | | | | | | |
|---|---|---|---|---|---|---|
| | 154,3315 | 8,7551 | 5 | 0.103 | | 27 |
| | 154,3490 | -0,2496 | 5 | 0.063 | | 27 |
| | 154,3803 | 48,9066 | 5 | 0.062 | | 27 |
| | 154,3877 | 48,9520 | 4 | 0.061 | | 27 |
| | 154,3896 | 6,0546 | 4 | 0.075 | | 27 |
| | 154,4343 | -0,1777 | 5 | 0.065 | | 27 |
| | 154,4470 | 3,3271 | 6 | 0.105 | | 27 |
| HCG 044 | 154,5000 | 21,8122 | 4 | 0,005 | ARP316 | 1 |
| | | | 4 | 0,004 | RSCG 35 | 9 |
| | | | 6 | 0,004 | UZCCG114 | 5 |
| | 154,5334 | 0,0068 | 6 | 0.064 | | 27 |
| | 154,5565 | 49,9260 | 4 | 0.146 | | 27 |
| | 154,5756 | 39,3141 | 4 | 0.054 | | 27 |
| | 154,6123 | 41,0632 | 4 | 0.093 | | 27 |
| | 154,6437 | 57,0592 | 4 | 0.125 | | 27 |
| | 154,6642 | 10,1113 | 4 | 0.077 | | 27 |
| | 154,6989 | 0,3283 | 5 | 0.063 | | 27 |
| | 154,6991 | 46,5143 | 6 | 0.03 | | 27 |
| | 154,7405 | 1,9496 | 4 | 0.046 | | 27 |
| | 154,7435 | 50,5206 | 4 | 0.056 | | 27 |
| HCG 045 | 154,7958 | 59,1097 | 4 | 0,073 | U5564 | 1 |
| | 154,8802 | 40,9422 | 4 | 0.092 | | 27 |
| | 154,9162 | 37,4111 | 4 | 0.048 | | 27 |
| | 154,9293 | 2,8841 | 6 | 0.046 | | 27 |
| | 154,9330 | -0,7071 | 6 | 0.093 | | 27 |
| | 154,9672 | 48,1625 | 4 | 0.051 | | 27 |
| | 154,9776 | 7,7851 | 4 | 0.045 | | 27 |
| | 154,9953 | 42,8004 | 4 | 0.085 | | 27 |
| LCCG 20 | 155,0042 | -3,8778 | 3 | 0,071 | | 2 |
| | 155,0255 | 41,0029 | 7 | 0.095 | | 27 |
| | 155,0303 | 41,0013 | 5 | 0.09 | | 27 |
| | 155,0581 | 0,4637 | 6 | 0.096 | | 27 |
| | 155,0713 | 64,0070 | 5 | 0.041 | | 27 |
| | 155,1076 | 55,5226 | 5 | 0.068 | | 27 |
| | 155,1580 | 46,8711 | 8 | 0.063 | | 27 |
| | 155,1642 | 45,0295 | 4 | 0.069 | | 27 |
| | 155,1813 | 4,4189 | 6 | 0.033 | | 27 |
| UZCCG115 | 155,1958 | 42,9928 | 3 | 0,023 | | 5 |
| | 155,2460 | 39,5801 | 5 | 0.119 | | 27 |
| | 155,2514 | 13,7106 | 4 | 0.143 | | 27 |
| | 155,2713 | -0,8081 | 4 | 0.053 | | 27 |
| | 155,2842 | 45,7223 | 4 | 0.087 | | 27 |
| | 155,2920 | 6,7527 | 7 | 0.111 | | 27 |
| | 155,4050 | 52,1859 | 6 | 0.069 | | 27 |
| | 155,4220 | 57,2592 | 8 | 0.026 | | 27 |
| | 155,4373 | 38,5616 | 8 | 0.052 | | 27 |

| | | | | | | | | |
|---|---|---|---|---|---|---|---|---|
| | 155,4408 | 48,0263 | 4 | 0.041 | | | 27 | |
| | 155,4832 | 38,5192 | 6 | 0.057 | | | 27 | |
| | 155,4928 | 38,6002 | 4 | 0.051 | | | 27 | |
| HCG 046 | 155,5083 | 17,8150 | 4 | 0,027 | | | 1 | |
| | 155,5286 | 8,2546 | 4 | 0.103 | | | 27 | |
| | 155,5449 | 1,9040 | 4 | 0.073 | | | 27 | |
| | 155,5513 | 58,3545 | 4 | 0.044 | | | 27 | |
| | 155,5613 | 38,4329 | 4 | 0.054 | | | 27 | |
| | 155,5740 | 13,6703 | 4 | 0.085 | | | 27 | |
| | 155,5765 | 2,2896 | 8 | 0.071 | | | 27 | |
| | 155,5969 | 38,4440 | 9 | 0.053 | | | 27 | |
| | 155,6158 | 5,8087 | 5 | 0.029 | | | 27 | |
| | 155,6225 | 13,0184 | 8 | 0.032 | | | 27 | |
| | 155,6501 | 13,7997 | 6 | 0.044 | | | 27 | |
| | 155,6772 | 7,5741 | 4 | 0.121 | | | 27 | |
| | 155,6919 | 6,9833 | 9 | 0.033 | | | 27 | |
| | 155,6922 | 12,3433 | 4 | 0.047 | | | 27 | |
| | 155,7468 | 53,9112 | 6 | 0.046 | | | 27 | |
| | 155,7482 | 37,6964 | 4 | 0.054 | | | 27 | |
| | 155,7516 | 6,9234 | 4 | 0.111 | | | 27 | |
| | 155,7571 | 9,8692 | 5 | 0.05 | | | 27 | |
| LCCG 21 | 155,8083 | -5,8766 | 3 | 0,089 | | | 2 | |
| | 155,8317 | 7,6854 | 5 | 0.045 | | | 27 | |
| | 155,8492 | 12,4200 | 8 | 0.045 | | | 27 | |
| | 155,9130 | 64,5353 | 5 | 0.119 | | | 27 | |
| | 155,9748 | 64,8466 | 4 | 0.04 | | | 27 | |
| | 155,9893 | 8,6637 | 4 | 0.106 | | | 27 | |
| | 156,0075 | 42,0703 | 4 | 0.081 | | | 27 | |
| | 156,0157 | 53,9922 | 8 | 0.032 | | | 27 | |
| | 156,0281 | 41,6775 | 5 | 0.068 | | | 27 | |
| | 156,0332 | 1,4947 | 4 | 0.096 | | | 27 | |
| | 156,0336 | 41,7901 | 4 | 0.046 | | | 27 | |
| | 156,0346 | 12,8409 | 7 | 0.031 | | | 27 | |
| | 156,0370 | 1,3037 | 4 | 0.098 | | | 27 | |
| | 156,0533 | 59,8431 | 4 | 0.073 | | | 27 | |
| | 156,0726 | 41,7540 | 8 | 0.043 | | | 27 | |
| | 156,1108 | 2,9399 | 4 | 0.073 | | | 27 | |
| | 156,1116 | 6,4595 | 6 | 0.044 | | | 27 | |
| | 156,1188 | 47,2841 | 4 | 0.062 | | | 27 | |
| | 156,1461 | 41,5095 | 4 | 0.045 | | | 27 | |
| | 156,1967 | 40,0944 | 5 | 0.067 | | | 27 | |
| | 156,2372 | 47,8287 | 4 | 0.061 | | | 27 | |
| | 156,2441 | 38,0782 | 4 | 0.127 | | | 27 | |
| | 156,2453 | 61,1979 | 4 | 0.074 | | | 27 | |
| PCG102512+091835 | 156,3000 | 9,3100 | 4 | | 0.1420 | B | 7 | group member/coord.from Ref.(6) |
| | 156,3198 | 8,6646 | 6 | 0.073 | | | 27 | |

| | | | | | | |
|---|---|---|---|---|---|---|
| | 156,3380 | 9,8288 | 6 | 0.046 | | 27 |
| | 156,3690 | 47,7264 | 4 | 0.064 | | 27 |
| | 156,3769 | 11,3319 | 4 | 0.062 | | 27 |
| SDSS 018 | 156,4083 | 0,5617 | 5 | 0,169 | | 8 |
| | 156,4110 | 6,5119 | 4 | 0.043 | | 27 |
| | 156,4138 | 38,3161 | 4 | 0.054 | | 27 |
| | 156,4403 | 40,2116 | 4 | 0.064 | | 27 |
| | 156,4404 | 45,3314 | 4 | 0.044 | | 27 |
| HCG 047 | 156,4500 | 13,7317 | 4 | 0,032 | U5644 | 1 |
| | 156,4888 | 13,7976 | 7 | 0.032 | | 27 |
| | 156,4918 | 41,0183 | 5 | 0.076 | | 27 |
| | 156,4929 | 46,9242 | 9 | 0.06 | | 27 |
| | 156,4989 | 58,8684 | 4 | 0.031 | | 27 |
| SDSS 020 | 156,5083 | 0,4914 | 5 | 0,172 | | 8 |
| SDSS 021 | 156,5542 | 0,0381 | 4 | 0,104 | | 8 |
| | 156,5947 | 0,4816 | 4 | 0.098 | | 27 |
| | 156,6330 | 5,3979 | 4 | 0.111 | | 27 |
| SDSS 022 | 156,6542 | 0,7308 | 4 | 0,106 | | 8 |
| | 156,6651 | 37,9319 | 5 | 0.055 | | 27 |
| | 156,7290 | 9,6539 | 4 | 0.092 | | 27 |
| | 156,7475 | 47,0118 | 6 | 0.059 | | 27 |
| UZCCG116 | 156,7708 | 16,0633 | 3 | 0,033 | | 5 |
| | 156,7783 | 9,4393 | 4 | 0.046 | | 27 |
| | 156,8067 | 8,5895 | 4 | 0.073 | | 27 |
| | 156,8178 | 10,9250 | 4 | 0.031 | | 27 |
| | 156,8213 | 53,2023 | 4 | 0.064 | | 27 |
| | 156,8354 | 45,6109 | 4 | 0.059 | | 27 |
| | 156,8591 | 46,2657 | 5 | 0.06 | | 27 |
| | 156,8691 | 55,5220 | 5 | 0.13 | | 27 |
| | 156,9070 | 10,8211 | 4 | 0.109 | | 27 |
| | 156,9245 | 10,5683 | 6 | 0.108 | | 27 |
| | 156,9362 | 10,5270 | 7 | 0.111 | | 27 |
| | 156,9506 | 40,7544 | 6 | 0.076 | | 27 |
| | 156,9831 | 1,5000 | 4 | 0.022 | | 27 |
| | 156,9838 | 4,1720 | 4 | 0.066 | | 27 |
| | 156,9964 | 10,5760 | 5 | 0.11 | | 27 |
| | 157,0052 | 43,6364 | 6 | 0.052 | | 27 |
| | 157,0280 | 37,8412 | 5 | 0.107 | | 27 |
| | 157,0644 | 10,1794 | 7 | 0.066 | | 27 |
| | 157,0679 | 8,5453 | 7 | 0.047 | | 27 |
| | 157,0830 | 47,3155 | 6 | 0.062 | | 27 |
| | 157,1301 | 9,7018 | 5 | 0.105 | | 27 |
| | 157,1361 | 9,6985 | 4 | 0.106 | | 27 |
| | 157,1407 | 53,8933 | 4 | 0.046 | | 27 |
| | 157,1454 | 3,8733 | 8 | 0.074 | | 27 |
| | 157,1631 | 40,7848 | 4 | 0.077 | | 27 |

| | | | | | |
|---|---|---|---|---|---|
| | 157,1857 | 40,2799 | 5 | 0.067 | 27 |
| | 157,1940 | 3,6162 | 8 | 0.073 | 27 |
| | 157,1947 | 51,5275 | 6 | 0.045 | 27 |
| | 157,2016 | 8,7176 | 5 | 0.048 | 27 |
| LCCG 22 | 157,2250 | -3,2400 | 3 | 0,036 | 2 |
| | 157,2265 | 3,7043 | 5 | 0.072 | 27 |
| | 157,2759 | 3,1386 | 8 | 0.073 | 27 |
| | 157,3047 | 51,6920 | 5 | 0.046 | 27 |
| | 157,3314 | 3,1211 | 5 | 0.074 | 27 |
| | 157,4027 | 47,2279 | 5 | 0.052 | 27 |
| | 157,4647 | 7,5491 | 7 | 0.035 | 27 |
| | 157,4855 | 41,1085 | 4 | 0.091 | 27 |
| | 157,5165 | 4,0303 | 5 | 0.067 | 27 |
| | 157,5218 | 1,0024 | 8 | 0.089 | 27 |
| | 157,5347 | 47,1501 | 4 | 0.058 | 27 |
| LCCG 23 | 157,5417 | -3,1678 | 3 | 0,037 | 2 |
| | 157,5491 | 54,3330 | 4 | 0.065 | 27 |
| | 157,5503 | 3,9897 | 5 | 0.068 | 27 |
| | 157,6041 | 7,1310 | 4 | 0.036 | 27 |
| | 157,6131 | 12,4086 | 4 | 0.109 | 27 |
| | 157,6160 | 4,0789 | 8 | 0.066 | 27 |
| | 157,6821 | 13,4262 | 4 | 0.118 | 27 |
| | 157,6896 | 54,6223 | 4 | 0.134 | 27 |
| | 157,7376 | 54,9712 | 4 | 0.065 | 27 |
| | 157,7591 | 0,4842 | 7 | 0.029 | 27 |
| | 157,7618 | 43,8818 | 9 | 0.028 | 27 |
| | 157,7720 | 11,1727 | 4 | 0.065 | 27 |
| | 157,7896 | 11,1238 | 4 | 0.066 | 27 |
| | 157,8138 | 7,5692 | 4 | 0.106 | 27 |
| | 157,8275 | 11,0681 | 6 | 0.064 | 27 |
| | 157,8564 | 38,5551 | 4 | 0.097 | 27 |
| | 157,8724 | 40,1579 | 6 | 0.069 | 27 |
| | 157,9195 | 7,9775 | 4 | 0.073 | 27 |
| | 157,9294 | 40,0946 | 5 | 0.07 | 27 |
| | 157,9498 | 58,1770 | 6 | 0.093 | 27 |
| | 157,9536 | 53,2742 | 4 | 0.135 | 27 |
| | 157,9851 | 40,2508 | 4 | 0.077 | 27 |
| | 157,9943 | 40,1517 | 8 | 0.068 | 27 |
| | 158,0117 | 40,1957 | 6 | 0.066 | 27 |
| | 158,0350 | 3,5180 | 5 | 0.068 | 27 |
| | 158,1087 | 53,3587 | 5 | 0.133 | 27 |
| | 158,1099 | 40,2837 | 4 | 0.075 | 27 |
| | 158,1226 | 40,2728 | 4 | 0.076 | 27 |
| | 158,1473 | 40,1544 | 5 | 0.065 | 27 |
| | 158,1859 | 40,2802 | 6 | 0.079 | 27 |
| | 158,1899 | 52,9556 | 9 | 0.064 | 27 |

| | | | | | | | | |
|---|---|---|---|---|---|---|---|---|
| | 158,1908 | 53,1063 | 5 | 0.137 | | | 27 | |
| | 158,2102 | 43,2334 | 4 | 0.158 | | | 27 | |
| | 158,2608 | 11,9396 | 7 | 0.033 | | | 27 | |
| | 158,2810 | 53,1174 | 4 | 0.063 | | | 27 | |
| PCG103308+090210 | 158,2833 | 9,0361 | 4 | | 0.2270 | B | 7 | group member/coord.from Ref.(6) |
| | 158,3403 | 56,8936 | 8 | 0.046 | | | 27 | |
| | 158,3492 | 56,5813 | 4 | 0.047 | | | 27 | |
| | 158,3528 | 47,1294 | 5 | 0.13 | | | 27 | |
| | 158,3559 | 48,9225 | 4 | 0.045 | | | 27 | |
| | 158,3858 | 11,9300 | 5 | 0.034 | | | 27 | |
| | 158,3960 | 52,8933 | 5 | 0.065 | | | 27 | |
| | 158,3978 | 48,3880 | 4 | 0.063 | | | 27 | |
| | 158,4144 | 61,5594 | 5 | 0.115 | | | 27 | |
| | 158,4466 | 47,2593 | 6 | 0.063 | | | 27 | |
| | 158,4541 | 60,4789 | 4 | 0.09 | | | 27 | |
| J1033+2253 | 158,4542 | 22,8900 | 4 | | 0.0630 | A | 4 | not CG / discordant redshifts |
| | | | | | 0.1099 | B | 4 | |
| | | | | | 0.1115 | C | 4 | |
| | | | | | 0.0629 | D | 4 | |
| SDSS 024 | 158,5000 | 0,4092 | 4 | 0,150 | | | 8 | |
| | 158,5364 | 44,7271 | 5 | 0.025 | | | 27 | |
| | 158,5575 | 7,5942 | 7 | 0.067 | | | 27 | |
| | 158,6150 | 4,7601 | 4 | 0.1 | | | 27 | |
| UZCCG117 | 158,6583 | 64,6856 | 3 | 0,006 | | | 5 | |
| | 158,6797 | 3,7307 | 4 | 0.085 | | | 27 | |
| | 158,6992 | 10,8570 | 5 | 0.051 | | | 27 | |
| | 158,7484 | 63,0893 | 4 | 0.117 | | | 27 | |
| | 158,7495 | 2,9269 | 4 | 0.072 | | | 27 | |
| UZCCG118 | 158,7500 | 44,6550 | 3 | 0,025 | | | 5 | |
| | 158,7675 | 44,8823 | 4 | 0.026 | | | 27 | |
| | 158,8614 | 57,1227 | 7 | 0.047 | | | 27 | |
| | 158,8750 | 54,6459 | 5 | 0.046 | | | 27 | |
| | 158,8896 | 42,6300 | 4 | 0.087 | | | 27 | |
| | 158,9098 | 39,4834 | 5 | 0.043 | | | 27 | |
| | 159,0354 | 44,7535 | 5 | 0.091 | | | 27 | |
| | 159,0826 | 4,3881 | 4 | 0.069 | | | 27 | |
| | 159,1024 | 50,0409 | 6 | 0.044 | | | 27 | |
| | 159,1871 | 44,8781 | 4 | 0.125 | | | 27 | |
| | 159,2435 | 50,1112 | 6 | 0.046 | | | 27 | |
| | 159,2663 | 2,1685 | 7 | 0.095 | | | 27 | |
| | 159,3057 | 39,8361 | 4 | 0.068 | | | 27 | |
| | 159,3190 | 10,9505 | 4 | 0.053 | | | 27 | |
| | 159,3307 | 11,2961 | 4 | 0.093 | | | 27 | |
| | 159,3413 | 45,3061 | 5 | 0.049 | | | 27 | |
| SDSS 026 | 159,3458 | 0,5156 | 4 | 0,114 | | | 8 | |
| | 159,3874 | 8,6951 | 6 | 0.067 | | | 27 | |
| | 159,3902 | 58,9732 | 4 | 0.028 | | | 27 | |
| | 159,4077 | 44,0629 | 6 | 0.126 | | | 27 | |
| | 159,4415 | 13,8965 | 4 | 0.066 | | | 27 | |
| | 159,4431 | 50,0838 | 8 | 0.045 | | | 27 | |
| | 159,5346 | 47,0173 | 5 | 0.063 | | | 27 | |
| | 159,5406 | 12,1052 | 4 | 0.094 | | | 27 | |
| | 159,5496 | 1,0149 | 4 | 0.059 | | | 27 | |
| | 159,5540 | 40,1127 | 5 | 0.093 | | | 27 | |
| | 159,5548 | 3,0443 | 6 | 0.076 | | | 27 | |
| | 159,6021 | 6,7433 | 4 | 0.067 | | | 27 | |
| | 159,6353 | 53,0728 | 4 | 0.065 | | | 27 | |
| | 159,6542 | 8,2046 | 7 | 0.067 | | | 27 | |
| | 159,6668 | 44,6092 | 4 | 0.049 | | | 27 | |
| | 159,6867 | 1,7999 | 4 | 0.056 | | | 27 | |
| | 159,6878 | 42,0481 | 4 | 0.091 | | | 27 | |
| | 159,7139 | 5,2501 | 8 | 0.067 | | | 27 | |
| | 159,7179 | 5,6425 | 5 | 0.028 | | | 27 | |
| | 159,7186 | 12,0650 | 5 | 0.092 | | | 27 | |
| | 159,7530 | 4,5414 | 8 | 0.067 | | | 27 | |
| | 159,7559 | 2,2823 | 4 | 0.083 | | | 27 | |
| PCG103901+051000 | 159,7583 | 5,1669 | 4 | | 0.0680 | A | 7 | group member/coord.from Ref.(6) |
| | 159,7648 | 12,6074 | 4 | 0.094 | | | 27 | |
| | 159,7707 | 4,6790 | 9 | 0.069 | | | 27 | |
| | 159,7887 | 40,0114 | 4 | 0.082 | | | 27 | |
| | 159,8316 | 12,0197 | 4 | 0.095 | | | 27 | |
| | 159,8434 | 5,2082 | 4 | 0.064 | | | 27 | |
| | 159,8464 | 5,1860 | 5 | 0.072 | | | 27 | |
| | 159,8468 | 2,3420 | 4 | 0.084 | | | 27 | |
| | 159,8654 | 5,0740 | 8 | 0.073 | | | 27 | |
| | 159,8965 | 4,5102 | 4 | 0.126 | | | 27 | |
| | 159,8990 | 2,1822 | 4 | 0.058 | | | 27 | |
| | 159,9034 | 5,1080 | 8 | 0.071 | | | 27 | |
| | 159,9501 | 5,3464 | 4 | 0.07 | | | 27 | |
| | 159,9757 | -0,6649 | 4 | 0.096 | | | 27 | |
| | 159,9944 | 13,5550 | 6 | 0.032 | | | 27 | |
| J1039+2749 | 159,9958 | 27,8297 | 4 | 0,099 | | | 4 | group coord.from ref.(3) |
| | 160,0069 | 11,0772 | 4 | 0.053 | | | 27 | |
| | 160,0320 | 2,9316 | 6 | 0.076 | | | 27 | |
| | 160,0441 | 58,2996 | 6 | 0.072 | | | 27 | |
| | 160,0618 | 43,7639 | 4 | 0.054 | | | 27 | |
| | 160,0973 | 6,7125 | 4 | 0.03 | | | 27 | |
| | 160,1000 | 8,7113 | 5 | 0.055 | | | 27 | |
| | 160,1073 | 45,5987 | 4 | 0.054 | | | 27 | |
| UZCCG119 | 160,1083 | 47,7039 | 3 | 0,005 | | | 5 | |
| | 160,1294 | 5,4893 | 5 | 0.027 | | | 27 | |
| | 160,1437 | 6,3522 | 4 | 0.034 | | | 27 | |
| | 160,1805 | 62,4040 | 7 | 0.052 | | | 27 | |
| | 160,1924 | 58,2271 | 8 | 0.073 | | | 27 | |
| SDSS 027 | 160,2500 | 0,9558 | 5 | 0,087 | | | 8 | |
| | 160,3346 | 0,8648 | 4 | 0.065 | | | 27 | |
| | 160,3813 | 40,1090 | 4 | 0.043 | | | 27 | |
| | 160,4180 | 1,3891 | 8 | 0.073 | | | 27 | |
| | 160,4566 | 3,4992 | 4 | 0.083 | | | 27 | |
| | 160,5179 | 2,6071 | 7 | 0.077 | | | 27 | |
| | 160,5234 | 3,7747 | 5 | 0.056 | | | 27 | |
| | 160,5309 | -0,2292 | 4 | 0.114 | | | 27 | |
| | 160,5499 | 42,1833 | 4 | 0.066 | | | 27 | |
| PCG104215+035811 | 160,5667 | 3,9700 | 4 | | 0.0980 | A | 7 | group member/coord.from Ref.(6) |
| | 160,6019 | 12,5638 | 7 | 0.064 | | | 27 | |
| | 160,6167 | 58,6999 | 7 | 0.031 | | | 27 | |
| | 160,6220 | 61,8695 | 5 | 0.132 | | | 27 | |
| | 160,6473 | -0,9559 | 4 | 0.026 | | | 27 | |
| | 160,6824 | 8,6318 | 4 | 0.046 | | | 27 | |
| | 160,6863 | 40,4401 | 5 | 0.055 | | | 27 | |
| | 160,7092 | 41,0357 | 5 | 0.074 | | | 27 | |
| | 160,7105 | 52,2034 | 4 | 0.025 | | | 27 | |
| | 160,8089 | 52,4959 | 5 | 0.065 | | | 27 | |
| | 160,8251 | 43,1156 | 5 | 0.104 | | | 27 | |
| | 160,8374 | 40,8670 | 5 | 0.03 | | | 27 | |
| | 160,8651 | 0,5768 | 4 | 0.124 | | | 27 | |
| | 160,8981 | -0,6754 | 6 | 0.039 | | | 27 | |
| J1043+2817 | 160,9083 | 28,2864 | 4 | | 0.0778 | A | 4 | not CG / discordant redshifts |
| | | | | | 0.0844 | B | 4 | |
| | | | | | 0.2454 | C | 4 | |
| | | | | | 0.2452 | D | 4 | |
| | 160,9178 | 3,1382 | 4 | 0.081 | | | 27 | |
| | 160,9247 | -0,3166 | 4 | 0.075 | | | 27 | |
| | 160,9268 | 10,6429 | 5 | 0.054 | | | 27 | |
| | 160,9594 | 54,8041 | 4 | 0.127 | | | 27 | |
| | 160,9814 | 60,7795 | 4 | 0.114 | | | 27 | |
| | 160,9831 | 11,0279 | 5 | 0.055 | | | 27 | |
| | 161,0064 | 3,7754 | 4 | 0.073 | | | 27 | |
| | 161,0557 | 52,0081 | 4 | 0.063 | | | 27 | |
| | 161,0703 | 60,7180 | 4 | 0.115 | | | 27 | |
| PCG104418+024814 | 161,0792 | 2,8039 | 4 | | 0.0690 | A | 7 | group member/coord.from Ref.(6) |
| | 161,0911 | 7,6502 | 4 | 0.069 | | | 27 | |
| | 161,1522 | 39,7212 | 4 | 0.069 | | | 27 | |
| PCG104450+353601 | 161,2083 | 35,6006 | 4 | | 0.0510 | A | 7 | group member/coord.from Ref.(6) |
| | 161,2212 | 5,0247 | 4 | 0.069 | | | 27 | |
| | 161,2603 | 40,5584 | 6 | 0.07 | | | 27 | |
| | 161,2659 | 2,1549 | 4 | 0.106 | | | 27 | |
| | 161,3514 | 53,9650 | 4 | 0.072 | | | 27 | |
| PCG104527+493118 | 161,3625 | 49,5219 | 4 | | 0.1680 | C | 7 | group member/coord.from Ref.(6) |
| J1045+1758 | 161,4083 | 17,9739 | 4 | | 0.0723 | A | 4 | not CG / discordant redshifts |
| | | | | | 0.1059 | B | 4 | |
| | | | | | 0.1056 | C | 4 | |
| | | | | | 0.1303 | D | 4 | |
| LCCG 24 | 161,5167 | -2,5755 | 3 | 0,108 | | | 2 | |
| | 161,5293 | 42,9142 | 5 | 0.074 | | | 27 | |
| UZCCG120 | 161,5583 | 25,9289 | 3 | 0,021 | | | 5 | |
| | 161,5924 | 7,5758 | 4 | 0.057 | | | 27 | |
| | 161,6449 | 1,4211 | 4 | 0.107 | | | 27 | |
| LCCG 25 | 161,6458 | -11,6305 | 3 | 0,054 | | | 2 | |
| | 161,7227 | -0,2001 | 4 | 0.039 | | | 27 | |
| | 161,7256 | 39,0239 | 4 | 0.037 | | | 27 | |
| PCG104724+051125 | 161,8500 | 5,1903 | 4 | | 0.0700 | B | 7 | group member/coord.from Ref.(6) |
| | 161,8664 | 59,4958 | 6 | 0.028 | | | 27 | |
| | 161,8699 | 12,0695 | 4 | 0.033 | | | 27 | |
| | 161,8807 | 57,8425 | 5 | 0.073 | | | 27 | |
| | 161,8988 | 2,0983 | 4 | 0.062 | | | 27 | |
| | 161,8995 | 45,5683 | 5 | 0.05 | | | 27 | |
| | 161,9545 | 41,5814 | 5 | 0.063 | | | 27 | |
| | 162,0225 | 43,6841 | 5 | 0.086 | | | 27 | |
| RSCG 36 | 162,0333 | 12,5611 | 3 | 0,003 | | | 9 | |
| RSCG 37 | 162,0417 | 14,1022 | 3 | 0,002 | | | 9 | |
| SHK 191 | 162,0458 | 31,5025 | 25 | | 0.1176 | 2 | 11 | group member/coord.from Ref.(10) |
| | 162,0458 | 31,5025 | 12 | 0,115 | | | 23 | mean value from NED data |
| | 162,0632 | 4,8912 | 6 | 0.026 | | | 27 | |
| | 162,1248 | 12,5215 | 5 | 0.054 | | | 27 | |
| | 162,1376 | 4,9212 | 5 | 0.072 | | | 27 | |
| | 162,1644 | 7,6728 | 8 | 0.053 | | | 27 | |
| | 162,2327 | 57,8449 | 4 | 0.074 | | | 27 | |
| | 162,3188 | 56,7232 | 7 | 0.046 | | | 27 | |
| SDSS 028 | 162,3333 | 0,2364 | 4 | 0,125 | | | 8 | |
| | 162,3339 | 11,0437 | 4 | 0.078 | | | 27 | |
| | 162,4223 | 0,3778 | 7 | 0.039 | | | 27 | |
| | 162,4768 | 59,5641 | 5 | 0.116 | | | 27 | |
| | 162,5900 | -0,8557 | 5 | 0.038 | | | 27 | |
| | 162,6021 | 6,7022 | 4 | 0.111 | | | 27 | |
| | 162,6255 | 11,5428 | 4 | 0.137 | | | 27 | |
| SDSS 029 | 162,6500 | 0,3681 | 6 | 0,150 | | | 8 | |
| UZCCG121 | 162,6500 | 8,7550 | 3 | 0,021 | | | 5 | |
| | 162,7531 | 0,3387 | 5 | 0.038 | | | 27 | |
| | 162,8127 | 47,9256 | 4 | 0.028 | | | 27 | |
| | 162,8526 | 50,9150 | 5 | 0.026 | | | 27 | |
| | 162,8941 | 2,5403 | 5 | 0.051 | | | 27 | |
| | 162,9352 | 55,6113 | 5 | 0.072 | | | 27 | |
| | 162,9445 | 66,7133 | 5 | 0.042 | | | 27 | |

| | | | | | | |
|---|---|---|---|---|---|---|
| RSCG 38 | 162,9458 | 32,8586 | 3 | 0,005 | | 9 |
| | | | 6 | 0,005 | UZCCG122 | 5 |
| | 162,9502 | 3,3117 | 4 | 0.041 | | 27 |
| | 163,0380 | 57,6695 | 6 | 0.073 | | 27 |
| | 163,1921 | 2,6941 | 5 | 0.036 | | 27 |
| SHK 282 | 163,2208 | -11,0078 | 11 | 0,142 | | 20 |
| | 163,2771 | 47,6950 | 4 | 0.119 | | 27 |
| UZCCG123 | 163,2833 | 34,1422 | 3 | 0,006 | | 5 |
| | 163,3373 | 56,2955 | 4 | 0.077 | | 27 |
| | 163,3657 | 55,5110 | 4 | 0.047 | | 27 |
| | 163,4156 | 54,8967 | 6 | 0.073 | | 27 |
| UZCCG124 | 163,4500 | 17,3603 | 9 | 0,004 | | 5 |
| | 163,4743 | 44,4238 | 4 | 0.049 | | 27 |
| | 163,4769 | 54,7845 | 4 | 0.069 | | 27 |
| | 163,4843 | 56,6009 | 5 | 0.076 | | 27 |
| | 163,5243 | 12,5640 | 5 | 0.086 | | 27 |
| | 163,5570 | 54,8518 | 8 | 0.071 | | 27 |
| | 163,5820 | 48,2502 | 7 | 0.044 | | 27 |
| | 163,6386 | 55,6683 | 5 | 0.048 | | 27 |
| | 163,6618 | -0,8893 | 5 | 0.074 | | 27 |
| UZCCG125 | 163,7042 | 57,0753 | 3 | 0,007 | | 5 |
| RSCG 39 | 163,7792 | 17,1747 | 4 | 0,004 | | 9 |
| | 163,8293 | 5,4809 | 6 | 0.043 | | 27 |
| | 163,9136 | 46,2155 | 7 | 0.063 | | 27 |
| | 163,9266 | 6,5693 | 4 | 0.071 | | 27 |
| HCG 048 | 163,9500 | -27,0875 | 4 | 0,009 | | 1 |
| | 163,9571 | 11,3680 | 4 | 0.088 | | 27 |
| | 164,0492 | 44,3385 | 7 | 0.037 | | 27 |
| | 164,2459 | 41,6581 | 4 | 0.134 | | 27 |
| | 164,2611 | 40,5760 | 5 | 0.119 | | 27 |
| | 164,3479 | 57,9131 | 4 | 0.077 | | 27 |
| | 164,3989 | 43,0184 | 4 | 0.078 | | 27 |
| | 164,4424 | 1,5684 | 8 | 0.039 | | 27 |
| | 164,4652 | 0,7523 | 5 | 0.039 | | 27 |
| | 164,5739 | 57,5294 | 5 | 0.081 | | 27 |
| | 164,6157 | 6,7328 | 4 | 0.035 | | 27 |
| | 164,6747 | 49,9697 | 4 | 0.024 | | 27 |
| | 164,7013 | 7,7724 | 4 | 0.109 | | 27 |
| | 164,7074 | 5,2566 | 5 | 0.035 | | 27 |
| | 164,7319 | 8,2420 | 7 | 0.035 | | 27 |
| | 164,7715 | 1,5219 | 6 | 0.041 | | 27 |
| | 164,7941 | 9,7120 | 4 | 0.064 | | 27 |
| | 164,8764 | 44,5347 | 5 | 0.08 | | 27 |
| | 164,9007 | 9,3878 | 4 | 0.063 | | 27 |
| | 164,9289 | 7,7332 | 4 | 0.106 | | 27 |
| | 164,9406 | 9,0291 | 4 | 0.062 | | 27 |

| | | | | | | | | |
|---|---|---|---|---|---|---|---|---|
| | 164,9468 | 56,9588 | 4 | 0.047 | | | 27 | |
| UZCCG126 | 164,9542 | 50,0189 | 3 | 0,024 | | | 5 | |
| | 164,9646 | 8,4594 | 4 | 0.073 | | | 27 | |
| | 164,9683 | 10,5578 | 4 | 0.038 | | | 27 | |
| PCG110002+082435 | 165,0125 | 8,4097 | 4 | | 0.0730 | C | 7 | group member/coord.from Ref.(6) |
| | 165,0255 | 45,1435 | 4 | 0.065 | | | 27 | |
| | 165,0559 | 53,8459 | 4 | 0.072 | | | 27 | |
| | 165,0896 | 4,2082 | 4 | 0.039 | | | 27 | |
| | 165,1202 | 8,8267 | 4 | 0.101 | | | 27 | |
| | 165,1697 | 4,6653 | 4 | 0.033 | | | 27 | |
| | 165,1705 | 7,9457 | 4 | 0.038 | | | 27 | |
| | 165,1984 | 9,9709 | 4 | 0.075 | | | 27 | |
| | 165,2004 | 55,6394 | 9 | 0.049 | | | 27 | |
| | 165,2285 | 41,6977 | 4 | 0.076 | | | 27 | |
| SDSS 030 | 165,2583 | 1,0981 | 4 | 0,185 | | | 8 | |
| | 165,2679 | 10,4896 | 4 | 0.033 | | | 27 | |
| | 165,3372 | 9,2356 | 4 | 0.035 | | | 27 | |
| | 165,3726 | 50,3969 | 4 | 0.023 | | | 27 | |
| | 165,4208 | 10,3540 | 6 | 0.036 | | | 27 | |
| | 165,4877 | 43,8809 | 4 | 0.026 | | | 27 | |
| SHK 237 | 165,4958 | 38,2308 | 7 | | 0.0301 | 1 | 11 | group member/coord.from Ref.(10) |
| | 165,5399 | -1,0258 | 4 | 0.085 | | | 27 | |
| | 165,6284 | 5,1012 | 5 | 0.052 | | | 27 | |
| | 165,6676 | 8,1040 | 4 | 0.072 | | | 27 | |
| | 165,6812 | 44,4528 | 7 | 0.025 | | | 27 | |
| SDSS 031 | 165,6833 | 0,6578 | 4 | 0,110 | | | 8 | |
| | 165,6957 | 7,5361 | 7 | 0.072 | | | 27 | |
| | 165,7167 | 43,8184 | 5 | 0.049 | | | 27 | |
| | 165,7169 | -0,2684 | 6 | 0.102 | | | 27 | |
| | 165,7396 | 7,5985 | 9 | 0.071 | | | 27 | |
| | 165,7446 | 44,5859 | 7 | 0.037 | | | 27 | |
| | 165,7631 | 7,6310 | 5 | 0.073 | | | 27 | |
| | 165,7683 | 41,6960 | 5 | 0.075 | | | 27 | |
| | 165,7819 | 0,6967 | 4 | 0.096 | | | 27 | |
| | 165,7927 | 50,6304 | 7 | 0.022 | | | 27 | |
| | 165,7979 | 0,1916 | 5 | 0.05 | | | 27 | |
| | 165,8780 | 6,7185 | 5 | 0.061 | | | 27 | |
| | 165,9191 | 52,3792 | 5 | 0.07 | | | 27 | |
| | 165,9372 | 59,2532 | 4 | 0.057 | | | 27 | |
| SDSS 032 | 165,9833 | 0,6350 | 4 | 0,096 | | | 8 | |
| | 166,0154 | 45,0840 | 5 | 0.021 | | | 27 | |
| | 166,0158 | 6,8169 | 4 | 0.06 | | | 27 | |
| SDSS 033 | 166,0208 | 1,0786 | 4 | 0,120 | | | 8 | |
| UZCCG127 | 166,0500 | 45,1742 | 4 | 0,022 | | | 5 | |
| | 166,0921 | 49,9299 | 4 | 0.024 | | | 27 | |
| SHK 120 | 166,1417 | 35,8803 | 7 | 0,071 | | | 25 | |
| | 166,1548 | 2,3698 | 4 | 0.101 | | | 27 | |
| | 166,1950 | 40,8581 | 4 | 0.033 | | | 27 | |
| UZCCG128 | 166,2042 | 4,2956 | 3 | 0,025 | | | 5 | |
| | 166,2123 | 42,9275 | 4 | 0.075 | | | 27 | |
| | 166,2747 | 7,7316 | 4 | 0.07 | | | 27 | |
| | 166,3058 | 59,6667 | 5 | 0.033 | | | 27 | |
| | 166,3087 | 57,1636 | 4 | 0.047 | | | 27 | |
| | 166,3214 | 42,7759 | 6 | 0.076 | | | 27 | |
| | 166,4375 | 7,4867 | 4 | 0.069 | | | 27 | |
| | 166,4833 | 6,7059 | 5 | 0.061 | | | 27 | |
| | 166,5743 | 2,0550 | 4 | 0.107 | | | 27 | |
| | 166,6092 | 1,6254 | 5 | 0.031 | | | 27 | |
| | 166,7396 | 40,0879 | 6 | 0.074 | | | 27 | |
| | 166,7435 | 43,0156 | 4 | 0.058 | | | 27 | |
| | 166,7934 | 1,1804 | 4 | 0.067 | | | 27 | |
| | 166,8336 | 7,7304 | 4 | 0.075 | | | 27 | |
| | 166,8678 | 57,2046 | 4 | 0.074 | | | 27 | |
| | 166,8691 | 43,4488 | 6 | 0.075 | | | 27 | |
| | 166,8926 | 8,1991 | 5 | 0.074 | | | 27 | |
| | 166,8931 | 2,8162 | 6 | 0.058 | | | 27 | |
| | 166,9183 | 44,4295 | 7 | 0.076 | | | 27 | |
| | 166,9327 | 0,0416 | 6 | 0.067 | | | 27 | |
| | 166,9532 | 43,3001 | 4 | 0.076 | | | 27 | |
| | 166,9624 | 2,6961 | 6 | 0.057 | | | 27 | |
| | 166,9639 | 5,9498 | 5 | 0.078 | | | 27 | |
| | 166,9658 | 42,2609 | 4 | 0.074 | | | 27 | |
| | 167,0074 | 45,5181 | 4 | 0.064 | | | 27 | |
| | 167,0503 | 43,9441 | 9 | 0.058 | | | 27 | |
| | 167,0599 | 45,0034 | 4 | 0.062 | | | 27 | |
| | 167,0842 | 6,1758 | 5 | 0.044 | | | 27 | |
| | 167,1589 | 40,4875 | 5 | 0.075 | | | 27 | |
| | 167,1815 | 63,8305 | 5 | 0.044 | | | 27 | |
| | 167,1901 | 44,0718 | 7 | 0.056 | | | 27 | |
| | 167,1972 | 40,6092 | 6 | 0.074 | | | 27 | |
| | 167,1981 | 43,9716 | 7 | 0.062 | | | 27 | |
| | 167,2077 | 43,8251 | 4 | 0.113 | | | 27 | |
| | 167,2136 | 41,5164 | 9 | 0.077 | | | 27 | |
| | 167,2150 | 8,4561 | 5 | 0.039 | | | 27 | |
| | 167,2250 | 41,5844 | 4 | 0.075 | | | 27 | |
| | 167,2428 | 0,5553 | 4 | 0.025 | | | 27 | |
| | 167,2436 | 44,4137 | 7 | 0.061 | | | 27 | |
| | 167,2623 | 43,9650 | 7 | 0.057 | | | 27 | |
| | 167,2766 | 48,7880 | 4 | 0.109 | | | 27 | |
| PCG110907+022442 | 167,2833 | 2,4117 | 4 | | 0.1350 | A | 7 | group member/coord.from Ref.(6) |
| | 167,2841 | 2,6824 | 4 | 0.036 | | | 27 | |
| | 167,3265 | 43,0109 | 4 | 0.062 | | | 27 | |
| | 167,3485 | 43,2301 | 5 | 0.075 | | | 27 | |
| | 167,3614 | 41,5475 | 4 | 0.078 | | | 27 | |
| | 167,3804 | 5,3465 | 7 | 0.076 | | | 27 | |
| PCG110933+413359 | 167,3917 | 41,5664 | 4 | | 0.0790 | B | 7 | group member/coord.from Ref.(6) |
| | 167,3956 | 55,9425 | 5 | 0.049 | | | 27 | |
| SDSS 034 | 167,4167 | 0,7597 | 6 | 0,332 | | | 8 | |
| J1109+2033 | 167,4208 | 20,5556 | 4 | 0,139 | | | 4 | group coord.from ref.(3) |
| | 167,4304 | 43,7758 | 4 | 0.067 | | | 27 | |
| | 167,4629 | 1,4075 | 5 | 0.097 | | | 27 | |
| | 167,4656 | 41,5180 | 7 | 0.076 | | | 27 | |
| | 167,4678 | 12,8345 | 4 | 0.043 | | | 27 | |
| UZCCG129 | 167,4792 | 10,5603 | 3 | 0,005 | | | 5 | |
| | 167,5252 | -0,5425 | 4 | 0.04 | | | 27 | |
| | 167,5293 | 1,4907 | 4 | 0.098 | | | 27 | |
| | 167,5599 | 47,9041 | 4 | 0.111 | | | 27 | |
| SHK 351 | 167,5708 | 4,7819 | 6 | 0,029 | | | 26 | |
| SHK 351 | 167,5708 | 4,7819 | | | 0.0303 | 6 | 11 | group member/coord.from Ref.(10) |
| | 167,5807 | 4,5877 | 5 | 0.029 | | | 27 | |
| | 167,6019 | 4,8433 | 6 | 0.031 | | | 27 | |
| UZCCG130 | 167,6333 | 4,0828 | 3 | 0,029 | | | 5 | |
| | 167,6925 | 4,0858 | 4 | 0.054 | | | 27 | |
| | 167,6930 | 4,8267 | 4 | 0.03 | | | 27 | |
| | 167,6955 | 10,1203 | 7 | 0.069 | | | 27 | |
| | 167,7063 | 55,1681 | 6 | 0.058 | | | 27 | |
| | 167,7159 | 3,1307 | 6 | 0.03 | | | 27 | |
| UZCCG131 | 167,7208 | 4,7694 | 3 | 0,022 | | | 5 | |
| | 167,7605 | -0,7681 | 8 | 0.064 | | | 27 | |
| SDSS 035 | 167,8000 | 1,1072 | 4 | 0,091 | | | 8 | |
| | 167,8083 | 43,8305 | 5 | 0.112 | | | 27 | |
| | 167,8565 | 7,4945 | 4 | 0.046 | | | 27 | |
| | 167,8695 | 57,0354 | 7 | 0.047 | | | 27 | |
| | 167,8862 | 44,1605 | 6 | 0.076 | | | 27 | |
| | 167,9018 | 44,8502 | 4 | 0.112 | | | 27 | |
| | 167,9116 | 41,1162 | 4 | 0.064 | | | 27 | |
| | 167,9313 | 43,9561 | 4 | 0.053 | | | 27 | |
| | 167,9421 | 40,8595 | 5 | 0.077 | | | 27 | |
| UZCCG132 | 167,9542 | 35,4261 | 3 | 0,025 | | | 5 | |
| | 167,9995 | 40,9822 | 4 | 0.072 | | | 27 | |
| UZCCG133 | 168,0125 | 3,1078 | 3 | 0,030 | | | 5 | |
| | 168,0398 | 40,5253 | 6 | 0.072 | | | 27 | |
| | 168,0638 | 40,7316 | 7 | 0.073 | | | 27 | |
| | 168,0651 | 8,7639 | 5 | 0.083 | | | 27 | |
| | 168,0657 | 9,7131 | 4 | 0.047 | | | 27 | |
| LCCG 26 | 168,0875 | -3,1583 | 3 | 0,083 | | | 2 | |
| | 168,1606 | 7,9654 | 4 | 0.074 | | | 27 | |
| | 168,1895 | 2,4287 | 4 | 0.079 | | | 27 | |
| PCG111250+132815 | 168,2083 | 13,4711 | 4 | | 0.1690 | A | 7 | group member/coord.from Ref.(6) |
| | 168,2112 | 4,9654 | 6 | 0.027 | | | 27 | |
| | 168,2498 | 57,0991 | 4 | 0.049 | | | 27 | |
| | 168,2670 | 63,4021 | 4 | 0.033 | | | 27 | |
| | 168,2903 | -0,1298 | 4 | 0.04 | | | 27 | |
| | 168,2981 | 56,9248 | 7 | 0.048 | | | 27 | |
| | 168,2985 | 48,6464 | 4 | 0.026 | | | 27 | |
| | 168,3111 | 41,4030 | 4 | 0.094 | | | 27 | |
| | 168,3155 | 64,0467 | 5 | 0.035 | | | 27 | |
| | 168,3397 | 2,1882 | 4 | 0.079 | | | 27 | |
| SDSS 036 | 168,4000 | 0,4814 | 5 | 0,100 | | | 8 | |
| | 168,4009 | 1,5959 | 5 | 0.104 | | | 27 | |
| | 168,4103 | 57,1457 | 4 | 0.045 | | | 27 | |
| | 168,4205 | -0,4491 | 4 | 0.1 | | | 27 | |
| | 168,4318 | 2,4340 | 8 | 0.074 | | | 27 | |
| | 168,4334 | 2,4840 | 4 | 0.072 | | | 27 | |
| | 168,4826 | 11,6827 | 5 | 0.038 | | | 27 | |
| | 168,5094 | 40,3379 | 5 | 0.073 | | | 27 | |
| | 168,5206 | 56,2366 | 6 | 0.05 | | | 27 | |
| | 168,5317 | 0,0103 | 4 | 0.073 | | | 27 | |
| | 168,5504 | 12,8961 | 5 | 0.112 | | | 27 | |
| | 168,5660 | 50,2790 | 5 | 0.047 | | | 27 | |
| | 168,6711 | 4,0593 | 7 | 0.039 | | | 27 | |
| | 168,6794 | 2,4310 | 4 | 0.076 | | | 27 | |
| LCCG 27 | 168,6958 | -6,3672 | 4 | 0,081 | | | 2 | |
| | 168,7025 | 56,6023 | 4 | 0.074 | | | 27 | |
| | 168,7426 | 53,2398 | 5 | 0.038 | | | 27 | |
| | 168,7899 | 4,2243 | 6 | 0.04 | | | 27 | |
| | 168,7942 | 50,4817 | 4 | 0.107 | | | 27 | |
| | 168,7957 | 41,3815 | 4 | 0.073 | | | 27 | |
| | 168,8353 | 2,7305 | 7 | 0.078 | | | 27 | |
| | 168,8361 | 54,3160 | 5 | 0.069 | | | 27 | |
| LCCG 28 | 168,8458 | -3,9072 | 5 | 0,081 | | | 2 | |
| | 168,8739 | 41,6546 | 4 | 0.066 | | | 27 | |
| | 168,8806 | 48,6961 | 6 | 0.074 | | | 27 | |
| | 168,8807 | 54,5500 | 5 | 0.072 | | | 27 | |
| | 168,8907 | 54,4582 | 5 | 0.067 | | | 27 | |
| | 168,9326 | 40,6880 | 7 | 0.08 | | | 27 | |
| | 168,9349 | 3,4611 | 4 | 0.076 | | | 27 | |
| | 168,9608 | 0,5035 | 4 | 0.078 | | | 27 | |
| LCCG 29 | 168,9625 | -11,5647 | 3 | 0,106 | | | 2 | |
| | 168,9711 | 2,8501 | 5 | 0.077 | | | 27 | |
| | 168,9854 | 47,4720 | 4 | 0.075 | | | 27 | |
| | 168,9942 | 61,9915 | 6 | 0.067 | | | 27 | |
| | 168,9993 | 61,1442 | 4 | 0.057 | | | 27 | |
| | 169,0046 | 3,4129 | 4 | 0.078 | | | 27 | |

| | | | | | | | | | |
|---|---|---|---|---|---|---|---|---|---|
| | 169,0110 | 47,9755 | 4 | 0.074 | | | | 27 | |
| PCG111605+042937 | 169,0250 | 4,4939 | 4 | | 0.1100 | A | | 7 | group member/coord.from Ref.(6) |
| | 169,0421 | 40,6499 | 7 | 0.079 | | | | 27 | |
| | 169,0422 | 3,3936 | 6 | 0.077 | | | | 27 | |
| RSCG 40 | 169,0667 | 17,9864 | 5 | 0,003 | | | | 9 | |
| | 169,0720 | 41,9858 | 4 | 0.082 | | | | 27 | |
| | 169,0922 | 41,0967 | 7 | 0.072 | | | | 27 | |
| | 169,0949 | 12,0220 | 4 | 0.048 | | | | 27 | |
| | 169,1090 | 54,8464 | 4 | 0.069 | | | | 27 | |
| | 169,1182 | 2,4639 | 8 | 0.076 | | | | 27 | |
| | 169,2041 | -0,9533 | 5 | 0.075 | | | | 27 | |
| | 169,2107 | 5,6543 | 4 | 0.048 | | | | 27 | |
| | 169,2263 | 52,6815 | 5 | 0.078 | | | | 27 | |
| | 169,2362 | 54,6052 | 4 | 0.112 | | | | 27 | |
| | 169,2577 | 3,9281 | 6 | 0.024 | | | | 27 | |
| HCG 050 | 169,2750 | 54,9186 | 5 | 0,139 | | | | 1 | |
| | | | 5 | 0,140 | | | SHK 005 | 11 | mean value |
| | 169,3278 | -0,2716 | 4 | 0.109 | | | | 27 | |
| UZCCG134 | 169,3417 | 4,5022 | 3 | 0,006 | | | | 5 | |
| PCG111728+074639 | 169,3708 | 7,7778 | 4 | | 0.1600 | C | | 7 | group member/coord.from Ref.(6) |
| | 169,4438 | 0,4331 | 4 | 0.107 | | | | 27 | |
| | 169,4727 | 44,5893 | 4 | 0.107 | | | | 27 | |
| | 169,4821 | 54,0643 | 5 | 0.034 | | | | 27 | |
| | 169,5831 | 51,7672 | 8 | 0.081 | | | | 27 | |
| | 169,6081 | 45,5703 | 4 | 0.113 | | | | 27 | |
| UZCCG135 | 169,6250 | 23,3989 | 3 | 0,022 | | | | 5 | |
| | 169,6289 | 55,6800 | 4 | 0.05 | | | | 27 | |
| | 169,6352 | 2,5484 | 5 | 0.03 | | | | 27 | |
| | 169,6447 | 53,9552 | 5 | 0.035 | | | | 27 | |
| | 169,6969 | 8,9218 | 4 | 0.076 | | | | 27 | |
| | 169,6996 | 0,9912 | 6 | 0.064 | | | | 27 | |
| | 169,8420 | 5,2187 | 4 | 0.049 | | | | 27 | |
| | 169,8628 | 12,0306 | 6 | 0.039 | | | | 27 | |
| UZCCG136 | 169,8750 | 57,8464 | 3 | 0,006 | | | | 5 | |
| | 169,8788 | 51,7706 | 4 | 0.08 | | | | 27 | |
| | 169,9105 | 54,4937 | 7 | 0.071 | | | | 27 | |
| | 169,9295 | 4,6738 | 4 | 0.032 | | | | 27 | |
| | 169,9501 | 54,8965 | 4 | 0.073 | | | | 27 | |
| | 169,9684 | 55,2795 | 5 | 0.107 | | | | 27 | |
| | 169,9992 | -0,1940 | 4 | 0.04 | | | | 27 | |
| | 170,0074 | 11,9465 | 8 | 0.04 | | | | 27 | |
| | 170,0169 | 0,2336 | 6 | 0.073 | | | | 27 | |
| | 170,0170 | 53,7054 | 4 | 0.104 | | | | 27 | |
| | 170,0425 | 53,6549 | 6 | 0.106 | | | | 27 | |
| | 170,0445 | 0,0929 | 6 | 0.025 | | | | 27 | |
| | 170,0762 | 59,6020 | 5 | 0.055 | | | | 27 | |

| | | | | | | |
|---|---|---|---|---|---|---|
| | 170,0870 | 56,7906 | 5 | 0.048 | | 27 |
| | 170,1426 | 42,2918 | 5 | 0.085 | | 27 |
| | 170,1480 | 47,1301 | 6 | 0.111 | | 27 |
| | 170,1581 | 47,1722 | 4 | 0.113 | | 27 |
| | 170,1648 | 53,6591 | 4 | 0.105 | | 27 |
| | 170,1820 | 53,7586 | 5 | 0.103 | | 27 |
| | 170,1898 | 4,2974 | 7 | 0.049 | | 27 |
| | 170,2104 | 46,7559 | 6 | 0.054 | | 27 |
| UZCCG137 | 170,2125 | 0,4908 | 3 | 0,024 | | 5 |
| | 170,2293 | 2,9616 | 6 | 0.048 | | 27 |
| | 170,2347 | 0,0095 | 7 | 0.102 | | 27 |
| | 170,2635 | 47,9915 | 5 | 0.112 | | 27 |
| | 170,2637 | -0,2059 | 4 | 0.098 | | 27 |
| LCCG 30 | 170,2667 | -2,7494 | 3 | 0,062 | | 2 |
| | 170,2840 | 2,8622 | 9 | 0.049 | | 27 |
| | 170,3003 | 44,3220 | 5 | 0.031 | | 27 |
| | 170,3566 | 48,4320 | 4 | 0.111 | | 27 |
| | 170,3770 | 2,8917 | 5 | 0.047 | | 27 |
| | 170,3789 | -0,2500 | 4 | 0.101 | | 27 |
| | 170,3843 | 2,8874 | 4 | 0.053 | | 27 |
| | 170,3982 | 41,5249 | 5 | 0.045 | | 27 |
| | 170,4058 | 60,8355 | 4 | 0.06 | | 27 |
| | 170,4136 | 0,4833 | 4 | 0.024 | | 27 |
| | 170,4475 | 56,6320 | 9 | 0.047 | | 27 |
| | 170,4566 | 42,7866 | 4 | 0.072 | | 27 |
| | 170,4763 | 46,4333 | 4 | 0.032 | | 27 |
| | 170,4979 | 12,4751 | 6 | 0.084 | | 27 |
| | 170,4994 | 66,2735 | 5 | 0.114 | | 27 |
| | 170,5496 | 49,9965 | 4 | 0.115 | | 27 |
| | 170,5565 | 53,5976 | 4 | 0.034 | | 27 |
| | 170,5598 | 1,0871 | 4 | 0.076 | | 27 |
| | 170,5647 | 58,2984 | 7 | 0.053 | | 27 |
| | 170,5785 | 43,8954 | 4 | 0.097 | | 27 |
| | 170,5864 | 9,8554 | 4 | 0.076 | | 27 |
| HCG 051 | 170,5875 | 24,2931 | 6 | 0,026 | N3651 | 1 |
| | | | 3 | 0,028 | UZCCG138 | 5 |
| UZCCG139 | 170,6208 | 59,0053 | 3 | 0,005 | | 5 |
| | 170,6342 | 49,3347 | 4 | 0.111 | | 27 |
| | 170,6432 | 0,4633 | 5 | 0.103 | | 27 |
| | 170,6652 | 59,3269 | 4 | 0.05 | | 27 |
| | 170,6790 | 40,8494 | 5 | 0.061 | | 27 |
| | 170,6894 | 12,2822 | 4 | 0.084 | | 27 |
| | 170,6985 | 42,3849 | 5 | 0.078 | | 27 |
| | 170,7176 | 54,9663 | 5 | 0.073 | | 27 |
| SHK 154 | 170,7250 | 1,1144 | 6 | 0,073 | | 19 |
| | 170,7413 | 0,7588 | 5 | 0.072 | | 27 |

| | | | | | | | | |
|---|---|---|---|---|---|---|---|---|
| | 170,7568 | 1,1039 | 9 | 0.075 | | | 27 | |
| | 170,7812 | 1,0883 | 4 | 0.077 | | | 27 | |
| | 170,7974 | 0,1264 | 4 | 0.068 | | | 27 | |
| | 170,8035 | 5,6646 | 4 | 0.038 | | | 27 | |
| | 170,8419 | -0,9142 | 7 | 0.04 | | | 27 | |
| | 170,8490 | 47,9409 | 5 | 0.126 | | | 27 | |
| | 170,8607 | 55,9473 | 4 | 0.095 | | | 27 | |
| | 170,9203 | 56,1146 | 5 | 0.068 | | | 27 | |
| | 170,9208 | 2,7558 | 4 | 0.049 | | | 27 | |
| | 170,9325 | 47,0074 | 5 | 0.053 | | | 27 | |
| | 170,9342 | 67,1934 | 4 | 0.056 | | | 27 | |
| | 170,9440 | 0,9601 | 5 | 0.1 | | | 27 | |
| | 170,9487 | 47,1626 | 4 | 0.054 | | | 27 | |
| PCG112357+212857 | 170,9917 | 21,4828 | 4 | | 0.1900 | A | 7 | group member/coord.from Ref.(6) |
| | 171,0728 | 2,6966 | 6 | 0.063 | | | 27 | |
| | 171,0870 | 5,3481 | 4 | 0.149 | | | 27 | |
| | 171,1887 | 57,7536 | 4 | 0.069 | | | 27 | |
| | 171,2067 | 46,9862 | 4 | 0.034 | | | 27 | |
| PCG112452+414012 | 171,2167 | 41,6703 | 4 | | 0.1910 | A | 7 | group member/coord.from Ref.(6) |
| | 171,2710 | 60,9852 | 4 | 0.108 | | | 27 | |
| | 171,3208 | 52,0889 | 7 | 0.034 | | | 27 | |
| | 171,3636 | 48,4821 | 5 | 0.107 | | | 27 | |
| UZCCG140 | 171,4792 | 63,5336 | 3 | 0,012 | | | 5 | |
| | 171,5109 | 10,4661 | 4 | 0.066 | | | 27 | |
| | 171,5352 | 56,9381 | 4 | 0.056 | | | 27 | |
| | 171,5741 | 55,7850 | 5 | 0.052 | | | 27 | |
| | 171,5746 | 55,5833 | 4 | 0.047 | | | 27 | |
| HCG 052 | 171,5792 | 21,0892 | 4 | 0,043 | | | 1 | |
| | 171,5995 | 5,1069 | 5 | 0.04 | | | 27 | |
| | 171,6298 | 44,4144 | 4 | 0.091 | | | 27 | |
| | 171,6404 | 57,1024 | 7 | 0.057 | | | 27 | |
| | 171,7275 | 54,6657 | 8 | 0.047 | | | 27 | |
| | 171,7980 | 1,3137 | 4 | 0.128 | | | 27 | |
| | 171,7989 | 47,5357 | 5 | 0.033 | | | 27 | |
| RSCG 41 | 171,8125 | 17,0183 | 4 | 0,004 | | | 9 | |
| | 171,8308 | 10,6117 | 8 | 0.114 | | | 27 | |
| | 171,8466 | 52,0896 | 4 | 0.033 | | | 27 | |
| | 171,8774 | 59,2048 | 8 | 0.06 | | | 27 | |
| | 171,8919 | 0,1643 | 6 | 0.13 | | | 27 | |
| | 171,9341 | 1,4465 | 6 | 0.072 | | | 27 | |
| | 171,9391 | -2,2547 | 5 | 0.098 | | | 27 | |
| | 171,9435 | 61,3733 | 6 | 0.047 | | | 27 | |
| | 171,9778 | 1,3407 | 4 | 0.073 | | | 27 | |
| | 171,9877 | 56,9487 | 4 | 0.058 | | | 27 | |
| | 172,0013 | 10,6285 | 4 | 0.113 | | | 27 | |
| | 172,0558 | 0,1374 | 4 | 0.05 | | | 27 | |

| | | | | | | |
|---|---|---|---|---|---|---|
| | 172,0673 | 50,6175 | 4 | 0.029 | | 27 |
| UZCCG141 | 172,1542 | 9,0883 | 6 | 0,021 | | 5 |
| | 172,1676 | 11,0785 | 4 | 0.114 | | 27 |
| | 172,2110 | 56,2775 | 9 | 0.059 | | 27 |
| | 172,2251 | 58,1112 | 4 | 0.06 | | 27 |
| | 172,2306 | 54,1345 | 5 | 0.068 | | 27 |
| HCG 053 | 172,2417 | 20,7764 | 4 | 0,021 | N3697 | 1 |
| | 172,2763 | 56,8360 | 5 | 0.059 | | 27 |
| | 172,3072 | 1,7923 | 5 | 0.098 | | 27 |
| HCG 054 | 172,3125 | 20,5786 | 4 | 0,005 | ROSE27 | 1 |
| | 172,3316 | 54,0628 | 8 | 0.071 | | 27 |
| | 172,3398 | 58,4201 | 4 | 0.041 | | 27 |
| | 172,3426 | 55,4513 | 5 | 0.069 | | 27 |
| | 172,3701 | 55,3564 | 7 | 0.066 | | 27 |
| SDSS 041 | 172,3792 | 0,5533 | 4 | 0,075 | | 8 |
| | 172,3922 | 55,5164 | 7 | 0.068 | | 27 |
| | 172,4189 | 51,0266 | 6 | 0.034 | | 27 |
| | 172,4580 | 55,3603 | 5 | 0.07 | | 27 |
| | 172,4938 | 42,0137 | 6 | 0.044 | | 27 |
| | 172,5528 | 4,5975 | 4 | 0.085 | | 27 |
| | 172,6047 | 4,1759 | 4 | 0.062 | | 27 |
| | 172,6494 | 59,3308 | 4 | 0.089 | | 27 |
| | 172,6540 | 2,1389 | 5 | 0.076 | | 27 |
| | 172,7246 | 46,2827 | 5 | 0.105 | | 27 |
| | 172,8092 | 47,3424 | 4 | 0.127 | | 27 |
| | 172,8487 | 67,2475 | 6 | 0.055 | | 27 |
| | 172,8863 | 55,9653 | 5 | 0.056 | | 27 |
| | 172,8943 | -1,2214 | 8 | 0.048 | | 27 |
| SDSS 042 | 172,9083 | 0,2153 | 6 | 0,132 | | 8 |
| | 172,9172 | 55,9718 | 7 | 0.051 | | 27 |
| | 172,9331 | 64,9700 | 4 | 0.038 | | 27 |
| | 172,9904 | -2,0802 | 4 | 0.041 | | 27 |
| | 173,0160 | 45,4549 | 4 | 0.045 | | 27 |
| | 173,0179 | 53,5561 | 8 | 0.047 | | 27 |
| HCG 055 | 173,0292 | 70,8119 | 5 | 0,053 | ARP329 | 1 |
| | 173,0322 | 55,9734 | 6 | 0.059 | | 27 |
| | 173,0544 | 1,7855 | 4 | 0.047 | | 27 |
| | 173,0663 | 66,3971 | 4 | 0.116 | | 27 |
| | 173,0683 | 55,7815 | 9 | 0.057 | | 27 |
| | 173,0847 | 3,9387 | 4 | 0.091 | | 27 |
| HCG 056 | 173,1333 | 52,9486 | 5 | 0,027 | ARP322 | 1 |
| | 173,1357 | 56,1342 | 4 | 0.075 | | 27 |
| | 173,1706 | 62,6140 | 4 | 0.047 | | 27 |
| | 173,1722 | 57,5756 | 5 | 0.068 | | 27 |
| | 173,2391 | 56,9206 | 4 | 0.059 | | 27 |
| | 173,2426 | 62,1898 | 4 | 0.033 | | 27 |

| | | | | | |
|---|---|---|---|---|---|
| | 173,2681 | 62,8188 | 5 | 0.048 | 27 |
| | 173,2788 | 66,3518 | 4 | 0.113 | 27 |
| | 173,2892 | 2,2830 | 5 | 0.099 | 27 |
| | 173,2937 | 56,0460 | 9 | 0.052 | 27 |
| SDSS 043 | 173,3000 | 0,4569 | 5 | 0,105 | 8 |
| | 173,3125 | -3,0107 | 4 | 0.146 | 27 |
| | 173,3239 | 56,1400 | 4 | 0.058 | 27 |
| | 173,3635 | -3,4564 | 4 | 0.052 | 27 |
| | 173,4203 | 51,2251 | 5 | 0.026 | 27 |
| | 173,4503 | 47,4447 | 5 | 0.046 | 27 |
| | 173,4788 | 1,4277 | 4 | 0.075 | 27 |
| | 173,4789 | 57,7840 | 6 | 0.067 | 27 |
| | 173,5284 | 47,2618 | 4 | 0.102 | 27 |
| | 173,5790 | 66,3224 | 4 | 0.117 | 27 |
| LCCG 31 | 173,7125 | -12,6403 | 3 | 0,117 | 2 |
| | 173,7160 | 41,9059 | 5 | 0.061 | 27 |
| | 173,7199 | 50,1144 | 4 | 0.087 | 27 |
| | 173,7205 | 48,9700 | 8 | 0.031 | 27 |
| | 173,7255 | 60,8345 | 4 | 0.062 | 27 |
| | 173,7376 | 53,8322 | 6 | 0.071 | 27 |
| | 173,7395 | 12,6690 | 4 | 0.065 | 27 |
| | 173,7420 | 48,8906 | 5 | 0.035 | 27 |
| | 173,7861 | 55,4493 | 9 | 0.065 | 27 |
| | 173,8214 | 55,7022 | 6 | 0.075 | 27 |
| | 173,8383 | 1,5160 | 5 | 0.094 | 27 |
| | 173,8576 | 55,9845 | 4 | 0.061 | 27 |
| | 173,8629 | 41,7119 | 4 | 0.074 | 27 |
| | 173,9598 | 49,1081 | 4 | 0.036 | 27 |
| | 173,9599 | 48,3698 | 7 | 0.055 | 27 |
| | 173,9612 | 56,9330 | 6 | 0.094 | 27 |
| | 173,9791 | 3,5477 | 5 | 0.105 | 27 |
| | 173,9913 | 43,4204 | 4 | 0.024 | 27 |
| | 174,0243 | 55,1291 | 5 | 0.057 | 27 |
| | 174,0244 | 50,8278 | 4 | 0.085 | 27 |
| | 174,0350 | 41,9356 | 4 | 0.07 | 27 |
| | 174,0871 | 55,1055 | 9 | 0.056 | 27 |
| | 174,0899 | 55,0817 | 5 | 0.055 | 27 |
| | 174,0904 | 55,0082 | 9 | 0.058 | 27 |
| RSCG 42 | 174,2125 | 19,9886 | 3 | 0,021 | 9 |
| | 174,2672 | 13,1748 | 5 | 0.08 | 27 |
| | 174,2676 | 47,5515 | 4 | 0.034 | 27 |
| | 174,2910 | -0,8483 | 4 | 0.046 | 27 |
| | 174,3056 | 57,4756 | 6 | 0.067 | 27 |
| | 174,3247 | 64,8960 | 5 | 0.043 | 27 |
| | 174,3348 | 2,4478 | 4 | 0.074 | 27 |
| | 174,3477 | 53,7590 | 8 | 0.072 | 27 |

| | | | | | | |
|---|---|---|---|---|---|---|
| | 174,4007 | 55,4256 | 8 | 0.064 | | 27 |
| | 174,4436 | 56,7073 | 8 | 0.08 | | 27 |
| HCG 057 | 174,4625 | 21,9850 | 8 | 0,030 | ARP320 | 1 |
| | | | 3 | 0,030 | RSCG 43 | 9 |
| | | | 4 | 0,030 | UZCCG142 | 5 |
| | 174,5362 | 13,1517 | 5 | 0.082 | | 27 |
| | 174,5892 | 5,1515 | 4 | 0.092 | | 27 |
| | 174,6067 | 11,3152 | 4 | 0.085 | | 27 |
| | 174,6317 | 56,2266 | 4 | 0.051 | | 27 |
| | 174,7030 | 55,6379 | 6 | 0.062 | | 27 |
| | 174,7479 | 55,7070 | 4 | 0.059 | | 27 |
| | 174,7570 | 55,6355 | 4 | 0.063 | | 27 |
| | 174,7774 | 54,9427 | 8 | 0.057 | | 27 |
| | 174,8453 | 54,3557 | 5 | 0.075 | | 27 |
| | 174,8498 | 1,4381 | 4 | 0.076 | | 27 |
| | 174,9261 | 47,9753 | 4 | 0.089 | | 27 |
| | 174,9277 | 4,3921 | 7 | 0.075 | | 27 |
| | 174,9492 | 54,9896 | 4 | 0.075 | | 27 |
| | 174,9569 | 13,0382 | 4 | 0.1 | | 27 |
| | 175,0146 | 4,3497 | 5 | 0.074 | | 27 |
| | 175,0382 | 10,7549 | 6 | 0.082 | | 27 |
| LCCG 32 | 175,0417 | -3,3594 | 3 | 0,049 | | 2 |
| | 175,0680 | 43,6815 | 5 | 0.102 | | 27 |
| | 175,0895 | 10,3615 | 5 | 0.104 | | 27 |
| | 175,0923 | 55,4252 | 5 | 0.06 | | 27 |
| | 175,1004 | 10,0453 | 5 | 0.082 | | 27 |
| | 175,1174 | 4,0147 | 6 | 0.099 | | 27 |
| | 175,1177 | 0,9308 | 4 | 0.075 | | 27 |
| | 175,1327 | 0,3086 | 4 | 0.107 | | 27 |
| | 175,1377 | 10,3648 | 4 | 0.107 | | 27 |
| | 175,1431 | 2,0052 | 4 | 0.075 | | 27 |
| | 175,1441 | 10,3312 | 5 | 0.105 | | 27 |
| | 175,1512 | 55,2956 | 4 | 0.064 | | 27 |
| | 175,1608 | 56,5839 | 7 | 0.066 | | 27 |
| | 175,1665 | 10,1490 | 5 | 0.021 | | 27 |
| | 175,1738 | 11,0300 | 4 | 0.107 | | 27 |
| | 175,1740 | 10,9575 | 5 | 0.08 | | 27 |
| | 175,2110 | 10,4456 | 4 | 0.112 | | 27 |
| UZCCG143 | 175,2500 | 24,7728 | 3 | 0,012 | | 5 |
| | 175,2611 | 5,6228 | 8 | 0.097 | | 27 |
| | 175,3139 | 65,1670 | 5 | 0.042 | | 27 |
| | 175,3233 | 10,6740 | 5 | 0.116 | | 27 |
| | 175,3289 | 5,5411 | 4 | 0.099 | | 27 |
| | 175,3810 | 4,6387 | 4 | 0.099 | | 27 |
| | 175,3851 | 5,8114 | 4 | 0.1 | | 27 |
| | 175,3986 | 8,7784 | 5 | 0.1 | | 27 |

| | | | | | | | | | |
|---|---|---|---|---|---|---|---|---|---|
| | 175,3990 | 56,3559 | 4 | 0.097 | | | | 27 | |
| | 175,4041 | -1,2037 | 4 | 0.078 | | | | 27 | |
| | 175,4179 | 57,8530 | 4 | 0.066 | | | | 27 | |
| | 175,4552 | 10,1946 | 5 | 0.119 | | | | 27 | |
| | 175,4606 | 9,6167 | 5 | 0.077 | | | | 27 | |
| | 175,4789 | 4,1947 | 4 | 0.127 | | | | 27 | |
| | 175,4897 | 10,4199 | 4 | 0.107 | | | | 27 | |
| | 175,4970 | 4,2732 | 5 | 0.126 | | | | 27 | |
| HCG 058 | 175,5500 | 10,3169 | 5 | 0,021 | | | N3825 | 1 | |
| | 175,5500 | 10,3169 | 4 | 0,021 | | | UZCCG144 | 5 | |
| | 175,5648 | 56,4058 | 4 | 0.098 | | | | 27 | |
| | 175,5738 | 54,2081 | 4 | 0.076 | | | | 27 | |
| | 175,5812 | 62,8234 | 4 | 0.098 | | | | 27 | |
| J1142+1407 | 175,6375 | 14,1272 | 4 | 0,125 | | | | 4 | group coord.from ref.(3) |
| UZCCG145 | 175,6500 | 8,8725 | 3 | 0,022 | | | | 5 | |
| | 175,7607 | 5,4585 | 7 | 0.074 | | | | 27 | |
| | 175,7646 | 10,2047 | 6 | 0.119 | | | | 27 | |
| | 175,8676 | -3,2815 | 5 | 0.118 | | | | 27 | |
| SHK 371 | 175,8708 | 21,9058 | 5 | 0,130 | | | | 26 | |
| | | | 5 | 0,132 | | | J1143+2153 | 4 | group coord.from ref.(3) |
| PCG114333+215356 | 175,8708 | 21,9058 | 5 | | 0.1300 | B | | 7 | group member/coord.from Ref.(6) |
| UZCCG146 | 175,8875 | 10,2508 | 3 | 0,020 | | | | 5 | |
| | 175,8922 | -1,7845 | 4 | 0.109 | | | | 27 | |
| | 175,8974 | 49,5624 | 4 | 0.024 | | | | 27 | |
| | 175,9051 | 54,7889 | 5 | 0.059 | | | | 27 | |
| UZCCG147 | 175,9083 | 31,5533 | 4 | 0,006 | | | | 5 | |
| | 175,9253 | -1,7749 | 4 | 0.105 | | | | 27 | |
| | 175,9523 | -1,7199 | 6 | 0.108 | | | | 27 | |
| | 175,9672 | 2,0538 | 7 | 0.076 | | | | 27 | |
| | 175,9706 | 60,1374 | 6 | 0.064 | | | | 27 | |
| RSCG 44 | 176,0042 | 19,9456 | 6 | 0,021 | | | | 9 | |
| | 176,0099 | 5,7824 | 5 | 0.099 | | | | 27 | |
| | 176,0557 | 5,7701 | 6 | 0.104 | | | | 27 | |
| | 176,0621 | 5,7738 | 4 | 0.102 | | | | 27 | |
| | 176,0700 | 5,8004 | 4 | 0.106 | | | | 27 | |
| RSCG 45 | 176,0833 | 19,8353 | 3 | 0,018 | | | | 9 | |
| | 176,0949 | 52,2826 | 4 | 0.131 | | | | 27 | |
| | 176,1165 | -1,7567 | 8 | 0.107 | | | | 27 | |
| | 176,1405 | -1,1761 | 4 | 0.078 | | | | 27 | |
| SHK 123 | 176,1708 | 57,5058 | 11 | 0,115 | | | | 16 | mean value |
| | 176,1913 | 61,6657 | 6 | 0.035 | | | | 27 | |
| | 176,2115 | -2,5274 | 4 | 0.047 | | | | 27 | |
| | 176,2307 | 4,5571 | 4 | 0.098 | | | | 27 | |
| UZCCG148 | 176,2375 | 19,8117 | 4 | 0,026 | | | | 5 | |
| | 176,2416 | 51,2549 | 4 | 0.128 | | | | 27 | |
| | 176,2704 | 64,5482 | 9 | 0.063 | | | | 27 | |
| | 176,3188 | 56,1695 | 4 | 0.072 | | | | 27 | |
| | 176,3283 | 9,7556 | 5 | 0.097 | | | | 27 | |
| | 176,3351 | 64,4579 | 6 | 0.064 | | | | 27 | |
| | 176,3849 | -1,3397 | 4 | 0.118 | | | | 27 | |
| | 176,4390 | 6,0605 | 5 | 0.103 | | | | 27 | |
| | 176,5046 | 55,4203 | 8 | 0.053 | | | | 27 | |
| | 176,5501 | 42,3304 | 5 | 0.061 | | | | 27 | |
| PCG114616+494908 | 176,5708 | 49,8192 | 4 | | 0.1920 | B | | 7 | group member/coord.from Ref.(6) |
| UZCCG149 | 176,5875 | 33,1453 | 4 | 0,033 | | | | 5 | |
| SDSS 045 | 176,6208 | -1,0239 | 4 | 0,080 | | | | 8 | |
| | 176,6490 | 54,5566 | 5 | 0.057 | | | | 27 | |
| UZCCG150 | 176,6583 | 13,7819 | 3 | 0,011 | | | | 5 | |
| | 176,6753 | -1,0893 | 4 | 0.08 | | | | 27 | |
| | 176,7488 | 55,7155 | 9 | 0.051 | | | | 27 | |
| | 176,7684 | 61,2114 | 5 | 0.067 | | | | 27 | |
| | 176,8129 | 4,5655 | 4 | 0.13 | | | | 27 | |
| | 176,8162 | 55,7476 | 5 | 0.056 | | | | 27 | |
| | 176,8174 | 42,4470 | 4 | 0.086 | | | | 27 | |
| | 176,8418 | -3,0880 | 6 | 0.108 | | | | 27 | |
| | 176,8847 | 10,5787 | 5 | 0.11 | | | | 27 | |
| | 176,8848 | 55,6970 | 8 | 0.05 | | | | 27 | |
| | 176,8993 | 59,2880 | 5 | 0.061 | | | | 27 | |
| | 176,9072 | 54,7135 | 8 | 0.061 | | | | 27 | |
| | 176,9749 | 49,9725 | 5 | 0.033 | | | | 27 | |
| | 176,9891 | 64,7554 | 4 | 0.042 | | | | 27 | |
| | 176,9902 | 54,6137 | 5 | 0.062 | | | | 27 | |
| | 177,0264 | 54,6695 | 5 | 0.058 | | | | 27 | |
| PCG114812+543852 | 177,0542 | 54,6481 | 4 | | 0.0600 | A | | 7 | group member/coord.from Ref.(6) |
| | 177,0555 | 55,7233 | 9 | 0.052 | | | | 27 | |
| | 177,0576 | 10,3391 | 5 | 0.117 | | | | 27 | |
| HCG 059 | 177,1083 | 12,7261 | 5 | 0,014 | | | ROSE7 | 1 | |
| | | | 3 | 0,014 | | | UZCCG151 | 5 | |
| | | | 4 | 0,013 | | | RSCG 46 | 9 | |
| | 177,1317 | 4,1916 | 6 | 0.075 | | | | 27 | |
| | 177,1378 | 9,9917 | 5 | 0.085 | | | | 27 | |
| | 177,1515 | 51,6057 | 4 | 0.132 | | | | 27 | |
| SDSS 046 | 177,1542 | 0,7853 | 5 | 0,126 | | | | 8 | |
| | 177,2015 | 56,5891 | 4 | 0.099 | | | | 27 | |
| | 177,2060 | 1,8550 | 4 | 0.107 | | | | 27 | |
| | 177,2280 | 10,0347 | 7 | 0.087 | | | | 27 | |
| | 177,2318 | 65,7779 | 4 | 0.063 | | | | 27 | |
| | 177,2413 | 11,0597 | 5 | 0.084 | | | | 27 | |
| | 177,2832 | 53,7789 | 5 | 0.062 | | | | 27 | |
| | 177,2872 | 9,8153 | 7 | 0.088 | | | | 27 | |
| | 177,2915 | 56,1991 | 5 | 0.08 | | | | 27 | |
| | 177,3192 | 2,8926 | 5 | 0.047 | | | | 27 | |

| | | | | | | | | |
|---|---|---|---|---|---|---|---|---|
| UZCCG152 | 177,3292 | -1,0908 | 3 | 0,020 | | | 5 | |
| | 177,3389 | -1,4172 | 5 | 0.081 | | | 27 | |
| UZCCG153 | 177,3875 | 26,7475 | 3 | 0,006 | | | 5 | |
| | 177,3894 | 53,6073 | 5 | 0.061 | | | 27 | |
| | 177,4125 | -3,3046 | 4 | 0.027 | | | 27 | |
| | 177,4947 | 5,6953 | 5 | 0.076 | | | 27 | |
| | 177,5113 | 10,8071 | 5 | 0.085 | | | 27 | |
| | 177,5125 | 5,6662 | 5 | 0.073 | | | 27 | |
| | 177,5478 | 5,7837 | 9 | 0.075 | | | 27 | |
| | 177,5552 | 53,6099 | 4 | 0.059 | | | 27 | |
| PCG115017+104955 | 177,5750 | 10,8322 | 4 | | 0.0850 | B | 7 | group member/coord.from Ref.(6) |
| | 177,6327 | 3,4175 | 4 | 0.047 | | | 27 | |
| | 177,6441 | 53,1484 | 4 | 0.067 | | | 27 | |
| | 177,6984 | 56,4928 | 9 | 0.066 | | | 27 | |
| | 177,7595 | -3,2099 | 6 | 0.077 | | | 27 | |
| | 177,7928 | 2,9829 | 4 | 0.076 | | | 27 | |
| LCCG 33 | 177,8000 | -3,0286 | 3 | 0,091 | | | 2 | |
| | 177,8197 | -3,0556 | 7 | 0.091 | | | 27 | |
| | 177,8198 | 56,4785 | 8 | 0.065 | | | 27 | |
| | 177,8216 | 53,8100 | 6 | 0.052 | | | 27 | |
| | 177,9059 | 46,8357 | 4 | 0.055 | | | 27 | |
| | 177,9077 | -2,8234 | 5 | 0.077 | | | 27 | |
| | 177,9225 | 4,9321 | 5 | 0.075 | | | 27 | |
| | 177,9286 | 6,1240 | 4 | 0.08 | | | 27 | |
| | 177,9365 | -0,1266 | 4 | 0.059 | | | 27 | |
| | 177,9389 | 55,1363 | 4 | 0.08 | | | 27 | |
| | 177,9391 | 5,0696 | 7 | 0.076 | | | 27 | |
| | 177,9640 | 3,5852 | 4 | 0.081 | | | 27 | |
| | 177,9836 | 53,4801 | 6 | 0.06 | | | 27 | |
| | 178,0433 | 3,3889 | 6 | 0.04 | | | 27 | |
| | 178,1213 | 59,4457 | 5 | 0.066 | | | 27 | |
| | 178,1481 | 4,3400 | 4 | 0.076 | | | 27 | |
| | 178,1771 | -3,5057 | 4 | 0.147 | | | 27 | |
| | 178,2120 | 67,9106 | 5 | 0.117 | | | 27 | |
| | 178,2394 | 49,2480 | 7 | 0.056 | | | 27 | |
| | 178,2571 | 4,3652 | 5 | 0.075 | | | 27 | |
| | 178,2779 | 43,5382 | 4 | 0.043 | | | 27 | |
| | 178,2849 | 55,5899 | 5 | 0.064 | | | 27 | |
| | 178,2955 | 5,3983 | 4 | 0.077 | | | 27 | |
| | 178,3504 | 10,4654 | 7 | 0.022 | | | 27 | |
| | 178,4719 | 57,0150 | 5 | 0.042 | | | 27 | |
| | 178,5740 | 43,1450 | 4 | 0.067 | | | 27 | |
| | 178,6082 | 52,5617 | 4 | 0.071 | | | 27 | |
| | 178,6572 | 59,3339 | 4 | 0.032 | | | 27 | |
| | 178,6923 | 49,2628 | 5 | 0.053 | | | 27 | |
| | 178,6975 | 48,5362 | 6 | 0.076 | | | 27 | |
| | 178,7057 | 9,8062 | 4 | 0.103 | | | 27 | |
| | 178,7154 | 50,9055 | 5 | 0.053 | | | 27 | |
| SDSS 048 | 178,7708 | 0,5836 | 5 | 0,131 | | | 8 | |
| | 178,7769 | 55,0466 | 8 | 0.052 | | | 27 | |
| | 178,8055 | 44,5709 | 9 | 0.071 | | | 27 | |
| | 178,8390 | 52,7348 | 4 | 0.07 | | | 27 | |
| | 178,8941 | 52,8298 | 5 | 0.069 | | | 27 | |
| | 178,9194 | 55,2855 | 9 | 0.051 | | | 27 | |
| SDSS 049 | 178,9208 | 0,9242 | 4 | 0,107 | | | 8 | |
| SDSS 050 | 178,9250 | 0,8858 | 4 | 0,208 | | | 8 | |
| | 178,9412 | 48,3136 | 6 | 0.032 | | | 27 | |
| | 179,0114 | 43,7227 | 4 | 0.07 | | | 27 | |
| | 179,0228 | 67,9795 | 4 | 0.063 | | | 27 | |
| PCG115606+021907 | 179,0250 | 2,3186 | 4 | | 0.0800 | A | 7 | group member/coord.from Ref.(6) |
| PCG115610+031802 | 179,0417 | 3,3006 | 4 | | 0.0720 | C | 7 | group member/coord.from Ref.(6) |
| | 179,0572 | 43,8923 | 6 | 0.072 | | | 27 | |
| | 179,0679 | 4,2037 | 4 | 0.074 | | | 27 | |
| | 179,0726 | 4,9696 | 8 | 0.074 | | | 27 | |
| | 179,0779 | 2,9449 | 5 | 0.081 | | | 27 | |
| | 179,0923 | 43,9113 | 5 | 0.071 | | | 27 | |
| | 179,1125 | 4,9648 | 4 | 0.073 | | | 27 | |
| | 179,1271 | 10,1854 | 4 | 0.066 | | | 27 | |
| SDSS 051 | 179,1542 | 1,1211 | 4 | 0,157 | | | 8 | |
| | 179,1578 | 52,6021 | 6 | 0.052 | | | 27 | |
| | 179,2191 | 3,9366 | 6 | 0.096 | | | 27 | |
| PCG115655+241600 | 179,2292 | 24,2669 | 5 | | 0.1420 | A | 7 | group member/coord.from Ref.(6) |
| | 179,2712 | 3,0538 | 6 | 0.097 | | | 27 | |
| | 179,2737 | 55,0876 | 5 | 0.064 | | | 27 | |
| | 179,2873 | 54,0513 | 5 | 0.08 | | | 27 | |
| | 179,2982 | 48,5345 | 4 | 0.071 | | | 27 | |
| | 179,3113 | -0,7710 | 5 | 0.126 | | | 27 | |
| | 179,3135 | 5,0593 | 8 | 0.077 | | | 27 | |
| | 179,3482 | 5,1969 | 4 | 0.078 | | | 27 | |
| | 179,3885 | 10,0696 | 4 | 0.09 | | | 27 | |
| RSCG 47 | 179,4042 | 32,3050 | 4 | 0,011 | | | 9 | |
| SDSS 052 | 179,4333 | 0,8736 | 4 | 0,132 | | | 8 | |
| | 179,4532 | 66,6214 | 5 | 0.046 | | | 27 | |
| | 179,4661 | 5,1059 | 6 | 0.075 | | | 27 | |
| | 179,4661 | -2,5530 | 7 | 0.081 | | | 27 | |
| | 179,4695 | 0,2216 | 5 | 0.108 | | | 27 | |
| | 179,4875 | 45,1873 | 7 | 0.054 | | | 27 | |
| | 179,5315 | -2,4932 | 4 | 0.083 | | | 27 | |
| LCCG 34 | 179,5417 | -2,5564 | 5 | 0,080 | | | 2 | |
| SDSS 053 | 179,5833 | 0,0872 | 4 | 0,106 | | | 8 | |
| | 179,5836 | -1,8764 | 6 | 0.02 | | | 27 | |
| | 179,5988 | 41,4594 | 4 | 0.067 | | | 27 | |

| | | | | | |
|---|---|---|---|---|---|
| | 179,6006 | -1,7660 | 8 | 0.021 | 27 |
| SDSS 054 | 179,6083 | 0,7114 | 4 | 0,048 | 8 |
| | 179,6685 | 54,0068 | 7 | 0.052 | 27 |
| | 179,6919 | 10,5593 | 4 | 0.089 | 27 |
| | 179,7071 | 10,6691 | 9 | 0.061 | 27 |
| | 179,8114 | 41,2204 | 4 | 0.078 | 27 |
| | 179,8228 | 56,0315 | 7 | 0.065 | 27 |
| | 179,8379 | 52,8849 | 7 | 0.062 | 27 |
| | 179,8398 | 54,8369 | 4 | 0.08 | 27 |
| | 179,8701 | -1,5477 | 4 | 0.052 | 27 |
| | 179,8863 | 55,4540 | 6 | 0.08 | 27 |
| | 179,8949 | 56,5494 | 4 | 0.052 | 27 |
| | 179,9029 | 49,9559 | 4 | 0.085 | 27 |
| | 179,9209 | 48,9963 | 5 | 0.079 | 27 |
| | 179,9378 | 56,1354 | 7 | 0.067 | 27 |
| | 180,0200 | -0,4106 | 4 | 0.08 | 27 |
| | 180,0220 | 45,5322 | 5 | 0.067 | 27 |
| SDSS 055 | 180,0375 | 0,6739 | 4 | 0,085 | 8 |
| SDSS 056 | 180,0417 | 0,5986 | 4 | 0,169 | 8 |
| SDSS 057 | 180,0542 | 0,4261 | 5 | 0,077 | 8 |
| | 180,0934 | 43,5681 | 4 | 0.073 | 27 |
| | 180,1304 | 3,4599 | 4 | 0.086 | 27 |
| UZCCG154 | 180,1333 | 31,8850 | 3 | 0,026 | 5 |
| | 180,1598 | 5,0009 | 4 | 0.078 | 27 |
| | 180,1651 | 51,8683 | 5 | 0.067 | 27 |
| | 180,1679 | 10,6698 | 9 | 0.062 | 27 |
| | 180,1886 | 43,6038 | 7 | 0.072 | 27 |
| | 180,2093 | 10,7026 | 7 | 0.063 | 27 |
| LCCG 35 | 180,2208 | -11,6816 | 3 | 0,084 | 2 |
| UZCCG155 | 180,2250 | 13,8564 | 4 | 0,005 | 5 |
| | 180,2920 | 56,3586 | 7 | 0.061 | 27 |
| | 180,2963 | 54,2026 | 4 | 0.056 | 27 |
| | 180,3090 | 58,0027 | 4 | 0.101 | 27 |
| | 180,3169 | 3,6499 | 4 | 0.091 | 27 |
| | 180,3255 | 2,2773 | 5 | 0.078 | 27 |
| | 180,3447 | 65,7789 | 4 | 0.079 | 27 |
| | 180,3682 | 51,5894 | 4 | 0.063 | 27 |
| | 180,3776 | 1,7892 | 6 | 0.078 | 27 |
| | 180,3790 | 58,0760 | 6 | 0.103 | 27 |
| | 180,3831 | 0,9882 | 6 | 0.085 | 27 |
| | 180,3911 | 46,4751 | 5 | 0.102 | 27 |
| | 180,4594 | 1,1896 | 4 | 0.082 | 27 |
| | 180,4978 | 58,0408 | 4 | 0.105 | 27 |
| | 180,5490 | 59,1990 | 4 | 0.033 | 27 |
| | 180,5911 | 5,4306 | 4 | 0.078 | 27 |
| | 180,6621 | 59,1684 | 4 | 0.055 | 27 |

| | | | | | | | | |
|---|---|---|---|---|---|---|---|---|
| | 180,6674 | 4,2966 | 4 | 0.021 | | | 27 | |
| | 180,6714 | 43,4962 | 7 | 0.054 | | | 27 | |
| | 180,7458 | 41,5175 | 4 | 0.053 | | | 27 | |
| HCG 060 | 180,7708 | 51,6931 | 4 | 0,063 | | | 1 | |
| | 180,8443 | 60,8971 | 5 | 0.064 | | | 27 | |
| | 180,8500 | 1,8977 | 9 | 0.021 | | | 27 | |
| | 180,8710 | 52,8811 | 4 | 0.063 | | | 27 | |
| | 180,9241 | 3,7800 | 4 | 0.113 | | | 27 | |
| | 180,9261 | 43,4671 | 4 | 0.101 | | | 27 | |
| | 180,9312 | 57,2985 | 4 | 0.065 | | | 27 | |
| | 180,9317 | 60,7432 | 6 | 0.066 | | | 27 | |
| | 180,9467 | 43,3658 | 6 | 0.073 | | | 27 | |
| | 180,9847 | 1,3116 | 9 | 0.083 | | | 27 | |
| | 180,9848 | 54,6782 | 9 | 0.049 | | | 27 | |
| | 180,9932 | 57,2540 | 4 | 0.064 | | | 27 | |
| | 181,0300 | 4,2945 | 4 | 0.137 | | | 27 | |
| UZCCG156 | 181,0625 | 20,2900 | 6 | 0,023 | | | 5 | |
| | 181,1191 | 54,6901 | 5 | 0.052 | | | 27 | |
| | 181,1315 | 52,3052 | 4 | 0.063 | | | 27 | |
| | 181,1453 | 42,9284 | 7 | 0.052 | | | 27 | |
| | 181,1779 | 42,5729 | 5 | 0.054 | | | 27 | |
| | 181,2624 | 54,2942 | 4 | 0.065 | | | 27 | |
| | 181,2820 | 61,9682 | 5 | 0.05 | | | 27 | |
| | 181,2983 | 10,4892 | 5 | 0.096 | | | 27 | |
| | 181,3098 | 4,1313 | 4 | 0.138 | | | 27 | |
| | 181,3346 | 54,0928 | 8 | 0.05 | | | 27 | |
| | 181,3414 | 10,9486 | 8 | 0.063 | | | 27 | |
| PCG120525+500032 | 181,3542 | 50,0092 | 4 | | 0.0950 | A | 7 | group member/coord.from Ref.(6) |
| | 181,3734 | 54,1236 | 4 | 0.081 | | | 27 | |
| | 181,4047 | 48,6967 | 9 | 0.064 | | | 27 | |
| | 181,4171 | 10,6997 | 4 | 0.033 | | | 27 | |
| | 181,4292 | 67,1289 | 4 | 0.117 | | | 27 | |
| | 181,4350 | 51,3936 | 8 | 0.085 | | | 27 | |
| UZCCG157 | 181,4583 | 20,5483 | 6 | 0,024 | | | 5 | |
| | 181,4642 | 47,2628 | 5 | 0.115 | | | 27 | |
| | 181,4664 | 54,1958 | 4 | 0.051 | | | 27 | |
| | 181,6107 | 0,9659 | 4 | 0.113 | | | 27 | |
| | 181,6113 | 50,1657 | 4 | 0.034 | | | 27 | |
| | 181,6336 | 51,5664 | 4 | 0.091 | | | 27 | |
| | 181,6410 | 42,7273 | 5 | 0.074 | | | 27 | |
| | 181,6730 | 68,5780 | 4 | 0.105 | | | 27 | |
| RSCG 48 | 181,7083 | 43,1300 | 3 | 0,003 | | | 9 | |
| UZCCG158 | 181,7208 | 67,2586 | 4 | 0,008 | | | 5 | |
| | 181,8033 | 62,9782 | 4 | 0.04 | | | 27 | |
| | 181,8145 | 57,4347 | 4 | 0.065 | | | 27 | |
| | 181,9101 | 50,7296 | 4 | 0.058 | | | 27 | |
| | 181,9109 | 64,4375 | 4 | 0.11 | | | 27 | |
| | 181,9348 | 9,7889 | 6 | 0.069 | | | 27 | |
| | 181,9429 | 44,0258 | 4 | 0.113 | | | 27 | |
| | 181,9527 | 10,3960 | 5 | 0.095 | | | 27 | |
| | 181,9720 | 56,9343 | 5 | 0.062 | | | 27 | |
| | 181,9965 | 53,3127 | 4 | 0.083 | | | 27 | |
| | 182,0256 | 57,8355 | 5 | 0.047 | | | 27 | |
| | 182,0673 | 56,4672 | 6 | 0.064 | | | 27 | |
| | 182,0841 | 56,4982 | 8 | 0.065 | | | 27 | |
| | 182,0917 | 52,0259 | 5 | 0.06 | | | 27 | |
| | 182,1109 | 2,6785 | 4 | 0.078 | | | 27 | |
| | 182,1116 | 61,3396 | 4 | 0.102 | | | 27 | |
| | 182,1306 | 3,8590 | 4 | 0.077 | | | 27 | |
| | 182,1468 | 2,0394 | 4 | 0.078 | | | 27 | |
| | 182,1675 | 57,6327 | 4 | 0.064 | | | 27 | |
| | 182,2050 | 54,6149 | 4 | 0.09 | | | 27 | |
| | 182,2081 | 53,2600 | 4 | 0.086 | | | 27 | |
| | 182,2472 | 10,0493 | 4 | 0.114 | | | 27 | |
| | 182,2482 | 52,4450 | 4 | 0.062 | | | 27 | |
| UZCCG159 | 182,2500 | 29,2439 | 3 | 0,013 | | | 5 | |
| | 182,2511 | 53,2971 | 9 | 0.082 | | | 27 | |
| | 182,3684 | 53,2651 | 5 | 0.081 | | | 27 | |
| | 182,3731 | 56,7500 | 9 | 0.063 | | | 27 | |
| | 182,3737 | 10,6871 | 7 | 0.068 | | | 27 | |
| | 182,3747 | 10,8013 | 4 | 0.069 | | | 27 | |
| SDSS 061 | 182,4667 | 0,5675 | 5 | 0,187 | | | 8 | |
| | 182,4832 | 47,2057 | 4 | 0.031 | | | 27 | |
| | 182,5130 | 56,7756 | 5 | 0.052 | | | 27 | |
| | 182,5306 | 53,9676 | 7 | 0.05 | | | 27 | |
| | 182,5899 | 5,3955 | 4 | 0.08 | | | 27 | |
| | 182,6180 | 5,2901 | 5 | 0.079 | | | 27 | |
| | 182,6401 | 5,2652 | 4 | 0.075 | | | 27 | |
| | 182,6476 | 0,7532 | 5 | 0.021 | | | 27 | |
| | 182,7081 | 0,5324 | 5 | 0.02 | | | 27 | |
| | 182,8164 | 0,1152 | 4 | 0.078 | | | 27 | |
| UZCCG160 | 182,8500 | 11,9961 | 3 | 0,006 | | | 5 | |
| | 182,8546 | 67,0467 | 5 | 0.059 | | | 27 | |
| | 182,8576 | 42,1183 | 5 | 0.023 | | | 27 | |
| | 182,9056 | 53,0483 | 7 | 0.049 | | | 27 | |
| | 182,9417 | 5,8993 | 4 | 0.086 | | | 27 | |
| | 182,9486 | 61,5354 | 4 | 0.05 | | | 27 | |
| | 182,9736 | 53,0147 | 4 | 0.082 | | | 27 | |
| J1211+1344 | 182,9875 | 13,7392 | 4 | | 0.0808 | A | 4 | not CG / discordant redshifts |
| | | | | | 0.1739 | B | 4 | |
| | | | | | 0.0686 | C | 4 | |
| | | | | | 0.2926 | D | 4 | |

| | | | | | | | | | |
|---|---|---|---|---|---|---|---|---|---|
| PCG121157+134421 | 182,9875 | 13,7392 | 4 | | 0.0970 | B | | 7 | group member/coord.from Ref.(6) |
| | 183,0034 | 50,4473 | 5 | 0.095 | | | | 27 | |
| | 183,0066 | 59,6137 | 4 | 0.045 | | | | 27 | |
| | 183,0131 | 5,3926 | 5 | 0.076 | | | | 27 | |
| | 183,0231 | 62,7581 | 4 | 0.051 | | | | 27 | |
| | 183,0253 | 54,1902 | 4 | 0.048 | | | | 27 | |
| UZCCG161 | 183,0792 | 15,3558 | 3 | 0,024 | | | | 5 | |
| | 183,0952 | 5,6014 | 5 | 0.112 | | | | 27 | |
| HCG 061 | 183,1000 | 29,1778 | 4 | 0,013 | | | ROSE10 | 1 | |
| | | | 3 | 0,013 | | | UZCCG162 | 5 | |
| | | | 3 | 0,013 | | | RSCG 49 | 9 | |
| | 183,1265 | 67,0071 | 4 | 0.032 | | | | 27 | |
| | 183,1701 | 45,6050 | 6 | 0.066 | | | | 27 | |
| J1212+2235 | 183,2167 | 22,5886 | 4 | 0,085 | | | | 4 | group coord.from ref.(3) |
| | 183,2175 | 61,4896 | 4 | 0.101 | | | | 27 | |
| | 183,2359 | 6,1216 | 5 | 0.138 | | | | 27 | |
| | 183,2392 | 3,0291 | 5 | 0.074 | | | | 27 | |
| | 183,2816 | 51,5019 | 4 | 0.086 | | | | 27 | |
| | 183,3031 | 5,6727 | 5 | 0.083 | | | | 27 | |
| | 183,3050 | 59,1881 | 5 | 0.098 | | | | 27 | |
| | 183,3303 | 42,4086 | 5 | 0.074 | | | | 27 | |
| UZCCG163 | 183,4125 | 6,9875 | 3 | 0,008 | | | | 5 | |
| | 183,4785 | 1,9686 | 4 | 0.079 | | | | 27 | |
| | 183,4886 | 5,9842 | 4 | 0.08 | | | | 27 | |
| PCG121359+015956 | 183,5000 | 1,9992 | 4 | | 0.0720 | B | | 7 | group member/coord.from Ref.(6) |
| | 183,5038 | 56,7344 | 5 | 0.052 | | | | 27 | |
| | 183,5269 | 59,9890 | 8 | 0.059 | | | | 27 | |
| | 183,5298 | 50,5574 | 7 | 0.083 | | | | 27 | |
| SDSS 062 | 183,5667 | 0,4497 | 4 | 0,245 | | | | 8 | |
| | 183,6331 | 41,4845 | 4 | 0.049 | | | | 27 | |
| | 183,6396 | 5,4331 | 5 | 0.078 | | | | 27 | |
| | 183,7029 | 9,1766 | 4 | 0.074 | | | | 27 | |
| | 183,7204 | 5,0526 | 4 | 0.077 | | | | 27 | |
| | 183,7429 | 5,0544 | 4 | 0.079 | | | | 27 | |
| RSCG 50 | 183,7500 | 13,1389 | 3 | 0,002 | | | | 9 | |
| | 183,7594 | -2,6893 | 4 | 0.11 | | | | 27 | |
| J1215+1534 | 183,8167 | 15,5667 | 4 | | 0.0983 | A | | 4 | not CG / discordant redshifts |
| | | | | | 0.1045 | B | | 4 | |
| | | | | | 0.0816 | C | | 4 | |
| | | | | | 0.0985 | D | | 4 | |
| UZCCG164 | 183,8542 | 23,9706 | 4 | 0,023 | | | | 5 | |
| | 183,8674 | 44,1408 | 4 | 0.113 | | | | 27 | |
| SDSS 063 | 183,8792 | 0,6425 | 4 | 0,075 | | | | 8 | |
| | 183,8996 | 5,7448 | 4 | 0.126 | | | | 27 | |
| | 183,9369 | 5,6375 | 4 | 0.079 | | | | 27 | |
| | 183,9504 | -0,4857 | 4 | 0.076 | | | | 27 | |

| | | | | | | | | |
|---|---|---|---|---|---|---|---|---|
| | 183,9742 | 60,5008 | 4 | 0.049 | | | 27 | |
| | 183,9863 | 3,3197 | 4 | 0.077 | | | 27 | |
| | 184,0094 | -0,5630 | 5 | 0.021 | | | 27 | |
| | 184,0260 | 44,6270 | 5 | 0.076 | | | 27 | |
| | 184,0414 | 1,2766 | 4 | 0.075 | | | 27 | |
| | 184,0658 | 60,1410 | 7 | 0.043 | | | 27 | |
| | 184,0688 | 5,4499 | 4 | 0.115 | | | 27 | |
| | 184,1425 | -3,2625 | 8 | 0.111 | | | 27 | |
| | 184,2234 | 57,9928 | 5 | 0.05 | | | 27 | |
| | 184,2236 | 44,1002 | 5 | 0.067 | | | 27 | |
| | 184,2314 | 3,6070 | 4 | 0.077 | | | 27 | |
| LCCG 36 | 184,2667 | -3,2000 | 5 | 0,107 | | | 2 | |
| SDSS 064 | 184,2708 | 0,9150 | 4 | 0,198 | | | 8 | |
| | 184,2728 | 60,0495 | 6 | 0.044 | | | 27 | |
| | 184,2767 | 5,3354 | 4 | 0.081 | | | 27 | |
| | 184,2780 | -0,7740 | 4 | 0.07 | | | 27 | |
| | 184,3470 | 3,7037 | 7 | 0.082 | | | 27 | |
| | 184,3703 | 3,7276 | 6 | 0.074 | | | 27 | |
| PCG121740+033933 | 184,4208 | 3,6594 | 4 | | 0.0770 | C | 7 | group member/coord.from Ref.(6) |
| | 184,4227 | 45,6230 | 5 | 0.069 | | | 27 | |
| | 184,4246 | 54,0572 | 4 | 0.089 | | | 27 | |
| | 184,4438 | 54,6746 | 4 | 0.102 | | | 27 | |
| | 184,4463 | 3,7190 | 7 | 0.073 | | | 27 | |
| | 184,4629 | 3,6213 | 7 | 0.075 | | | 27 | |
| | 184,4684 | 3,6889 | 7 | 0.079 | | | 27 | |
| | 184,4747 | 4,1562 | 4 | 0.079 | | | 27 | |
| | 184,5112 | 5,2499 | 8 | 0.077 | | | 27 | |
| | 184,5151 | 5,3243 | 7 | 0.076 | | | 27 | |
| | 184,5163 | 43,2538 | 4 | 0.038 | | | 27 | |
| | 184,5379 | 64,2881 | 5 | 0.097 | | | 27 | |
| | 184,5947 | 2,4161 | 5 | 0.075 | | | 27 | |
| | 184,6506 | -1,5112 | 4 | 0.069 | | | 27 | |
| | 184,6654 | 52,1526 | 4 | 0.057 | | | 27 | |
| | 184,6800 | 5,3214 | 4 | 0.073 | | | 27 | |
| | 184,6960 | 48,8601 | 6 | 0.045 | | | 27 | |
| | 184,7182 | 63,5287 | 4 | 0.108 | | | 27 | |
| | 184,7186 | 0,0858 | 4 | 0.117 | | | 27 | |
| | 184,8039 | 11,9585 | 7 | 0.026 | | | 27 | |
| | 184,8157 | 3,0938 | 7 | 0.077 | | | 27 | |
| | 184,8368 | 12,0010 | 4 | 0.065 | | | 27 | |
| | 184,8612 | 47,8137 | 8 | 0.059 | | | 27 | |
| | 184,8916 | 12,3038 | 5 | 0.027 | | | 27 | |
| SHK 202 | 184,9042 | 28,3892 | 16 | 0,028 | | | 11 | mean value/coord.from Ref.(10) |
| SHK 202 | 184,9042 | 28,3892 | | 0,027 | | | 17 | |
| | 184,9145 | 42,4259 | 5 | 0.075 | | | 27 | |
| UZCCG165 | 184,9250 | 28,7908 | 3 | 0,026 | | | 5 | |

| | | | | | | | | | |
|---|---|---|---|---|---|---|---|---|---|
| | 184,9388 | 48,5986 | 4 | 0.117 | | | | 27 | |
| | 184,9946 | 4,3334 | 5 | 0.078 | | | | 27 | |
| | 184,9967 | 46,1484 | 5 | 0.039 | | | | 27 | |
| | 185,0320 | 50,9007 | 6 | 0.047 | | | | 27 | |
| | 185,0511 | -1,1069 | 4 | 0.118 | | | | 27 | |
| | 185,0602 | 3,6340 | 5 | 0.077 | | | | 27 | |
| | 185,0739 | 49,1427 | 4 | 0.119 | | | | 27 | |
| | 185,1203 | 42,5881 | 6 | 0.048 | | | | 27 | |
| | 185,1605 | 55,6182 | 5 | 0.031 | | | | 27 | |
| | 185,2081 | 56,0337 | 6 | 0.034 | | | | 27 | |
| | 185,2498 | 46,6932 | 4 | 0.063 | | | | 27 | |
| RSCG 51 | 185,2500 | 12,3083 | 7 | 0,002 | | | | 9 | |
| RSCG 52 | 185,2500 | 39,8811 | 3 | 0,022 | | | | 9 | |
| | | | 3 | 0,023 | | | UZCCG166 | 5 | |
| | 185,2798 | 0,9198 | 4 | 0.099 | | | | 27 | |
| | 185,3497 | 50,4888 | 7 | 0.089 | | | | 27 | |
| | 185,3542 | -2,9707 | 6 | 0.072 | | | | 27 | |
| PCG122142+554821 | 185,4250 | 55,8061 | 4 | | 0.0340 | C | | 7 | group member/coord.from Ref.(6) |
| | 185,4503 | 9,6464 | 4 | 0.084 | | | | 27 | |
| RSCG 53 | 185,4833 | 14,6811 | 3 | 0,004 | | | | 9 | |
| | 185,5379 | 49,6182 | 4 | 0.085 | | | | 27 | |
| | 185,5388 | 1,4237 | 6 | 0.079 | | | | 27 | |
| | 185,5948 | 51,0915 | 6 | 0.031 | | | | 27 | |
| | 185,6446 | 10,0649 | 5 | 0.067 | | | | 27 | |
| | 185,6447 | -2,9290 | 5 | 0.113 | | | | 27 | |
| | 185,6467 | 1,0410 | 4 | 0.074 | | | | 27 | |
| SDSS 065 | 185,6667 | 0,9806 | 4 | 0,175 | | | | 8 | |
| | 185,7083 | 12,3539 | 4 | 0.092 | | | | 27 | |
| | 185,7539 | 10,6770 | 6 | 0.026 | | | | 27 | |
| | 185,7633 | 10,6350 | 6 | 0.025 | | | | 27 | |
| | 185,7702 | 57,1503 | 4 | 0.068 | | | | 27 | |
| | 185,8018 | 10,3894 | 6 | 0.068 | | | | 27 | |
| | 185,9138 | 55,1083 | 5 | 0.034 | | | | 27 | |
| | 185,9374 | 44,2379 | 5 | 0.068 | | | | 27 | |
| UZCCG167 | 185,9375 | 58,3969 | 3 | 0,015 | | | | 5 | |
| | 185,9872 | 61,2049 | 5 | 0.051 | | | | 27 | |
| | 185,9879 | 66,5720 | 5 | 0.087 | | | | 27 | |
| | 186,0051 | 63,2749 | 4 | 0.106 | | | | 27 | |
| UZCCG168 | 186,0292 | 9,3114 | 6 | 0,024 | | | | 5 | |
| | 186,0778 | 63,3169 | 4 | 0.104 | | | | 27 | |
| | 186,1610 | 61,5050 | 7 | 0.07 | | | | 27 | |
| | 186,1921 | 2,7005 | 4 | 0.026 | | | | 27 | |
| SHK 245 | 186,2208 | 31,9564 | 5 | 0,063 | | | | 25 | |
| | 186,3071 | 9,0103 | 4 | 0.023 | | | | 27 | |
| SDSS 066 | 186,3083 | 0,1808 | 4 | 0,287 | | | | 8 | |
| | 186,3152 | 64,2501 | 5 | 0.111 | | | | 27 | |
| | 186,3165 | 43,1358 | 5 | 0.074 | | | | 27 | |
| RSCG 54 | 186,3417 | 18,1981 | 3 | 0,003 | | | | 9 | |
| | 186,3598 | 45,4888 | 8 | 0.024 | | | | 27 | |
| | 186,3963 | 66,4061 | 7 | 0.086 | | | | 27 | |
| | 186,4939 | 9,1288 | 9 | 0.025 | | | | 27 | |
| | 186,5328 | 61,5299 | 8 | 0.071 | | | | 27 | |
| UZCCG169 | 186,6417 | 9,0306 | 4 | 0,025 | | | R 55 | 5 | |
| RSCG 55 | 186,6542 | 9,0342 | 4 | 0,025 | | | | 9 | |
| UZCCG170 | 186,6792 | -1,0294 | 3 | 0,007 | | | | 5 | |
| | 186,7056 | 49,8702 | 4 | 0.086 | | | | 27 | |
| | 186,7106 | 43,3908 | 4 | 0.066 | | | | 27 | |
| | 186,7586 | 50,6203 | 5 | 0.041 | | | | 27 | |
| | 186,8593 | 8,8179 | 4 | 0.094 | | | | 27 | |
| RSCG 56 | 186,8667 | 12,4231 | 5 | 0,003 | | | | 9 | |
| | 186,8896 | 8,8781 | 6 | 0.09 | | | | 27 | |
| | 186,9015 | 52,5247 | 5 | 0.084 | | | | 27 | |
| | 186,9018 | -0,4469 | 4 | 0.114 | | | | 27 | |
| | 187,0010 | 0,0776 | 4 | 0.023 | | | | 27 | |
| | 187,0250 | 48,1135 | 4 | 0.026 | | | | 27 | |
| | 187,0341 | 1,6107 | 5 | 0.077 | | | | 27 | |
| | 187,0657 | 9,2778 | 4 | 0.09 | | | | 27 | |
| RSCG 57 | 187,1083 | 9,6564 | 6 | 0,002 | | | | 9 | |
| RSCG 58 | 187,1167 | 13,0900 | 3 | 0,002 | | | | 9 | |
| | 187,1457 | 56,5663 | 4 | 0.053 | | | | 27 | |
| | 187,1617 | 52,1845 | 4 | 0.083 | | | | 27 | |
| PCG122850-010938 | 187,2125 | -1,1608 | 4 | | 0.1150 | A | | 7 | group member/coord.from Ref.(6) |
| | 187,2404 | 53,4769 | 4 | 0.036 | | | | 27 | |
| | 187,2707 | 63,9698 | 5 | 0.111 | | | | 27 | |
| | 187,2708 | 51,2677 | 4 | 0.086 | | | | 27 | |
| PCG122905+083949 | 187,2750 | 8,6636 | 4 | | 0.0910 | A | | 7 | group member/coord.from Ref.(6) |
| | 187,3324 | -1,9760 | 4 | 0.083 | | | | 27 | |
| PCG122925+114733 | 187,3542 | 11,7928 | 4 | | 0.0860 | A | | 7 | group member/coord.from Ref.(6) |
| RSCG 59 | 187,3667 | 14,0211 | 3 | 0,004 | | | | 9 | |
| RSCG 60 | 187,3750 | 8,8639 | 4 | 0,002 | | | | 9 | |
| | 187,4122 | 65,4112 | 4 | 0.05 | | | | 27 | |
| | 187,4969 | 2,9697 | 4 | 0.08 | | | | 27 | |
| | 187,5294 | 5,6072 | 4 | 0.069 | | | | 27 | |
| RSCG 61 | 187,6333 | 12,3603 | 5 | 0,005 | | | | 9 | |
| | 187,6373 | 9,9137 | 5 | 0.085 | | | | 27 | |
| | 187,6658 | 50,7762 | 4 | 0.031 | | | | 27 | |
| | 187,6857 | 51,4612 | 5 | 0.044 | | | | 27 | |
| | 187,6917 | 3,0453 | 6 | 0.081 | | | | 27 | |
| | 187,6919 | 47,1854 | 9 | 0.039 | | | | 27 | |
| | 187,6973 | 3,5828 | 4 | 0.098 | | | | 27 | |
| | 187,7031 | 65,7236 | 4 | 0.082 | | | | 27 | |
| | 187,7150 | 47,1045 | 4 | 0.04 | | | | 27 | |

| | | | | | | | | |
|---|---|---|---|---|---|---|---|---|
| | 187,7595 | 5,1458 | 6 | 0.08 | | | 27 | |
| | 187,7964 | 50,4338 | 8 | 0.03 | | | 27 | |
| | 187,8371 | 9,9221 | 4 | 0.129 | | | 27 | |
| | 187,8587 | 0,9058 | 4 | 0.023 | | | 27 | |
| RSCG 62 | 187,8708 | 11,6897 | 5 | 0,003 | | | 9 | |
| | 187,8908 | 10,6651 | 7 | 0.086 | | | 27 | |
| SDSS 067 | 187,9167 | 0,5683 | 4 | 0,202 | | | 8 | |
| UZCCG171 | 187,9333 | 64,0208 | 4 | 0,008 | | | 5 | |
| | 188,0881 | 66,3260 | 4 | 0.047 | | | 27 | |
| UZCCG172 | 188,1250 | 15,1131 | 3 | 0,008 | | | 5 | |
| | 188,1288 | 64,0339 | 4 | 0.079 | | | 27 | |
| | 188,1354 | 52,9639 | 5 | 0.119 | | | 27 | |
| UZCCG173 | 188,1750 | 0,2147 | 3 | 0,005 | | | 5 | |
| | 188,3422 | 67,1640 | 4 | 0.104 | | | 27 | |
| | 188,3637 | -3,3550 | 4 | 0.063 | | | 27 | |
| | 188,4484 | 62,2578 | 4 | 0.075 | | | 27 | |
| | 188,4488 | 6,1077 | 4 | 0.084 | | | 27 | |
| SDSS 068 | 188,4542 | 0,1589 | 5 | 0,135 | | | 8 | |
| | 188,5096 | 56,7718 | 4 | 0.151 | | | 27 | |
| | 188,5308 | 67,0903 | 4 | 0.105 | | | 27 | |
| | 188,5442 | 5,5439 | 8 | 0.081 | | | 27 | |
| | 188,5651 | 50,7037 | 6 | 0.039 | | | 27 | |
| | 188,6071 | 50,8686 | 7 | 0.041 | | | 27 | |
| | 188,6225 | 3,8915 | 6 | 0.085 | | | 27 | |
| | 188,6585 | 47,5101 | 9 | 0.031 | | | 27 | |
| PCG123439+234951 | 188,6625 | 23,8311 | 5 | | 0.1320 | D | 7 | group member/coord.from Ref.(6) |
| | 188,6911 | 61,7044 | 4 | 0.115 | | | 27 | |
| | 188,7140 | 58,9588 | 8 | 0.043 | | | 27 | |
| | 188,7218 | 1,8228 | 7 | 0.077 | | | 27 | |
| | 188,7575 | 50,7627 | 4 | 0.04 | | | 27 | |
| | 188,7891 | 1,7778 | 5 | 0.082 | | | 27 | |
| | 188,7916 | 1,7996 | 4 | 0.079 | | | 27 | |
| | 188,7994 | 2,0009 | 6 | 0.08 | | | 27 | |
| PCG123512+014705 | 188,8000 | 1,7847 | 4 | | 0.0790 | B | 7 | group member/coord.from Ref.(6) |
| | 188,8044 | 4,7529 | 4 | 0.085 | | | 27 | |
| | 188,8349 | -1,7049 | 4 | 0.023 | | | 27 | |
| SHK 205 | 188,8458 | 27,5747 | 14 | 0,096 | | | 11 | mean value/coord.from Ref.(10) |
| | 188,8591 | -0,0239 | 4 | 0.023 | | | 27 | |
| | 188,8622 | 1,8466 | 5 | 0.081 | | | 27 | |
| RSCG 63 | 188,8750 | 12,4872 | 4 | 0,002 | | | 9 | |
| | 188,9114 | -3,6530 | 4 | 0.062 | | | 27 | |
| | 188,9139 | 4,9875 | 4 | 0.065 | | | 27 | |
| | 188,9149 | 1,8327 | 8 | 0.078 | | | 27 | |
| | 189,0039 | 55,9488 | 5 | 0.102 | | | 27 | |
| | 189,0824 | 4,5447 | 4 | 0.085 | | | 27 | |
| | 189,1167 | 5,5295 | 8 | 0.094 | | | 27 | |

| | | | | | | | | | |
|---|---|---|---|---|---|---|---|---|---|
| SDSS 070 | 189,2125 | 0,5858 | 4 | 0,009 | | | | 8 | |
| | 189,3442 | 57,1995 | 4 | 0.106 | | | | 27 | |
| | 189,4325 | 64,5014 | 5 | 0.102 | | | | 27 | |
| SDSS 071 | 189,4583 | 0,3003 | 4 | 0,141 | | | | 8 | |
| | 189,4661 | 3,3141 | 9 | 0.085 | | | | 27 | |
| | 189,4912 | 10,2182 | 4 | 0.042 | | | | 27 | |
| | 189,5706 | 42,3741 | 4 | 0.066 | | | | 27 | |
| | 189,6178 | 59,4143 | 4 | 0.042 | | | | 27 | |
| | 189,7371 | -3,1300 | 5 | 0.065 | | | | 27 | |
| | 189,7665 | 6,1306 | 8 | 0.074 | | | | 27 | |
| | 189,7856 | 0,2936 | 4 | 0.09 | | | | 27 | |
| | 189,8091 | 6,0524 | 4 | 0.073 | | | | 27 | |
| | 189,8368 | 53,3680 | 5 | 0.065 | | | | 27 | |
| | 189,9059 | 1,6794 | 4 | 0.074 | | | | 27 | |
| | 189,9063 | 9,4857 | 4 | 0.083 | | | | 27 | |
| | 189,9073 | 53,2589 | 6 | 0.067 | | | | 27 | |
| | 189,9889 | 9,7002 | 4 | 0.028 | | | | 27 | |
| | 190,1167 | 5,9460 | 6 | 0.076 | | | | 27 | |
| | 190,1482 | 65,1524 | 4 | 0.102 | | | | 27 | |
| | 190,2054 | 61,0179 | 4 | 0.06 | | | | 27 | |
| | 190,2247 | -2,0525 | 5 | 0.089 | | | | 27 | |
| | 190,2980 | 0,8339 | 4 | 0.09 | | | | 27 | |
| RSCG 64 | 190,3667 | 26,0508 | 3 | 0,016 | | | | 9 | |
| | | | 3 | 0,016 | | | UZCCG174 | 5 | |
| | 190,3725 | 41,5134 | 5 | 0.065 | | | | 27 | |
| PCG124140+332431 | 190,4167 | 33,4086 | 4 | | 0.0220 | A | | 7 | group member/coord.from Ref.(6) |
| | 190,4276 | 3,1245 | 5 | 0.093 | | | | 27 | |
| | 190,5131 | -2,1112 | 9 | 0.084 | | | | 27 | |
| | 190,5137 | 52,9054 | 4 | 0.054 | | | | 27 | |
| | 190,5162 | 4,2462 | 5 | 0.047 | | | | 27 | |
| | 190,5354 | -1,9480 | 4 | 0.085 | | | | 27 | |
| | 190,5640 | 68,2694 | 4 | 0.077 | | | | 27 | |
| RSCG 65 | 190,5792 | 11,3683 | 7 | 0,004 | | | | 9 | |
| | 190,6310 | 42,2403 | 8 | 0.071 | | | | 27 | |
| PCG124233-025231 | 190,6375 | -2,8755 | 4 | | 0.0480 | A | | 7 | group member/coord.from Ref.(6) |
| | 190,6614 | -2,2792 | 4 | 0.082 | | | | 27 | |
| | 190,6935 | 63,2382 | 4 | 0.062 | | | | 27 | |
| | 190,7006 | 2,8369 | 5 | 0.086 | | | | 27 | |
| | 190,7080 | 0,5585 | 4 | 0.08 | | | | 27 | |
| SDSS 072 | 190,7875 | 0,7806 | 4 | 0,143 | | | | 8 | |
| LCCG 37 | 190,8000 | -12,2311 | 3 | 0,096 | | | | 2 | |
| RSCG 66 | 190,8250 | 13,1964 | 3 | 0,003 | | | | 9 | |
| | 190,8663 | 2,4000 | 4 | 0.085 | | | | 27 | |
| SDSS 073 | 190,8833 | -1,1161 | 4 | 0,166 | | | | 8 | |
| | 190,9019 | 42,1468 | 6 | 0.074 | | | | 27 | |
| | 190,9520 | 64,4896 | 6 | 0.042 | | | | 27 | |

| | | | | | | | | |
|---|---|---|---|---|---|---|---|---|
| | 190,9531 | -1,9010 | 4 | 0.086 | | | 27 | |
| UZCCG175 | 191,0167 | 0,4567 | 5 | 0,007 | | | 5 | |
| | 191,0864 | 0,7298 | 4 | 0.082 | | | 27 | |
| | 191,1008 | 3,6472 | 5 | 0.048 | | | 27 | |
| | 191,1422 | 50,3706 | 4 | 0.095 | | | 27 | |
| | 191,1449 | 49,8467 | 7 | 0.039 | | | 27 | |
| | 191,1482 | 1,1267 | 4 | 0.09 | | | 27 | |
| | 191,1747 | 53,1145 | 4 | 0.053 | | | 27 | |
| | 191,1766 | -1,6817 | 4 | 0.047 | | | 27 | |
| | 191,1885 | 4,6058 | 5 | 0.087 | | | 27 | |
| SDSS 074 | 191,1917 | -1,0244 | 7 | 0,147 | | | 8 | |
| | 191,1970 | 50,3380 | 4 | 0.094 | | | 27 | |
| | 191,2080 | 2,1783 | 4 | 0.085 | | | 27 | |
| | 191,2363 | 1,7150 | 6 | 0.047 | | | 27 | |
| | 191,2571 | 0,8947 | 4 | 0.09 | | | 27 | |
| | 191,3924 | 0,2369 | 4 | 0.048 | | | 27 | |
| | 191,5260 | 43,5247 | 5 | 0.041 | | | 27 | |
| PCG124615+050223 | 191,5667 | 5,0400 | 4 | | 0.0740 | B | 7 | group member/coord.from Ref.(6) |
| | 191,5822 | 52,2093 | 5 | 0.067 | | | 27 | |
| | 191,6063 | 42,9430 | 4 | 0.111 | | | 27 | |
| UZCCG176 | 191,6500 | 71,4639 | 4 | 0,005 | | | 5 | |
| | 191,6947 | 0,1907 | 4 | 0.091 | | | 27 | |
| | 191,7626 | 0,2381 | 7 | 0.087 | | | 27 | |
| | 191,7795 | 0,2897 | 9 | 0.089 | | | 27 | |
| | 191,7956 | 0,2663 | 4 | 0.09 | | | 27 | |
| | 191,8023 | 54,9930 | 7 | 0.084 | | | 27 | |
| | 191,8221 | 4,8211 | 5 | 0.065 | | | 27 | |
| | 191,8309 | 55,0878 | 4 | 0.081 | | | 27 | |
| | 191,8681 | -2,9685 | 5 | 0.089 | | | 27 | |
| | 191,9368 | 50,0486 | 4 | 0.095 | | | 27 | |
| | 191,9432 | 50,1407 | 4 | 0.096 | | | 27 | |
| | 191,9548 | -0,2270 | 7 | 0.09 | | | 27 | |
| | 191,9559 | -0,1995 | 4 | 0.089 | | | 27 | |
| | 191,9615 | 45,0417 | 6 | 0.077 | | | 27 | |
| | 192,0537 | 48,1900 | 6 | 0.031 | | | 27 | |
| | 192,0629 | 41,7189 | 4 | 0.039 | | | 27 | |
| | 192,0865 | 62,6345 | 4 | 0.106 | | | 27 | |
| | 192,1349 | -1,0665 | 4 | 0.084 | | | 27 | |
| | 192,1386 | -1,2754 | 4 | 0.083 | | | 27 | |
| | 192,1617 | 62,6658 | 4 | 0.105 | | | 27 | |
| | 192,1645 | 62,7296 | 4 | 0.102 | | | 27 | |
| | 192,1874 | 55,0445 | 4 | 0.085 | | | 27 | |
| | 192,1969 | 62,7053 | 4 | 0.103 | | | 27 | |
| | 192,2747 | 46,8722 | 4 | 0.061 | | | 27 | |
| | 192,3604 | 47,3567 | 4 | 0.114 | | | 27 | |
| | 192,3880 | -1,6972 | 9 | 0.084 | | | 27 | |

| | | | | | |
|---|---|---|---|---|---|
| LCCG 38 | 192,4000 | -6,3497 | 3 | 0,125 | 2 |
| | 192,4215 | -3,5847 | 4 | 0.09 | 27 |
| | 192,4485 | -1,4661 | 5 | 0.081 | 27 |
| | 192,5138 | -1,6181 | 7 | 0.082 | 27 |
| | 192,5218 | -2,5869 | 4 | 0.071 | 27 |
| | 192,5442 | 51,0049 | 4 | 0.123 | 27 |
| | 192,5586 | 55,0884 | 8 | 0.083 | 27 |
| SDSS 076 | 192,7417 | 0,7481 | 5 | 0,257 | 8 |
| | 192,8520 | -0,6559 | 4 | 0.083 | 27 |
| | 192,9460 | 45,0907 | 4 | 0.078 | 27 |
| | 192,9738 | 4,5613 | 6 | 0.064 | 27 |
| | 192,9914 | 4,5778 | 5 | 0.065 | 27 |
| | 193,0032 | 47,6536 | 9 | 0.032 | 27 |
| | 193,0439 | 63,2131 | 6 | 0.105 | 27 |
| | 193,0551 | 51,6921 | 4 | 0.036 | 27 |
| | 193,1387 | 50,8819 | 6 | 0.122 | 27 |
| | 193,1478 | 45,0667 | 6 | 0.079 | 27 |
| | 193,1553 | 55,6574 | 6 | 0.064 | 27 |
| | 193,1911 | 0,5006 | 4 | 0.082 | 27 |
| | 193,2253 | 44,9490 | 4 | 0.076 | 27 |
| LCCG 39 | 193,2542 | -12,0836 | 3 | 0,102 | 2 |
| HCG 062 | 193,2833 | -9,2241 | 4 | 0,014 | 1 |
| | 193,2860 | -2,4235 | 4 | 0.047 | 27 |
| | 193,2953 | 48,1088 | 9 | 0.029 | 27 |
| | 193,3261 | 0,6653 | 4 | 0.105 | 27 |
| | 193,3275 | 62,2345 | 6 | 0.061 | 27 |
| SDSS 077 | 193,3458 | 0,4244 | 5 | 0,047 | 8 |
| | 193,3512 | 0,4556 | 5 | 0.081 | 27 |
| | 193,3706 | 44,9893 | 5 | 0.077 | 27 |
| | 193,4566 | 62,1838 | 6 | 0.038 | 27 |
| | 193,5578 | 3,8712 | 8 | 0.049 | 27 |
| | 193,6264 | 46,5748 | 4 | 0.061 | 27 |
| | 193,6368 | 5,0663 | 6 | 0.065 | 27 |
| SDSS 078 | 193,6500 | 0,1789 | 4 | 0,082 | 8 |
| | 193,6827 | 53,0400 | 5 | 0.055 | 27 |
| | 193,6907 | 53,0854 | 5 | 0.054 | 27 |
| | 193,7418 | 62,9256 | 5 | 0.052 | 27 |
| UZCCG177 | 193,7458 | 73,2486 | 3 | 0,005 | 5 |
| SDSS 079 | 193,7875 | -1,0894 | 5 | 0,123 | 8 |
| | 193,8085 | 3,4265 | 4 | 0.082 | 27 |
| | 193,8112 | 63,8155 | 5 | 0.128 | 27 |
| | 193,9184 | -1,4145 | 4 | 0.085 | 27 |
| | 193,9733 | 62,1454 | 4 | 0.105 | 27 |
| | 193,9997 | 52,1837 | 4 | 0.038 | 27 |
| SDSS 080 | 194,0708 | 0,6397 | 4 | 0,135 | 8 |
| | 194,1113 | 6,0499 | 6 | 0.092 | 27 |

| | | | | | | | |
|---|---|---|---|---|---|---|---|
| | 194,1317 | 68,3044 | 4 | 0.07 | | 27 | |
| | 194,1616 | 63,6656 | 4 | 0.132 | | 27 | |
| | 194,1679 | 46,6160 | 6 | 0.05 | | 27 | |
| | 194,2626 | 47,9964 | 6 | 0.07 | | 27 | |
| | 194,3615 | 48,2163 | 7 | 0.03 | | 27 | |
| | 194,4287 | 41,8555 | 5 | 0.027 | | 27 | |
| | 194,4385 | 48,9859 | 5 | 0.037 | | 27 | |
| | 194,4409 | 2,0345 | 4 | 0.088 | | 27 | |
| | 194,4862 | 4,8413 | 4 | 0.138 | | 27 | |
| LCCG 40 | 194,5208 | -12,1516 | 4 | 0,113 | | 2 | |
| | 194,5282 | 50,2601 | 7 | 0.1 | | 27 | |
| | 194,6278 | -1,6440 | 4 | 0.081 | | 27 | |
| | 194,6381 | -1,7543 | 5 | 0.085 | | 27 | |
| | 194,6498 | 4,2808 | 5 | 0.048 | | 27 | |
| | 194,6531 | -1,5168 | 4 | 0.082 | | 27 | |
| | 194,6972 | 62,3671 | 5 | 0.089 | | 27 | |
| | 194,7313 | -1,7206 | 7 | 0.084 | | 27 | |
| | 194,7561 | 4,9221 | 4 | 0.024 | | 27 | |
| | 194,7758 | -2,7467 | 4 | 0.082 | | 27 | |
| | 194,8197 | -1,8195 | 9 | 0.083 | | 27 | |
| UZCCG178 | 194,8458 | 14,1328 | 3 | 0,007 | | 5 | |
| | 194,8847 | 50,2776 | 4 | 0.036 | | 27 | |
| RSCG 67 | 194,8875 | 27,9575 | 3 | 0,024 | | 9 | |
| RSCG 68 | 195,0458 | 27,9714 | 3 | 0,022 | | 9 | |
| | 195,0527 | 58,9645 | 4 | 0.046 | | 27 | |
| | 195,0911 | 55,4128 | 4 | 0.064 | | 27 | |
| | 195,1092 | 52,7952 | 4 | 0.053 | | 27 | |
| | 195,1198 | 60,1353 | 4 | 0.07 | | 27 | |
| | 195,1350 | -3,5357 | 6 | 0.088 | | 27 | |
| | 195,1390 | 50,7658 | 6 | 0.113 | | 27 | |
| UZCCG179 | 195,1625 | 80,0733 | 3 | 0,033 | | 5 | |
| | 195,2183 | 57,7655 | 4 | 0.07 | | 27 | |
| | 195,2199 | 1,1087 | 5 | 0.082 | | 27 | |
| | 195,2293 | 5,1177 | 7 | 0.048 | | 27 | |
| | 195,2800 | 4,7110 | 4 | 0.111 | | 27 | |
| | 195,3139 | 62,5183 | 4 | 0.075 | | 27 | |
| | 195,3219 | -3,5231 | 6 | 0.084 | | 27 | |
| | 195,3675 | -3,5218 | 5 | 0.086 | | 27 | |
| | 195,4371 | -3,5302 | 4 | 0.085 | | 27 | |
| | 195,4720 | 62,6883 | 4 | 0.077 | | 27 | |
| J1301+1915 | 195,4875 | 19,2531 | 4 | 0,079 | | 4 | group coord.from ref.(3) |
| | 195,4960 | 5,7321 | 4 | 0.062 | | 27 | |
| | 195,5022 | -0,8477 | 5 | 0.085 | | 27 | |
| | 195,5234 | -2,4350 | 4 | 0.079 | | 27 | |
| HCG 063 | 195,5417 | -32,7680 | 4 | 0,031 | | 1 | |
| | | | | 0,031 | HCG 063 | 12 | |

| | | | | | | | | |
|---|---|---|---|---|---|---|---|---|
| | 195,5638 | 3,2301 | 5 | 0.048 | | | 27 | |
| | 195,5757 | -2,7070 | 7 | 0.082 | | | 27 | |
| | 195,6152 | 12,6304 | 4 | 0.065 | | | 27 | |
| | 195,6163 | 65,1924 | 4 | 0.081 | | | 27 | |
| | 195,6262 | -2,4923 | 7 | 0.084 | | | 27 | |
| | 195,6567 | 0,5167 | 9 | 0.068 | | | 27 | |
| | 195,7016 | 3,2865 | 4 | 0.07 | | | 27 | |
| | 195,7018 | -2,4542 | 5 | 0.086 | | | 27 | |
| | 195,7270 | 3,2884 | 5 | 0.071 | | | 27 | |
| | 195,7365 | 62,3405 | 4 | 0.076 | | | 27 | |
| PCG130257+053112 | 195,7375 | 5,5200 | 4 | | 0.0690 | A | 7 | group member/coord.from Ref.(6) |
| | 195,7745 | -2,4063 | 4 | 0.082 | | | 27 | |
| PCG130308-022207 | 195,7875 | -2,3689 | 4 | | 0.0850 | B | 7 | group member/coord.from Ref.(6) |
| | 195,8415 | -2,4921 | 5 | 0.085 | | | 27 | |
| | 195,8699 | 47,2074 | 4 | 0.024 | | | 27 | |
| | 195,9110 | 47,3818 | 6 | 0.025 | | | 27 | |
| | 196,0027 | 55,1579 | 4 | 0.09 | | | 27 | |
| | 196,0135 | 2,7069 | 5 | 0.072 | | | 27 | |
| | 196,0287 | 4,7853 | 4 | 0.064 | | | 27 | |
| | 196,0534 | 55,4796 | 4 | 0.028 | | | 27 | |
| | 196,0633 | 54,3883 | 4 | 0.064 | | | 27 | |
| | 196,0829 | 3,7032 | 8 | 0.04 | | | 27 | |
| | 196,0895 | 42,6844 | 4 | 0.038 | | | 27 | |
| | 196,1041 | 61,2566 | 9 | 0.052 | | | 27 | |
| | 196,1393 | 49,5123 | 5 | 0.033 | | | 27 | |
| | 196,1565 | 2,4423 | 5 | 0.042 | | | 27 | |
| | 196,1656 | 51,6309 | 4 | 0.102 | | | 27 | |
| | 196,1798 | 55,6369 | 5 | 0.065 | | | 27 | |
| | 196,1874 | 60,0774 | 4 | 0.06 | | | 27 | |
| | 196,1925 | 55,2148 | 6 | 0.089 | | | 27 | |
| | 196,2491 | 11,8160 | 4 | 0.08 | | | 27 | |
| | 196,3211 | 43,6066 | 6 | 0.038 | | | 27 | |
| | 196,3266 | 55,0373 | 7 | 0.064 | | | 27 | |
| UZCCG180 | 196,3833 | 53,6408 | 4 | 0,030 | | | 5 | |
| | 196,4159 | 49,4186 | 8 | 0.034 | | | 27 | |
| SDSS 082 | 196,4250 | 0,3978 | 4 | 0,223 | | | 8 | |
| | 196,4395 | 61,1274 | 5 | 0.053 | | | 27 | |
| | 196,4563 | 53,5782 | 5 | 0.029 | | | 27 | |
| | 196,5237 | 0,2974 | 4 | 0.09 | | | 27 | |
| | 196,5546 | -0,7672 | 4 | 0.084 | | | 27 | |
| | 196,6541 | -2,5784 | 4 | 0.114 | | | 27 | |
| SDSS 083 | 196,7125 | 0,5578 | 4 | 0,128 | | | 8 | |
| | 196,7388 | 43,5540 | 7 | 0.035 | | | 27 | |
| | 196,7397 | 14,9144 | 4 | 0.085 | | | 27 | |
| | 196,7515 | 59,7931 | 5 | 0.06 | | | 27 | |
| | 196,7781 | 61,0803 | 5 | 0.107 | | | 27 | |
| | 196,8093 | -1,5048 | 4 | 0.08 | | | 27 | |
| | 196,8542 | 52,0647 | 4 | 0.03 | | | 27 | |
| | 196,9092 | 50,1477 | 7 | 0.055 | | | 27 | |
| | 196,9113 | 60,1936 | 8 | 0.028 | | | 27 | |
| | 196,9296 | 15,6432 | 5 | 0.084 | | | 27 | |
| | 197,0870 | 3,1246 | 6 | 0.038 | | | 27 | |
| | 197,1036 | 13,1067 | 5 | 0.086 | | | 27 | |
| | 197,1315 | -1,3001 | 4 | 0.085 | | | 27 | |
| | 197,1590 | 13,7509 | 4 | 0.062 | | | 27 | |
| | 197,1745 | 0,1433 | 7 | 0.081 | | | 27 | |
| | 197,2237 | -1,6986 | 8 | 0.084 | | | 27 | |
| UZCCG181 | 197,2333 | 0,7950 | 3 | 0,018 | | | 5 | |
| | 197,3354 | -1,6439 | 5 | 0.082 | | | 27 | |
| J1309+1553 | 197,3583 | 15,8994 | 4 | 0,149 | | | 4 | group coord.from ref.(3) |
| | 197,3858 | 49,9319 | 4 | 0.048 | | | 27 | |
| | 197,4540 | 51,4122 | 5 | 0.056 | | | 27 | |
| | 197,5293 | 5,1944 | 4 | 0.047 | | | 27 | |
| | 197,5320 | 1,2834 | 5 | 0.083 | | | 27 | |
| | 197,5321 | 42,1957 | 4 | 0.048 | | | 27 | |
| | 197,5890 | 13,9720 | 4 | 0.085 | | | 27 | |
| SDSS 084 | 197,6667 | 0,6656 | 5 | 0,081 | | | 8 | |
| | 197,7085 | 54,5759 | 9 | 0.064 | | | 27 | |
| | 197,7113 | 13,7549 | 4 | 0.086 | | | 27 | |
| | 197,7372 | 0,1525 | 4 | 0.095 | | | 27 | |
| UZCCG182 | 197,7375 | 29,6414 | 3 | 0,023 | | | 5 | |
| | 197,7451 | 2,3998 | 5 | 0.067 | | | 27 | |
| | 197,7753 | 15,3379 | 4 | 0.087 | | | 27 | |
| | 197,8397 | -0,1018 | 5 | 0.041 | | | 27 | |
| | 197,8799 | 49,0092 | 8 | 0.055 | | | 27 | |
| PCG131132-011944 | 197,8833 | -1,3289 | 4 | | 0.1800 | D | 7 | group member/coord.from Ref.(6) |
| | 197,8875 | -0,5221 | 4 | 0.09 | | | 27 | |
| | 197,9480 | 46,7817 | 6 | 0.027 | | | 27 | |
| | 197,9570 | 43,1671 | 8 | 0.028 | | | 27 | |
| | 197,9947 | 53,9246 | 5 | 0.064 | | | 27 | |
| | 198,0118 | 55,3535 | 4 | 0.082 | | | 27 | |
| SDSS 085 | 198,0208 | 0,5383 | 4 | 0,121 | | | 8 | |
| | 198,0432 | 44,8594 | 4 | 0.037 | | | 27 | |
| PCG131211+071828 | 198,0500 | 7,3078 | 4 | | 0.0920 | B | 7 | group member/coord.from Ref.(6) |
| SHK 355 | 198,0542 | 7,3047 | 6 | 0,093 | | | 11 | mean value/coord.from Ref.(10) |
| | 198,0799 | 40,8156 | 5 | 0.068 | | | 27 | |
| | 198,0863 | -1,0883 | 4 | 0.086 | | | 27 | |
| | 198,0897 | -1,1517 | 6 | 0.085 | | | 27 | |
| SDSS 086 | 198,1042 | -1,1036 | 4 | 0,109 | | | 8 | |
| | 198,1155 | 4,9919 | 6 | 0.048 | | | 27 | |
| | 198,1261 | -0,8560 | 5 | 0.083 | | | 27 | |
| | 198,1915 | 46,9169 | 7 | 0.028 | | | 27 | |

| | | | | | | | | | |
|---|---|---|---|---|---|---|---|---|---|
| | 198,1980 | 0,9844 | 7 | 0.073 | | | | 27 | |
| | 198,2080 | 1,0199 | 4 | 0.074 | | | | 27 | |
| | 198,2393 | 56,9841 | 8 | 0.079 | | | | 27 | |
| | 198,3121 | 47,6660 | 5 | 0.056 | | | | 27 | |
| | 198,4146 | 3,8867 | 4 | 0.077 | | | | 27 | |
| | 198,4495 | 60,9763 | 7 | 0.125 | | | | 27 | |
| | 198,4499 | 4,2783 | 4 | 0.076 | | | | 27 | |
| UZCCG183 | 198,4542 | 6,9717 | 3 | 0,022 | | | | 5 | |
| | 198,4901 | 5,1158 | 4 | 0.111 | | | | 27 | |
| | 198,5542 | 53,3084 | 6 | 0.062 | | | | 27 | |
| | 198,5644 | 13,5133 | 6 | 0.08 | | | | 27 | |
| | 198,6077 | 48,8359 | 4 | 0.056 | | | | 27 | |
| UZCCG184 | 198,7250 | 3,0344 | 3 | 0,022 | | | | 5 | |
| SDSS 088 | 198,7583 | -1,0683 | 4 | 0,215 | | | | 8 | |
| | 198,8833 | 41,3427 | 5 | 0.075 | | | | 27 | |
| | 198,9233 | 46,4063 | 5 | 0.057 | | | | 27 | |
| | 198,9514 | 0,0049 | 4 | 0.082 | | | | 27 | |
| | 198,9582 | 62,5968 | 4 | 0.041 | | | | 27 | |
| | 198,9611 | 63,7320 | 4 | 0.037 | | | | 27 | |
| | 199,0137 | 43,4818 | 5 | 0.086 | | | | 27 | |
| | 199,0224 | -0,8500 | 4 | 0.112 | | | | 27 | |
| | 199,1217 | 62,2776 | 4 | 0.079 | | | | 27 | |
| | 199,1402 | 0,9100 | 4 | 0.08 | | | | 27 | |
| | 199,2029 | 5,3692 | 4 | 0.096 | | | | 27 | |
| UZCCG185 | 199,2083 | 31,0292 | 4 | 0,019 | | | | 5 | |
| | 199,2185 | 0,7148 | 4 | 0.024 | | | | 27 | |
| UZCCG186 | 199,2375 | -2,1844 | 3 | 0,019 | | | | 5 | |
| | 199,2541 | 44,0897 | 5 | 0.052 | | | | 27 | |
| | 199,2699 | 44,9631 | 4 | 0.092 | | | | 27 | |
| | 199,3071 | 42,5310 | 6 | 0.074 | | | | 27 | |
| | 199,3381 | 44,8005 | 7 | 0.061 | | | | 27 | |
| RSCG 69 | 199,3458 | 20,6239 | 3 | 0,022 | | | | 9 | |
| | | | 3 | 0,022 | | | UZCCG187 | 5 | |
| PCG131725-014820 | 199,3542 | -1,8058 | 4 | | 0.0580 | C | | 7 | group member/coord.from Ref.(6) |
| PCG131730-031041 | 199,3750 | -3,1780 | 4 | | 0.0180 | B | | 7 | group member/coord.from Ref.(6) |
| | 199,4097 | 42,4046 | 8 | 0.075 | | | | 27 | |
| | 199,4145 | 48,1258 | 4 | 0.057 | | | | 27 | |
| | 199,4489 | -1,7850 | 5 | 0.048 | | | | 27 | |
| | 199,4529 | 41,2192 | 4 | 0.067 | | | | 27 | |
| | 199,4812 | 4,0950 | 4 | 0.046 | | | | 27 | |
| | 199,4899 | 43,3352 | 4 | 0.035 | | | | 27 | |
| | 199,5605 | -0,8708 | 4 | 0.084 | | | | 27 | |
| SDSS 089 | 199,5625 | 0,3272 | 4 | 0,082 | | | | 8 | |
| UZCCG188 | 199,5708 | 4,4514 | 3 | 0,021 | | | | 5 | |
| | 199,5967 | 66,4106 | 7 | 0.066 | | | | 27 | |
| | 199,5993 | 66,4529 | 7 | 0.067 | | | | 27 | |

| | | | | | |
|---|---|---|---|---|---|
| | 199,6402 | 47,2171 | 4 | 0.058 | 27 |
| | 199,6456 | 41,2568 | 4 | 0.065 | 27 |
| | 199,6611 | 60,6572 | 5 | 0.064 | 27 |
| SDSS 090 | 199,6875 | 0,7297 | 4 | 0,086 | 8 |
| SDSS 091 | 199,6958 | 0,7586 | 4 | 0,087 | 8 |
| | 199,6975 | 53,9673 | 8 | 0.033 | 27 |
| | 199,7304 | 43,8602 | 5 | 0.028 | 27 |
| | 199,7788 | -1,0841 | 4 | 0.086 | 27 |
| | 199,8009 | 55,8539 | 4 | 0.136 | 27 |
| | 199,8267 | -1,0629 | 4 | 0.083 | 27 |
| | 199,8305 | 57,9746 | 4 | 0.098 | 27 |
| | 199,8311 | -0,9579 | 7 | 0.082 | 27 |
| | 199,8535 | 0,0399 | 4 | 0.109 | 27 |
| SDSS 092 | 199,8875 | -1,0961 | 4 | 0,080 | 8 |
| | 199,8911 | -0,9224 | 6 | 0.084 | 27 |
| UZCCG189 | 199,9500 | 57,6261 | 3 | 0,008 | 5 |
| UZCCG190 | 199,9708 | 52,0608 | 3 | 0,016 | 5 |
| | 200,0609 | 41,3458 | 5 | 0.098 | 27 |
| | 200,0798 | 60,6627 | 4 | 0.071 | 27 |
| | 200,1163 | 42,5916 | 4 | 0.074 | 27 |
| | 200,1306 | 42,6169 | 4 | 0.073 | 27 |
| | 200,1747 | 57,3032 | 4 | 0.117 | 27 |
| | 200,2865 | 53,6379 | 4 | 0.121 | 27 |
| | 200,3587 | 3,5156 | 4 | 0.082 | 27 |
| | 200,3909 | 61,4992 | 5 | 0.134 | 27 |
| | 200,3951 | -0,8374 | 4 | 0.106 | 27 |
| | 200,4133 | 59,0356 | 7 | 0.075 | 27 |
| | 200,4346 | 57,5717 | 5 | 0.118 | 27 |
| | 200,4753 | 58,9540 | 5 | 0.074 | 27 |
| | 200,4867 | 57,5116 | 6 | 0.119 | 27 |
| | 200,6004 | 46,5785 | 4 | 0.061 | 27 |
| | 200,6057 | 45,4843 | 4 | 0.057 | 27 |
| | 200,7228 | -2,3973 | 4 | 0.09 | 27 |
| | 200,7378 | -1,5815 | 4 | 0.086 | 27 |
| | 200,7680 | -1,8919 | 5 | 0.085 | 27 |
| | 200,7979 | -0,2806 | 6 | 0.082 | 27 |
| | 200,7982 | -1,6220 | 5 | 0.083 | 27 |
| | 200,8307 | 43,5052 | 4 | 0.135 | 27 |
| | 200,9021 | -1,6171 | 4 | 0.084 | 27 |
| | 200,9187 | 42,1570 | 4 | 0.076 | 27 |
| | 200,9503 | 55,2988 | 5 | 0.066 | 27 |
| | 200,9644 | 1,1386 | 5 | 0.108 | 27 |
| | 200,9648 | 1,2734 | 8 | 0.109 | 27 |
| | 200,9709 | 62,4130 | 4 | 0.077 | 27 |
| | 200,9753 | 4,8376 | 4 | 0.133 | 27 |
| | 200,9959 | 57,6721 | 4 | 0.115 | 27 |

| | | | | | | | | | |
|---|---|---|---|---|---|---|---|---|---|
| | 201,0034 | 59,0034 | 6 | 0.073 | | | | 27 | |
| SDSS 093 | 201,1625 | 0,3678 | 4 | 0,108 | | | | 8 | |
| SDSS 094 | 201,2042 | 0,0119 | 4 | 0,082 | | | | 8 | |
| RSCG 70 | 201,2417 | 36,4067 | 3 | 0,018 | | | HCG 68 | 9 | |
| | | | 3 | 0,018 | | | UZCCG191 | 5 | |
| | 201,2771 | 0,9005 | 4 | 0.079 | | | | 27 | |
| SDSS 095 | 201,2833 | 0,6419 | 4 | 0,085 | | | | 8 | |
| | 201,3071 | 2,2927 | 4 | 0.102 | | | | 27 | |
| | 201,3115 | -1,3407 | 5 | 0.083 | | | | 27 | |
| | 201,3682 | 43,9869 | 6 | 0.028 | | | | 27 | |
| | 201,3885 | 61,7804 | 7 | 0.036 | | | | 27 | |
| HCG 064 | 201,4292 | -3,8578 | 4 | 0,036 | | | | 1 | |
| | 201,4787 | 60,8039 | 7 | 0.071 | | | | 27 | |
| | 201,5086 | 58,9942 | 7 | 0.074 | | | | 27 | |
| | 201,6034 | 46,8518 | 5 | 0.091 | | | | 27 | |
| | 201,6663 | 0,3692 | 4 | 0.084 | | | | 27 | |
| | 201,6898 | 53,0375 | 6 | 0.03 | | | | 27 | |
| | 201,7042 | 4,3462 | 6 | 0.134 | | | | 27 | |
| | 201,7361 | 1,2714 | 4 | 0.082 | | | | 27 | |
| | 201,7460 | -2,0787 | 4 | 0.087 | | | | 27 | |
| | 201,7866 | -0,9582 | 5 | 0.074 | | | | 27 | |
| | 201,8287 | 1,4279 | 4 | 0.078 | | | | 27 | |
| | 201,8393 | 52,8704 | 4 | 0.104 | | | | 27 | |
| | 201,8408 | -2,6360 | 5 | 0.085 | | | | 27 | |
| | 201,8419 | 1,3285 | 7 | 0.079 | | | | 27 | |
| | 201,9123 | 0,9767 | 5 | 0.083 | | | | 27 | |
| | 201,9336 | 59,9719 | 4 | 0.119 | | | | 27 | |
| | 202,0972 | 61,8586 | 4 | 0.097 | | | | 27 | |
| | 202,1067 | 53,3004 | 5 | 0.029 | | | | 27 | |
| PCG132826+012636 | 202,1125 | 1,4436 | 4 | | 0.0780 | A | | 7 | group member/coord.from Ref.(6) |
| SHK 019 | 202,1125 | 15,8417 | 5 | 0,067 | | | MCG 3-34-39 | 21 | mean value from NED data |
| UZCCG192 | 202,2083 | 46,6247 | 5 | 0,009 | | | | 5 | |
| PCG132857+541852 | 202,2417 | 54,3147 | 4 | | 0.1040 | B | | 7 | group member/coord.from Ref.(6) |
| | 202,2693 | 44,2810 | 4 | 0.134 | | | | 27 | |
| | 202,2696 | 59,1546 | 5 | 0.072 | | | | 27 | |
| | 202,3008 | 54,5187 | 4 | 0.102 | | | | 27 | |
| | 202,3333 | 45,3330 | 4 | 0.161 | | | | 27 | |
| | 202,3339 | 44,9774 | 4 | 0.061 | | | | 27 | |
| UZCCG193 | 202,3625 | 11,7217 | 4 | 0,023 | | | | 5 | |
| | 202,4448 | 43,5846 | 5 | 0.068 | | | | 27 | |
| HCG 065 | 202,4750 | -29,4997 | 5 | 0,048 | | | | 1 | |
| | 202,5343 | -2,1044 | 6 | 0.087 | | | | 27 | |
| | 202,5427 | 54,0443 | 6 | 0.107 | | | | 27 | |
| | 202,5610 | 2,4961 | 4 | 0.083 | | | | 27 | |
| | 202,6292 | 41,9990 | 5 | 0.061 | | | | 27 | |
| | 202,6606 | -1,9742 | 6 | 0.085 | | | | 27 | |

| | | | | | | | | |
|---|---|---|---|---|---|---|---|---|
| PCG133042-003302 | 202,6750 | 0,5508 | 4 | | 0.1680 | A | 7 | group member/coord.from Ref.(6) |
| | 202,6978 | 4,6736 | 5 | 0.081 | | | 27 | |
| | 202,7071 | -1,1385 | 4 | 0.087 | | | 27 | |
| | 202,7074 | 60,2730 | 7 | 0.07 | | | 27 | |
| | 202,7085 | -1,6349 | 6 | 0.083 | | | 27 | |
| | 202,7524 | -1,8371 | 9 | 0.084 | | | 27 | |
| | 202,7927 | -1,9829 | 9 | 0.088 | | | 27 | |
| | 202,7931 | -1,6734 | 7 | 0.082 | | | 27 | |
| | 202,8421 | -2,1883 | 8 | 0.021 | | | 27 | |
| | 202,8774 | -1,6342 | 4 | 0.085 | | | 27 | |
| | 202,8954 | -1,7556 | 4 | 0.075 | | | 27 | |
| | 202,9077 | 43,9052 | 4 | 0.074 | | | 27 | |
| | 202,9104 | 40,4526 | 5 | 0.027 | | | 27 | |
| | 202,9565 | 3,1589 | 4 | 0.178 | | | 27 | |
| | 202,9766 | 47,5647 | 6 | 0.061 | | | 27 | |
| | 202,9882 | 40,0919 | 6 | 0.081 | | | 27 | |
| | 203,0242 | 2,1688 | 5 | 0.057 | | | 27 | |
| | 203,0303 | -1,6847 | 4 | 0.087 | | | 27 | |
| | 203,0403 | -1,7478 | 4 | 0.076 | | | 27 | |
| | 203,0855 | 54,0001 | 4 | 0.04 | | | 27 | |
| | 203,1069 | -1,8266 | 6 | 0.084 | | | 27 | |
| | 203,1476 | 5,4317 | 4 | 0.024 | | | 27 | |
| | 203,1665 | 57,1189 | 5 | 0.069 | | | 27 | |
| UZCCG194 | 203,1792 | 7,3017 | 6 | 0,023 | | | 5 | |
| | 203,2295 | 0,7111 | 4 | 0.079 | | | 27 | |
| | 203,2359 | 66,6791 | 5 | 0.094 | | | 27 | |
| UZCCG195 | 203,2417 | -1,1514 | 3 | 0,012 | | | 5 | |
| | 203,2515 | 4,4747 | 5 | 0.101 | | | 27 | |
| | 203,2767 | 54,4553 | 4 | 0.108 | | | 27 | |
| | 203,3553 | 4,6754 | 8 | 0.022 | | | 27 | |
| | 203,3765 | 53,3034 | 4 | 0.102 | | | 27 | |
| | 203,3917 | 59,5194 | 4 | 0.074 | | | 27 | |
| | 203,4163 | 56,9246 | 4 | 0.024 | | | 27 | |
| | 203,4187 | -2,3934 | 6 | 0.108 | | | 27 | |
| | 203,4669 | -1,9148 | 4 | 0.091 | | | 27 | |
| | 203,4877 | 5,2366 | 4 | 0.082 | | | 27 | |
| | 203,5546 | 2,7884 | 4 | 0.078 | | | 27 | |
| | 203,5751 | 54,8577 | 6 | 0.067 | | | 27 | |
| | 203,6468 | -1,4999 | 4 | 0.087 | | | 27 | |
| SDSS 099 | 203,6500 | -1,0572 | 4 | 0,077 | | | 8 | |
| | 203,6671 | 47,5440 | 4 | 0.089 | | | 27 | |
| SDSS 100 | 203,7750 | 0,4725 | 4 | 0,087 | | | 8 | |
| UZCCG196 | 203,7833 | 13,7500 | 3 | 0,023 | | | 5 | |
| | 203,7879 | -2,3346 | 5 | 0.047 | | | 27 | |
| | 203,8158 | -3,4883 | 5 | 0.108 | | | 27 | |
| | 203,8415 | 59,0475 | 6 | 0.067 | | | 27 | |

| | | | | | | |
|---|---|---|---|---|---|---|
| | 203,8545 | 0,2788 | 4 | 0.086 | | 27 |
| | 203,8570 | -0,9048 | 4 | 0.072 | | 27 |
| | 203,9249 | 42,7491 | 5 | 0.062 | | 27 |
| | 203,9952 | 59,3095 | 5 | 0.066 | | 27 |
| | 204,0298 | 42,9052 | 5 | 0.063 | | 27 |
| | 204,0307 | 56,0414 | 4 | 0.068 | | 27 |
| | 204,0401 | 59,8872 | 5 | 0.042 | | 27 |
| | 204,0557 | 42,0362 | 4 | 0.062 | | 27 |
| | 204,0726 | 60,3619 | 4 | 0.072 | | 27 |
| | 204,0825 | 4,7453 | 4 | 0.034 | | 27 |
| | 204,0858 | 59,2739 | 9 | 0.068 | | 27 |
| | 204,0917 | 44,8764 | 5 | 0.079 | | 27 |
| | 204,0997 | 59,2583 | 4 | 0.074 | | 27 |
| | 204,1354 | 54,8565 | 7 | 0.107 | | 27 |
| | 204,1789 | 43,9286 | 5 | 0.062 | | 27 |
| | 204,1835 | 43,9251 | 6 | 0.063 | | 27 |
| SHK 251 | 204,2167 | 36,8322 | 7 | 0,061 | | 24 |
| | 204,2414 | 43,4817 | 5 | 0.053 | | 27 |
| | 204,2815 | 59,2977 | 5 | 0.069 | | 27 |
| | 204,3228 | 3,4965 | 4 | 0.079 | | 27 |
| | 204,3299 | 62,8456 | 4 | 0.068 | | 27 |
| | 204,3648 | 1,7374 | 4 | 0.077 | | 27 |
| | 204,3744 | 1,7086 | 5 | 0.078 | | 27 |
| | 204,3964 | 2,1505 | 4 | 0.114 | | 27 |
| | 204,4185 | -1,0988 | 4 | 0.071 | | 27 |
| | 204,4389 | 5,7555 | 4 | 0.082 | | 27 |
| | 204,4779 | 61,4440 | 4 | 0.036 | | 27 |
| | 204,5069 | 45,2587 | 6 | 0.062 | | 27 |
| | 204,5133 | 2,2816 | 4 | 0.116 | | 27 |
| SDSS 101 | 204,5208 | 0,9656 | 4 | 0,143 | | 8 |
| | 204,5770 | 57,2610 | 4 | 0.066 | | 27 |
| | 204,5964 | 54,9428 | 5 | 0.108 | | 27 |
| HCG 066 | 204,6417 | 57,3044 | 4 | 0,070 | VV135 | 1 |
| | 204,6546 | 61,8634 | 8 | 0.031 | | 27 |
| SDSS 102 | 204,6750 | 0,7428 | 4 | 0,346 | | 8 |
| | 204,6820 | 63,2952 | 4 | 0.102 | | 27 |
| SDSS 103 | 204,6875 | 0,2761 | 4 | 0,129 | | 8 |
| | 204,6918 | 1,4254 | 4 | 0.079 | | 27 |
| UZCCG197 | 204,7667 | 31,3114 | 3 | 0,016 | | 5 |
| | 204,7776 | 57,2301 | 4 | 0.069 | | 27 |
| | 204,8438 | 4,6192 | 5 | 0.085 | | 27 |
| | 204,9441 | -2,6593 | 5 | 0.094 | | 27 |
| | 205,0806 | 5,1019 | 4 | 0.076 | | 27 |
| | 205,1747 | 44,4623 | 8 | 0.063 | | 27 |
| | 205,1781 | 3,1725 | 4 | 0.115 | | 27 |
| | 205,1961 | 3,0803 | 5 | 0.114 | | 27 |

| | | | | | | | | |
|---|---|---|---|---|---|---|---|---|
| | 205,2126 | 4,4296 | 9 | 0.023 | | | 27 | |
| | 205,2154 | 56,3659 | 6 | 0.071 | | | 27 | |
| | 205,2295 | 56,1788 | 6 | 0.07 | | | 27 | |
| | 205,2308 | 3,6825 | 4 | 0.079 | | | 27 | |
| | 205,2359 | -0,5451 | 5 | 0.087 | | | 27 | |
| SDSS 106 | 205,2458 | 0,7589 | 4 | 0,166 | | | 8 | |
| SDSS 107 | 205,2750 | 0,0264 | 4 | 0,100 | | | 8 | |
| | 205,2789 | 2,3349 | 4 | 0.076 | | | 27 | |
| | 205,3400 | 1,8962 | 4 | 0.08 | | | 27 | |
| UZCCG198 | 205,3417 | 5,0594 | 3 | 0,024 | | | 5 | |
| | 205,3498 | 58,5849 | 4 | 0.083 | | | 27 | |
| | 205,3579 | 4,4650 | 5 | 0.076 | | | 27 | |
| SDSS 108 | 205,3625 | -1,1725 | 4 | 0,089 | | | 8 | |
| LCCG 41 | 205,4000 | -12,1005 | 3 | 0,051 | | | 2 | |
| PCG134137+371637 | 205,4042 | 37,2769 | 4 | | 0.1700 | A | 7 | group member/coord.from Ref.(6) |
| | 205,4255 | 0,9151 | 4 | 0.087 | | | 27 | |
| | 205,4721 | 56,5960 | 4 | 0.068 | | | 27 | |
| | 205,4783 | 2,0267 | 6 | 0.076 | | | 27 | |
| | 205,4891 | 54,8140 | 7 | 0.065 | | | 27 | |
| | 205,5117 | 0,5466 | 6 | 0.074 | | | 27 | |
| | 205,5490 | -0,2336 | 7 | 0.089 | | | 27 | |
| | 205,5631 | 2,2752 | 6 | 0.075 | | | 27 | |
| | 205,5640 | -1,0922 | 4 | 0.088 | | | 27 | |
| | 205,6028 | 2,2261 | 9 | 0.079 | | | 27 | |
| SDSS 110 | 205,6250 | 0,3553 | 4 | 0,243 | | | 8 | |
| | 205,6571 | 2,3503 | 8 | 0.077 | | | 27 | |
| | 205,6769 | -1,0250 | 4 | 0.089 | | | 27 | |
| SDSS 111 | 205,6833 | 0,7617 | 4 | 0,073 | | | 8 | |
| | 205,6938 | 2,2924 | 6 | 0.08 | | | 27 | |
| | 205,7139 | 3,3503 | 7 | 0.078 | | | 27 | |
| | 205,7186 | 2,4151 | 6 | 0.078 | | | 27 | |
| | 205,7494 | 61,7002 | 4 | 0.033 | | | 27 | |
| | 205,7548 | 49,0481 | 8 | 0.063 | | | 27 | |
| | 205,7828 | 2,4522 | 6 | 0.076 | | | 27 | |
| | 205,8139 | 55,6905 | 7 | 0.07 | | | 27 | |
| | 205,8245 | 55,9643 | 7 | 0.067 | | | 27 | |
| | 205,8483 | 2,7927 | 4 | 0.079 | | | 27 | |
| | 205,8734 | 58,1361 | 4 | 0.071 | | | 27 | |
| PCG134338+624933 | 205,9083 | 62,8261 | 4 | | 0.1150 | C | 7 | group member/coord.from Ref.(6) |
| | 205,9406 | 55,6374 | 5 | 0.069 | | | 27 | |
| UZCCG199 | 205,9583 | 3,8969 | 3 | 0,023 | | | 5 | |
| | 205,9749 | 57,1024 | 4 | 0.096 | | | 27 | |
| | 205,9909 | 45,6354 | 4 | 0.058 | | | 27 | |
| | 206,1168 | 3,7299 | 4 | 0.079 | | | 27 | |
| | 206,1345 | 61,4238 | 5 | 0.071 | | | 27 | |
| | 206,1428 | 40,4531 | 4 | 0.076 | | | 27 | |

| | | | | | | | | | |
|---|---|---|---|---|---|---|---|---|---|
| | 206,2276 | 2,8989 | 6 | 0.077 | | | | 27 | |
| | 206,2324 | 2,9433 | 7 | 0.078 | | | | 27 | |
| | 206,2403 | 0,2360 | 5 | 0.09 | | | | 27 | |
| | 206,2509 | 5,1229 | 4 | 0.079 | | | | 27 | |
| | 206,2704 | 53,5159 | 5 | 0.068 | | | | 27 | |
| | 206,2928 | 3,0119 | 7 | 0.076 | | | | 27 | |
| | 206,3058 | 2,0752 | 8 | 0.072 | | | | 27 | |
| SHK 213 | 206,3542 | 26,8833 | 6 | | 0.0582 | 1 | | 11 | group member/coord.from Ref.(10) |
| | 206,3672 | 2,7844 | 5 | 0.079 | | | | 27 | |
| | 206,3786 | -0,5412 | 4 | 0.077 | | | | 27 | |
| | 206,3942 | -0,0057 | 7 | 0.088 | | | | 27 | |
| | 206,4154 | -3,2921 | 4 | 0.092 | | | | 27 | |
| | 206,4632 | 44,8348 | 4 | 0.063 | | | | 27 | |
| SDSS 113 | 206,4667 | 0,0211 | 4 | 0,091 | | | | 8 | |
| | 206,4718 | 44,8732 | 7 | 0.062 | | | | 27 | |
| | 206,4888 | 56,5080 | 6 | 0.035 | | | | 27 | |
| | 206,4911 | 45,6093 | 4 | 0.063 | | | | 27 | |
| | 206,5438 | 61,2589 | 5 | 0.035 | | | | 27 | |
| | 206,6098 | 55,1033 | 4 | 0.04 | | | | 27 | |
| | 206,6218 | 3,1994 | 5 | 0.078 | | | | 27 | |
| | 206,6779 | 5,1279 | 4 | 0.031 | | | | 27 | |
| | 206,6780 | 60,3428 | 5 | 0.072 | | | | 27 | |
| | 206,6824 | -1,7325 | 4 | 0.089 | | | | 27 | |
| | 206,7054 | 58,3077 | 4 | 0.06 | | | | 27 | |
| | 206,7062 | 45,7179 | 8 | 0.064 | | | | 27 | |
| | 206,7629 | 48,4215 | 4 | 0.17 | | | | 27 | |
| UZCCG200 | 206,7792 | 3,7194 | 3 | 0,023 | | | | 5 | |
| | 206,7907 | 40,7830 | 7 | 0.075 | | | | 27 | |
| | 206,9347 | 60,3784 | 4 | 0.073 | | | | 27 | |
| UZCCG201 | 206,9417 | 3,3983 | 3 | 0,023 | | | | 5 | |
| | 206,9945 | 3,3604 | 6 | 0.024 | | | | 27 | |
| | 207,0835 | 56,6977 | 7 | 0.034 | | | | 27 | |
| | 207,1199 | 57,2197 | 4 | 0.097 | | | | 27 | |
| | 207,1358 | 5,0751 | 6 | 0.078 | | | | 27 | |
| | 207,2019 | 46,8462 | 4 | 0.088 | | | | 27 | |
| HCG 067 | 207,2667 | -7,2055 | 4 | 0,025 | | | | 1 | |
| | | | 4 | 0,026 | | | HCG 067 | 12 | |
| PCG134932+280017 | 207,3875 | 28,0047 | 5 | | 0.0700 | A | | 7 | group member/coord.from Ref.(6) |
| UZCCG202 | 207,4708 | 59,9636 | 3 | 0,007 | | | | 5 | |
| SDSS 114 | 207,4833 | 0,4689 | 4 | 0,260 | | | | 8 | |
| | 207,5031 | 56,0086 | 8 | 0.068 | | | | 27 | |
| | 207,5926 | 43,4374 | 5 | 0.093 | | | | 27 | |
| UZCCG203 | 207,5958 | 33,6531 | 3 | 0,015 | | | | 5 | |
| | 207,6742 | 56,0029 | 6 | 0.035 | | | | 27 | |
| UZCCG204 | 207,7083 | 2,3189 | 3 | 0,033 | | | | 5 | |
| SDSS 115 | 207,7125 | 0,5172 | 4 | 0,150 | | | | 8 | |
| | 207,7941 | 2,6501 | 4 | 0.042 | | | | 27 | |
| | 207,8162 | 46,3467 | 5 | 0.065 | | | | 27 | |
| SDSS 116 | 207,8833 | 0,7375 | 4 | 0,088 | | | | 8 | |
| | 207,8969 | 55,9919 | 7 | 0.069 | | | | 27 | |
| | 207,9018 | 46,2549 | 5 | 0.06 | | | | 27 | |
| | 207,9113 | 2,1705 | 4 | 0.033 | | | | 27 | |
| | 208,0202 | 2,7341 | 4 | 0.069 | | | | 27 | |
| | 208,0497 | 47,3306 | 4 | 0.062 | | | | 27 | |
| | 208,0863 | 5,1077 | 7 | 0.082 | | | | 27 | |
| SHK 253 | 208,1167 | 37,5206 | 13 | | 0.073 | 3 | | 11 | group member/coord.from Ref.(10) |
| SDSS 117 | 208,1458 | -1,0289 | 4 | 0,141 | | | | 8 | |
| | 208,2471 | 58,6833 | 4 | 0.094 | | | | 27 | |
| | 208,2556 | 5,1044 | 5 | 0.079 | | | | 27 | |
| | 208,2786 | 5,0830 | 9 | 0.078 | | | | 27 | |
| | 208,2829 | -2,5672 | 5 | 0.076 | | | | 27 | |
| | 208,2845 | 5,2141 | 4 | 0.076 | | | | 27 | |
| | 208,3086 | 5,1764 | 5 | 0.084 | | | | 27 | |
| | 208,3212 | 5,1428 | 6 | 0.081 | | | | 27 | |
| | 208,3439 | 40,0086 | 4 | 0.106 | | | | 27 | |
| | 208,3520 | 46,3970 | 9 | 0.061 | | | | 27 | |
| | 208,4116 | 44,7543 | 6 | 0.092 | | | | 27 | |
| HCG 068 | 208,4208 | 40,3281 | 5 | 0,008 | | | N5353 | 1 | |
| | | | 5 | 0,008 | | | RSCG 71 | 9 | |
| | 208,4304 | 54,8883 | 4 | 0.07 | | | | 27 | |
| | 208,6189 | 60,6051 | 8 | 0.075 | | | | 27 | |
| | 208,6517 | 62,0739 | 4 | 0.073 | | | | 27 | |
| | 208,7647 | 3,3891 | 5 | 0.069 | | | | 27 | |
| HCG 069 | 208,8792 | 25,0628 | 4 | 0,029 | | | VV281 | 1 | |
| | | | 4 | 0,030 | | | UZCCG205 | 5 | |
| SDSS 119 | 208,9917 | 0,2564 | 5 | 0,133 | | | | 8 | |
| | 209,0150 | 39,1059 | 4 | 0.094 | | | | 27 | |
| SHK 254 | 209,0708 | 35,1561 | 11 | 0,064 | | | | 26 | |
| SHK 376 | 209,1333 | 23,3561 | 4 | 0,067 | | | | 18 | |
| | 209,2116 | 52,9531 | 4 | 0.094 | | | | 27 | |
| | 209,2630 | 48,5121 | 4 | 0.055 | | | | 27 | |
| RSCG 72 | 209,3000 | 12,0064 | 3 | 0,021 | | | | 9 | |
| | 209,3250 | 36,4478 | 7 | 0.062 | | | | 27 | |
| | 209,3374 | 5,3748 | 6 | 0.052 | | | | 27 | |
| UZCCG206 | 209,4083 | 17,4756 | 3 | 0,022 | | | | 5 | |
| UZCCG207 | 209,4208 | 37,5814 | 4 | 0,011 | | | | 5 | |
| | 209,4357 | 52,9163 | 6 | 0.068 | | | | 27 | |
| | 209,4925 | 53,0333 | 4 | 0.069 | | | | 27 | |
| | 209,4996 | 51,3762 | 4 | 0.126 | | | | 27 | |
| | 209,5614 | 44,7196 | 4 | 0.126 | | | | 27 | |
| | 209,5843 | 58,5267 | 5 | 0.061 | | | | 27 | |
| | 209,6705 | 58,9842 | 5 | 0.072 | | | | 27 | |

| | | | | | | | | |
|---|---|---|---|---|---|---|---|---|
| | 209,6952 | 53,9500 | 4 | 0.075 | | | 27 | |
| | 209,7249 | 3,2617 | 5 | 0.078 | | | 27 | |
| | 209,7799 | 39,5903 | 5 | 0.08 | | | 27 | |
| | 209,7822 | 0,2858 | 4 | 0.033 | | | 27 | |
| | 209,8729 | 58,1015 | 4 | 0.059 | | | 27 | |
| | 209,9043 | 39,6080 | 4 | 0.079 | | | 27 | |
| | 209,9224 | 44,6569 | 4 | 0.125 | | | 27 | |
| | 209,9470 | 55,6523 | 4 | 0.041 | | | 27 | |
| | 210,0158 | -2,9898 | 4 | 0.026 | | | 27 | |
| PCG140026+053457 | 210,1125 | 5,5828 | 4 | | 0.0350 | B | 7 | group member/coord.from Ref.(6) |
| UZCCG208 | 210,1542 | 54,7903 | 3 | 0,007 | | | 5 | |
| | 210,1795 | -2,7363 | 6 | 0.025 | | | 27 | |
| | 210,2538 | 51,6815 | 4 | 0.07 | | | 27 | |
| | 210,2677 | 63,5941 | 5 | 0.106 | | | 27 | |
| | 210,2768 | 41,7921 | 5 | 0.096 | | | 27 | |
| | 210,2832 | 38,8322 | 4 | 0.079 | | | 27 | |
| | 210,3733 | 59,0206 | 4 | 0.062 | | | 27 | |
| | 210,3777 | 3,8186 | 4 | 0.035 | | | 27 | |
| LCCG 42 | 210,4375 | -11,6414 | 3 | 0,037 | | | 2 | |
| LCCG 44 | 210,4375 | -3,7453 | 3 | 0,077 | | | 2 | |
| LCCG 43 | 210,4500 | -11,5708 | 4 | 0,038 | | | 2 | |
| LCCG 45 | 210,5833 | -12,2508 | 3 | 0,060 | | | 2 | |
| | 210,6701 | 3,0435 | 4 | 0.077 | | | 27 | |
| | 210,6871 | 2,9225 | 4 | 0.076 | | | 27 | |
| RSCG 73 | 210,6917 | 9,3597 | 3 | 0,019 | | | 9 | |
| | 210,7141 | 59,8540 | 4 | 0.098 | | | 27 | |
| | 210,7438 | 58,2344 | 5 | 0.04 | | | 27 | |
| UZCCG209 | 210,7792 | 34,9300 | 4 | 0,013 | | | 5 | |
| | 210,7912 | 38,9199 | 4 | 0.097 | | | 27 | |
| PCG140314+413636 | 210,8083 | 41,6103 | 4 | | 0.0340 | A | 7 | group member/coord.from Ref.(6) |
| | 210,9508 | 0,9677 | 4 | 0.043 | | | 27 | |
| | 210,9546 | 43,3065 | 4 | 0.116 | | | 27 | |
| | 211,0012 | 46,2526 | 4 | 0.062 | | | 27 | |
| HCG 070 | 211,0542 | 33,3278 | 6 | 0,064 | | | 1 | |
| | 211,0577 | 46,4828 | 4 | 0.067 | | | 27 | |
| PCG140430+102224 | 211,1292 | 10,3736 | 4 | | 0.1020 | A | 7 | group member/coord.from Ref.(6) |
| | 211,2469 | 37,8112 | 4 | 0.065 | | | 27 | |
| | 211,2489 | -1,7453 | 6 | 0.054 | | | 27 | |
| | 211,2753 | 39,7226 | 5 | 0.08 | | | 27 | |
| | 211,2782 | 38,6504 | 4 | 0.064 | | | 27 | |
| SDSS 120 | 211,2833 | 0,1497 | 4 | 0,243 | | | 8 | |
| | 211,3396 | -1,3367 | 4 | 0.053 | | | 27 | |
| | 211,4241 | 4,5986 | 4 | 0.025 | | | 27 | |
| | 211,4282 | 43,8336 | 4 | 0.034 | | | 27 | |
| | 211,5128 | 65,6529 | 5 | 0.032 | | | 27 | |
| | 211,8516 | 54,1152 | 4 | 0.079 | | | 27 | |

| | | | | | | |
|---|---|---|---|---|---|---|
| LCCG 46 | 211,9083 | -2,5930 | 3 | 0,139 | | 2 |
| | 211,9140 | 40,0337 | 4 | 0.08 | | 27 |
| | 211,9444 | 5,3369 | 7 | 0.086 | | 27 |
| | 211,9539 | 5,4093 | 4 | 0.09 | | 27 |
| | 211,9657 | -0,8073 | 5 | 0.053 | | 27 |
| | 211,9819 | 55,5070 | 7 | 0.076 | | 27 |
| | 211,9858 | 60,1760 | 4 | 0.037 | | 27 |
| | 211,9914 | 52,9621 | 4 | 0.043 | | 27 |
| | 211,9917 | -0,8033 | 5 | 0.052 | | 27 |
| | 212,0170 | 53,8681 | 4 | 0.078 | | 27 |
| | 212,0686 | 5,4494 | 4 | 0.091 | | 27 |
| | 212,0940 | 5,3679 | 4 | 0.085 | | 27 |
| | 212,1750 | 52,6521 | 5 | 0.082 | | 27 |
| | 212,1996 | 55,4335 | 5 | 0.072 | | 27 |
| | 212,2199 | 55,5545 | 8 | 0.074 | | 27 |
| | 212,2414 | 5,4219 | 4 | 0.088 | | 27 |
| | 212,2427 | 37,9332 | 4 | 0.067 | | 27 |
| | 212,2549 | -0,9868 | 5 | 0.083 | | 27 |
| | 212,2946 | 54,0073 | 6 | 0.042 | | 27 |
| | 212,3547 | 0,0791 | 4 | 0.054 | | 27 |
| | 212,3750 | 56,4228 | 4 | 0.042 | | 27 |
| | 212,4029 | 65,8883 | 4 | 0.067 | | 27 |
| | 212,4449 | 41,7739 | 8 | 0.094 | | 27 |
| | 212,4738 | -1,4755 | 4 | 0.115 | | 27 |
| UZCCG210 | 212,5125 | 17,6317 | 3 | 0,018 | | 5 |
| | 212,5173 | 58,1009 | 4 | 0.074 | | 27 |
| | 212,5798 | 51,1704 | 4 | 0.074 | | 27 |
| | 212,6640 | 54,8230 | 6 | 0.04 | | 27 |
| | 212,6840 | 38,8858 | 6 | 0.026 | | 27 |
| | 212,7283 | 40,1367 | 4 | 0.082 | | 27 |
| SDSS 122 | 212,7292 | 0,9850 | 4 | 0,180 | | 8 |
| | 212,7333 | 64,3948 | 5 | 0.068 | | 27 |
| HCG 071 | 212,7708 | 25,4850 | 4 | 0,030 | I4381 | 1 |
| | 212,7834 | 40,2438 | 4 | 0.081 | | 27 |
| | 212,8851 | 4,9280 | 6 | 0.094 | | 27 |
| | 212,9762 | 1,2311 | 4 | 0.025 | | 27 |
| SDSS 123 | 213,0250 | -1,0578 | 4 | 0,182 | | 8 |
| | 213,0534 | 0,0523 | 4 | 0.054 | | 27 |
| | 213,0647 | 5,1459 | 6 | 0.092 | | 27 |
| RSCG 74 | 213,0708 | 15,8575 | 3 | 0,017 | | 9 |
| | 213,1237 | 41,1324 | 4 | 0.103 | | 27 |
| | 213,1365 | 40,5899 | 5 | 0.082 | | 27 |
| | 213,2080 | 54,2485 | 5 | 0.075 | | 27 |
| | 213,2399 | 1,5850 | 9 | 0.026 | | 27 |
| | 213,2763 | 39,4213 | 4 | 0.063 | | 27 |
| | 213,3544 | 43,6955 | 4 | 0.089 | | 27 |

| | | | | | | | | |
|---|---|---|---|---|---|---|---|---|
| | 213,3807 | 1,0574 | 4 | 0.054 | | | 27 | |
| | 213,4728 | 53,8903 | 6 | 0.041 | | | 27 | |
| | 213,4799 | -2,1103 | 4 | 0.038 | | | 27 | |
| SDSS 124 | 213,5000 | 0,3956 | 4 | 0,189 | | | 8 | |
| | 213,5734 | 2,9880 | 4 | 0,025 | | | 27 | |
| UZCCG211 | 213,6292 | 3,1822 | 3 | 0,026 | | | 5 | |
| | 213,6444 | 1,7665 | 7 | 0.055 | | | 27 | |
| | 213,6503 | 1,9609 | 8 | 0.053 | | | 27 | |
| | 213,6684 | 50,3822 | 4 | 0.076 | | | 27 | |
| | 213,7184 | 4,7341 | 4 | 0.054 | | | 27 | |
| | 213,7294 | -0,3629 | 6 | 0.14 | | | 27 | |
| | 213,7329 | 3,2599 | 4 | 0.027 | | | 27 | |
| | 213,7396 | 39,7058 | 5 | 0.043 | | | 27 | |
| SDSS 126 | 213,7667 | 0,3761 | 4 | 0,143 | | | 8 | |
| | 213,7782 | 52,3767 | 5 | 0.074 | | | 27 | |
| | 213,7833 | 0,9262 | 4 | 0.123 | | | 27 | |
| | 213,8030 | 44,4276 | 4 | 0.137 | | | 27 | |
| | 213,8033 | -1,9187 | 9 | 0.048 | | | 27 | |
| | 213,8730 | 50,3129 | 4 | 0.075 | | | 27 | |
| | 213,8791 | 4,9577 | 9 | 0.08 | | | 27 | |
| | 213,8904 | 3,5537 | 4 | 0.056 | | | 27 | |
| | 213,8931 | 58,2227 | 4 | 0.04 | | | 27 | |
| PCG141535+292026 | 213,8958 | 29,3408 | 4 | | 0.0670 | D | 7 | group member/coord.from Ref.(6) |
| | 213,9454 | 4,5573 | 6 | 0.08 | | | 27 | |
| | 213,9668 | -2,7999 | 4 | 0.032 | | | 27 | |
| SDSS 127 | 213,9750 | 0,2678 | 4 | 0,126 | | | 8 | |
| | 214,0002 | 50,4027 | 5 | 0.074 | | | 27 | |
| | 214,0332 | -1,2315 | 5 | 0.05 | | | 27 | |
| | 214,0605 | 3,6403 | 6 | 0.055 | | | 27 | |
| | 214,0708 | 3,6773 | 5 | 0.054 | | | 27 | |
| | 214,1274 | 53,6999 | 8 | 0.075 | | | 27 | |
| | 214,1821 | -1,6171 | 4 | 0.126 | | | 27 | |
| | 214,1875 | 1,7706 | 7 | 0.056 | | | 27 | |
| | 214,2230 | 40,9325 | 4 | 0.068 | | | 27 | |
| | 214,2430 | 54,8757 | 5 | 0.041 | | | 27 | |
| | 214,2807 | 1,2056 | 5 | 0.054 | | | 27 | |
| | 214,4456 | 43,3843 | 5 | 0.106 | | | 27 | |
| | 214,4612 | 0,4296 | 5 | 0.05 | | | 27 | |
| | 214,4681 | 0,9481 | 4 | 0.026 | | | 27 | |
| UZCCG212 | 214,4875 | 7,5572 | 4 | 0,024 | | | 5 | |
| | 214,5154 | 1,6945 | 4 | 0.083 | | | 27 | |
| | 214,5798 | 41,0275 | 4 | 0.082 | | | 27 | |
| | 214,5951 | 44,4124 | 4 | 0.141 | | | 27 | |
| | 214,6137 | 2,2883 | 5 | 0.072 | | | 27 | |
| | 214,6474 | -1,7096 | 5 | 0.12 | | | 27 | |
| | 214,7009 | 37,8338 | 4 | 0.137 | | | 27 | |
| | 214,7057 | 52,9283 | 6 | 0.082 | | | 27 | |
| | 214,7309 | 55,9094 | 4 | 0.077 | | | 27 | |
| PCG141902+582218 | 214,7625 | 58,3717 | 4 | | 0.0630 | A | 7 | group member/coord.from Ref.(6) |
| UZCCG213 | 214,9542 | 17,7928 | 4 | 0,019 | | | 5 | |
| | 214,9641 | 5,0614 | 6 | 0.088 | | | 27 | |
| | 215,0014 | 56,5083 | 4 | 0.074 | | | 27 | |
| | 215,0562 | 2,6861 | 4 | 0.09 | | | 27 | |
| SHK 328 | 215,1292 | -9,3391 | 9 | 0,046 | | | 19 | |
| | 215,1550 | 40,3747 | 4 | 0.082 | | | 27 | |
| UZCCG214 | 215,1667 | 3,6700 | 6 | 0,005 | | | 5 | |
| | 215,2818 | 4,3566 | 4 | 0.027 | | | 27 | |
| SHK 74 E | 215,2833 | 43,0661 | 8 | 0,104 | | | 24 | |
| SHK 74 W | 215,2833 | 43,0661 | 3 | 0,219 | | | 24 | |
| UZCCG215 | 215,2917 | 35,2211 | 3 | 0,011 | | | 5 | |
| | 215,3051 | 44,6072 | 8 | 0.093 | | | 27 | |
| | 215,4467 | 61,6856 | 4 | 0.037 | | | 27 | |
| SDSS 128 | 215,5083 | 0,9631 | 4 | 0,103 | | | 8 | |
| | 215,5360 | 38,7728 | 4 | 0.137 | | | 27 | |
| | 215,6244 | 4,7541 | 4 | 0.055 | | | 27 | |
| | 215,6271 | 46,3761 | 4 | 0.034 | | | 27 | |
| | 215,6652 | 0,9228 | 4 | 0.04 | | | 27 | |
| UZCCG216 | 215,7042 | 40,3714 | 3 | 0,019 | | | 5 | |
| UZCCG217 | 215,7917 | 14,2800 | 3 | 0,008 | | | 5 | |
| | 215,8571 | -3,1818 | 4 | 0.078 | | | 27 | |
| | 215,8726 | 4,2387 | 4 | 0.054 | | | 27 | |
| | 215,9271 | 46,0399 | 9 | 0.072 | | | 27 | |
| | 215,9557 | 40,1887 | 5 | 0.082 | | | 27 | |
| | 215,9798 | 2,7157 | 6 | 0.056 | | | 27 | |
| | 215,9813 | 4,6171 | 8 | 0.026 | | | 27 | |
| | 216,0273 | 2,7682 | 5 | 0.052 | | | 27 | |
| SDSS 129 | 216,0708 | 0,1503 | 4 | 0,174 | | | 8 | |
| | 216,0789 | 40,3697 | 6 | 0.081 | | | 27 | |
| | 216,2914 | 3,2038 | 4 | 0.036 | | | 27 | |
| SHK 014 | 216,3125 | 47,2417 | 6 | 0,075 | | | 21 | mean value from NED data |
| | 216,3153 | 4,6057 | 4 | 0.056 | | | 27 | |
| SDSS 130 | 216,3167 | 0,4964 | 4 | 0,108 | | | 8 | |
| SDSS 131 | 216,3708 | 0,3422 | 4 | 0,134 | | | 8 | |
| SDSS 132 | 216,3917 | 0,3272 | 4 | 0,085 | | | 8 | |
| | 216,4481 | 3,9810 | 4 | 0.088 | | | 27 | |
| | 216,4697 | 3,1721 | 4 | 0.035 | | | 27 | |
| | 216,4875 | 37,8434 | 4 | 0.167 | | | 27 | |
| | 216,5012 | -1,4088 | 7 | 0.055 | | | 27 | |
| | 216,6123 | 37,4634 | 5 | 0.032 | | | 27 | |
| | 216,6558 | 42,0346 | 4 | 0.099 | | | 27 | |
| | 216,6765 | 55,7978 | 6 | 0.131 | | | 27 | |
| LCCG 47 | 216,7292 | -3,4966 | 3 | 0,079 | | | 2 | |

| | | | | | |
|---|---|---|---|---|---|
| | 216,7844 | 43,7560 | 4 | 0.072 | 27 |
| | 216,8483 | -0,6774 | 4 | 0.073 | 27 |
| | 216,8492 | 4,4444 | 4 | 0.055 | 27 |
| | 216,8781 | 37,4158 | 5 | 0.033 | 27 |
| | 216,8814 | 4,4803 | 4 | 0.057 | 27 |
| | 216,9185 | -1,5314 | 4 | 0.055 | 27 |
| | 216,9313 | 46,3713 | 6 | 0.074 | 27 |
| | 217,0020 | 56,9143 | 6 | 0.106 | 27 |
| | 217,0039 | 55,8708 | 4 | 0.132 | 27 |
| | 217,0500 | -2,4550 | 5 | 0.052 | 27 |
| | 217,0593 | 45,9784 | 5 | 0.075 | 27 |
| | 217,0733 | 56,7931 | 4 | 0.108 | 27 |
| UZCCG218 | 217,1083 | 11,3853 | 3 | 0,026 | 5 |
| | 217,1104 | 64,5719 | 5 | 0.037 | 27 |
| | 217,1164 | 54,1552 | 4 | 0.042 | 27 |
| | 217,1183 | 0,8563 | 4 | 0.126 | 27 |
| | 217,1756 | 0,6205 | 5 | 0.057 | 27 |
| | 217,1913 | 45,8512 | 4 | 0.074 | 27 |
| UZCCG219 | 217,2292 | 30,6264 | 3 | 0,013 | 5 |
| | 217,2999 | -0,5091 | 4 | 0.03 | 27 |
| | 217,3018 | 56,6428 | 5 | 0.105 | 27 |
| | 217,3377 | 1,9919 | 4 | 0.034 | 27 |
| | 217,3759 | 54,7239 | 4 | 0.124 | 27 |
| SDSS 133 | 217,3875 | 0,4975 | 4 | 0,055 | 8 |
| | 217,4017 | 63,1948 | 4 | 0.124 | 27 |
| SDSS 134 | 217,4125 | 0,3689 | 5 | 0,055 | 8 |
| | 217,4220 | 0,6270 | 5 | 0.054 | 27 |
| | 217,4866 | 0,2344 | 8 | 0.055 | 27 |
| | 217,5787 | -1,7045 | 8 | 0.056 | 27 |
| | 217,6404 | 57,3767 | 7 | 0.106 | 27 |
| | 217,6768 | 59,8555 | 4 | 0.06 | 27 |
| | 217,6779 | 3,4343 | 4 | 0.033 | 27 |
| | 217,6917 | 2,5400 | 4 | 0.138 | 27 |
| | 217,7240 | -1,7695 | 4 | 0.055 | 27 |
| UZCCG220 | 217,7250 | 25,4528 | 3 | 0,015 | 5 |
| | 217,7804 | -1,7906 | 5 | 0.053 | 27 |
| | 217,8108 | 42,5337 | 4 | 0.087 | 27 |
| | 217,8405 | -1,8501 | 4 | 0.052 | 27 |
| | 217,8470 | 0,6807 | 4 | 0.095 | 27 |
| | 217,8474 | 4,3599 | 4 | 0.11 | 27 |
| UZCCG221 | 217,9125 | 49,7356 | 4 | 0,007 | 5 |
| | 217,9346 | -1,4901 | 4 | 0.091 | 27 |
| | 217,9426 | 43,9311 | 4 | 0.096 | 27 |
| | 217,9775 | 1,5217 | 9 | 0.03 | 27 |
| UZCCG222 | 217,9875 | 6,1994 | 3 | 0,008 | 5 |
| | 218,0328 | 4,9828 | 5 | 0.085 | 27 |
| SDSS 136 | 218,1083 | 0,0528 | 5 | 0,093 | 8 |
| | 218,1420 | 41,7195 | 4 | 0.099 | 27 |
| | 218,1879 | 53,4342 | 4 | 0.07 | 27 |
| | 218,2190 | 2,0497 | 5 | 0.112 | 27 |
| | 218,2400 | 56,4057 | 5 | 0.069 | 27 |
| | 218,3013 | 57,4621 | 5 | 0.105 | 27 |
| | 218,3273 | 45,8552 | 4 | 0.036 | 27 |
| | 218,3546 | 53,0578 | 6 | 0.044 | 27 |
| SDSS 137 | 218,3667 | 0,1178 | 4 | 0,034 | 8 |
| | 218,4042 | -0,8166 | 4 | 0.056 | 27 |
| | 218,4133 | 3,5752 | 4 | 0.031 | 27 |
| | 218,4321 | 56,6429 | 4 | 0.075 | 27 |
| UZCCG223 | 218,4458 | 3,7142 | 3 | 0,029 | 5 |
| SDSS 138 | 218,4875 | 0,5833 | 4 | 0,222 | 8 |
| | 218,5289 | 61,4758 | 4 | 0.11 | 27 |
| | 218,5512 | 55,7238 | 7 | 0.042 | 27 |
| | 218,6554 | 54,7985 | 4 | 0.14 | 27 |
| | 218,6809 | 0,9750 | 4 | 0.138 | 27 |
| LCCG 48 | 218,6917 | -11,8294 | 3 | 0,059 | 2 |
| | 218,7476 | 42,8330 | 6 | 0.071 | 27 |
| | 218,7610 | 58,5977 | 4 | 0.044 | 27 |
| | 218,8010 | 43,1243 | 6 | 0.083 | 27 |
| | 218,8136 | 1,0286 | 4 | 0.125 | 27 |
| | 218,8606 | 49,1147 | 4 | 0.075 | 27 |
| | 218,8978 | 60,3048 | 9 | 0.049 | 27 |
| | 218,8981 | 0,8507 | 4 | 0.03 | 27 |
| | 218,9223 | 55,1770 | 6 | 0.141 | 27 |
| SDSS 139 | 218,9500 | 0,1672 | 4 | 0,100 | 8 |
| | 218,9553 | -1,3627 | 5 | 0.056 | 27 |
| UZCCG224 | 218,9667 | 48,6356 | 4 | 0,008 | 5 |
| | 218,9984 | 60,4539 | 7 | 0.05 | 27 |
| | 219,1050 | 51,6947 | 7 | 0.044 | 27 |
| | 219,2369 | 44,7496 | 4 | 0.027 | 27 |
| SDSS 140 | 219,2625 | 0,5981 | 4 | 0,214 | 8 |
| LCCG 49 | 219,2958 | -3,7689 | 3 | 0,082 | 2 |
| | 219,3007 | 0,8364 | 4 | 0.078 | 27 |
| SDSS 141 | 219,3958 | 0,5558 | 4 | 0,180 | 8 |
| | 219,4642 | 4,2682 | 5 | 0.061 | 27 |
| | 219,5164 | 53,3612 | 4 | 0.126 | 27 |
| SDSS 142 | 219,5167 | 0,3244 | 4 | 0,247 | 8 |
| | 219,5268 | 61,0507 | 5 | 0.049 | 27 |
| | 219,6124 | 55,9551 | 6 | 0.043 | 27 |
| | 219,6308 | 62,0553 | 4 | 0.111 | 27 |
| | 219,7094 | 45,2455 | 4 | 0.092 | 27 |
| | 219,7559 | -0,9182 | 4 | 0.072 | 27 |
| | 220,0881 | 3,6544 | 9 | 0.029 | 27 |

| | | | | | | | | | |
|---|---|---|---|---|---|---|---|---|---|
| | 220,0974 | 36,5652 | 4 | 0.099 | | | | 27 | |
| | 220,1665 | 1,2474 | 4 | 0.078 | | | | 27 | |
| | 220,1697 | 3,4371 | 5 | 0.024 | | | | 27 | |
| UZCCG225 | 220,1792 | 3,4325 | 4 | 0,027 | | | | 5 | |
| | 220,2299 | 50,3871 | 4 | 0.067 | | | | 27 | |
| | 220,2500 | 54,0997 | 4 | 0.049 | | | | 27 | |
| | 220,3124 | 53,4535 | 8 | 0.038 | | | | 27 | |
| | 220,3445 | 43,4928 | 5 | 0.09 | | | | 27 | |
| UZCCG226 | 220,5417 | 3,3178 | 3 | 0,026 | | | | 5 | |
| LCCG 50 | 220,6292 | -3,9316 | 4 | 0,059 | | | | 2 | |
| | 220,7364 | 55,1443 | 4 | 0.104 | | | | 27 | |
| | 220,7402 | 55,2213 | 5 | 0.105 | | | | 27 | |
| SDSS 144 | 220,8250 | 0,8019 | 4 | 0,147 | | | | 8 | |
| UZCCG227 | 220,9708 | 1,6669 | 4 | 0,006 | | | | 5 | |
| | 221,2271 | 4,7728 | 6 | 0.061 | | | | 27 | |
| | 221,3912 | 3,8097 | 4 | 0.061 | | | | 27 | |
| | 221,4355 | 48,7966 | 5 | 0.157 | | | | 27 | |
| | 221,4949 | 58,3134 | 5 | 0.029 | | | | 27 | |
| | 221,5576 | 1,6885 | 5 | 0.034 | | | | 27 | |
| | 221,6012 | 55,5971 | 9 | 0.075 | | | | 27 | |
| | 221,6046 | 37,5614 | 4 | 0.071 | | | | 27 | |
| SHK 257 | 221,7250 | 37,5581 | 6 | 0,073 | | | | 26 | |
| UZCCG228 | 221,7375 | 13,6661 | 3 | 0,030 | | | | 5 | |
| RSCG 75 | 221,7792 | 11,5914 | 3 | 0,029 | | | | 9 | |
| | 221,8056 | 35,8226 | 5 | 0.091 | | | | 27 | |
| SDSS 148 | 221,8125 | 0,9244 | 4 | 0,137 | | | | 8 | |
| UZCCG229 | 221,9542 | 9,5411 | 3 | 0,029 | | | | 5 | |
| HCG 072 | 221,9792 | 19,0594 | 5 | 0,042 | | | ARP328 | 1 | |
| | 222,0155 | 3,9453 | 4 | 0.095 | | | | 27 | |
| PCG144814+314430 | 222,0583 | 31,7419 | 4 | | 0.0450 | A | | 7 | group member/coord.from Ref.(6) |
| | 222,0817 | 38,4607 | 9 | 0.032 | | | | 27 | |
| | 222,0968 | 57,4819 | 4 | 0.105 | | | | 27 | |
| LCCG 51 | 222,1208 | -3,2553 | 4 | 0,071 | | | | 2 | |
| | 222,1491 | 55,0787 | 5 | 0.04 | | | | 27 | |
| | 222,1496 | 2,1692 | 4 | 0.121 | | | | 27 | |
| SDSS 149 | 222,1542 | 0,6483 | 4 | 0,141 | | | | 8 | |
| | 222,2138 | -0,7330 | 4 | 0.028 | | | | 27 | |
| | 222,2520 | -2,3411 | 5 | 0.043 | | | | 27 | |
| | 222,3978 | 36,8234 | 4 | 0.07 | | | | 27 | |
| PCG144948+261243 | 222,4542 | 26,2119 | 4 | | 0.1170 | B | | 7 | group member/coord.from Ref.(6) |
| | 222,7173 | 62,6846 | 4 | 0.076 | | | | 27 | |
| | 222,7513 | 49,0652 | 6 | 0.076 | | | | 27 | |
| | 222,8515 | 55,4396 | 4 | 0.038 | | | | 27 | |
| | 222,9222 | 2,7441 | 4 | 0.034 | | | | 27 | |
| | 223,0395 | 3,0013 | 8 | 0.029 | | | | 27 | |
| | 223,0858 | -0,0992 | 6 | 0.044 | | | | 27 | |

| | | | | | | | | |
|---|---|---|---|---|---|---|---|---|
| | 223,0971 | 36,1212 | 4 | 0.071 | | | 27 | |
| | 223,1322 | 38,9155 | 4 | 0.061 | | | 27 | |
| J1452+2759 | 223,1625 | 27,9847 | 4 | 0,126 | | | 4 | group coord.from ref.(3) |
| | 223,1781 | 3,1206 | 4 | 0.028 | | | 27 | |
| | 223,2056 | 49,9810 | 4 | 0.095 | | | 27 | |
| | 223,2421 | 55,0883 | 7 | 0.075 | | | 27 | |
| | 223,2523 | 50,2325 | 6 | 0.096 | | | 27 | |
| | 223,2651 | 53,7848 | 4 | 0.093 | | | 27 | |
| SDSS 150 | 223,2917 | 0,9725 | 4 | 0,119 | | | 8 | |
| SDSS 151 | 223,3333 | 0,6256 | 4 | 0,186 | | | 8 | |
| LCCG 52 | 223,3875 | -2,9094 | 3 | 0,074 | | | 2 | |
| UZCCG230 | 223,3917 | 3,5736 | 5 | 0,005 | | | 5 | |
| UZCCG231 | 223,4000 | 3,0267 | 3 | 0,029 | | | 5 | |
| | 223,4813 | 59,8259 | 4 | 0.07 | | | 27 | |
| | 223,5158 | 53,3051 | 4 | 0.095 | | | 27 | |
| | 223,5452 | 4,7589 | 5 | 0.028 | | | 27 | |
| | 223,5519 | 54,2946 | 4 | 0.101 | | | 27 | |
| | 223,5547 | 2,7991 | 4 | 0.027 | | | 27 | |
| | 223,6013 | 54,2624 | 9 | 0.1 | | | 27 | |
| | 223,7501 | 49,6158 | 8 | 0.037 | | | 27 | |
| | 223,8019 | 0,8243 | 4 | 0.043 | | | 27 | |
| | 223,8049 | 51,2971 | 4 | 0.037 | | | 27 | |
| | 223,8076 | 0,8259 | 7 | 0.044 | | | 27 | |
| | 223,8688 | -0,3577 | 5 | 0.083 | | | 27 | |
| | 223,8729 | -0,4581 | 4 | 0.084 | | | 27 | |
| | 223,9643 | 52,9676 | 4 | 0.086 | | | 27 | |
| SDSS 153 | 224,0667 | 0,3319 | 5 | 0,139 | | | 8 | |
| UZCCG232 | 224,1167 | 49,6969 | 3 | 0,014 | | | 5 | |
| SDSS 154 | 224,1292 | 0,2264 | 4 | 0,185 | | | 8 | |
| | 224,2249 | -2,4956 | 4 | 0.057 | | | 27 | |
| | 224,2311 | 34,7865 | 5 | 0.078 | | | 27 | |
| | 224,2664 | 62,9236 | 5 | 0.043 | | | 27 | |
| SDSS 155 | 224,4083 | 0,0772 | 4 | 0,042 | | | 8 | |
| | 224,4842 | 38,4525 | 6 | 0.073 | | | 27 | |
| | 224,5095 | 54,4115 | 6 | 0.1 | | | 27 | |
| | 224,5671 | 47,4899 | 5 | 0.087 | | | 27 | |
| UZCCG233 | 224,6333 | 53,8647 | 3 | 0,011 | | | 5 | |
| | 224,6464 | 47,6161 | 4 | 0.086 | | | 27 | |
| | 224,6522 | 48,3712 | 5 | 0.037 | | | 27 | |
| | 224,6644 | 2,9885 | 4 | 0.083 | | | 27 | |
| PCG145847+375935 | 224,7000 | 37,9931 | 4 | | 0.0780 | B | 7 | group member/coord.from Ref.(6) |
| PCG145853-014235 | 224,7208 | -1,7097 | 4 | | 0.1230 | A | 7 | group member/coord.from Ref.(6) |
| | 224,7429 | 53,7486 | 4 | 0.076 | | | 27 | |
| | 224,7538 | 38,7659 | 4 | 0.091 | | | 27 | |
| | 224,8274 | 48,0963 | 8 | 0.031 | | | 27 | |
| | 224,8478 | 37,6030 | 4 | 0.062 | | | 27 | |

| | | | | | | | | | |
|---|---|---|---|---|---|---|---|---|---|
| PCG145952+602856 | 224,9708 | 60,4822 | 4 | | 0.1560 | A | | 7 | group member/coord.from Ref.(6) |
| | 224,9969 | 55,6366 | 4 | 0.026 | | | | 27 | |
| | 225,0321 | 4,6769 | 5 | 0.063 | | | | 27 | |
| | 225,1426 | 36,4767 | 5 | 0.092 | | | | 27 | |
| | 225,1802 | 40,5819 | 4 | 0.063 | | | | 27 | |
| | 225,2502 | 52,8719 | 4 | 0.076 | | | | 27 | |
| | 225,2555 | 47,3201 | 8 | 0.087 | | | | 27 | |
| | 225,2750 | 46,8272 | 5 | 0.105 | | | | 27 | |
| | 225,2843 | -0,2371 | 5 | 0.041 | | | | 27 | |
| | 225,3855 | 47,3297 | 6 | 0.085 | | | | 27 | |
| | 225,4528 | 40,9942 | 5 | 0.169 | | | | 27 | |
| | 225,4657 | 55,5600 | 4 | 0.037 | | | | 27 | |
| HCG 073 | 225,6667 | 23,3536 | 5 | 0,045 | | | ARP42 | 1 | |
| | 225,6888 | 34,9353 | 4 | 0.056 | | | | 27 | |
| | 225,6937 | 47,2652 | 4 | 0.088 | | | | 27 | |
| | 225,7068 | 37,6818 | 6 | 0.116 | | | | 27 | |
| | 225,7094 | -2,2504 | 5 | 0.04 | | | | 27 | |
| | 225,7296 | 48,9115 | 8 | 0.027 | | | | 27 | |
| | 225,7807 | 48,6182 | 4 | 0.026 | | | | 27 | |
| | 225,7977 | 36,1043 | 4 | 0.073 | | | | 27 | |
| | 225,8219 | 54,5841 | 6 | 0.039 | | | | 27 | |
| | 225,8856 | 35,7912 | 5 | 0.049 | | | | 27 | |
| | 225,9186 | 37,7337 | 5 | 0.12 | | | | 27 | |
| | 226,0097 | 56,0669 | 5 | 0.065 | | | | 27 | |
| | 226,0114 | 56,0317 | 4 | 0.066 | | | | 27 | |
| PCG150410+565535 | 226,0417 | 56,9264 | 4 | | 0.1480 | A | | 7 | group member/coord.from Ref.(6) |
| | 226,1204 | -2,1096 | 6 | 0.034 | | | | 27 | |
| | 226,1829 | 35,9884 | 4 | 0.047 | | | | 27 | |
| | 226,2103 | 35,9373 | 5 | 0.048 | | | | 27 | |
| LCCG 53 | 226,2375 | -11,4883 | 3 | 0,081 | | | | 2 | |
| | 226,2861 | 37,9511 | 4 | 0.124 | | | | 27 | |
| | 226,4383 | 3,6864 | 4 | 0.036 | | | | 27 | |
| PCG150556+360303 | 226,4875 | 36,0508 | 4 | | 0.0970 | A | | 7 | group member/coord.from Ref.(6) |
| | 226,5478 | 56,5482 | 4 | 0.03 | | | | 27 | |
| | 226,5479 | 36,5947 | 6 | 0.055 | | | | 27 | |
| | 226,5607 | -1,2973 | 4 | 0.071 | | | | 27 | |
| | 226,7156 | -1,2006 | 8 | 0.072 | | | | 27 | |
| | 226,7207 | 3,5516 | 7 | 0.049 | | | | 27 | |
| RSCG 76 | 226,7208 | 12,8581 | 3 | 0,022 | | | | 9 | |
| | | | 3 | 0,022 | | | UZCCG234 | 5 | |
| UZCCG235 | 226,7375 | 12,6061 | 3 | 0,022 | | | | 5 | |
| | 226,8357 | 39,5524 | 4 | 0.037 | | | | 27 | |
| | 226,8429 | -1,0130 | 4 | 0.07 | | | | 27 | |
| SDSS 157 | 226,9083 | 0,0411 | 4 | 0,231 | | | | 8 | |
| | 226,9119 | 4,1911 | 4 | 0.035 | | | | 27 | |
| PCG150741+525000 | 226,9208 | 52,8336 | 4 | | 0.1060 | C | | 7 | group member/coord.from Ref.(6) |

| | | | | | | | | |
|---|---|---|---|---|---|---|---|---|
| | 226,9216 | -0,4063 | 4 | 0.153 | | | 27 | |
| SDSS 158 | 226,9458 | 0,5014 | 4 | 0,153 | | | 8 | |
| | 227,0027 | -0,3111 | 7 | 0.092 | | | 27 | |
| | 227,0320 | 36,0461 | 4 | 0.047 | | | 27 | |
| | 227,0856 | -2,4181 | 4 | 0.077 | | | 27 | |
| | 227,1007 | -1,3309 | 7 | 0.055 | | | 27 | |
| | 227,1142 | 37,6740 | 4 | 0.03 | | | 27 | |
| | 227,1226 | -0,2674 | 4 | 0.087 | | | 27 | |
| SDSS 159 | 227,1250 | 0,2442 | 4 | 0,095 | | | 8 | |
| | 227,1296 | -0,2620 | 4 | 0.089 | | | 27 | |
| PCG150834+461317 | 227,1417 | 46,2214 | 4 | | 0.0930 | B | 7 | group member/coord.from Ref.(6) |
| | 227,1668 | -1,5414 | 5 | 0.072 | | | 27 | |
| | 227,1759 | 46,3462 | 5 | 0.037 | | | 27 | |
| | 227,1856 | -0,1910 | 7 | 0.09 | | | 27 | |
| | 227,1899 | -2,3138 | 4 | 0.059 | | | 27 | |
| | 227,2129 | 3,0273 | 4 | 0.045 | | | 27 | |
| | 227,2141 | 1,1283 | 4 | 0.093 | | | 27 | |
| | 227,2356 | 36,0849 | 4 | 0.112 | | | 27 | |
| SDSS 161 | 227,4083 | 0,8014 | 4 | 0,083 | | | 8 | |
| | 227,4320 | 33,3184 | 6 | 0.112 | | | 27 | |
| | 227,4680 | 2,9937 | 7 | 0.092 | | | 27 | |
| | 227,4792 | 46,2067 | 4 | 0.038 | | | 27 | |
| | 227,4843 | 33,4350 | 4 | 0.117 | | | 27 | |
| | 227,5281 | 33,4846 | 6 | 0.109 | | | 27 | |
| | 227,5675 | 48,6814 | 4 | 0.074 | | | 27 | |
| | 227,5704 | 47,2186 | 6 | 0.087 | | | 27 | |
| | 227,5884 | 59,0332 | 8 | 0.071 | | | 27 | |
| | 227,6261 | 54,3587 | 4 | 0.096 | | | 27 | |
| | 227,6318 | -2,0187 | 7 | 0.037 | | | 27 | |
| | 227,6395 | 33,8622 | 5 | 0.114 | | | 27 | |
| | 227,6515 | 3,3424 | 8 | 0.078 | | | 27 | |
| | 227,7393 | 2,8984 | 4 | 0.041 | | | 27 | |
| PCG151057+031443 | 227,7417 | 3,2453 | 4 | | 0.1750 | B | 7 | group member/coord.from Ref.(6) |
| | 227,7859 | -0,0498 | 6 | 0.09 | | | 27 | |
| | 227,7981 | 0,1988 | 4 | 0.05 | | | 27 | |
| | 227,8173 | -0,0767 | 4 | 0.089 | | | 27 | |
| | 227,8264 | 1,7307 | 5 | 0.041 | | | 27 | |
| | 227,8508 | -0,1023 | 9 | 0.092 | | | 27 | |
| | 227,8605 | 1,8139 | 4 | 0.039 | | | 27 | |
| | 227,8616 | 37,1231 | 6 | 0.055 | | | 27 | |
| | 227,9021 | 41,0692 | 4 | 0.028 | | | 27 | |
| | 227,9321 | 4,3550 | 4 | 0.036 | | | 27 | |
| | 227,9388 | 35,9243 | 5 | 0.042 | | | 27 | |
| | 228,0007 | 4,4889 | 5 | 0.035 | | | 27 | |
| | 228,1027 | 46,0666 | 4 | 0.096 | | | 27 | |
| SDSS 162 | 228,1417 | 0,3261 | 5 | 0,117 | | | 8 | |
| UZCCG236 | 228,1417 | 1,4889 | 3 | 0,029 | | | 5 | |
| | 228,1630 | 35,2654 | 4 | 0.075 | | | 27 | |
| | 228,1631 | 2,6775 | 4 | 0.039 | | | 27 | |
| | 228,1742 | 4,3564 | 5 | 0.037 | | | 27 | |
| | 228,2413 | 4,3359 | 6 | 0.079 | | | 27 | |
| | 228,2642 | 4,4211 | 8 | 0.038 | | | 27 | |
| | 228,3236 | 2,0308 | 5 | 0.038 | | | 27 | |
| PCG151329+025509 | 228,3708 | 2,9192 | 4 | | 0.1340 | C | 7 | group member/coord.from Ref.(6) |
| PCG151352+275615 | 228,4708 | 27,9375 | 4 | | 0.0900 | A | 7 | group member/coord.from Ref.(6) |
| PCG151409+413606 | 228,5417 | 41,6019 | 4 | | 0.1490 | C | 7 | group member/coord.from Ref.(6) |
| | 228,6391 | 46,1782 | 5 | 0.037 | | | 27 | |
| | 228,6897 | 4,0942 | 9 | 0.097 | | | 27 | |
| SDSS 164 | 228,6917 | 0,0539 | 4 | 0,072 | | | 8 | |
| SDSS 165 | 228,7000 | 0,1522 | 4 | 0,100 | | | 8 | |
| LCCG 54 | 228,7708 | -2,7750 | 3 | 0,114 | | | 2 | |
| | 228,7737 | 3,9080 | 5 | 0.079 | | | 27 | |
| | 228,7754 | -1,5792 | 4 | 0.093 | | | 27 | |
| | 228,7898 | 4,3164 | 6 | 0.095 | | | 27 | |
| | 228,7956 | 4,3222 | 5 | 0.101 | | | 27 | |
| | 228,8199 | 4,2914 | 4 | 0.099 | | | 27 | |
| | 228,8436 | 32,6903 | 4 | 0.06 | | | 27 | |
| | 228,8835 | 4,3209 | 5 | 0.1 | | | 27 | |
| | 228,9187 | 57,8790 | 7 | 0.059 | | | 27 | |
| | 228,9381 | 49,7632 | 4 | 0.074 | | | 27 | |
| | 229,0152 | -0,3010 | 4 | 0.051 | | | 27 | |
| SDSS 166 | 229,0458 | 0,4644 | 5 | 0,108 | | | 8 | |
| | 229,0490 | -0,0591 | 4 | 0.121 | | | 27 | |
| | 229,0520 | 45,1898 | 5 | 0.064 | | | 27 | |
| | 229,0919 | -1,2168 | 6 | 0.119 | | | 27 | |
| SDSS 167 | 229,0958 | 0,2925 | 4 | 0,086 | | | 8 | |
| SDSS 168 | 229,1000 | 0,1008 | 4 | 0,115 | | | 8 | |
| PCG151624+025757 | 229,1042 | 2,9658 | 4 | | 0.1130 | B | 7 | group member/coord.from Ref.(6) |
| | 229,1107 | 33,5501 | 5 | 0.113 | | | 27 | |
| | 229,1541 | 51,7062 | 4 | 0.057 | | | 27 | |
| | 229,1550 | -0,8381 | 5 | 0.12 | | | 27 | |
| SDSS 169 | 229,2000 | 0,8089 | 4 | 0,119 | | | 8 | |
| | 229,2289 | -1,0342 | 4 | 0.119 | | | 27 | |
| | 229,2739 | -0,6574 | 6 | 0.116 | | | 27 | |
| | 229,2825 | 39,9069 | 4 | 0.063 | | | 27 | |
| | 229,3046 | 54,4549 | 4 | 0.073 | | | 27 | |
| | 229,3091 | 57,0297 | 4 | 0.067 | | | 27 | |
| SDSS 170 | 229,3458 | 0,0039 | 5 | 0,138 | | | 8 | |
| | 229,3855 | 38,7791 | 4 | 0.063 | | | 27 | |
| | 229,5220 | 56,9790 | 5 | 0.068 | | | 27 | |
| UZCCG237 | 229,5625 | 12,9578 | 3 | 0,028 | | | 5 | |
| | 229,6264 | 4,0561 | 9 | 0.036 | | | 27 | |

| | | | | | | | | | |
|---|---|---|---|---|---|---|---|---|---|
| PCG151833-013726 | 229,6375 | -1,6241 | 4 | | 0.0630 | B | | 7 | group member/coord.from Ref.(6) |
| | 229,6400 | 49,2468 | 5 | 0.037 | | | | 27 | |
| | 229,6457 | 2,8755 | 4 | 0.094 | | | | 27 | |
| | 229,6816 | 4,2493 | 5 | 0.048 | | | | 27 | |
| | 229,7242 | 33,5115 | 7 | 0.084 | | | | 27 | |
| | 229,7483 | 4,2118 | 4 | 0.035 | | | | 27 | |
| | 229,7978 | 45,1320 | 4 | 0.065 | | | | 27 | |
| | 229,8174 | 4,2156 | 7 | 0.046 | | | | 27 | |
| | 229,8175 | 3,3138 | 4 | 0.047 | | | | 27 | |
| | 229,8252 | -0,4448 | 4 | 0.037 | | | | 27 | |
| | 229,8303 | 33,2785 | 4 | 0.081 | | | | 27 | |
| | 229,8365 | 4,1130 | 4 | 0.101 | | | | 27 | |
| | 229,8385 | 3,5927 | 9 | 0.048 | | | | 27 | |
| | 229,8464 | 4,1137 | 4 | 0.052 | | | | 27 | |
| | 229,8498 | 4,2141 | 5 | 0.103 | | | | 27 | |
| HCG 074 | 229,8667 | 20,8936 | 5 | 0,040 | | | VV139 | 1 | |
| | 229,8690 | 45,0572 | 5 | 0.062 | | | | 27 | |
| | 229,8750 | 47,6977 | 5 | 0.058 | | | | 27 | |
| | 229,9207 | 51,2764 | 4 | 0.077 | | | | 27 | |
| | 229,9974 | 51,1053 | 6 | 0.079 | | | | 27 | |
| | 230,0158 | 32,8446 | 9 | 0.08 | | | | 27 | |
| | 230,0239 | 48,7820 | 8 | 0.074 | | | | 27 | |
| | 230,0328 | 45,0664 | 5 | 0.064 | | | | 27 | |
| | 230,0569 | -0,2971 | 5 | 0.076 | | | | 27 | |
| | 230,0702 | 33,3909 | 5 | 0.082 | | | | 27 | |
| | 230,0887 | 33,4244 | 5 | 0.084 | | | | 27 | |
| | 230,1078 | 33,4142 | 4 | 0.08 | | | | 27 | |
| | 230,1784 | 48,6583 | 4 | 0.076 | | | | 27 | |
| | 230,2582 | 48,7070 | 7 | 0.073 | | | | 27 | |
| | 230,3410 | 4,2725 | 4 | 0.052 | | | | 27 | |
| | 230,3701 | 32,0876 | 9 | 0.111 | | | | 27 | |
| HCG 075 | 230,3917 | 21,1833 | 6 | 0,042 | | | | 1 | |
| | 230,3964 | 32,0977 | 6 | 0.11 | | | | 27 | |
| | 230,4052 | 3,4766 | 6 | 0.104 | | | | 27 | |
| | 230,4103 | 32,1232 | 4 | 0.113 | | | | 27 | |
| | 230,4842 | 44,8231 | 9 | 0.067 | | | | 27 | |
| | 230,5393 | -1,2806 | 4 | 0.055 | | | | 27 | |
| | 230,5545 | 33,5071 | 5 | 0.062 | | | | 27 | |
| | 230,5812 | 52,9989 | 5 | 0.086 | | | | 27 | |
| | 230,6182 | -0,4784 | 5 | 0.131 | | | | 27 | |
| | 230,6424 | 40,5948 | 4 | 0.044 | | | | 27 | |
| | 230,6608 | 31,5933 | 4 | 0.115 | | | | 27 | |
| SDSS 171 | 230,6792 | 0,6889 | 4 | 0,076 | | | | 8 | |
| | 230,6997 | 48,4666 | 4 | 0.075 | | | | 27 | |
| | 230,7072 | -0,8524 | 5 | 0.028 | | | | 27 | |
| | 230,7166 | 46,1243 | 4 | 0.037 | | | | 27 | |

| | | | | | | | | |
|---|---|---|---|---|---|---|---|---|
| | 230,7636 | 44,5776 | 5 | 0.063 | | | 27 | |
| PCG152313+310047 | 230,8042 | 31,0131 | 5 | | 0.1110 | C | 7 | group member/coord.from Ref.(6) |
| | 230,8983 | 32,6229 | 4 | 0.08 | | | 27 | |
| | 230,9918 | 31,6118 | 5 | 0.074 | | | 27 | |
| | 231,0868 | 3,9264 | 5 | 0.037 | | | 27 | |
| | 231,1054 | 32,8396 | 5 | 0.069 | | | 27 | |
| | 231,2482 | 48,6159 | 7 | 0.038 | | | 27 | |
| | 231,3792 | 3,7724 | 4 | 0.052 | | | 27 | |
| | 231,4048 | 47,6808 | 5 | 0.073 | | | 27 | |
| PCG152551+282232 | 231,4667 | 28,3758 | 4 | | 0.0740 | A | 7 | group member/coord.from Ref.(6) |
| | 231,4891 | 32,0738 | 4 | 0.079 | | | 27 | |
| | 231,4971 | 48,4443 | 7 | 0.037 | | | 27 | |
| | 231,5262 | 33,0968 | 7 | 0.068 | | | 27 | |
| | 231,5686 | 32,9205 | 4 | 0.031 | | | 27 | |
| | 231,6021 | 3,9623 | 4 | 0.086 | | | 27 | |
| | 231,6275 | 0,7279 | 4 | 0.051 | | | 27 | |
| | 231,7671 | -0,8134 | 4 | 0.125 | | | 27 | |
| | 231,8325 | 49,3294 | 4 | 0.074 | | | 27 | |
| | 231,9493 | 32,6341 | 6 | 0.078 | | | 27 | |
| | 231,9557 | 57,3097 | 4 | 0.071 | | | 27 | |
| | 231,9628 | 3,4627 | 4 | 0.084 | | | 27 | |
| | 232,0062 | 3,4571 | 4 | 0.085 | | | 27 | |
| | 232,0388 | 52,7351 | 4 | 0.074 | | | 27 | |
| | 232,1125 | 45,7059 | 7 | 0.065 | | | 27 | |
| | 232,1270 | 31,2706 | 5 | 0.078 | | | 27 | |
| | 232,2200 | 3,3196 | 5 | 0.075 | | | 27 | |
| PCG152854+030011 | 232,2250 | 3,0031 | 4 | | 0.2980 | B | 7 | group member/coord.from Ref.(6) |
| PCG152909-001049 | 232,2917 | 0,1803 | 4 | | 0.0870 | B | 7 | group member/coord.from Ref.(6) |
| | 232,3045 | 52,8239 | 7 | 0.071 | | | 27 | |
| | 232,3387 | 50,8958 | 4 | 0.049 | | | 27 | |
| | 232,4032 | 44,8228 | 6 | 0.071 | | | 27 | |
| | 232,4233 | 32,5730 | 4 | 0.078 | | | 27 | |
| | 232,4491 | 30,7578 | 4 | 0.115 | | | 27 | |
| | 232,4560 | 35,3455 | 5 | 0.063 | | | 27 | |
| | 232,5076 | 51,5428 | 5 | 0.051 | | | 27 | |
| | 232,5121 | 54,2760 | 4 | 0.074 | | | 27 | |
| | 232,5273 | 3,7832 | 4 | 0.039 | | | 27 | |
| | 232,5541 | 43,8181 | 4 | 0.147 | | | 27 | |
| | 232,5950 | 3,4876 | 9 | 0.086 | | | 27 | |
| | 232,6010 | -0,1002 | 5 | 0.071 | | | 27 | |
| | 232,6126 | -1,3576 | 7 | 0.087 | | | 27 | |
| | 232,6159 | -1,6777 | 6 | 0.037 | | | 27 | |
| | 232,6529 | 31,7457 | 4 | 0.069 | | | 27 | |
| | 232,6665 | -1,5156 | 9 | 0.086 | | | 27 | |
| | 232,6829 | 45,3252 | 4 | 0.071 | | | 27 | |
| | 232,7476 | 3,4424 | 6 | 0.088 | | | 27 | |
| | 232,7669 | 50,3032 | 5 | 0.07 | | | 27 | |
| | 232,7724 | 45,5761 | 4 | 0.072 | | | 27 | |
| | 232,8627 | 0,3841 | 4 | 0.08 | | | 27 | |
| | 232,9115 | -0,4728 | 4 | 0.059 | | | 27 | |
| SDSS 173 | 232,9167 | 0,5203 | 4 | 0,080 | | | 8 | |
| HCG 076 | 232,9250 | 7,3081 | 5 | 0,034 | | | 1 | |
| PCG153148+384452 | 232,9542 | 38,7478 | 4 | | 0.2000 | C | 7 | group member/coord.from Ref.(6) |
| | 232,9833 | 48,8876 | 5 | 0.075 | | | 27 | |
| | 232,9852 | 31,3811 | 5 | 0.106 | | | 27 | |
| | 233,1334 | -0,6797 | 7 | 0.083 | | | 27 | |
| | 233,1621 | 30,4834 | 4 | 0.066 | | | 27 | |
| | 233,2357 | 0,0456 | 9 | 0.038 | | | 27 | |
| | 233,2473 | 31,0779 | 4 | 0.069 | | | 27 | |
| | 233,3225 | 52,9482 | 6 | 0.073 | | | 27 | |
| PCG153322+493858 | 233,3458 | 49,6494 | 5 | | 0.1130 | A | 7 | group member/coord.from Ref.(6) |
| | 233,3586 | 45,1074 | 7 | 0.072 | | | 27 | |
| | 233,3755 | 31,1572 | 4 | 0.067 | | | 27 | |
| | 233,4219 | 0,2332 | 6 | 0.083 | | | 27 | |
| | 233,4462 | 31,2158 | 6 | 0.068 | | | 27 | |
| | 233,4725 | 31,4844 | 4 | 0.066 | | | 27 | |
| | 233,5117 | 31,3704 | 9 | 0.065 | | | 27 | |
| | 233,5939 | 30,8828 | 6 | 0.067 | | | 27 | |
| | 233,7079 | 32,3013 | 5 | 0.067 | | | 27 | |
| | 233,7652 | -0,5558 | 4 | 0.066 | | | 27 | |
| | 233,7654 | 31,7169 | 4 | 0.106 | | | 27 | |
| | 233,8464 | -0,0080 | 4 | 0.137 | | | 27 | |
| UZCCG238 | 233,8625 | 30,8661 | 3 | 0,006 | | | 5 | |
| | 233,8825 | 32,1922 | 4 | 0.068 | | | 27 | |
| | 233,9297 | 47,0170 | 6 | 0.037 | | | 27 | |
| UZCCG239 | 234,0208 | 16,5325 | 3 | 0,006 | | | 5 | |
| | 234,0394 | 0,5464 | 4 | 0.095 | | | 27 | |
| | 234,1371 | 31,1692 | 4 | 0.1 | | | 27 | |
| | 234,1718 | 37,6412 | 4 | 0.152 | | | 27 | |
| | 234,1838 | -2,0837 | 6 | 0.144 | | | 27 | |
| | 234,2070 | 47,1000 | 4 | 0.072 | | | 27 | |
| PCG153705+433504 | 234,2708 | 43,5847 | 4 | | 0.0770 | C | 7 | group member/coord.from Ref.(6) |
| | 234,2736 | 52,8397 | 6 | 0.072 | | | 27 | |
| PCG153711+450759 | 234,3000 | 45,1331 | 4 | | 0.0720 | B | 7 | group member/coord.from Ref.(6) |
| | 234,3173 | 30,8507 | 4 | 0.097 | | | 27 | |
| | 234,3204 | 3,2349 | 4 | 0.086 | | | 27 | |
| | 234,4050 | 41,9362 | 4 | 0.055 | | | 27 | |
| | 234,4467 | -1,9527 | 5 | 0.098 | | | 27 | |
| | 234,4528 | 35,0925 | 4 | 0.077 | | | 27 | |
| | 234,4901 | 30,7741 | 8 | 0.056 | | | 27 | |
| UZCCG240 | 234,5167 | 59,4114 | 4 | 0,009 | | | 5 | |
| | 234,5349 | 31,1701 | 4 | 0.055 | | | 27 | |
| | 234,5624 | 38,1998 | 6 | 0.065 | | | 27 | |
| | 234,5696 | 30,5736 | 4 | 0.097 | | | 27 | |
| | 234,5852 | 31,5571 | 4 | 0.053 | | | 27 | |
| | 234,6252 | -1,2909 | 4 | 0.12 | | | 27 | |
| | 234,6385 | -0,3402 | 5 | 0.079 | | | 27 | |
| | 234,6613 | 44,8061 | 4 | 0.037 | | | 27 | |
| PCG153909+462407 | 234,7917 | 46,4022 | 4 | | 0.1490 | A | 7 | group member/coord.from Ref.(6) |
| | 234,8057 | 36,1469 | 4 | 0.065 | | | 27 | |
| | 235,0136 | 3,1270 | 4 | 0.085 | | | 27 | |
| | 235,0370 | 46,0418 | 6 | 0.037 | | | 27 | |
| | 235,0934 | 45,4452 | 4 | 0.095 | | | 27 | |
| | 235,1424 | 34,6698 | 7 | 0.057 | | | 27 | |
| | 235,2195 | 48,6616 | 5 | 0.067 | | | 27 | |
| PCG154114+034610 | 235,3083 | 3,7694 | 4 | | 0.1090 | B | 7 | group member/coord.from Ref.(6) |
| | 235,3199 | 47,5543 | 4 | 0.075 | | | 27 | |
| PCG154126+030747 | 235,3583 | 3,1300 | 4 | | 0.0880 | D | 7 | group member/coord.from Ref.(6) |
| SHK 360 | 235,3583 | 4,7322 | 15 | 0,108 | | | 19 | |
| | | | | 0,108 | | | 11 | mean value |
| | 235,3630 | 48,0109 | 5 | 0.072 | | | 27 | |
| | 235,3717 | -0,4971 | 4 | 0.096 | | | 27 | |
| | 235,3798 | 36,2887 | 5 | 0.068 | | | 27 | |
| | 235,4600 | 42,6539 | 4 | 0.076 | | | 27 | |
| | 235,4960 | -0,0913 | 4 | 0.078 | | | 27 | |
| | 235,5387 | 29,9072 | 4 | 0.082 | | | 27 | |
| | 235,6161 | 29,9528 | 6 | 0.059 | | | 27 | |
| | 235,6450 | 45,1680 | 6 | 0.037 | | | 27 | |
| UZCCG241 | 235,6500 | 23,7586 | 4 | 0,023 | | | 5 | |
| | 235,6663 | 0,5040 | 4 | 0.093 | | | 27 | |
| | 235,7034 | -1,8182 | 4 | 0.056 | | | 27 | |
| | 235,7140 | 35,2735 | 4 | 0.07 | | | 27 | |
| | 235,8209 | 52,2539 | 6 | 0.068 | | | 27 | |
| | 235,8317 | 48,5160 | 5 | 0.038 | | | 27 | |
| | 235,8410 | 35,0452 | 4 | 0.068 | | | 27 | |
| | 235,8599 | -2,1352 | 4 | 0.098 | | | 27 | |
| | 235,8667 | 42,3856 | 4 | 0.101 | | | 27 | |
| | 235,9091 | 43,9447 | 8 | 0.037 | | | 27 | |
| | 235,9400 | 36,4129 | 7 | 0.065 | | | 27 | |
| | 235,9919 | -1,9643 | 4 | 0.099 | | | 27 | |
| | 236,0425 | 44,2003 | 4 | 0.12 | | | 27 | |
| | 236,0482 | 54,7345 | 8 | 0.067 | | | 27 | |
| | 236,0609 | 51,0839 | 4 | 0.035 | | | 27 | |
| | 236,0751 | 42,6677 | 4 | 0.075 | | | 27 | |
| | 236,0797 | -1,8506 | 4 | 0.098 | | | 27 | |
| | 236,1024 | 36,0611 | 9 | 0.069 | | | 27 | |
| | 236,1200 | 36,1192 | 4 | 0.068 | | | 27 | |
| | 236,1261 | 34,7465 | 4 | 0.07 | | | 27 | |

| | | | | | | | | | |
|---|---|---|---|---|---|---|---|---|---|
| | 236,1362 | 47,8254 | 4 | 0.129 | | | | 27 | |
| | 236,1481 | 41,5433 | 4 | 0.141 | | | | 27 | |
| | 236,1506 | 46,9576 | 4 | 0.037 | | | | 27 | |
| | 236,1825 | 36,3254 | 5 | 0.066 | | | | 27 | |
| | 236,1926 | 42,4174 | 4 | 0.076 | | | | 27 | |
| | 236,2110 | 45,1969 | 5 | 0.038 | | | | 27 | |
| | 236,2302 | 36,1379 | 5 | 0.07 | | | | 27 | |
| | 236,2531 | 48,2714 | 4 | 0.076 | | | | 27 | |
| | 236,3077 | 36,0315 | 8 | 0.064 | | | | 27 | |
| | 236,3080 | 46,2620 | 4 | 0.06 | | | | 27 | |
| SHK 022 | 236,4333 | 55,1289 | 13 | 0,082 | | | | 21 | mean value from NED data |
| | 236,4656 | 36,8704 | 5 | 0.074 | | | | 27 | |
| UZCCG242 | 236,5042 | 2,5131 | 3 | 0,013 | | | | 5 | |
| | 236,5249 | -1,6963 | 4 | 0.121 | | | | 27 | |
| PCG154629+005120 | 236,6250 | 0,8556 | 5 | | 0.0630 | C | | 7 | group member/coord.from Ref.(6) |
| | 236,6666 | 29,3965 | 6 | 0.096 | | | | 27 | |
| | 236,6691 | 54,9395 | 4 | 0.109 | | | | 27 | |
| | 236,7214 | 43,5723 | 4 | 0.034 | | | | 27 | |
| | 236,7282 | 3,2938 | 4 | 0.096 | | | | 27 | |
| | 236,8179 | 46,5422 | 5 | 0.066 | | | | 27 | |
| | 236,8233 | 3,1774 | 7 | 0.095 | | | | 27 | |
| PCG154728+405909 | 236,8708 | 40,9858 | 4 | | 0.1470 | A | | 7 | group member/coord.from Ref.(6) |
| | 236,8803 | 54,6961 | 4 | 0.108 | | | | 27 | |
| PCG154802+030416 | 237,0125 | 3,0714 | 4 | | 0.1390 | B | | 7 | group member/coord.from Ref.(6) |
| UZCCG243 | 237,0875 | 68,1750 | 3 | 0,029 | | | | 5 | |
| | 237,0875 | 68,1750 | 4 | | 0.0287 | A | HCG 078 | 1 | not CG / discordant redshifts |
| | 237,0875 | 68,1750 | | | 0.0318 | B | HCG 078 | 1 | |
| | 237,0875 | 68,1750 | | | 0.0607 | C | HCG 078 | 1 | |
| | 237,0875 | 68,1750 | | | 0.0334 | D | HCG 078 | 1 | |
| | 237,0952 | 0,2909 | 9 | 0.033 | | | | 27 | |
| UZCCG244 | 237,3042 | 71,2244 | 4 | 0,025 | | | | 5 | |
| | 237,3176 | 41,9783 | 4 | 0.035 | | | | 27 | |
| HCG 077 | 237,3208 | 21,8283 | 4 | | 0.0351 | A | | 1 | not CG / discordant redshifts |
| | | | | | 0.0357 | B | | 1 | |
| | | | | | 0.0073 | C | | 1 | |
| | | | | | 0.0075 | D | | 1 | |
| UZCCG245 | 237,3542 | 0,2147 | 3 | 0,032 | | | | 5 | |
| J1549+2756 | 237,3750 | 27,9436 | 4 | | 0.0764 | A | | 4 | not CG / discordant redshifts |
| | 237,3750 | 27,9436 | | | 0.1216 | B | | 4 | |
| | 237,3750 | 27,9436 | | | 0.0947 | C | | 4 | |
| | 237,3750 | 27,9436 | | | 0.0764 | D | | 4 | |
| | 237,4275 | 55,2656 | 5 | 0.066 | | | | 27 | |
| | 237,4401 | 0,2414 | 6 | 0.032 | | | | 27 | |
| | 237,4471 | 52,8895 | 7 | 0.05 | | | | 27 | |
| SHK 223 | 237,4583 | 29,1492 | 10 | 0,083 | | | | 25 | |
| | 237,4606 | 28,6051 | 7 | 0.085 | | | | 27 | |

| | | | | | | | | |
|---|---|---|---|---|---|---|---|---|
| | 237,4709 | 29,1240 | 5 | 0.083 | | | 27 | |
| | 237,5105 | 29,0905 | 6 | 0.084 | | | 27 | |
| | 237,5577 | 52,9476 | 5 | 0.049 | | | 27 | |
| PCG155024+290119 | 237,6000 | 29,0219 | 4 | | 0.0750 | A | 7 | group member/coord.from Ref.(6) |
| | 237,6292 | 28,7098 | 4 | 0.031 | | | 27 | |
| | 237,6730 | 53,4552 | 4 | 0.066 | | | 27 | |
| | 237,6983 | 42,6611 | 4 | 0.034 | | | 27 | |
| | 237,7189 | 53,4742 | 6 | 0.065 | | | 27 | |
| | 237,7761 | 53,0660 | 6 | 0.048 | | | 27 | |
| | 237,8297 | 45,5574 | 5 | 0.119 | | | 27 | |
| | 237,8376 | 52,8356 | 4 | 0.066 | | | 27 | |
| | 237,8437 | 53,3763 | 5 | 0.064 | | | 27 | |
| | 237,9160 | 43,4876 | 6 | 0.041 | | | 27 | |
| | 237,9208 | 27,9850 | 7 | 0.082 | | | 27 | |
| | 238,1503 | 43,1733 | 4 | 0.196 | | | 27 | |
| | 238,2457 | 27,7332 | 4 | 0.082 | | | 27 | |
| | 238,2496 | 47,7803 | 4 | 0.08 | | | 27 | |
| | 238,3511 | 39,6769 | 4 | 0.077 | | | 27 | |
| | 238,3807 | 46,8737 | 6 | 0.059 | | | 27 | |
| | 238,5338 | 32,1720 | 5 | 0.071 | | | 27 | |
| | 238,5389 | 52,9701 | 6 | 0.047 | | | 27 | |
| | 238,6472 | 46,1463 | 4 | 0.04 | | | 27 | |
| | 238,6634 | 28,5171 | 5 | 0.079 | | | 27 | |
| | 238,7131 | 2,4487 | 4 | 0.087 | | | 27 | |
| | 238,7216 | 53,0930 | 4 | 0.066 | | | 27 | |
| | 238,7382 | 28,0194 | 5 | 0.091 | | | 27 | |
| | 238,8128 | 29,0584 | 4 | 0.077 | | | 27 | |
| PCG155521+460427 | 238,8417 | 46,0742 | 5 | | 0.1180 | B | 7 | group member/coord.from Ref.(6) |
| PCG155535+290953 | 238,9000 | 29,1650 | 4 | | 0.0770 | D | 7 | group member/coord.from Ref.(6) |
| | 238,9015 | 28,9698 | 4 | 0.08 | | | 27 | |
| | 238,9046 | 41,5196 | 9 | 0.034 | | | 27 | |
| | 238,9245 | 28,7291 | 8 | 0.076 | | | 27 | |
| | 238,9569 | 40,1078 | 7 | 0.071 | | | 27 | |
| | 238,9606 | 28,9035 | 4 | 0.079 | | | 27 | |
| | 239,0577 | 39,0793 | 4 | 0.077 | | | 27 | |
| | 239,0683 | 45,4557 | 6 | 0.039 | | | 27 | |
| | 239,1167 | 28,7752 | 4 | 0.077 | | | 27 | |
| | 239,1594 | 32,3180 | 4 | 0.057 | | | 27 | |
| | 239,1618 | 54,6399 | 4 | 0.046 | | | 27 | |
| UZCCG246 | 239,1667 | 48,1247 | 3 | 0,020 | | | 5 | |
| | 239,2318 | 27,6941 | 6 | 0.088 | | | 27 | |
| | 239,2876 | 27,3837 | 4 | 0.088 | | | 27 | |
| UZCCG247 | 239,3375 | 63,9211 | 3 | 0,031 | | | 5 | |
| | 239,4430 | 53,4036 | 6 | 0.043 | | | 27 | |
| | 239,4505 | 27,2213 | 4 | 0.086 | | | 27 | |
| | 239,4509 | 48,2382 | 6 | 0.021 | | | 27 | |

| | | | | | | | | | |
|---|---|---|---|---|---|---|---|---|---|
| | 239,4754 | 28,7478 | 4 | 0.075 | | | | 27 | |
| | 239,5296 | 27,1672 | 5 | 0.084 | | | | 27 | |
| | 239,5847 | 28,1672 | 4 | 0.091 | | | | 27 | |
| | 239,6014 | 27,0762 | 5 | 0.085 | | | | 27 | |
| | 239,6031 | 27,3995 | 6 | 0.095 | | | | 27 | |
| | 239,6139 | 27,2317 | 4 | 0.093 | | | | 27 | |
| | 239,7003 | 40,8044 | 5 | 0.083 | | | | 27 | |
| | 239,7203 | 52,2826 | 4 | 0.069 | | | | 27 | |
| | 239,7302 | 2,8902 | 4 | 0.048 | | | | 27 | |
| | 239,7306 | 44,7758 | 4 | 0.041 | | | | 27 | |
| | 239,7558 | 49,7596 | 4 | 0.044 | | | | 27 | |
| | 239,7749 | 2,7723 | 4 | 0.047 | | | | 27 | |
| | 239,7923 | 2,7649 | 4 | 0.046 | | | | 27 | |
| HCG 080 | 239,8000 | 65,2258 | 4 | 0,031 | | | | 1 | |
| HCG 079 | 239,8042 | 20,7517 | 5 | 0,015 | | | VV115 | 1 | |
| | | | 3 | 0,014 | | | RSCG 77 | 9 | |
| | 239,8458 | 27,5101 | 6 | 0.032 | | | | 27 | |
| | 239,8823 | 46,9818 | 6 | 0.056 | | | | 27 | |
| | 239,9327 | 2,6744 | 5 | 0.033 | | | | 27 | |
| | 239,9594 | 26,9923 | 4 | 0.092 | | | | 27 | |
| | 240,0567 | 31,8119 | 8 | 0.055 | | | | 27 | |
| | 240,0694 | 27,7326 | 4 | 0.052 | | | | 27 | |
| | 240,0724 | 3,1820 | 5 | 0.041 | | | | 27 | |
| | 240,0797 | 50,4875 | 5 | 0.071 | | | | 27 | |
| | 240,1126 | 52,3757 | 4 | 0.034 | | | | 27 | |
| | 240,3013 | 45,3740 | 6 | 0.043 | | | | 27 | |
| | 240,3056 | 31,0869 | 4 | 0.072 | | | | 27 | |
| | 240,3167 | 46,9901 | 7 | 0.043 | | | | 27 | |
| | 240,3373 | 51,8476 | 9 | 0.043 | | | | 27 | |
| | 240,3618 | 53,9280 | 4 | 0.107 | | | | 27 | |
| | 240,3627 | 45,3016 | 9 | 0.042 | | | | 27 | |
| | 240,3924 | 39,7012 | 6 | 0.031 | | | | 27 | |
| | 240,3931 | 53,6568 | 7 | 0.11 | | | | 27 | |
| | 240,4060 | 54,0001 | 5 | 0.108 | | | | 27 | |
| | 240,5012 | 53,8446 | 4 | 0.066 | | | | 27 | |
| | 240,5236 | 41,9955 | 4 | 0.118 | | | | 27 | |
| | 240,6177 | 47,3444 | 5 | 0.042 | | | | 27 | |
| | 240,6305 | 39,9537 | 6 | 0.071 | | | | 27 | |
| | 240,6419 | 47,7949 | 8 | 0.043 | | | | 27 | |
| | 240,8154 | 26,8457 | 6 | 0.045 | | | | 27 | |
| | 240,8529 | 27,0244 | 9 | 0.09 | | | | 27 | |
| | 240,8530 | 26,9173 | 9 | 0.089 | | | | 27 | |
| SHK 008 | 240,8750 | 52,3639 | 7 | 0,107 | | | | 21 | mean value from NED data |
| PCG160340+522107 | 240,8750 | 52,3639 | 4 | | 0.1100 | A | | 7 | group member/coord.from Ref.(6) |
| | 241,1833 | 31,3209 | 4 | 0.054 | | | | 27 | |
| UZCCG248 | 241,2000 | 34,6267 | 3 | 0,031 | | | | 5 | |

| | | | | | | | | |
|---|---|---|---|---|---|---|---|---|
| | 241,3207 | 53,1904 | 4 | 0.063 | | | 27 | |
| | 241,3524 | 37,2877 | 4 | 0.067 | | | 27 | |
| | 241,3881 | 31,4574 | 4 | 0.03 | | | 27 | |
| | 241,4496 | 48,0817 | 4 | 0.085 | | | 27 | |
| | 241,4546 | 27,2434 | 4 | 0.045 | | | 27 | |
| | 241,5787 | 27,1589 | 5 | 0.067 | | | 27 | |
| | 241,6712 | 39,1936 | 6 | 0.038 | | | 27 | |
| | 241,6791 | 27,4539 | 9 | 0.066 | | | 27 | |
| | 241,6799 | 51,0333 | 4 | 0.049 | | | 27 | |
| | 241,6861 | 27,1486 | 5 | 0.046 | | | 27 | |
| | 241,7267 | 51,3067 | 4 | 0.05 | | | 27 | |
| PCG160705+371209 | 241,7750 | 37,2028 | 4 | | 0.0660 | C | 7 | group member/coord.from Ref.(6) |
| | 241,7997 | 37,1013 | 4 | 0.066 | | | 27 | |
| | 241,9160 | 47,0452 | 5 | 0.046 | | | 27 | |
| | 241,9201 | 49,2036 | 6 | 0.059 | | | 27 | |
| | 241,9293 | 51,0674 | 4 | 0.051 | | | 27 | |
| | 242,0034 | 49,2294 | 4 | 0.06 | | | 27 | |
| | 242,0506 | 43,1067 | 5 | 0.085 | | | 27 | |
| | 242,0584 | 50,3517 | 5 | 0.054 | | | 27 | |
| | 242,1125 | 30,7019 | 4 | 0.05 | | | 27 | |
| | 242,2205 | 30,0587 | 8 | 0.048 | | | 27 | |
| PCG160905+311101 | 242,2708 | 31,1836 | 5 | | 0.1240 | B | 7 | group member/coord.from Ref.(6) |
| | 242,2726 | 53,7727 | 7 | 0.064 | | | 27 | |
| | 242,3443 | 53,8733 | 4 | 0.063 | | | 27 | |
| | 242,4143 | 53,7969 | 4 | 0.062 | | | 27 | |
| | 242,4164 | 25,5370 | 4 | 0.042 | | | 27 | |
| | 242,4956 | 52,9644 | 6 | 0.065 | | | 27 | |
| | 242,5478 | 37,3992 | 4 | 0.1 | | | 27 | |
| | 242,7076 | 47,6241 | 4 | 0.091 | | | 27 | |
| | 242,7311 | 52,4297 | 5 | 0.062 | | | 27 | |
| | 242,8097 | 29,6482 | 5 | 0.093 | | | 27 | |
| | 242,8145 | 36,1347 | 6 | 0.101 | | | 27 | |
| | 242,8355 | 36,5693 | 7 | 0.032 | | | 27 | |
| | 242,8519 | 26,2135 | 5 | 0.076 | | | 27 | |
| | 242,9019 | 31,0142 | 4 | 0.095 | | | 27 | |
| | 242,9100 | 37,0767 | 6 | 0.068 | | | 27 | |
| | 242,9299 | 36,9086 | 6 | 0.067 | | | 27 | |
| | 242,9961 | 41,8498 | 4 | 0.038 | | | 27 | |
| | 243,0558 | 29,5216 | 4 | 0.053 | | | 27 | |
| | 243,0640 | 29,5126 | 7 | 0.033 | | | 27 | |
| | 243,0790 | 29,4479 | 5 | 0.031 | | | 27 | |
| | 243,0839 | 29,7556 | 4 | 0.05 | | | 27 | |
| | 243,1266 | 30,0211 | 5 | 0.053 | | | 27 | |
| | 243,1428 | 30,8948 | 5 | 0.05 | | | 27 | |
| | 243,1582 | 29,4736 | 7 | 0.032 | | | 27 | |
| | 243,2067 | 29,9845 | 6 | 0.062 | | | 27 | |

| | | | | | | | | | |
|---|---|---|---|---|---|---|---|---|---|
| | 243,2452 | 30,5016 | 7 | 0.049 | | | | 27 | |
| | 243,3051 | 39,4970 | 7 | 0.032 | | | | 27 | |
| | 243,3068 | 49,0881 | 7 | 0.056 | | | | 27 | |
| | 243,4225 | 49,0530 | 8 | 0.057 | | | | 27 | |
| | 243,5239 | 39,5356 | 4 | 0.076 | | | | 27 | |
| | 243,5381 | 45,9664 | 4 | 0.047 | | | | 27 | |
| PCG161411+475621 | 243,5458 | 47,9394 | 4 | | 0.1720 | A | | 7 | group member/coord.from Ref.(6) |
| | 243,5539 | 50,5491 | 5 | 0.06 | | | | 27 | |
| | 243,6038 | 29,9957 | 4 | 0.034 | | | | 27 | |
| | 243,6141 | 49,2407 | 5 | 0.06 | | | | 27 | |
| | 243,6213 | 37,0456 | 4 | 0.066 | | | | 27 | |
| PCG161435+375001 | 243,6500 | 37,8336 | 4 | | 0.0590 | A | | 7 | group member/coord.from Ref.(6) |
| | 243,6878 | 30,1250 | 4 | 0.054 | | | | 27 | |
| | 243,7182 | 52,0373 | 6 | 0.089 | | | | 27 | |
| | 244,0773 | 28,5506 | 4 | 0.063 | | | | 27 | |
| | 244,1036 | 37,5594 | 5 | 0.066 | | | | 27 | |
| | 244,1178 | 51,9593 | 4 | 0.033 | | | | 27 | |
| PCG161650+383419 | 244,2125 | 38,5719 | 4 | | 0.0760 | A | | 7 | group member/coord.from Ref.(6) |
| | 244,3537 | 38,0972 | 7 | 0.066 | | | | 27 | |
| | 244,3759 | 43,1678 | 4 | 0.059 | | | | 27 | |
| | 244,4112 | 35,0025 | 8 | 0.031 | | | | 27 | |
| | 244,4509 | 35,0950 | 4 | 0.029 | | | | 27 | |
| | 244,4539 | 42,6998 | 6 | 0.061 | | | | 27 | |
| | 244,4631 | 41,4028 | 5 | 0.062 | | | | 27 | |
| | 244,4749 | 39,0842 | 4 | 0.064 | | | | 27 | |
| J1617+2758 | 244,4750 | 27,9761 | 4 | 0,126 | | | | 4 | group coord.from ref.(3) |
| | 244,4798 | 35,2599 | 7 | 0.03 | | | | 27 | |
| HCG 081 | 244,5542 | 12,7942 | 4 | 0,050 | | | U10319 | 1 | |
| | 244,6196 | 41,4573 | 4 | 0.061 | | | | 27 | |
| | 244,6823 | 39,7039 | 6 | 0.032 | | | | 27 | |
| | 244,7163 | 51,5450 | 4 | 0.057 | | | | 27 | |
| | 244,7342 | 37,6065 | 4 | 0.069 | | | | 27 | |
| | 244,7695 | 41,0793 | 4 | 0.038 | | | | 27 | |
| | 244,7997 | 40,7759 | 5 | 0.062 | | | | 27 | |
| | 244,8552 | 35,0932 | 7 | 0.033 | | | | 27 | |
| | 244,9261 | 42,8452 | 5 | 0.06 | | | | 27 | |
| SHK 004 | 244,9708 | 61,7153 | 8 | 0,099 | | | | 11 | mean value/coord.from Ref.(10) |
| PCG161951+614241 | 244,9708 | 61,7153 | 4 | | 0.0970 | B | | 7 | group member/coord.from Ref.(6) |
| | 245,1160 | 38,6846 | 4 | 0.064 | | | | 27 | |
| | 245,1395 | 38,1048 | 4 | 0.067 | | | | 27 | |
| | 245,1405 | 40,3993 | 5 | 0.033 | | | | 27 | |
| | 245,2372 | 49,3375 | 4 | 0.058 | | | | 27 | |
| | 245,2766 | 43,8936 | 4 | 0.06 | | | | 27 | |
| | 245,3913 | 40,7264 | 4 | 0.072 | | | | 27 | |
| | 245,4642 | 51,4890 | 6 | 0.056 | | | | 27 | |
| | 245,4857 | 48,3810 | 4 | 0.058 | | | | 27 | |
| | 245,5052 | 43,5186 | 4 | 0.032 | | | | 27 | |
| | 245,5203 | 40,2474 | 7 | 0.062 | | | | 27 | |
| | 245,5785 | 46,6504 | 5 | 0.048 | | | | 27 | |
| | 245,5949 | 44,3738 | 4 | 0.134 | | | | 27 | |
| | 245,7151 | 37,9136 | 6 | 0.069 | | | | 27 | |
| | 245,7462 | 37,9766 | 7 | 0.032 | | | | 27 | |
| PCG162327+272912 | 245,8625 | 27,4869 | 6 | | 0.0790 | B | | 7 | group member/coord.from Ref.(6) |
| | 245,8881 | 28,1961 | 4 | 0.114 | | | | 27 | |
| | 245,9278 | 37,9952 | 7 | 0.034 | | | | 27 | |
| | 246,0706 | 38,0926 | 5 | 0.028 | | | | 27 | |
| | 246,0992 | 33,6154 | 4 | 0.12 | | | | 27 | |
| | 246,1311 | 45,0884 | 4 | 0.115 | | | | 27 | |
| | 246,1525 | 27,5420 | 4 | 0.097 | | | | 27 | |
| | 246,1984 | 43,5560 | 7 | 0.033 | | | | 27 | |
| | 246,2381 | 33,3606 | 4 | 0.118 | | | | 27 | |
| | 246,3152 | 40,5441 | 4 | 0.075 | | | | 27 | |
| | 246,4458 | 44,0242 | 4 | 0.032 | | | | 27 | |
| | 246,5187 | 41,8398 | 6 | 0.062 | | | | 27 | |
| | 246,5209 | 39,2529 | 5 | 0.077 | | | | 27 | |
| | 246,5468 | 48,1752 | 5 | 0.051 | | | | 27 | |
| | 246,5855 | 28,1621 | 5 | 0.097 | | | | 27 | |
| PCG162629+371127 | 246,6250 | 37,1911 | 4 | | 0.1730 | A | | 7 | group member/coord.from Ref.(6) |
| | 246,6415 | 42,4746 | 9 | 0.031 | | | | 27 | |
| | 246,6601 | 37,5202 | 4 | 0.118 | | | | 27 | |
| | 246,6754 | 27,4847 | 4 | 0.112 | | | | 27 | |
| | 246,8217 | 42,7231 | 5 | 0.03 | | | | 27 | |
| PCG162743+261229 | 246,9333 | 26,2081 | 4 | | 0.2220 | C | | 7 | group member/coord.from Ref.(6) |
| | 247,0331 | 39,5353 | 5 | 0.035 | | | | 27 | |
| | 247,0737 | 27,2042 | 4 | 0.077 | | | | 27 | |
| HCG 082 | 247,0917 | 32,8236 | 4 | 0,036 | | | N6162 | 1 | |
| | 247,1337 | 42,7818 | 9 | 0.032 | | | | 27 | |
| | 247,3198 | 32,9614 | 8 | 0.037 | | | | 27 | |
| | 247,3396 | 48,8022 | 7 | 0.035 | | | | 27 | |
| PCG163018+265636 | 247,5750 | 26,9433 | 5 | | 0.0870 | A | | 7 | group member/coord.from Ref.(6) |
| | 247,6437 | 41,2283 | 4 | 0.093 | | | | 27 | |
| | 247,6679 | 41,2582 | 4 | 0.134 | | | | 27 | |
| | 247,6767 | 26,0590 | 4 | 0.049 | | | | 27 | |
| | 247,6923 | 40,4796 | 5 | 0.027 | | | | 27 | |
| | 247,7641 | 34,0491 | 5 | 0.123 | | | | 27 | |
| | 247,8002 | 34,6563 | 4 | 0.135 | | | | 27 | |
| | 247,9026 | 42,3866 | 4 | 0.071 | | | | 27 | |
| | 247,9257 | 25,8373 | 7 | 0.065 | | | | 27 | |
| | 247,9263 | 26,3691 | 4 | 0.105 | | | | 27 | |
| | 248,0047 | 25,5967 | 6 | 0.064 | | | | 27 | |
| | 248,2395 | 26,1743 | 5 | 0.066 | | | | 27 | |
| | 248,2828 | 36,3649 | 5 | 0.031 | | | | 27 | |

| | | | | | | | | |
|---|---|---|---|---|---|---|---|---|
| | 248,2919 | 32,0699 | 4 | 0.059 | | | 27 | |
| | 248,3081 | 26,8523 | 6 | 0.071 | | | 27 | |
| | 248,3123 | 41,0644 | 4 | 0.137 | | | 27 | |
| | 248,3140 | 34,8694 | 5 | 0.035 | | | 27 | |
| | 248,3218 | 26,9001 | 5 | 0.072 | | | 27 | |
| | 248,3797 | 25,8550 | 5 | 0.144 | | | 27 | |
| | 248,4083 | 43,2108 | 7 | 0.072 | | | 27 | |
| | 248,4298 | 26,0554 | 6 | 0.065 | | | 27 | |
| | 248,5759 | 37,4255 | 4 | 0.1 | | | 27 | |
| | 248,6764 | 26,0071 | 7 | 0.051 | | | 27 | |
| | 248,7682 | 26,5460 | 9 | 0.07 | | | 27 | |
| | 248,8101 | 26,5096 | 9 | 0.071 | | | 27 | |
| | 248,8117 | 26,2646 | 8 | 0.045 | | | 27 | |
| | 248,8320 | 26,2220 | 6 | 0.065 | | | 27 | |
| PCG163531+361150 | 248,8792 | 36,1975 | 4 | | 0.1020 | B | 7 | group member/coord.from Ref.(6) |
| | 248,8990 | 46,2739 | 4 | 0.031 | | | 27 | |
| | 248,9174 | 36,1187 | 4 | 0.061 | | | 27 | |
| HCG 083 | 248,9208 | 6,2700 | 4 | 0,053 | | | 1 | |
| | 248,9689 | 25,9848 | 6 | 0.052 | | | 27 | |
| | 248,9948 | 31,3892 | 4 | 0.099 | | | 27 | |
| | 249,0109 | 42,1220 | 5 | 0.031 | | | 27 | |
| | 249,0473 | 44,2707 | 8 | 0.057 | | | 27 | |
| | 249,0878 | 40,3295 | 6 | 0.034 | | | 27 | |
| | 249,1780 | 44,3048 | 9 | 0.031 | | | 27 | |
| | 249,1959 | 36,6660 | 6 | 0.074 | | | 27 | |
| | 249,2299 | 25,3536 | 5 | 0.057 | | | 27 | |
| | 249,2445 | 35,7812 | 4 | 0.038 | | | 27 | |
| | 249,2725 | 37,5731 | 5 | 0.099 | | | 27 | |
| | 249,2783 | 25,8673 | 6 | 0.05 | | | 27 | |
| | 249,3279 | 25,3273 | 5 | 0.058 | | | 27 | |
| UZCCG249 | 249,4750 | 36,0611 | 3 | 0,032 | | | 5 | |
| | 249,5440 | 44,2914 | 6 | 0.072 | | | 27 | |
| | 249,6941 | 31,8544 | 4 | 0.064 | | | 27 | |
| PCG163904+344212 | 249,7708 | 34,7033 | 4 | | 0.0790 | A | 7 | group member/coord.from Ref.(6) |
| | 249,7709 | 32,5145 | 5 | 0.052 | | | 27 | |
| | 249,8667 | 39,4450 | 9 | 0.031 | | | 27 | |
| | 249,8887 | 37,4804 | 6 | 0.06 | | | 27 | |
| | 250,0146 | 30,1378 | 4 | 0.056 | | | 27 | |
| | 250,0400 | 44,0497 | 4 | 0.03 | | | 27 | |
| | 250,1822 | 45,9253 | 4 | 0.072 | | | 27 | |
| | 250,2483 | 32,0588 | 4 | 0.064 | | | 27 | |
| | 250,2720 | 39,5989 | 7 | 0.03 | | | 27 | |
| | 250,3702 | 37,6839 | 6 | 0.061 | | | 27 | |
| | 250,4154 | 40,2836 | 6 | 0.032 | | | 27 | |
| | 250,4199 | 37,0208 | 4 | 0.031 | | | 27 | |
| | 250,4244 | 40,2562 | 4 | 0.033 | | | 27 | |
| | 250,5755 | 38,5352 | 5 | 0.062 | | | 27 | |
| | 250,6696 | 30,5991 | 4 | 0.104 | | | 27 | |
| HCG 084 | 251,0333 | 77,8361 | 6 | 0,056 | | | 1 | |
| | 251,0496 | 32,5587 | 4 | 0.04 | | | 27 | |
| | 251,1107 | 37,4998 | 4 | 0.061 | | | 27 | |
| | 251,2369 | 35,5800 | 4 | 0.115 | | | 27 | |
| | 251,2434 | 36,4114 | 5 | 0.098 | | | 27 | |
| | 251,2708 | 43,8857 | 4 | 0.034 | | | 27 | |
| | 251,3585 | 31,5375 | 4 | 0.055 | | | 27 | |
| PCG164608+482100 | 251,5375 | 48,3503 | 4 | | 0.1320 | A | 7 | group member/coord.from Ref.(6) |
| | 251,5684 | 31,0362 | 4 | 0.054 | | | 27 | |
| | 251,7107 | 45,0586 | 5 | 0.06 | | | 27 | |
| | 251,9374 | 29,2022 | 4 | 0.134 | | | 27 | |
| | 251,9482 | 24,0321 | 4 | 0.046 | | | 27 | |
| | 251,9786 | 29,8913 | 4 | 0.102 | | | 27 | |
| | 251,9984 | 33,3921 | 4 | 0.068 | | | 27 | |
| | 252,0387 | 35,9482 | 4 | 0.115 | | | 27 | |
| UZCCG250 | 252,1083 | 35,8944 | 3 | 0,032 | | | 5 | |
| | 252,1608 | 32,9219 | 6 | 0.086 | | | 27 | |
| SHK 016 | 252,2833 | 53,4314 | 14 | 0,030 | | | 11 | mean value/coord.from Ref.(10) |
| | 252,3642 | 64,1493 | 4 | 0.085 | | | 27 | |
| | 252,5295 | 32,2333 | 7 | 0.098 | | | 27 | |
| | 252,5344 | 24,5550 | 4 | 0.036 | | | 27 | |
| | 252,6167 | 38,7784 | 4 | 0.062 | | | 27 | |
| | 252,6655 | 39,9101 | 5 | 0.03 | | | 27 | |
| | 252,7016 | 32,8351 | 4 | 0.085 | | | 27 | |
| | 252,7401 | 32,8199 | 4 | 0.087 | | | 27 | |
| | 253,0539 | 40,1156 | 5 | 0.053 | | | 27 | |
| | 253,0555 | 31,4033 | 4 | 0.094 | | | 27 | |
| | 253,0597 | 24,5537 | 4 | 0.103 | | | 27 | |
| SHK 166 | 253,2000 | 81,6317 | 11 | 0,040 | | | 19 | |
| | 253,2518 | 40,8091 | 6 | 0.029 | | | 27 | |
| | 253,2669 | 24,3764 | 5 | 0.035 | | | 27 | |
| | 253,2816 | 23,8795 | 4 | 0.047 | | | 27 | |
| | 253,3158 | 33,0418 | 5 | 0.085 | | | 27 | |
| | 253,4262 | 23,6605 | 6 | 0.046 | | | 27 | |
| | 253,4578 | 24,3724 | 4 | 0.036 | | | 27 | |
| | 253,5269 | 23,1607 | 6 | 0.06 | | | 27 | |
| | 253,5726 | 31,4535 | 4 | 0.099 | | | 27 | |
| | 253,6040 | 29,9767 | 5 | 0.071 | | | 27 | |
| | 253,8086 | 23,4289 | 5 | 0.055 | | | 27 | |
| | 253,8115 | 32,1776 | 6 | 0.089 | | | 27 | |
| | 253,8404 | 39,2454 | 6 | 0.061 | | | 27 | |
| | 253,8819 | 37,9410 | 4 | 0.063 | | | 27 | |
| | 253,9083 | 23,5335 | 4 | 0.047 | | | 27 | |
| | 253,9973 | 33,0605 | 4 | 0.087 | | | 27 | |
| | 254,0577 | 38,3040 | 4 | 0.116 | | | 27 | |
| | 254,0740 | 39,5422 | 6 | 0.033 | | | 27 | |
| | 254,1588 | 39,2767 | 4 | 0.063 | | | 27 | |
| UZCCG251 | 254,4000 | 40,7214 | 3 | 0,030 | | | 5 | |
| | 254,4537 | 40,7544 | 4 | 0.03 | | | 27 | |
| | 254,4740 | 31,4118 | 4 | 0.093 | | | 27 | |
| | 254,4822 | 31,4887 | 4 | 0.092 | | | 27 | |
| UZCCG252 | 254,6167 | 59,0042 | 3 | 0,018 | | | 5 | |
| | 254,6207 | 32,5475 | 5 | 0.062 | | | 27 | |
| | 254,6883 | 32,5211 | 4 | 0.061 | | | 27 | |
| | 254,9175 | 32,3743 | 6 | 0.106 | | | 27 | |
| | 254,9719 | 32,5325 | 5 | 0.096 | | | 27 | |
| | 255,0115 | 40,6405 | 4 | 0.031 | | | 27 | |
| | 255,0356 | 32,6312 | 4 | 0.102 | | | 27 | |
| | 255,1548 | 33,3475 | 4 | 0.088 | | | 27 | |
| | 255,4207 | 35,1296 | 4 | 0.106 | | | 27 | |
| | 255,5017 | 35,0339 | 4 | 0.104 | | | 27 | |
| | 255,5562 | 33,0743 | 5 | 0.083 | | | 27 | |
| | 255,5753 | 33,5241 | 7 | 0.088 | | | 27 | |
| | 255,5858 | 33,4218 | 4 | 0.092 | | | 27 | |
| | 255,5948 | 33,5570 | 6 | 0.085 | | | 27 | |
| | 255,6057 | 34,0278 | 9 | 0.095 | | | 27 | |
| | 255,6201 | 33,7051 | 6 | 0.037 | | | 27 | |
| | 255,6351 | 34,0824 | 6 | 0.102 | | | 27 | |
| | 255,6611 | 34,1175 | 4 | 0.101 | | | 27 | |
| | 255,6669 | 34,0263 | 5 | 0.1 | | | 27 | |
| | 255,6670 | 34,0818 | 7 | 0.097 | | | 27 | |
| | 255,6746 | 33,5088 | 4 | 0.087 | | | 27 | |
| | 255,6828 | 33,4335 | 5 | 0.09 | | | 27 | |
| | 255,6836 | 34,0468 | 4 | 0.096 | | | 27 | |
| | 255,6978 | 33,5973 | 4 | 0.084 | | | 27 | |
| | 255,7104 | 33,8574 | 4 | 0.098 | | | 27 | |
| | 255,7370 | 31,6510 | 4 | 0.033 | | | 27 | |
| | 255,7603 | 33,4547 | 5 | 0.091 | | | 27 | |
| | 255,7730 | 34,5977 | 4 | 0.109 | | | 27 | |
| | 255,9152 | 20,8632 | 4 | 0.08 | | | 27 | |
| | 256,0373 | 30,9210 | 9 | 0.034 | | | 27 | |
| | 256,0419 | 39,2488 | 4 | 0.135 | | | 27 | |
| | 256,0464 | 35,7030 | 5 | 0.061 | | | 27 | |
| | 256,0617 | 38,5476 | 6 | 0.064 | | | 27 | |
| J1704+2818 | 256,2375 | 28,3094 | 4 | | 0.1639 | A | 4 | not CG / discordant redshifts |
| | | | | | 0.0811 | C | 4 | |
| | | | | | 0.1649 | D | 4 | |
| | 256,3007 | 31,5570 | 7 | 0.035 | | | 27 | |
| | 256,4172 | 61,0672 | 5 | 0.068 | | | 27 | |
| | 256,4527 | 33,7871 | 5 | 0.084 | | | 27 | |

| | | | | | |
|---|---|---|---|---|---|
| | 256,4566 | 33,5258 | 5 | 0.03 | 27 |
| | 256,5789 | 30,2928 | 4 | 0.082 | 27 |
| | 256,5991 | 31,4989 | 8 | 0.083 | 27 |
| | 256,6104 | 31,5392 | 7 | 0.082 | 27 |
| | 256,6214 | 33,5836 | 4 | 0.082 | 27 |
| | 256,6259 | 31,3738 | 4 | 0.084 | 27 |
| UZCCG253 | 256,9375 | 60,8231 | 3 | 0,011 | 5 |
| | 256,9391 | 31,5586 | 8 | 0.031 | 27 |
| | 256,9753 | 64,7672 | 5 | 0.08 | 27 |
| | 256,9878 | 30,8846 | 4 | 0.082 | 27 |
| | 257,0356 | 34,0775 | 9 | 0.031 | 27 |
| | 257,0536 | 62,9728 | 5 | 0.079 | 27 |
| | 257,0579 | 58,4731 | 4 | 0.079 | 27 |
| | 257,1907 | 60,1923 | 4 | 0.067 | 27 |
| | 257,1960 | 33,4233 | 5 | 0.103 | 27 |
| | 257,2324 | 34,3819 | 4 | 0.088 | 27 |
| | 257,2798 | 34,4192 | 5 | 0.085 | 27 |
| | 257,2904 | 34,1497 | 4 | 0.086 | 27 |
| | 257,3942 | 31,7585 | 4 | 0.055 | 27 |
| | 257,4645 | 30,7487 | 4 | 0.054 | 27 |
| | 257,6328 | 31,6300 | 4 | 0.082 | 27 |
| | 258,0139 | 31,5956 | 5 | 0.119 | 27 |
| | 258,1001 | 64,0648 | 4 | 0.074 | 27 |
| | 258,2362 | 58,5465 | 5 | 0.08 | 27 |
| | 258,2689 | 64,0592 | 6 | 0.078 | 27 |
| | 258,4311 | 64,2523 | 6 | 0.076 | 27 |
| | 258,4647 | 28,1732 | 4 | 0.071 | 27 |
| UZCCG254 | 258,6042 | 23,0717 | 3 | 0,029 | 5 |
| | 258,6291 | 59,5595 | 5 | 0.066 | 27 |
| | 258,6540 | 31,8591 | 5 | 0.08 | 27 |
| | 258,7165 | 31,9154 | 5 | 0.079 | 27 |
| | 258,8914 | 64,5192 | 4 | 0.08 | 27 |
| | 258,9071 | 57,3481 | 6 | 0.029 | 27 |
| | 259,0435 | 30,8183 | 4 | 0.108 | 27 |
| | 259,1718 | 57,6059 | 4 | 0.031 | 27 |
| | 259,2322 | 57,8105 | 4 | 0.033 | 27 |
| | 259,2912 | 57,3942 | 6 | 0.032 | 27 |
| | 259,3254 | 56,7385 | 4 | 0.11 | 27 |
| UZCCG255 | 259,3500 | 7,6994 | 3 | 0,022 | 5 |
| | 259,6006 | 29,4381 | 8 | 0.029 | 27 |
| | 259,8018 | 55,1292 | 4 | 0.063 | 27 |
| | 259,9185 | 28,5884 | 4 | 0.107 | 27 |
| | 259,9476 | 63,5375 | 5 | 0.079 | 27 |
| CG J1720-67.8 | 260,1167 | -67,7764 | 11 | 0,045 | 15 |
| | 260,1682 | 28,4809 | 4 | 0.048 | 27 |
| | 260,2522 | 64,8241 | 8 | 0.082 | 27 |

| | | | | | | |
|---|---|---|---|---|---|---|
| | 260,3991 | 59,2066 | 4 | 0.066 | | 27 |
| | 260,4738 | 27,5499 | 4 | 0.083 | | 27 |
| | 260,6061 | 58,6321 | 4 | 0.08 | | 27 |
| | 260,6957 | 56,8183 | 4 | 0.026 | | 27 |
| | 260,7126 | 27,8176 | 4 | 0.028 | | 27 |
| | 260,8016 | 27,3304 | 5 | 0.111 | | 27 |
| | 260,8383 | 58,6293 | 5 | 0.079 | | 27 |
| | 260,9288 | 27,4003 | 4 | 0.104 | | 27 |
| | 260,9786 | 62,3904 | 6 | 0.028 | | 27 |
| | 261,2145 | 54,6525 | 5 | 0.03 | | 27 |
| | 261,5328 | 60,2140 | 4 | 0.027 | | 27 |
| | 261,8321 | 58,5195 | 4 | 0.025 | | 27 |
| UZCCG256 | 261,9500 | 26,8700 | 3 | 0,029 | | 5 |
| | 262,0184 | 58,7082 | 7 | 0.026 | | 27 |
| | 262,1935 | 57,6175 | 4 | 0.079 | | 27 |
| | 262,5267 | 55,8146 | 4 | 0.113 | | 27 |
| | 262,8462 | 60,0314 | 7 | 0.029 | | 27 |
| | 264,7884 | 56,9675 | 6 | 0.084 | | 27 |
| | 265,4449 | 54,1926 | 4 | 0.089 | | 27 |
| | 265,4757 | 55,6615 | 4 | 0.103 | | 27 |
| UZCCG257 | 267,7458 | 23,0539 | 3 | 0,013 | | 5 |
| UZCCG258 | 268,1750 | 24,5144 | 4 | 0,011 | | 5 |
| UZCCG259 | 268,7792 | 18,3469 | 3 | 0,010 | | 5 |
| UZCCG260 | 273,0167 | 21,4694 | 4 | 0,017 | | 5 |
| UZCCG261 | 273,1333 | 25,5158 | 4 | 0,016 | | 5 |
| UZCCG262 | 275,1917 | 38,1439 | 3 | 0,028 | | 5 |
| UZCCG263 | 277,2625 | 22,8836 | 3 | 0,014 | | 5 |
| UZCCG264 | 277,4208 | 39,8472 | 3 | 0,019 | | 5 |
| UZCCG265 | 279,4000 | 73,7661 | 3 | 0,005 | | 5 |
| UZCCG266 | 279,4667 | 37,3542 | 3 | 0,021 | | 5 |
| HCG 085 | 282,5917 | 73,3500 | 4 | 0,039 | | 1 |
| UZCCG267 | 295,0125 | 50,8958 | 3 | 0,027 | | 5 |
| UZCCG268 | 296,8125 | 59,8331 | 3 | 0,011 | | 5 |
| HCG 086 | 297,9958 | -30,8261 | 4 | 0,020 | | 1 |
| UZCCG269 | 307,2083 | 10,7256 | 3 | 0,015 | | 5 |
| UZCCG270 | 309,8792 | 2,0453 | 3 | 0,013 | | 5 |
| UZCCG271 | 311,0417 | 12,4758 | 3 | 0,015 | | 5 |
| RSCG 78 | 311,8042 | 0,3678 | 4 | 0,013 | | 9 |
| | | | 6 | 0,013 | UZCCG272 | 5 |
| HCG 087 | 312,0500 | -19,8405 | 4 | 0,030 | | 1 |
| UZCCG273 | 312,5292 | 9,8906 | 3 | 0,015 | | 5 |
| | 312,9061 | -0,0399 | 5 | 0.109 | | 27 |
| | 312,9860 | -0,9677 | 6 | 0.055 | | 27 |
| HCG 088 | 313,0917 | -5,7578 | | 0,020 | | 12 |
| | 313,4116 | -0,1826 | 4 | 0.03 | | 27 |
| | 313,4280 | -5,0836 | 4 | 0.133 | | 27 |
| | 313,6825 | -6,9826 | 4 | 0.106 | | 27 |
| | 314,0767 | -5,3622 | 4 | 0.075 | | 27 |
| | 314,5971 | -0,3509 | 4 | 0.053 | | 27 |
| | 314,6139 | -7,4587 | 4 | 0.082 | | 27 |
| | 315,1224 | -6,4241 | 4 | 0.136 | | 27 |
| | 315,1298 | 0,3807 | 4 | 0.079 | | 27 |
| | 315,7583 | 0,3644 | 5 | 0.052 | | 27 |
| | 315,7769 | -0,9892 | 4 | 0.161 | | 27 |
| | 315,8564 | 10,8790 | 4 | 0.043 | | 27 |
| | 315,8693 | -6,7107 | 6 | 0.064 | | 27 |
| | 316,2788 | -7,5472 | 4 | 0.073 | | 27 |
| | 316,5552 | -0,9961 | 5 | 0.097 | | 27 |
| | 316,7692 | -0,2815 | 5 | 0.067 | | 27 |
| | 316,8034 | 10,6934 | 4 | 0.096 | | 27 |
| | 317,4641 | 9,9542 | 4 | 0.096 | | 27 |
| | 317,6515 | 9,5267 | 7 | 0.097 | | 27 |
| UZCCG274 | 317,7875 | -2,0175 | 3 | 0,032 | | 5 |
| RSCG 79 | 317,9208 | -23,1625 | 3 | 0,036 | | 9 |
| | 318,5567 | 0,2461 | 6 | 0.06 | | 27 |
| | 318,7492 | -7,8587 | 4 | 0.028 | | 27 |
| | 319,3185 | -0,3901 | 6 | 0.058 | | 27 |
| | 319,3314 | -7,0970 | 5 | 0.088 | | 27 |
| | 319,5157 | 0,9750 | 4 | 0.121 | | 27 |
| LCCG 55 | 319,7500 | -41,8203 | 3 | 0,109 | | 2 |
| | 319,9584 | -6,4753 | 4 | 0.045 | | 27 |
| HCG 089 | 320,0458 | -3,9089 | 4 | 0,030 | | 1 |
| | | | 4 | 0,030 | HCG 089 | 12 |
| | 320,0955 | 0,6192 | 4 | 0.035 | | 27 |
| | 320,1190 | 0,8429 | 5 | 0.034 | | 27 |
| | 320,3583 | 10,3361 | 5 | 0.077 | | 27 |
| LCCG 56 | 320,4542 | -45,4889 | 6 | 0,083 | | 2 |
| LCCG 57 | 320,6708 | -39,5203 | 3 | 0,060 | | 2 |
| | 320,6938 | -6,5418 | 4 | 0.088 | | 27 |
| | 320,8804 | -0,5341 | 4 | 0.051 | | 27 |
| | 320,8931 | -8,2938 | 4 | 0.119 | | 27 |
| | 321,1151 | -7,1317 | 4 | 0.116 | | 27 |
| | 321,2397 | 1,0013 | 9 | 0.05 | | 27 |
| | 321,2703 | -6,8877 | 6 | 0.119 | | 27 |
| | 321,3675 | -6,8890 | 4 | 0.118 | | 27 |
| | 321,4263 | 0,1498 | 7 | 0.063 | | 27 |
| | 321,4267 | 0,9467 | 5 | 0.137 | | 27 |
| | 321,5139 | -7,2255 | 5 | 0.088 | | 27 |
| | 321,5803 | 10,0894 | 4 | 0.094 | | 27 |
| | 321,6188 | -6,7177 | 4 | 0.116 | | 27 |
| | 321,7518 | -0,7370 | 5 | 0.052 | | 27 |
| | 322,0074 | 0,2177 | 6 | 0.052 | | 27 |

| | | | | | |
|---|---|---|---|---|---|
| | 322,0672 | 10,8033 | 4 | 0.094 | 27 |
| | 322,2086 | 9,9489 | 8 | 0.072 | 27 |
| | 322,2989 | 11,1522 | 5 | 0.088 | 27 |
| | 322,3948 | 0,1859 | 5 | 0.031 | 27 |
| | 322,4351 | -0,2571 | 5 | 0.136 | 27 |
| | 322,4932 | -0,1610 | 9 | 0.03 | 27 |
| | 322,6672 | 10,6628 | 4 | 0.061 | 27 |
| | 322,8742 | 10,1948 | 4 | 0.077 | 27 |
| | 323,1711 | -8,4041 | 4 | 0.096 | 27 |
| | 323,2974 | 1,1100 | 4 | 0.124 | 27 |
| | 323,3132 | 10,6342 | 4 | 0.089 | 27 |
| | 323,3823 | 10,3659 | 5 | 0.078 | 27 |
| | 323,4473 | -8,5843 | 6 | 0.095 | 27 |
| | 323,4634 | -8,5273 | 4 | 0.096 | 27 |
| | 323,5266 | 10,4119 | 5 | 0.077 | 27 |
| | 323,6813 | 0,9749 | 7 | 0.062 | 27 |
| | 323,9773 | -8,3482 | 5 | 0.087 | 27 |
| | 324,0010 | 0,5411 | 5 | 0.051 | 27 |
| | 324,1659 | 0,6042 | 4 | 0.029 | 27 |
| | 324,3761 | 0,4368 | 4 | 0.051 | 27 |
| | 324,4808 | 1,0316 | 4 | 0.105 | 27 |
| | 324,5387 | -7,8225 | 5 | 0.053 | 27 |
| | 324,6269 | 11,3798 | 4 | 0.078 | 27 |
| | 324,7603 | 10,5637 | 4 | 0.077 | 27 |
| | 324,8366 | 11,9768 | 4 | 0.08 | 27 |
| | 324,9661 | 10,4136 | 4 | 0.045 | 27 |
| | 324,9846 | -0,0134 | 4 | 0.061 | 27 |
| | 325,3094 | -8,3888 | 4 | 0.129 | 27 |
| | 325,3641 | -8,4108 | 5 | 0.128 | 27 |
| | 325,4375 | -0,6465 | 6 | 0.065 | 27 |
| | 325,6111 | -6,5784 | 4 | 0.088 | 27 |
| | 325,6760 | -6,8982 | 6 | 0.053 | 27 |
| | 325,6895 | -8,0624 | 4 | 0.13 | 27 |
| | 325,6936 | -6,8711 | 7 | 0.051 | 27 |
| | 325,7074 | -7,3133 | 6 | 0.07 | 27 |
| | 325,7130 | -6,9158 | 6 | 0.055 | 27 |
| | 325,7373 | -8,3075 | 4 | 0.069 | 27 |
| | 325,8160 | -6,7956 | 5 | 0.054 | 27 |
| LCCG 58 | 325,8167 | -45,4703 | 3 | 0,057 | 2 |
| | 326,1034 | -8,4674 | 4 | 0.051 | 27 |
| | 326,1893 | -7,1661 | 4 | 0.086 | 27 |
| | 326,3620 | -0,2941 | 4 | 0.076 | 27 |
| | 326,7973 | -8,4880 | 7 | 0.06 | 27 |
| | 326,8639 | -8,5077 | 7 | 0.061 | 27 |
| | 327,3617 | 0,2429 | 5 | 0.062 | 27 |
| | 327,6982 | -7,8888 | 4 | 0.121 | 27 |

| | | | | | | | | | |
|---|---|---|---|---|---|---|---|---|---|
| | 327,9336 | -7,0802 | 4 | 0.059 | | | | 27 | |
| | 328,0490 | -7,0760 | 9 | 0.058 | | | | 27 | |
| LCCG 59 | 328,1333 | -41,9280 | 3 | 0,064 | | | | 2 | |
| | 328,1638 | -7,0476 | 6 | 0.07 | | | | 27 | |
| LCCG 60 | 328,2250 | -41,9364 | 3 | 0,064 | | | | 2 | |
| | 328,3496 | -8,7935 | 4 | 0.078 | | | | 27 | |
| | 328,4140 | -7,2239 | 4 | 0.059 | | | | 27 | |
| | 329,1370 | -6,9643 | 8 | 0.065 | | | | 27 | |
| | 329,2080 | -6,9888 | 4 | 0.063 | | | | 27 | |
| | 329,2872 | -7,8175 | 4 | 0.062 | | | | 27 | |
| | 329,3300 | -7,7498 | 9 | 0.06 | | | | 27 | |
| | 329,3482 | -7,1007 | 6 | 0.085 | | | | 27 | |
| | 329,3881 | -7,8450 | 9 | 0.058 | | | | 27 | |
| | 329,7436 | -0,0176 | 4 | 0.109 | | | | 27 | |
| | 329,9717 | -0,5112 | 4 | 0.127 | | | | 27 | |
| | 330,0194 | -7,4787 | 4 | 0.064 | | | | 27 | |
| LCCG 61 | 330,0542 | -44,2589 | 3 | 0,138 | | | | 2 | |
| LCCG 62 | 330,0833 | -39,1739 | 3 | 0,061 | | | | 2 | |
| | 330,4385 | -8,0828 | 4 | 0.098 | | | | 27 | |
| | 330,4589 | 0,0187 | 4 | 0.099 | | | | 27 | |
| HCG 090 | 330,5250 | -31,9666 | 4 | 0,009 | | | N7172 | 1 | |
| | | | 3 | 0,009 | | | RSCG 80 | 9 | |
| | 330,8002 | 12,5772 | 4 | 0.026 | | | | 27 | |
| | 330,8309 | -8,5020 | 9 | 0.065 | | | | 27 | |
| | 331,0746 | -7,6400 | 4 | 0.059 | | | | 27 | |
| | 331,3642 | -9,0833 | 6 | 0.064 | | | | 27 | |
| LCCG 63 | 331,5125 | -38,1000 | 3 | 0,112 | | | | 2 | |
| | 331,5922 | -0,4450 | 4 | 0.083 | | | | 27 | |
| | 331,6517 | 12,3574 | 4 | 0.122 | | | | 27 | |
| | 331,7248 | -7,5430 | 4 | 0.061 | | | | 27 | |
| | 331,8263 | -7,6046 | 5 | 0.062 | | | | 27 | |
| | 331,8569 | -7,5089 | 8 | 0.059 | | | | 27 | |
| | 331,9216 | -7,5163 | 5 | 0.06 | | | | 27 | |
| PCG220748-004159 | 331,9542 | 0,6997 | 4 | | 0.1100 | A | | 7 | group member/coord.from Ref.(6) |
| LCCG 64 | 332,0917 | -44,0836 | 3 | 0,075 | | | | 2 | |
| | 332,1826 | -7,4909 | 4 | 0.056 | | | | 27 | |
| | 332,2603 | -9,1542 | 4 | 0.081 | | | | 27 | |
| | 332,2790 | -9,2669 | 6 | 0.082 | | | | 27 | |
| HCG 091 | 332,3000 | -27,7758 | 4 | 0,024 | | | | 1 | |
| | | | 3 | 0,023 | | | RSCG 81 | 9 | |
| | 332,4892 | -7,7787 | 4 | 0.116 | | | | 27 | |
| UZCCG275 | 332,5125 | 40,9664 | 3 | 0,015 | | | | 5 | |
| | 332,5541 | -9,3683 | 4 | 0.07 | | | | 27 | |
| LCCG 65 | 332,8167 | -45,4486 | 3 | 0,058 | | | | 2 | |
| | 332,8203 | -7,4697 | 4 | 0.116 | | | | 27 | |
| | 332,9710 | -9,3839 | 7 | 0.083 | | | | 27 | |
| UZCCG276 | 332,9875 | 38,7003 | 3 | 0,022 | | | | 5 | |
| | 333,1003 | 0,1467 | 4 | 0.057 | | | | 27 | |
| | 333,3052 | -0,2309 | 4 | 0.09 | | | | 27 | |
| | 333,5111 | -8,6028 | 4 | 0.085 | | | | 27 | |
| | 333,5397 | -8,4902 | 4 | 0.084 | | | | 27 | |
| PCG221414+002203 | 333,5625 | 0,3675 | 5 | | 0.1270 | A | | 7 | group member/coord.from Ref.(6) |
| | 333,5725 | 13,8629 | 7 | 0.028 | | | | 27 | |
| | 333,6251 | -8,1440 | 4 | 0.085 | | | | 27 | |
| | 333,8097 | 13,8564 | 4 | 0.025 | | | | 27 | |
| UZCCG277 | 333,8333 | 37,3000 | 3 | 0,019 | | | | 5 | |
| | 334,0217 | -9,2245 | 4 | 0.083 | | | | 27 | |
| | 334,0550 | -9,2411 | 4 | 0.085 | | | | 27 | |
| | 334,0952 | -9,3274 | 4 | 0.084 | | | | 27 | |
| | 334,1084 | -9,7609 | 5 | 0.092 | | | | 27 | |
| | 334,1185 | -9,3031 | 7 | 0.082 | | | | 27 | |
| | 334,1516 | -0,9214 | 4 | 0.112 | | | | 27 | |
| | 334,1540 | -9,3681 | 6 | 0.086 | | | | 27 | |
| | 334,2219 | -8,1528 | 4 | 0.071 | | | | 27 | |
| | 334,2468 | -9,6430 | 6 | 0.096 | | | | 27 | |
| | 334,2898 | -7,5701 | 4 | 0.077 | | | | 27 | |
| | 334,4219 | 12,6755 | 6 | 0.081 | | | | 27 | |
| | 334,5840 | 13,9830 | 5 | 0.119 | | | | 27 | |
| | 334,8711 | -8,7563 | 4 | 0.093 | | | | 27 | |
| PCG222111-010504 | 335,3000 | -1,0847 | 4 | | 0.1070 | A | | 7 | group member/coord.from Ref.(6) |
| PCG222121+002743 | 335,3417 | 0,4619 | 4 | | 0.0590 | B | | 7 | group member/coord.from Ref.(6) |
| | 335,5743 | -1,1579 | 4 | 0.101 | | | | 27 | |
| | 335,7910 | -8,9899 | 8 | 0.085 | | | | 27 | |
| UZCCG278 | 336,0500 | 36,1378 | 3 | 0,019 | | | | 5 | |
| | 336,2222 | -8,3154 | 5 | 0.1 | | | | 27 | |
| | 336,4740 | -8,7642 | 4 | 0.083 | | | | 27 | |
| | 336,5362 | -8,6899 | 4 | 0.081 | | | | 27 | |
| | 336,5649 | 13,3314 | 4 | 0.103 | | | | 27 | |
| | 336,6583 | 12,9324 | 4 | 0.131 | | | | 27 | |
| | 337,0257 | -8,5549 | 4 | 0.06 | | | | 27 | |
| | 337,1328 | -9,6354 | 4 | 0.085 | | | | 27 | |
| | 337,1474 | -9,6808 | 5 | 0.083 | | | | 27 | |
| | 337,1493 | 13,7892 | 4 | 0.101 | | | | 27 | |
| | 337,2166 | -8,1745 | 4 | 0.105 | | | | 27 | |
| | 337,3638 | 0,0107 | 8 | 0.058 | | | | 27 | |
| | 337,7603 | -8,5586 | 6 | 0.082 | | | | 27 | |
| | 338,0299 | -7,9305 | 4 | 0.073 | | | | 27 | |
| | 338,1388 | -9,4570 | 9 | 0.035 | | | | 27 | |
| | 338,2310 | -8,8169 | 4 | 0.024 | | | | 27 | |
| | 338,8596 | -1,1484 | 4 | 0.089 | | | | 27 | |
| LCCG 66 | 338,9458 | -41,4969 | 3 | 0,076 | | | | 2 | |
| HCG 092 | 339,0042 | 33,9658 | 5 | 0,022 | | | ARP319 | 1 | |
| | | | 3 | 0,021 | | | UZCCG279 | 5 | |
| | | | 4 | 0,022 | | | RSCG 82 | 9 | |
| | 339,0152 | -9,9252 | 4 | 0.062 | | | | 27 | |
| | 339,0399 | 13,0085 | 5 | 0.088 | | | | 27 | |
| | 339,0976 | 14,2419 | 4 | 0.038 | | | | 27 | |
| | 339,1447 | 13,4376 | 4 | 0.088 | | | | 27 | |
| | 339,2028 | 13,2404 | 4 | 0.09 | | | | 27 | |
| | 339,4382 | -8,6175 | 4 | 0.081 | | | | 27 | |
| | 339,4648 | -8,3739 | 8 | 0.08 | | | | 27 | |
| UZCCG280 | 339,6458 | 35,4200 | 6 | 0,028 | | | | 5 | |
| | 339,6636 | 13,1929 | 5 | 0.063 | | | | 27 | |
| LCCG 67 | 339,7208 | -41,4230 | 4 | 0,094 | | | | 2 | |
| | 339,7981 | -0,8799 | 5 | 0.052 | | | | 27 | |
| PCG223922-005611 | 339,8417 | 0,9364 | 4 | | 0.1340 | C | | 7 | group member/coord.from Ref.(6) |
| | 339,9213 | -9,2266 | 8 | 0.063 | | | | 27 | |
| | 340,1183 | -8,4318 | 4 | 0.079 | | | | 27 | |
| | 340,2637 | -8,7825 | 4 | 0.071 | | | | 27 | |
| | 340,4096 | 12,6392 | 4 | 0.101 | | | | 27 | |
| | 340,5077 | -7,9064 | 4 | 0.077 | | | | 27 | |
| | 340,5783 | -8,5143 | 5 | 0.083 | | | | 27 | |
| | 340,6216 | 0,8891 | 6 | 0.074 | | | | 27 | |
| | 340,8458 | 0,4889 | 4 | 0.057 | | | | 27 | |
| | 340,8778 | 0,5346 | 4 | 0.058 | | | | 27 | |
| | 340,9760 | 0,3157 | 4 | 0.059 | | | | 27 | |
| | 341,2554 | -8,3606 | 4 | 0.075 | | | | 27 | |
| | 341,2657 | 14,8646 | 4 | 0.089 | | | | 27 | |
| | 341,3525 | -10,1940 | 4 | 0.033 | | | | 27 | |
| LCCG 68 | 341,3875 | -45,0686 | 4 | 0,089 | | | | 2 | |
| | 341,4811 | -8,5819 | 9 | 0.081 | | | | 27 | |
| | 341,5618 | -7,9720 | 4 | 0.077 | | | | 27 | |
| | 341,7821 | -8,7480 | 6 | 0.066 | | | | 27 | |
| | 341,8615 | -8,7419 | 6 | 0.065 | | | | 27 | |
| | 342,4485 | -8,2559 | 4 | 0.076 | | | | 27 | |
| UZCCG281 | 342,5167 | 11,5983 | 6 | 0,026 | | | | 5 | |
| LCCG 69 | 342,5542 | -41,9386 | 3 | 0,087 | | | | 2 | |
| LCCG 70 | 342,6667 | -45,3628 | 4 | 0,052 | | | | 2 | |
| | 343,7101 | -0,4436 | 4 | 0.067 | | | | 27 | |
| | 343,9953 | -0,6678 | 4 | 0.112 | | | | 27 | |
| | 344,0782 | -0,4647 | 6 | 0.11 | | | | 27 | |
| | 344,1005 | -0,5063 | 4 | 0.109 | | | | 27 | |
| | 344,2989 | -8,7995 | 4 | 0.079 | | | | 27 | |
| | 344,3370 | 12,9634 | 5 | 0.026 | | | | 27 | |
| UZCCG282 | 344,4875 | 26,1175 | 4 | 0,025 | | | | 5 | |
| | 344,5239 | 14,1500 | 7 | 0.093 | | | | 27 | |
| PCG225807+011101 | 344,5292 | 1,1836 | 4 | | 0.1030 | A | | 7 | group member/coord.from Ref.(6) |
| | 344,5838 | -9,2990 | 5 | 0.081 | | | | 27 | |
| | 344,7189 | -9,3843 | 7 | 0.08 | | | | 27 | |
| PCG225902+132934 | 344,7625 | 13,4928 | 4 | | 0.1290 | A | | 7 | group member/coord.from Ref.(6) |
| | 344,9838 | -9,8180 | 7 | 0.08 | | | | 27 | |
| | 345,0906 | -8,3355 | 6 | 0.078 | | | | 27 | |
| | 345,0924 | 0,0522 | 4 | 0.11 | | | | 27 | |
| | 345,0951 | -9,8499 | 4 | 0.081 | | | | 27 | |
| | 345,1211 | 13,0183 | 4 | 0.029 | | | | 27 | |
| | 345,1479 | -8,3613 | 7 | 0.079 | | | | 27 | |
| | 345,4752 | -8,4362 | 4 | 0.056 | | | | 27 | |
| RSCG 83 | 345,4875 | 15,9758 | 3 | 0,007 | | | | 9 | |
| LCCG 71 | 345,5125 | -41,2828 | 3 | 0,046 | | | | 2 | |
| | 345,5385 | -8,4570 | 4 | 0.104 | | | | 27 | |
| SHK 361 | 345,8125 | 17,6847 | 11 | | 0.0785 | 2 | | 11 | group member/coord.from Ref.(10) |
| | | | | | 0.0797 | 3 | | 11 | group member/coord.from Ref.(10) |
| | 345,9372 | -10,6325 | 4 | 0.079 | | | | 27 | |
| | 345,9705 | -10,5764 | 4 | 0.078 | | | | 27 | |
| | 346,0388 | -10,2686 | 7 | 0.079 | | | | 27 | |
| | 346,3923 | 0,7876 | 4 | 0.07 | | | | 27 | |
| | 346,7215 | 13,7089 | 4 | 0.036 | | | | 27 | |
| | 346,9526 | 13,1696 | 4 | 0.096 | | | | 27 | |
| | 347,4408 | -0,6148 | 6 | 0.027 | | | | 27 | |
| | 347,8956 | 15,1226 | 4 | 0.04 | | | | 27 | |
| | 347,9635 | 14,4866 | 6 | 0.068 | | | | 27 | |
| | 347,9959 | -10,6120 | 4 | 0.089 | | | | 27 | |
| | 348,3569 | 13,9136 | 7 | 0.04 | | | | 27 | |
| RSCG 84 | 348,6042 | -2,6861 | 3 | 0,012 | | | | 9 | |
| UZCCG283 | 348,6375 | 28,4431 | 3 | 0,023 | | | | 5 | |
| | 348,7077 | -9,5607 | 4 | 0.071 | | | | 27 | |
| | 348,8282 | 13,7244 | 7 | 0.039 | | | | 27 | |
| HCG 093 | 348,8500 | 18,9831 | 5 | 0,017 | | | ARP99 | 1 | |
| | | | 3 | 0,017 | | | UZCCG284 | 5 | |
| | 348,9198 | 14,0022 | 7 | 0.083 | | | | 27 | |
| | 349,0429 | 15,1655 | 4 | 0.05 | | | | 27 | |
| | 349,2801 | -9,0211 | 8 | 0.084 | | | | 27 | |
| | 349,3093 | -8,9109 | 4 | 0.085 | | | | 27 | |
| HCG 094 | 349,3167 | 18,7197 | 7 | 0,042 | | | ARP170 | 1 | |
| | 349,6271 | 14,7573 | 4 | 0.094 | | | | 27 | |
| UZCCG285 | 349,6667 | 6,9161 | 3 | 0,014 | | | | 5 | |
| | 349,7702 | -10,3780 | 6 | 0.032 | | | | 27 | |
| LCCG 72 | 349,8417 | -41,9841 | 4 | 0,051 | | | | 2 | |
| HCG 095 | 349,8833 | 9,4919 | 4 | 0,040 | | | ARP150 | 1 | |
| | 349,8918 | 0,3220 | 5 | 0.034 | | | | 27 | |
| | 350,0583 | 1,1997 | 4 | 0.03 | | | | 27 | |
| LCCG 73 | 350,1417 | -41,7958 | 3 | 0,057 | | | | 2 | |
| RSCG 85 | 350,4042 | 27,0992 | 3 | 0,020 | | | | 9 | |
| | | | 3 | 0,020 | | | UZCCG286 | 5 | |
| UZCCG287 | 350,4292 | 8,9847 | 3 | 0,011 | | | | 5 | |
| UZCCG288 | 350,7333 | 29,1669 | 3 | 0,020 | | | | 5 | |
| | 350,9901 | -9,9404 | 4 | 0.068 | | | | 27 | |
| | 351,0466 | 14,4454 | 5 | 0.039 | | | | 27 | |
| | 351,0467 | 14,6264 | 4 | 0.037 | | | | 27 | |
| | 351,1055 | -9,8601 | 4 | 0.071 | | | | 27 | |
| | 351,1395 | 14,4911 | 7 | 0.04 | | | | 27 | |
| | 351,1421 | 14,6114 | 5 | 0.038 | | | | 27 | |
| PCG232445+005110 | 351,1875 | 0,8528 | 4 | | 0.1190 | D | | 7 | group member/coord.from Ref.(6) |
| | 351,2387 | -8,8430 | 4 | 0.098 | | | | 27 | |
| | 351,2566 | 15,2907 | 4 | 0.04 | | | | 27 | |
| | 351,2995 | 13,4176 | 4 | 0.046 | | | | 27 | |
| | 351,6176 | 13,4755 | 4 | 0.042 | | | | 27 | |
| HCG 096 | 351,9917 | 8,7742 | 4 | 0,029 | | | ARP182 | 1 | |
| | 352,1474 | -9,7096 | 4 | 0.066 | | | | 27 | |
| | 352,1655 | -9,5092 | 6 | 0.097 | | | | 27 | |
| | 352,2623 | 14,0092 | 4 | 0.091 | | | | 27 | |
| | 352,5633 | 13,7867 | 4 | 0.092 | | | | 27 | |
| SHK 362 | 353,1542 | 19,5297 | 14 | 0,082 | | | | 22 | mean value from NED data |
| | 353,6337 | -8,6793 | 5 | 0.092 | | | | 27 | |
| UZCCG289 | 353,6708 | 4,9067 | 5 | 0,019 | | | | 5 | |
| PCG233446+003743 | 353,6917 | 0,6286 | 4 | | 0.0860 | C | | 7 | group member/coord.from Ref.(6) |
| | 353,7022 | 14,9573 | 4 | 0.105 | | | | 27 | |
| | 353,7816 | 1,0268 | 6 | 0.085 | | | | 27 | |
| | 353,8593 | 1,0525 | 5 | 0.084 | | | | 27 | |
| | 353,8611 | -9,1580 | 4 | 0.06 | | | | 27 | |
| LCCG 74 | 354,0083 | -38,5805 | 3 | 0,065 | | | | 2 | |
| | 354,4784 | 15,6885 | 4 | 0.057 | | | | 27 | |
| | 354,5016 | -0,4748 | 5 | 0.036 | | | | 27 | |
| | 354,5263 | 15,8280 | 7 | 0.067 | | | | 27 | |
| | 354,5440 | 15,5709 | 4 | 0.069 | | | | 27 | |
| LCCG 75 | 354,5667 | -38,4697 | 3 | 0,056 | | | | 2 | |
| RSCG 86 | 354,6417 | 27,0233 | 4 | 0,029 | | | | 9 | |
| | 354,6628 | 15,2901 | 5 | 0.068 | | | | 27 | |
| | 354,6904 | -8,9167 | 4 | 0.091 | | | | 27 | |
| | 355,0335 | 0,2531 | 4 | 0.068 | | | | 27 | |
| PCG234100+000450 | 355,2500 | 0,0806 | 4 | | 0.1850 | A | | 7 | group member/coord.from Ref.(6) |
| | 355,2704 | -10,5512 | 7 | 0.087 | | | | 27 | |
| | 355,3339 | 14,1318 | 4 | 0.068 | | | | 27 | |
| | 355,3351 | -10,7109 | 4 | 0.089 | | | | 27 | |
| | 355,6162 | 15,0839 | 8 | 0.066 | | | | 27 | |
| | 355,9206 | -1,2127 | 4 | 0.066 | | | | 27 | |
| RSCG 87 | 356,8458 | -2,3258 | 3 | 0,021 | | | HCG 97 | 9 | |
| HCG 097 | 356,8625 | -2,3055 | 5 | 0,022 | | | | 1 | |
| RSCG 88 | 356,8708 | -28,1225 | 3 | 0,028 | | | | 9 | |
| | 357,0801 | -9,0061 | 4 | 0.085 | | | | 27 | |
| PCG235015+143723 | 357,5625 | 14,6233 | 4 | | 0.2010 | D | | 7 | group member/coord.from Ref.(6) |
| | 357,5864 | -0,9082 | 5 | 0.072 | | | | 27 | |
| UZCCG290 | 357,8208 | 20,1192 | 3 | 0,014 | | | | 5 | |
| | 357,9023 | -11,0353 | 4 | 0.076 | | | | 27 | |
| | 358,0756 | 14,0564 | 4 | 0.105 | | | | 27 | |
| PCG235225+143103 | 358,1042 | 14,5175 | 4 | | 0.1400 | C | | 7 | group member/coord.from Ref.(6) |
| | 358,2454 | -10,9652 | 5 | 0.076 | | | | 27 | |
| | 358,2618 | -8,9085 | 4 | 0.085 | | | | 27 | |
| | 358,3584 | -10,4198 | 6 | 0.075 | | | | 27 | |
| | 358,3835 | -10,5085 | 8 | 0.073 | | | | 27 | |
| | 358,4873 | -10,5593 | 4 | 0.08 | | | | 27 | |
| | 358,4993 | -10,3342 | 4 | 0.079 | | | | 27 | |
| HCG 098 | 358,5542 | 0,3733 | 4 | 0,027 | | | ARP323 | 1 | |
| | 358,5918 | 15,8256 | 7 | 0.075 | | | | 27 | |
| | 358,5990 | -10,3986 | 8 | 0.078 | | | | 27 | |
| | 358,6006 | -10,2855 | 6 | 0.076 | | | | 27 | |
| | 358,6437 | -9,6183 | 5 | 0.075 | | | | 27 | |
| | 358,7001 | -10,3049 | 5 | 0.075 | | | | 27 | |
| | 358,7297 | -9,4150 | 4 | 0.074 | | | | 27 | |
| | 358,7394 | -9,1950 | 5 | 0.075 | | | | 27 | |
| | 358,7788 | -9,3631 | 4 | 0.073 | | | | 27 | |
| | 358,9395 | -9,0639 | 4 | 0.077 | | | | 27 | |
| | 358,9542 | -9,3706 | 9 | 0.076 | | | | 27 | |
| | 358,9553 | -9,6270 | 5 | 0.078 | | | | 27 | |
| | 358,9967 | -9,5195 | 4 | 0.077 | | | | 27 | |
| | 359,0488 | -10,9863 | 6 | 0.044 | | | | 27 | |
| | 359,0962 | -10,0885 | 4 | 0.077 | | | | 27 | |
| | 359,5991 | -1,0826 | 4 | 0.064 | | | | 27 | |
| PCG235849+160051 | 359,7042 | 16,0142 | 4 | | 0.1150 | B | | 7 | group member/coord.from Ref.(6) |
| UZCCG291 | 359,9500 | 46,9131 | 3 | 0,017 | | | | 5 | |

References: (1) Hickson et al. 1992, (2) Allam & Tucker 2000, (3) Iovino et al. 2003, (4) Pompei et al. 2006, (5) Focardi & Kelm 2002, (6) and (7) De Carvalho et al. 2005, (8) Lee et al. 2004, (9) Barton et al. 1996, (10) Shakhbazian 1973 –1979, (11) Stoll et al. 1993-1997, (12) Stevens et al. 2004, (13) Bremer et al. 2002, (14) Temporin & Ciroi 2003, (15])Temporin et al. 2003, (16) Mirzoyan et al. 1975, (17) Kodaira et al. 1988, (18) Tovmassian et al. 2003a, (19) Tiersch et al. 2002, (20) Tovmassian et al. 2003b, (21) Tovmassian et al. 2005a, (22) Tovmassian et al. 2004, (23) Tovmassian et al. 2005b, (24) Tovmassian et al. 2005c, (25) Tovmassian et al. 2006 a, (26) Tovmassian et al. 2006 b, (27) Deng 2006